\documentclass[apsrev4-1,twocolumn,superscriptaddress,floatfix,longbibliography]{revtex4-1}

\usepackage{times}
\usepackage{float}
\usepackage{multirow}
\usepackage[dvipsnames]{xcolor}
\usepackage{amsmath}
\usepackage{amsthm}
\usepackage{amssymb}
\usepackage{amsbsy}
\usepackage{xfrac}
\usepackage{enumitem}
\usepackage{wasysym}
\usepackage[english]{babel}
\usepackage[T1]{fontenc}
\usepackage[utf8]{inputenc} 
\usepackage{graphicx}
\usepackage[colorlinks,bookmarks=false,citecolor=blue,linkcolor=red,urlcolor=blue]{hyperref}
\usepackage[normalem]{ulem}
\usepackage{pstricks}
\usepackage{rotating}			       
\usepackage{tabularx,hhline}	
\usepackage[caption=false]{subfig}		
\usepackage{hyperref}
\usepackage{appendix}
\usepackage{bigdelim}
\newcolumntype{P}[1]{>{\centering\arraybackslash}p{#1}}

\graphicspath{{./figs/}} 		

\makeindex

\newcommand{\Ham}{{\cal H}}
\newcommand{\V}[1]{\textbf{#1}}
\renewcommand{\exp}[1]{e^{#1}}

\newcommand{\oph}[1]{{\hat{\mathrm{#1}}}}
\newcommand{\ophb}[1]{{\boldsymbol{\hat{\mathrm{#1}}}}}

\newcommand{\opa}[3]{{\hat{\mathcal A}_{#1~#3}^{#2}}}
\newcommand{\opaA}[3]{{{\mathcal A}_{#1~#3}^{#2}}}
\newcommand{\ops}[2]{{\hat{\rm{S}}_{#1}^{#2}}}
\newcommand{\opsb}[1]{{\boldsymbol{\hat{\rm{S}}_{#1}}}}
\newcommand{\opq}[2]{{\hat{\rm{Q}}_{#1}^{#2}}}
\newcommand{\opqb}[1]{{\boldsymbol{\hat{\rm{Q}}_{#1}}}}
\newcommand{\op}[2]{{\hat{\rm{d}}_{ #1#2}}}
\newcommand{\opdD}[2]{{{\rm{d}}_{#1}^{\dagger #2}}}

\newcommand{\opx}[2]{{\boldsymbol{\hat{\rm{#1}}}_{#2}}}
\newcommand{\opdx}[2]{{\boldsymbol{\hat{\rm{ #1}}_{#2}^{\dagger}}}}

\newcommand{\ba}[1]{{\hat{\rm{a}}_{#1}}}
\newcommand{\bad}[1]{{\hat{\rm{a}}_{#1}^{\dagger}}}
\newcommand{\bb}[1]{{\hat{\rm{b}}_{#1}}}
\newcommand{\bbd}[1]{{\hat{\rm{b}}_{#1}^{\dagger}}}

\newcommand{\matU}[3]{{U_{#1}}_{#2}^{~#3}}
\newcommand{\matUinv}[3]{{U^{-1}_{#1}}_{#2}^{~#3}}
\newcommand{\matUd}[3]{{U^{\dagger ~#2}_{#1 #3}}}
\newcommand{\matUdinv}[3]{{U^{\dagger -1}_{#1}}_{#2}^{~#3}}

\newcommand{\opw}[3]{{\hat{\rm{#1}}_{#2#3}}}
\newcommand{\opdw}[3]{{\hat{\rm{ #1}}^{\dagger #3}_{#2}}}
\newcommand{\balpha}[1]{{\hat{\alpha}_{#1}}}
\newcommand{\balphad}[1]{{\hat{\alpha}_{#1}^{\dagger}}}
\newcommand{\brho}[1]{{\hat{\rho}_{#1}}}
\newcommand{\brhod}[1]{{\hat{\rho}_{#1}^{\dagger}}}
\newcommand{\bbeta}[1]{{\hat{\beta}_{#1}}}
\newcommand{\bbetad}[1]{{\hat{\beta}_{#1}^{\dagger}}}
\newcommand{\bsigma}[1]{{\hat{\sigma}_{#1}}}
\newcommand{\bsigmad}[1]{{\hat{\sigma}_{#1}^{\dagger}}}
\def \k{\V{k}}
\def \K{\V{K}}  
\def \r{\V{r}} 
\def \q{\V{q}} 

\def \l1{\overline{1}} 

\renewcommand{\O}[0]{\mathcal{O}}
\newcommand{\one}{\mathbb{I}}

\newcommand\numberthis{\addtocounter{equation}{1}\tag{\theequation}}

\def \dd  {{\rm d}}
\def \ddp  {{\partial}}
\def \diag {{\rm diag}}

\def \O {\mathcal O}

\def \st {\textsuperscript{st}}
\def \nd {\textsuperscript{nd}}

\def \th {\textsuperscript{th}}

\newcommand{\nn}[1]{		
	{ \langle #1 \rangle} }

\newcommand{\ket}[1]{
	{ \left| #1 \right\rangle}}

\newcommand{\bra}[1]{
	{ \left\langle #1\right|}}

\newcommand{\SP}[2]{
	{ \langle #1 | #2 \rangle}}

\newcommand{\com}[1]{ 
	{ \left[ #1 \right]} }
\newcommand{\norm}[1]{ 
	{ \left| #1 \right|} }

\newcommand{\Av}[1]{
	{ \langle #1 \rangle}}

\def \be {\begin{equation}}
\def \ee {\end{equation}}
\def \bee{\begin{equation*}}
\def \eee{\end{equation*}}
\def \bes {\begin{subequations}}
\def \ees {\end{subequations}}

\def \dd  {{\rm d}}
\def \ddp  {{\partial}}
\def \diag {{\rm diag}}
  
\def \st {\textsuperscript{st}}
\def \nd {\textsuperscript{nd}}

\def \th {\textsuperscript{th}} 
 
\def \k{\V{k}}
\def \K{\V{K}}  
\def \r{\V{r}} 
\def \q{\V{q}} 

\newcommand{\nic}{\textcolor{black}}
\newcommand{\kim}{\textcolor{black}}

\renewcommand*{\sectionautorefname}{Section}
\renewcommand*{\subsectionautorefname}{\sectionautorefname}
\let\subsectionautorefname\sectionautorefname

\def\chapterautorefname~#1\null{Chapter~#1\null}
\def\sectionautorefname~#1\null{Section~#1\null}
\def\subsectionautorefname~#1\null{Section~#1\null}
\def\figureautorefname~#1\null{Fig.~#1\null}
\def\tableautorefname~#1\null{Table~#1\null}
\def\equationautorefname~#1\null{Eq.~(#1)\null}

\newcommand{\Autoref}[1]{%
	\begingroup%
	\def\chapterautorefname~##1\null{Chapter~(##1)\null}%
	\def\sectionautorefname~##1\null{Section~##1\null}%
	\def\appendixautorefname~##1\null{\hyperref[~##1]{Appendix~}\null}%
	\def\subsectionautorefname~##1\null{Sub--Section~##1\null}%
	\def\figureautorefname~##1\null{Fig.~##1\null}%
	\def\tableautorefname~##1\null{Table~(##1)\null}%
	\def\equationautorefname~##1\null{Eq.~(##1)\null}%
	\autoref{#1}%
	\endgroup%
}
\begin{document}

\title{Semi--classical simulation of spin-1 magnets}

\author{Kimberly Remund}
\email{kimberly.remund@oist.jp}
\affiliation{Theory of Quantum Matter Unit, Okinawa Institute of Science and 
Technology Graduate University, Onna-son, Okinawa 904-0412, Japan}

\author{Rico Pohle}
\affiliation{Department of Applied Physics, University of Tokyo, Hongo,  Bunkyo-ku,
Tokyo, 113-8656, Japan}
\affiliation{Department of Applied Physics, Waseda University, Okubo, Shinjuku-ku, 
Tokyo 169-8555, Japan}

\author{Yutaka Akagi}
\affiliation{Department of Physics, Graduate School of Science, 
The University of Tokyo, Hongo, Tokyo 113-0033, Tokyo}

\author{Judit Romh\'anyi}
\affiliation{Department of Physics and Astronomy, University of California, Irvine, California 92697, USA}
\affiliation{Theory of Quantum Matter Unit, Okinawa Institute of Science and 
Technology Graduate University, Onna-son, Okinawa 904-0412, Japan}

\author{Nic Shannon}
\affiliation{Theory of Quantum Matter Unit, Okinawa Institute of Science and 
Technology Graduate University, Onna-son, Okinawa 904-0412, Japan}

\date{\today}

\begin{abstract}

Theoretical studies of magnets have traditionally concentrated on either 
classical spins, or the extreme quantum limit of spin--1/2. 
However, magnets built of \mbox{spin--1} moments are also intrinsically interesting, 
not least because they can support quadrupole, as well as dipole moments, 
on a single site. 
For this reason, \mbox{spin--1} models have been extensively studied as 
prototypes for quadrupolar (spin--nematic) order in magnetic insulators, 
and Fe--based superconductors. 
At the same time, because of the presence of quadrupoles, the classical limit of a \mbox{spin--1} 
moment is not an $O(3)$ vector, a fact which must be taken into account in 
describing their properties.
%
%
In this Article we develop a method to simulate spin-1 magnets based on a $u(3)$ algebra 
which treats both dipole and quadrupole moments on equal footing.
This approach is amenable to both classical and quantum calculations, and we develop
the techniques needed to calculate thermodynamic properties 
through Monte Carlo simulations and classical low--temperature expansion, 
and dynamical properties, through ``molecular dynamics'' simulations and a multiple--boson 
expansion.   
As a case study, we present detailed analytic and numerical results for the thermodynamic 
properties of ferroquadrupolar order on the triangular lattice, and its associated dynamics.
At low temperatures, we show that it is possible to ``correct'' for the effects of classical 
statistics in simulations, and extrapolate to the zero--temperature quantum 
results found in flavour--wave theory.
%

\end{abstract}


\maketitle

\section{Introduction}
\label{section.introduction}


Textbook discussions of magnetism usually begin either with classical spins, 
or with the \mbox{spin--1/2} moment of an individual electron.    
However, magnetic ions exist in many different forms, each of which requires its own mathematical representation \cite{Abragam1961-OUP,Abragam1970-OUP,Fazekas1999-WorldScientific,Khomskii2014-CUP}.
And this can have profound consequences, even for simple models.
One celebrated example is the gap found in integer--spin quantum antiferromagnets  
in one dimension \cite{Haldane1983-PhysLettA93,Haldane1983-PRL50,Affleck1987}, 
while half--integer systems remain gapless  \cite{Lieb1961}. 
The principles which underpin this gap 
are now well known \cite{Gu2009,Pollmann2012}, 
%
%
but much remains to be understood about higher--spin moments in general.  
For example, it is not widely appreciated that an $O(3)$ vector only provides 
an appropriate (semi--)classical limit for a quantum spin in the case 
of \mbox{spin--1/2} moments, a fact which has implications 
for both ground states and excitations.
In particular, the usual classical mean--field approximations, 
and ``large--S'' treatments of spin--wave excitations \cite{Holstein1940,Anderson1952}, 
both break down for spins larger than 1/2, 
because they do not adequately describe multipole moments \cite{Matveev1973,Papanicolaou1988}.


\begin{figure}[ht!]
	\centering
	\includegraphics[width=0.55\textwidth]{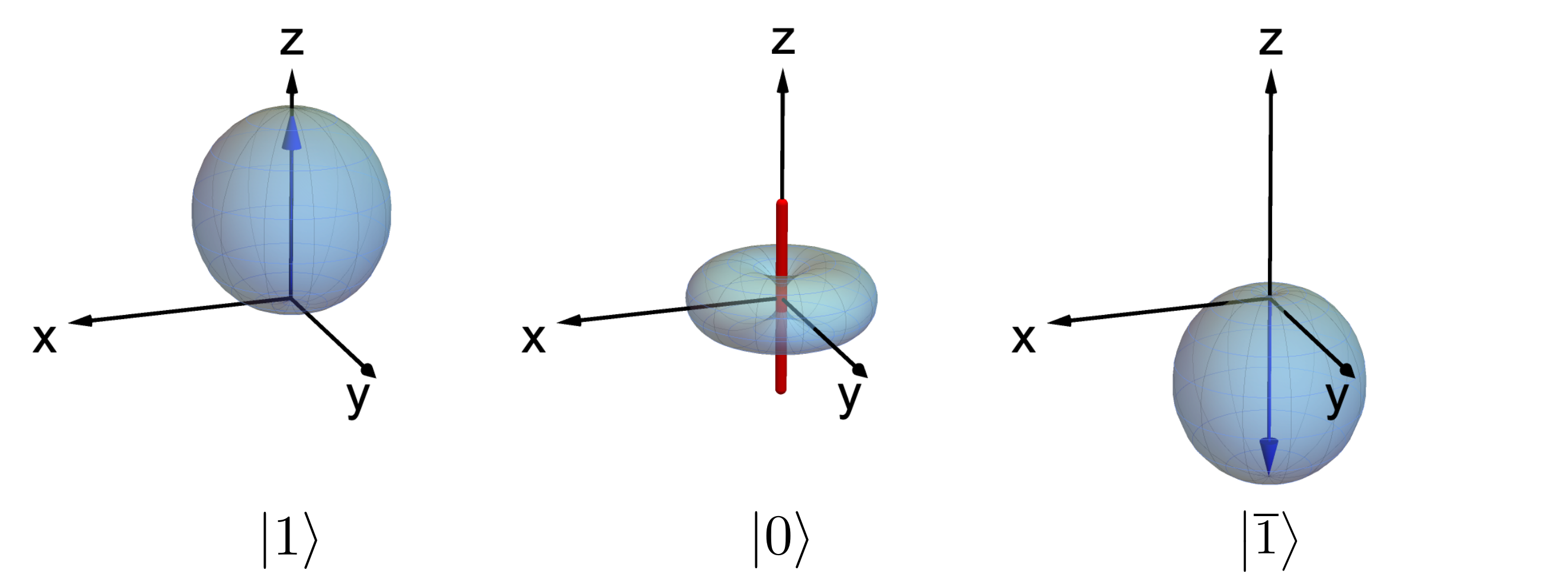}
	\caption{
		Usual, ``magnetic'' basis for a \mbox{spin--1} moment, formed by eigenstates of $S^z$. 
		States with $S^z = \pm 1$, labeled $| 1 \rangle$ and $| \overline{1} \rangle$, 
		break time--reversal symmetry and have a finite spin--dipole moment (blue arrow).
		Meanwhile, the state with $S^z = 0$, labeled $| 0 \rangle$, has 
		a quadrupolar magnetic moment, which can be represented through 
		a director (red bar), perpendicular to the plane of the quadrupole.
		States are represented as probability surfaces in spin--space, as described in \protect\hyperref[app:Spin_Fluct_Prob]{Appendix~}\ref{app:Spin_Fluct_Prob}.
	}
	\label{fig:magnetic.basis}
\end{figure}


Spin--1 magnets provide a natural focus for such questions.
Spin--1 moments differ from both classical vectors and quantum \mbox{spin--1/2} 
moments in that they can support a quadrupole on a single site 
[\Autoref{fig:magnetic.basis}].
This leads to both new types of ground state, and new kinds of excitation \cite{Matveev1973,Papanicolaou1988}.  
Spin--1 systems also support new forms of interaction, relative to a spin--1/2 moment, including single--ion anisotropies and bi--quadratic  
interactions, which can originate in exchange \cite{Fazekas1999-WorldScientific}, 
or be generated by coupling to the lattice \cite{Barma1975}.
As a consequence, the range of phases predicted to occur in \mbox{spin--1} magnets is very rich, 
including 
quadrupolar (spin--nematic) phases \cite{Matveev1973,Andreev1984,Papanicolaou1988,Harada2002,Tsunetsugu2006,Lauchli2006,Smerald2013}, 
and diverse forms of quantum spin liquid \cite{Serbyn2011,Bieri2012,Xu2012,Chen2012,Hwang2013,Buessen2018}, 
as well conventional, dipolar, magnetic order.


Further strong motivation to study \mbox{spin--1} magnets comes from experiment.
A well--studied example is provided by NiGa$_2$S$_4$, a triangular--lattice magnet  
which evades conventional magnetic order \cite{Nakatsuji2005, Nambu2006, Bhattacharjee2006}, and may realise a spin--nematic phase \cite{Tsunetsugu2006,Lauchli2006,Valentine2020}. 
Spin--1 magnets on the pyrochlore lattice have also recently come into 
focus \cite{Gao2020}.
Among these, NaCaNi$_2$F$_7$ is particularly interesting, showing spin--liquid--like properties (above a spin--freezing temperature) \cite{Plumb2019}, 
which cannot be explained within a framework based on $O(3)$ moments \cite{Zhang2019}.
Other \mbox{spin--1} materials discussed as candidate spin liquids include 
NiRh$_2$O$_4$, whose moments inhabit a diamond lattice \cite{Chamorro2018}, and YCa$_3$(VO)$_3$(BO$_3$)$_4$, which realises a Kagome lattice \cite{Miiller2011}, and the triangular--lattice system Ba$_3$NiSb$_2$O$_9$ \cite{Cheng2011,Quilliam2016,Fak2017}.
And in recent years, \mbox{spin--1} models have also been intensively studied as a way of understanding nematic phases in both Fe--based superconductors \cite{Fernandes2014,Luo2016,Wang2016,Gong2017,Lai2017}, 
and systems of cold atoms \cite{Demler2002,Imambekov2003,Kurn2013,deForgesdeParny2014,Zibold2016}.



Given this abundance of riches, there is clearly need for good theoretical tools to 
study \mbox{spin--1} magnets.
But the very things which make \mbox{spin--1} moments interesting, 
also make them difficult to simulate numerically.
Classical Monte Carlo (MC) simulations, 
based on $O(3)$ vector spins, fail to describe grounds states
or excitations built of local quadrupole moments.   
Exact diagonalisation does not suffer from this drawback, but the rapid growth 
of the Hilbert space typically restricts calculations to systems of 20 
sites or less \cite{Lauchli2006,deForgesdeParny2014}.
Variational calculations, based on matrix-- or tensor--product wave functions 
give a good account of dynamics in 1D \cite{White2008}, but cannot easily be extended beyond 
the calculation of ground--state properties in higher dimension \cite{Zhao2012,Niesen2017}.
And while Quantum Monte Carlo (QMC) simulation has yielded insights into 
both quadrupolar order \cite{Harada2002,Kaul2012}, and the associated dynamics \cite{Voell2015}, 
its use is restricted to a relatively small number of cases which do not suffer 
from a sign problem.
Moreover, within QMC, dynamics are only accessible for relatively small systems, 
through analytic continuation, which may be problematic for systems with 
complicated excitations.
As a consequence, much of what we know about the exotic properties 
of \mbox{spin--1} models is restricted to mean field theory (MFT), 
and the linear expansion of fluctuations about it, leaving many important questions 
out of reach.


In this Article, we develop a method of simulating spin-1 magnets, which treats both dipole 
and quadrupole moments on equal footing.
Our approach is based on embedding the algebra $su(3)$ describing a \mbox{spin--1} moment 
in the algebra $u(3)$, as discussed in earlier work of Papanicolaou \cite{Papanicolaou1988}.
This approach makes it possible to treat quantum aspects 
of the problem exactly, at the level of a single site.
We arrive at a formulation in terms of the generators of $U(3)$, which is suitable 
for both analytic and numerical approaches to \mbox{spin--1} magnets.
In particular, the uncluttered structure of the algebra $u(3)$ makes it 
possible to derive very compact  equations of motion (EoM), which can be integrated 
numerically to evaluate dynamics in cases with both conventional and 
unconventional forms of order.


\begin{figure}[t]
	\centering
	\subfloat[Mean Field phase diagram at T=0. \label{fig:phase.diagram}]{
		\includegraphics[width=0.45\textwidth]{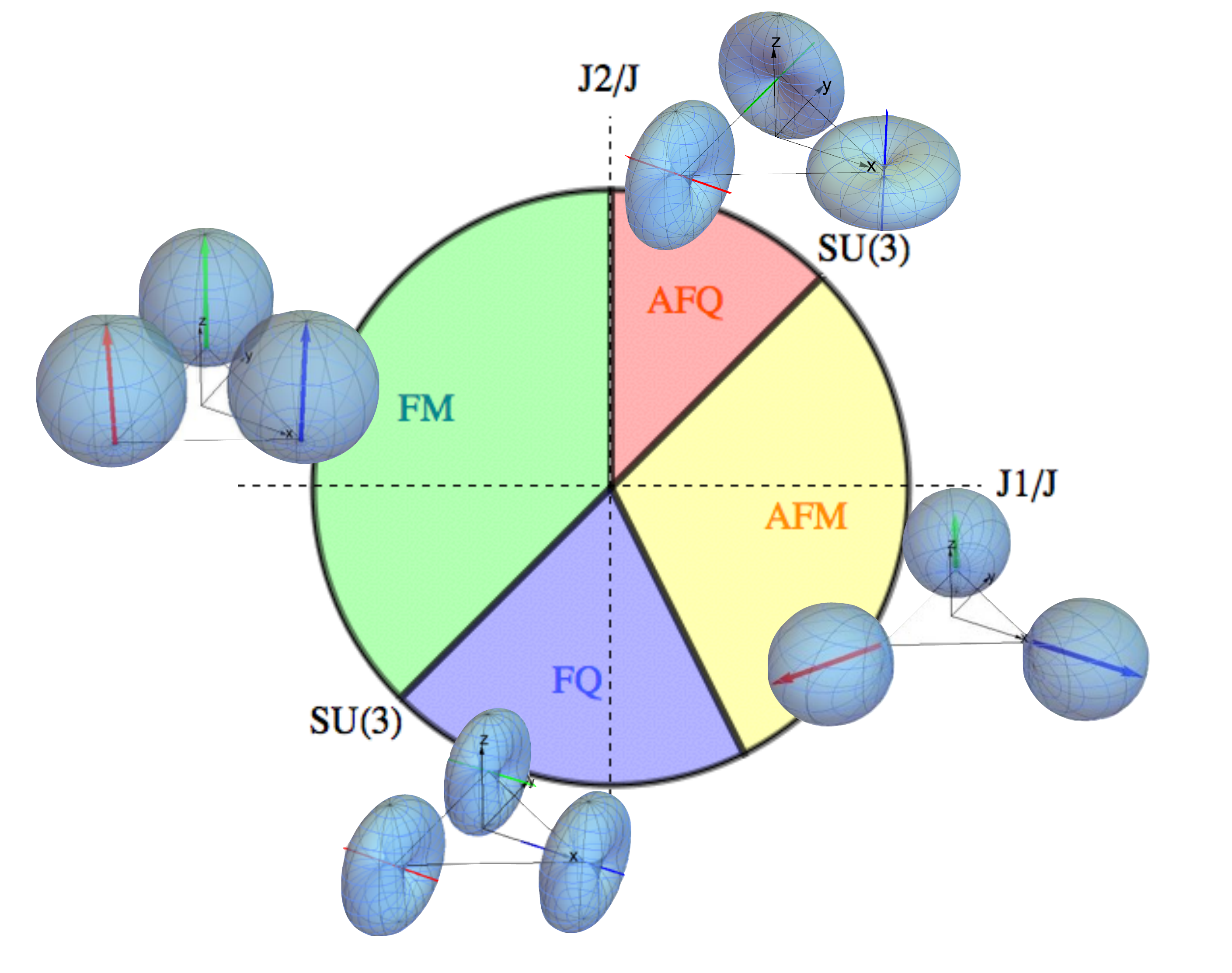}}\\
	\caption{
	Mean--field phase diagram of the \mbox{spin--1} bilinear--biquadratic 
	(BBQ) model on the triangular lattice [\protect\Autoref{eq:BBQ.model}]. 
	at $T=0$, following \cite{Lauchli2006,Smerald2013}. 
	The model shows four distinct ordered ground states: 
	ferromagnet (FM); 
	three--sublattice antiferromagnet (AFM);
	ferroquadrupolar (FQ);  
	and three--sublattice antiferroquadrupolar (AFQ). 
	For $J_1=J_2$ the model exhibits an enlarged, $SU(3)$ symmetry.
	}
	\label{fig:MF.phase.diagram}
\end{figure}


We illustrate our method by applying it to a \mbox{spin--1} model with the most general form 
of interactions allowed by $SU(2)$ symmetry, the bilinear--biquadratic (BBQ) Hamiltonian 
	\be
	\Ham_{\sf BBQ} = \sum_{\nn{i,j}} 
		\left[ J_1\opsb{i} \cdot \opsb{j} + J_2( \opsb{i} \cdot \opsb{j} )^2 \right] \; ,
	\label{eq:BBQ.model}
	\ee
on a triangular lattice, reproducing, and in many cases extending, known results for 
its low--temperature phases \cite{Lauchli2006,Smerald2013,Stoudenmire2009}.  
We pay particular attention to the simplest phase to exhibit all the new features 
of a \mbox{spin--1} moment, the ferroquadrupolar (FQ) order found for $J_2 < 0$ 
[\Autoref{fig:MF.phase.diagram}].  
In particular, we demonstrate that our approach reproduces known results for the 
dynamics of FQ order, where both QMC simulations  \cite{Voell2015} and analytic 
``flavour--wave'' calculations \cite{Lauchli2006} are available for comparison.


At the heart of this analysis is the need to introduce an extended set of operators 
to describe a single \mbox{spin--1} moment.
Any state of spin--1/2 moment can be represented as a point on a Bloch sphere,  
characterised by two polar angles \cite{Bloch1946}. 
By extension, any two states of a spin--1/2 moment can be connected by an $su(2)$ 
rotation, using two of the three generators of $su(2)$.
This simple geometrical picture provides a natural classical limit of a 
spin--1/2 moment, as an $O(3)$ vector carrying a finite dipole moment
\begin{eqnarray}
	\opsb{i}=\begin{pmatrix}
		\ops{i}{x}\\
		\ops{i}{y}\\
		\ops{i}{z}
	\end{pmatrix} \; . 
\label{eq:Sdipole}
\end{eqnarray}
In contrast, the usual magnetic basis for a spin--1 moment, \Autoref{fig:magnetic.basis}, 
includes states with both dipole and quadrupole moments.
There is no $su(2)$ rotation which connects dipoles with quadrupoles, and it 
follows that a general state of a spin--1 moment cannot be expressed in terms 
of two real angles.
As a consequence, its classical limit cannot be an $O(3)$ vector.


To properly characterise a spin--1 moment, 
we therefore need to seek an algebra which encompasses quadrupole moments of spin
\be
\V{Q}_i=\begin{pmatrix}
	Q^{x^2-y^2}\\
	Q^{3z^2-r^2}\\
	Q^{xy}\\
	Q^{yz}\\
	Q^{xz}
\end{pmatrix}_i
=
\begin{pmatrix}
	(S^x)^2-(S^y)^2\\
	\frac{1}{\sqrt{3}}(3((S^z)^2-S(S+1))\\
	S^xS^y+S^yS^x\\
	S^yS^z+S^zS^y\\
	S^xS^z+S^zS^x
\end{pmatrix}_i\; . 
\label{eq:def_Qvec}
\ee
The smallest algebra which can completely do so is $su(3)$, with a total of 8 generators 
\cite{Papanicolaou1988,Penc2011-Springer,Smerald2013}.  
In terms of these operators, the BBQ Hamiltonian [\Autoref{eq:BBQ.model}] 
can be expressed as 
\begin{align*}
	\Ham_{BBQ}=\sum_{\nn{i,j}}&\left(J_1-\frac{J_2}{2}\right)\opsb{i}\cdot\opsb{j}+\frac{J_2}{2}\left.\bf{\hat{Q}_{i}}\cdot\bf{\hat{Q}_{j}}\right.\\
	&+\frac{J_2}{3}s^2(s+1)^2\; , 
	\numberthis \label{eq:BBQ.modelQ}
\end{align*}
where it will prove convenient to write 
\be
\begin{matrix}
	J_1 = J \cos \theta & ,  \; J_2 = J \sin \theta
\end{matrix} \; .
\label{eq:J.from.theta}
\ee


Although the algebra $su(3)$ provides a complete portrait of a \mbox{spin--1} 
moment, its structure constants are very complicated \cite{Itzykson1980-McGraw-Hill}.
This makes $su(3)$ a challenging starting point for descriptions of 
dynamics \cite{Balla2014-Thesis,Remund2015-Thesis,Zhang2021}.
Happily, by adding just one more generator, the spin--length $\opsb{i}^2$,  
and subsequently imposing the constraint 
\begin{eqnarray}
\opsb{i}^2 = s(s+1) = 2 \; ,
\end{eqnarray} 
it is possible to transition to a description of \mbox{spin--1} moment in terms of the 
much simpler algebra $u(3)$ \cite{Papanicolaou1988}.   
This approach is illustrated schematically in \Autoref{eq:u3}.
%
%
\begin{widetext}
	\be
	\parbox{2cm}{$su(3)$ algebra}
	\begin{array}{l@{\quad} c l@{\quad} c}
		& \multirow{3}{3cm}{Spin length\\}&  & \\
		& &  \\
		& \multirow{2}{3cm}{3 linearly independent dipole components}& 	\ldelim \{ {3}{0.5cm}\\
		\ldelim \{ {5}{0.5cm} & & \\
		& & \multirow{5}{0.5cm}{ {$\left\lbrace  \right.$} }  \\
		&\multirow{5}{3cm}{5 linearly independent
			quadrupole components} & 	\ldelim \{ {5}{0.5cm}\\
		& &\\
		& & \\
		& & \\
		& & \\
		& & \\
	\end{array}~~
	\begin{pmatrix}
		\opsb{i}^2\\
		\ops{i}{x}\\
		\ops{i}{y}\\
		\ops{i}{z}\\
		\opq{i}{x^2-y^2}\\
		\opq{i}{3z^2-s^2}\\
		\opq{i}{xy}\\
		\opq{i}{xz}\\
		\opq{i}{yz}\\
	\end{pmatrix}\xrightarrow{\textrm{basis change}}
	\begin{pmatrix}
		\opa{i}{x}{x}\\
		\opa{i}{x}{y}\\
		\opa{i}{x}{z}\\
		\opa{i}{y}{x}\\
		\opa{i}{y}{y}\\
		\opa{i}{y}{z}\\
		\opa{i}{z}{x}\\
		\opa{i}{z}{y}\\
		\opa{i}{z}{z}\\
	\end{pmatrix}\parbox{3cm}{~~$u(3)$ algebra}
	\label{eq:u3}
	\ee
\end{widetext}


A convenient basis for $u(3)$ is provided by the tensors $\opa{i}{}{}$, a set of real, $3\times 3$ matrices with only a single non--vanishing matrix element \cite{Papanicolaou1988}, 
and commutation relations
\be
\begin{tabular}{rcl}
$\com{\opa{i}{\alpha}{\beta},\opa{i}{\gamma}{\eta}}
$&$=$&${\delta^{\gamma}}_{\beta}\opa{i}{\alpha}{\eta}-{\delta^{\alpha}}_{\eta}\opa{i}{\gamma}{\beta}$\; ,\\
& &\\
$\com{\opa{i}{\alpha}{\beta},\opa{j}{\gamma}{\eta}}$&=&0\; .
\end{tabular}\label{eq:ComRelAop}
\ee
Written in terms of these matrices, the BBQ Hamiltonian [\Autoref{eq:BBQ.model}] 
takes on the quadratic form  
\begin{align*}
\Ham_{\sf BBQ} = \sum_{\nn{i,j}}\left[ \right. &J_1\opa{i}{\alpha}{\beta}\opa{j}{\beta}{\alpha}+(J_2-J_1)\opa{i}{\alpha}{\beta}\opa{j}{\alpha}{\beta}\\
+&\frac{J_2}{4}s^2(s+1)^2\left.\right] \; , 
\label{eq:HamA1} \numberthis
\end{align*}
where we adopt the Einstein convention of summing over repeated indices.
Projection into states with \mbox{spin--1} can be accomplished by 
enforcing the constraint 
\be
\opa{i}{\alpha}{\alpha}  
= \frac{1}{2}s(s+1) = 1 \; ,
\label{eq:fix.spin.length}
\ee
on each site in the lattice.

\begin{figure*}[t]
	\centering
	\subfloat[Finite-T phase diagram from u3MC simulation]{\includegraphics[width=0.95\textwidth]{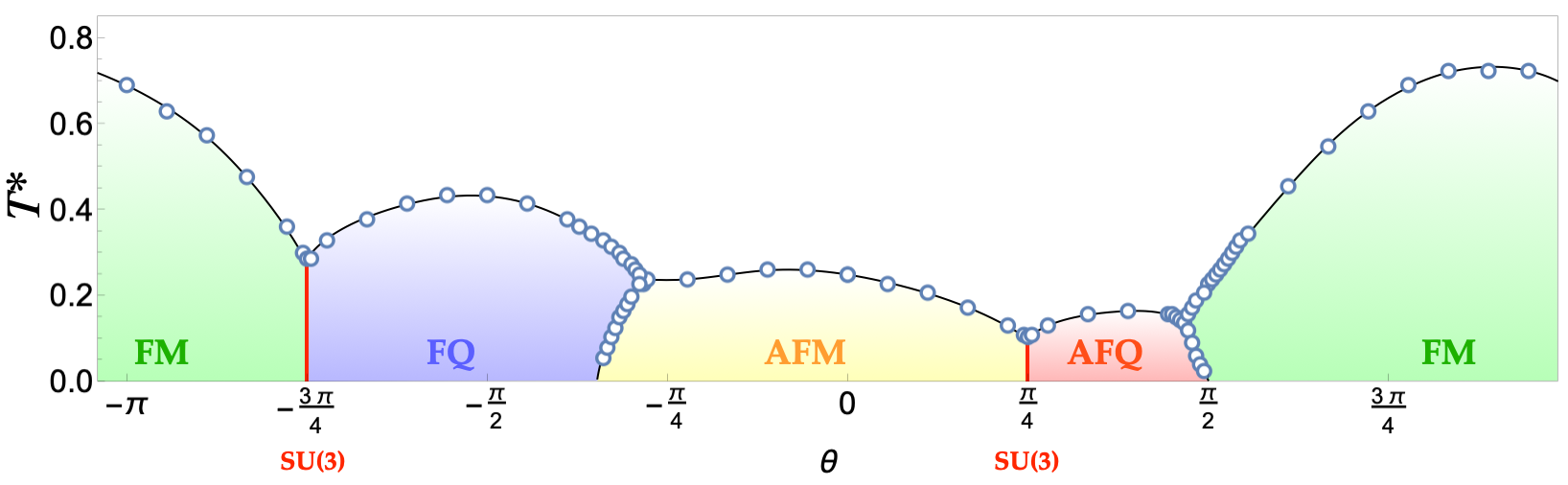}}\\
	\caption{
	Finite--temperature phase diagram of the \mbox{spin--1} bilinear--biquadratic 
	(BBQ) model on the triangular lattice, obtained from Monte Carlo simulation 
	of $\Ham_{\sf BBQ}$ [\protect\Autoref{eq:HamA1}] in the space of $u(3)$ matrices (u3MC).  
	Points show the location of peaks in heat capacity; 
	phases are labelled according to their dominant correlations 
	(cf. \protect\Autoref{fig:MF.phase.diagram} and \protect\Autoref{fig:PD.StructureFactor}).
	Details of simulation are provided in \protect\Autoref{sec:phase.diagram}.
	}
	\label{fig:MC.phase.diagram}
\end{figure*}

From this starting point, we can carry out classical Monte Carlo (MC) simulations 
of the BBQ model in the basis of $\opa{i}{}{}$, treating dipole and quadrupole 
moments of a single site on an equal footing.
We will refer to this approach as ``u3MC''.
Results for the finite--temperature 
phase diagram, obtained using u3MC,  
are shown in \Autoref{fig:MC.phase.diagram}.
At the level of thermodynamics, this approach is equivalent to 
the ``semi--classical $SU(3)$'' simulations of
Stoudenmire {\it et al.} \cite{Stoudenmire2009}, 
and yields identical results. 


However the real advantages of working with a $u(3)$ representation become 
apparent when considering dynamics.
Considering the Heisenberg equation of motion for $\opa{i}{}{}$, we find
\begin{align*}
\ddp_t \opa{i}{\gamma}{\eta}=&-i\com{\opa{{i}}{\gamma}{\eta},\Ham_{\sf BBQ}}\\
=&-i \sum_{\delta} \left[ \right. 
J_1(\opa{{i}}{\gamma}{\alpha}\opa{{i+\delta}}{\alpha}{\eta}-\opa{{i}}{\alpha}{\eta}\opa{{i+\delta}}{\gamma}{\alpha})\\
&+(J_2-J_1)(\opa{{i}}{\gamma}{\alpha}\opa{{i+\delta}}{\eta}{\alpha}-\opa{{i}}{\alpha}{\eta}\opa{{i+\delta}}{\alpha}{\gamma}) 
\left.\right] \; ,\label{eq:DA2}\numberthis
\end{align*}
a simple form which automatically preserves the length of the spin [\Autoref{eq:fix.spin.length}], 
and is well--suited to numerical integration.

\begin{figure}[t]
	\centering
	\subfloat[u3MD simulation at $T = 0.05$ \label{fig:illustration.raw.results}]{
		\includegraphics[width=0.45\textwidth]{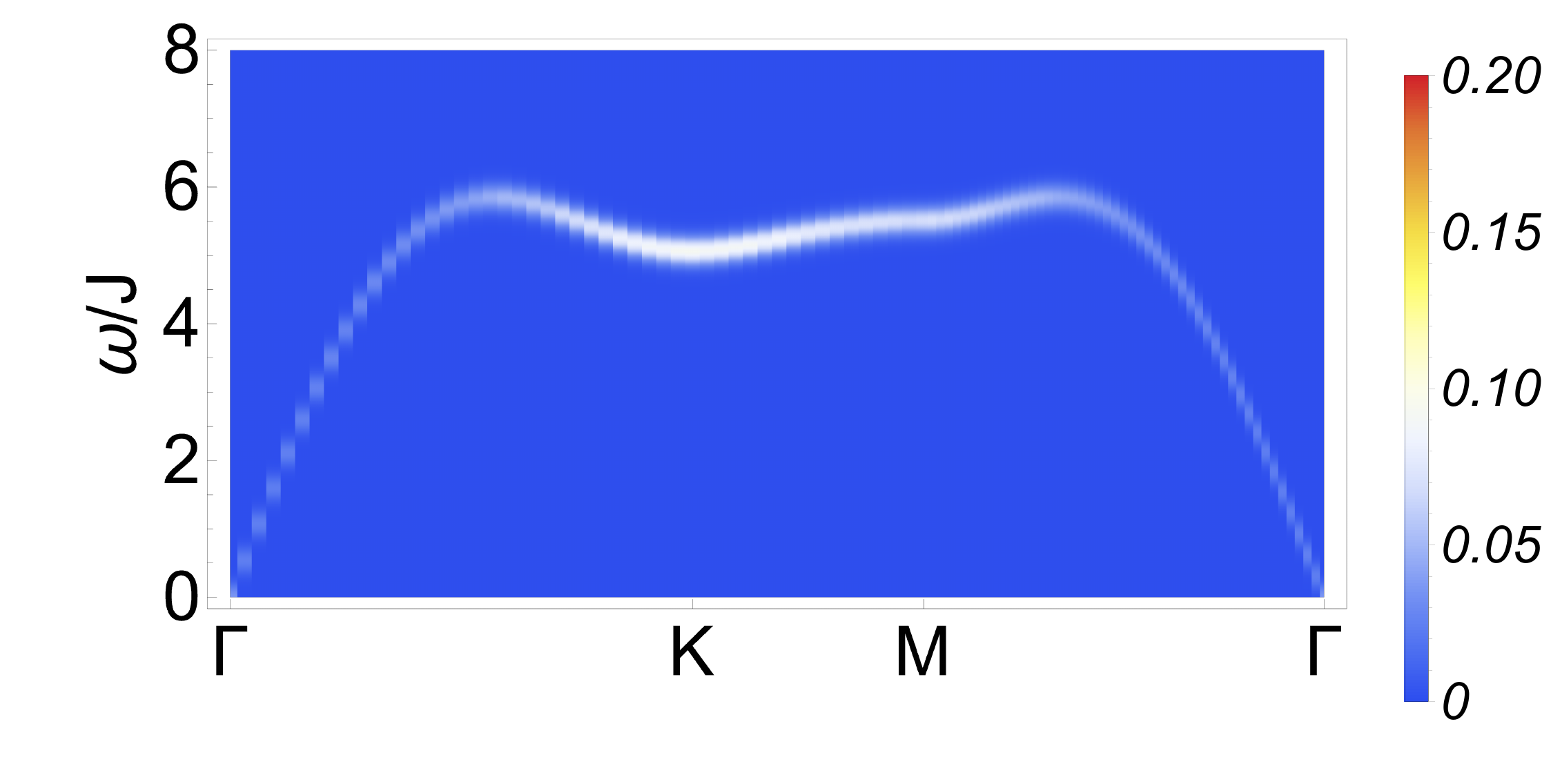}} \\
	\subfloat[u3MD, corrected for classical statistics  \label{fig:illustration.temperature.corrected}]{
		\includegraphics[width=0.45\textwidth]{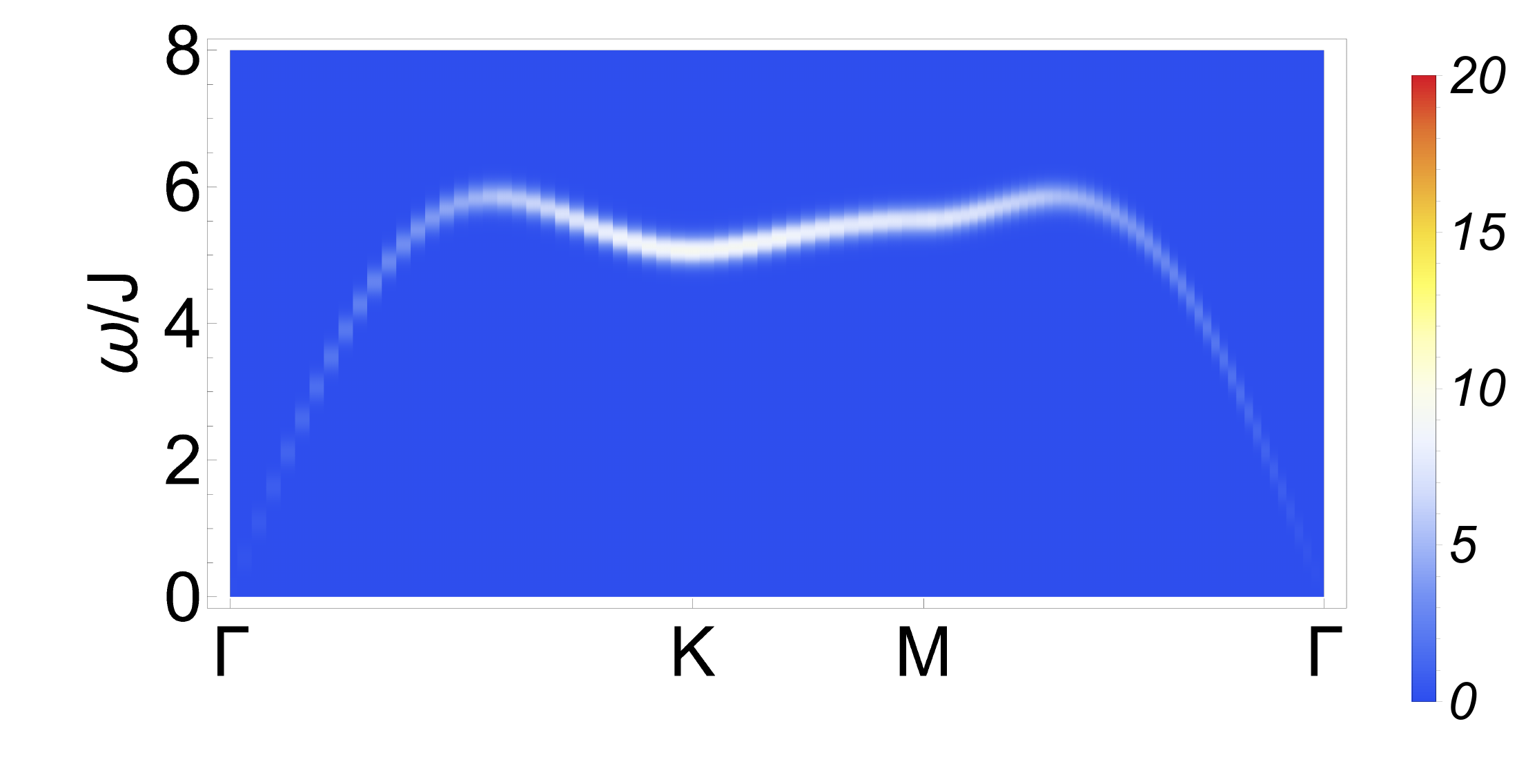}} \\
	\subfloat[Quantum flavour--wave theory at \mbox{$T=0$} \label{fig:illustration.flavour.wave}]{
		\includegraphics[width=0.45\textwidth]{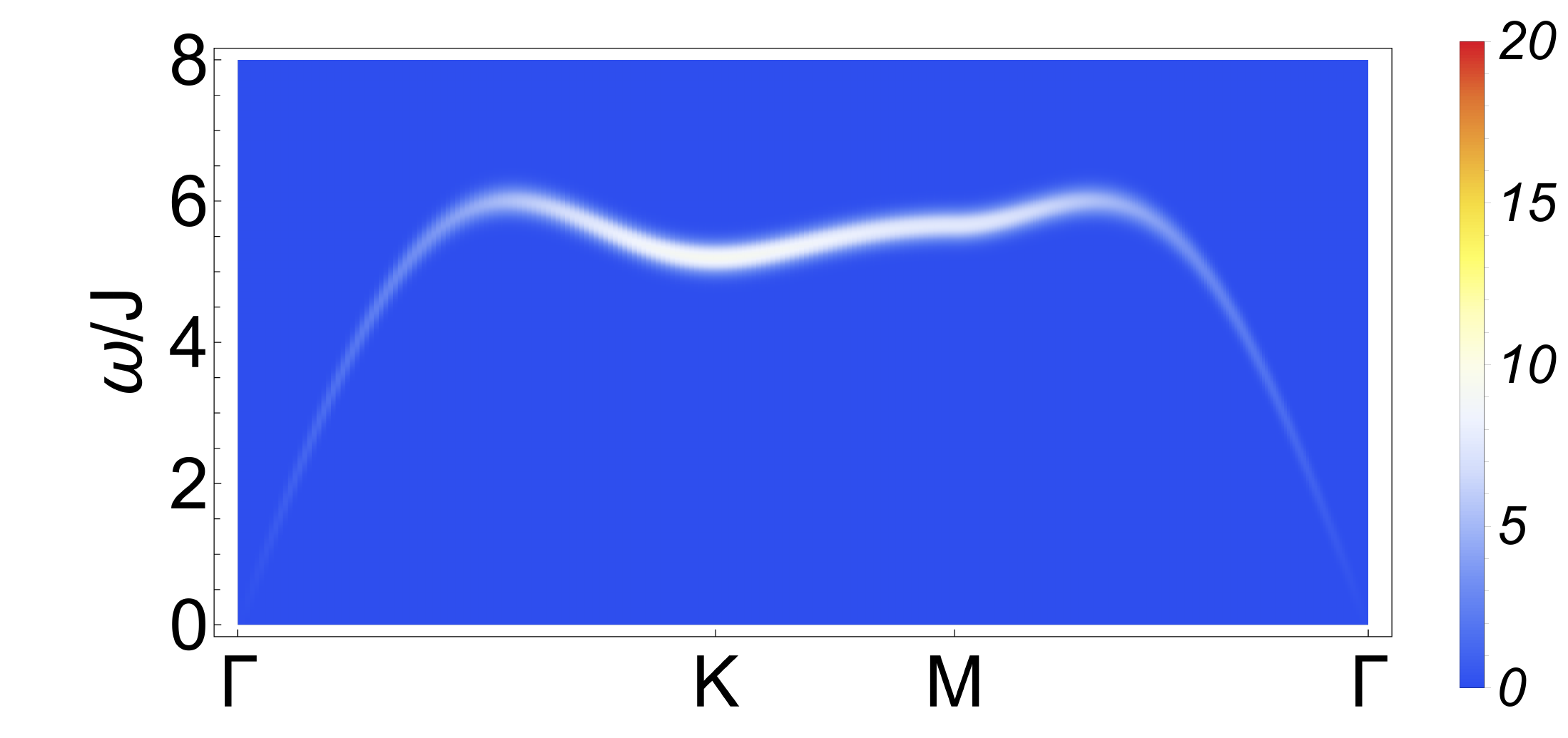}} \;
\caption{
Dynamical structure factor $S_{\rm{S}}(\q, \omega)$ 
for the ferroquadrupolar (FQ) phase of the BBQ model on the triangular lattice.
\protect\subref{fig:illustration.raw.results} Results of ``molecular dynamics'' (u3MD)
simulation at finite $T$, showing a dispersing band of excitations which are the 
Goldstone modes of the FQ order.
The spectral weight in these excitations is controlled by the classical statistics of 
the associated Monte Carlo (u3MC) simulations.
\protect\subref{fig:illustration.temperature.corrected} MD results corrected 
for the effects of classical statistics, for comparison with quantum results at $T=0$. 
\protect\subref{fig:illustration.flavour.wave} Predictions of quantum flavour--wave 
theory at $T=0$, showing agreement with the corrected results of MD simulation, 
as detailed in \protect\Autoref{fig:T-scaling_Sqw}.
Results are shown for \protect\Autoref{eq:HamA1} with $J_1 = 0.0; J_2 = -1.0$. %
Details of these calculations are provided in \protect\Autoref{section:MD.deconstructed}.
}
\label{fig:illustration.of.results}
\end{figure}

By combining classical MC simulation with numerical integration of the  
equation of motion, \Autoref{eq:DA2}, we obtain an approach to dynamics analogous 
to ``molecular dynamics'' (MD) simulation, which can be used to calculate 
dynamical structure factors.
We dub this ``u3MD''.
At low temperatures, we find it is possible to correct for the 
effect of classical statistics by multiplying structure factors 
by a temperature--dependent prefactor
\begin{eqnarray}
S^{\sf QM} ({\bf q}, \omega, T=0) = \lim_{T \to 0} \frac{\hbar\omega}{k_B T} S^{\sf MD} ({\bf q}, \omega, T) \; ,
\label{eq:corrected.Sqw}
\end{eqnarray}
obtaining results in agreement with semi--classical quantum results at $T=0$.
Results for u3MD simulations of the FQ phase are summarised in \Autoref{fig:illustration.of.results}.


In remainder of this paper we describe these results in some detail, developing both 
analytic and numerical approaches based on the $u(3)$ formalism.
We benchmark simulations against these analytic results, and 
published numerical results from other approaches.
We also also explore some of the further ramifications of the $u(3)$ approach, 
particularly with respect to anisotropic exchange interactions, 
and the extent to which quantum results for dynamics can be inferred from (semi--)classical simulations.


In order to keep the paper self--contained, we provide a detailed account 
of derivations, and review all of the necessary mathematical formalism.
However the paper is also constructed in such a way that readers uninterested 
in technical development of the method can skip directly to the results 
provided in \Autoref{section:numerics},  \Autoref{section:quantum.vs.classical} 
and  \Autoref{section:anisotropy}.
Additional technical details are provided in a series of Appendices.


The paper is structured as follows:


\Autoref{section:maths.for.spin.1} reviews the mathematical formalism needed to analyze 
spin--1 magnets in terms of a $u(3)$ algebra.
A single \mbox{spin--1} moment is analysed within a basis of time--reversal invariant states, 
where the most general possible spin configuration can be expressed in terms of a
complex vector ${\bf d}$.
The group $U(3)$ is shown to provide a convenient basis for all possible 
operations on \mbox{spin--1} moments.
Expressions are given for both the dipole and quadrupole associated with a 
spin--1 moment, in terms of the vector ${\bf d}$, and the $3\times3$ 
matrices $\opa{}{\alpha}{\beta}$, which provide a suitable representation of $u(3)$.
The BBQ model [\Autoref{sec:u3_HBBQ}], and corresponding 
Heisenberg EoM [\Autoref{sec:u3.EoM}], are also developed in 
terms of $\opa{}{\alpha}{\beta}$, in a forms suitable for numerical simulation.


\Autoref{section:numerical.method} introduces numerical simulation 
methods for \mbox{spin--1} magnets.   
In \Autoref{sec:MC}, the $u(3)$ algebra described 
in \Autoref{section:maths.for.spin.1} is shown to provide a convenient basis for 
\mbox{(semi--)classical} Monte Carlo simulations of \mbox{spin--1} moments.
A suitable MC update is developed in the basis of $u(3)$ matrices $\opa{}{\alpha}{\beta}$, 
and shown to be equivalent to earlier "sSU(3)" simulations of the \mbox{spin--1} BBQ model 
in the basis of the complex vector ${\bf d}$. 
These calculations are extended to general $J_1$, $J_2$, providing a
finite--temperature phase diagram for the \mbox{spin--1} BBQ model 
on a triangular lattice [\Autoref{fig:MC.phase.diagram}].
In \Autoref{sec:u3MD}, a MD simulation scheme is 
developed for \mbox{spin--1} moments, based on the EoM for the matrices
$\opa{}{\alpha}{\beta}$.
The technical implementation of this MD update is described.


In \Autoref{section:classical.theory} we develop an analytic theory of classical 
fluctuations about a ferroquadrupolar (FQ) ground state of the \mbox{spin--1} 
BBQ model on a triangular lattice, starting from the $u(3)$ representation 
introduced in \Autoref{section:maths.for.spin.1}.
In \Autoref{section:small.fluctuations}, small fluctuations about FQ order 
are recast in terms of the the matrices $\opa{}{\alpha}{\beta}$. 
In \Autoref{section:low.T} these are shown to provide a natural basis 
for a classical, low--temperature expansion, within which it is possible to 
calculate thermodynamic properties.
Finally, in \Autoref{section:theory.classical.thermodynamics} we use this theory 
to make explicit predictions for thermodynamic properties, including structure factors, 
for later comparison with MC simulation.


In \Autoref{section:quantum.theory}, we develop an equivalent, 
zero--temperature, quantum theory of fluctuations about a FQ ground state.
First, in \Autoref{section:quantization.of.fluctuations} we quantize the fluctuations 
introduced in \Autoref{section:small.fluctuations}, and show that the resulting 
multiple--Boson expansion is equivalent to earlier ``flavor--wave'' theory. 
Then, in \Autoref{sec:quantum.structure.factors}, we use this quantum theory to make 
explicit predictions for dynamical structure factors, for later comparison with MD simulation.


In \Autoref{section:numerics}, the numerical methods 
developed in \Autoref{section:numerical.method} are used to obtain 
a detailed portrait of FQ order in the \mbox{spin--1} BBQ model 
on a triangular lattice.
Monte Carlo simulation results for 
heat capacity [\Autoref{section:heat.capacity}], 
FQ order parameter [\Autoref{section:ordered.moment}], 
and equal--time structure factors [\Autoref{section:S.of.q}], 
are compared explicitly with the analytic theory developed 
in \Autoref{section:theory.classical.thermodynamics}.
The implications of the Mermin--Wagner Theorem are discussed, and the results 
of simulations at low temperatures are shown to conform exactly to the predictions 
of theory for a finite--size cluster.
In \Autoref{section:numerics.dynamics}, numerical results are presented for the 
dynamics of the FQ state, based on MD simulations.
The dynamical structure factors found in simulation are 
compared explicitly with the predictions of \Autoref{sec:quantum.structure.factors}.
Excitations are found to display the expected dispersion, but with intensities 
which, for \mbox{$T \to 0$}  differ from the analytic theory.


\Autoref{section:quantum.vs.classical} resolves this paradox.  
By combining the low--temperature and 
multiple--Boson expansions developed in \Autoref{section:classical.theory}, we show that MD 
results can be understood in terms of a semi--classical dynamics, with spectral weight 
determined by a factor coming from classical statistics.
Low--temperature MD results are corrected for this classical bias, and shown 
to agree exactly with the analytic theory of \Autoref{section:quantum.theory}, and equivalent 
``flavour--wave'' calculations, 
in the limit \mbox{$T \to 0$}. 


In \Autoref{section:anisotropy}, we address the generalisation of simulation 
to which are anisotropic in spin--space.
The EoM approach developed in \Autoref{sec:u3.EoM} and \Autoref{sec:u3MD} is shown to be 
robust against spin--anisotropy.
Concrete results analytic and numerical results are provided for FQ order in the presence 
of single--ion anisotropy.


The paper concludes in \Autoref{section:conclusions} with 
a brief summary of results, and discussion of potential future applications 
of the $u(3)$ approach.


A number of technical results are developed in Appendices.


\hyperref[app:Spin_Fluct_Prob]{Appendix~}\ref{app:Spin_Fluct_Prob}
provides a framework for visualising the quantum states 
individual \mbox{spin--1} moments starting from a coherent--state representation.


\hyperref[sec:appA]{Appendix~}\ref{sec:appA} details mathematical properties of the 
\mbox{tensors $\opa{}{\alpha}{\beta}$}, 
which act as generators of the group $U(3)$.


\hyperref[sec:conv_trig_lat]{Appendix~}\ref{sec:conv_trig_lat} sets out the conventions used in describing the triangular lattice.


\hyperref[sec:structure_factors_classical]{Appendix~}\ref{sec:structure_factors_classical} provides technical details of the calculation 
of equal--time structure factors within a classical low--T expansion.


\hyperref[sec:bogolioubov_transfomation]{Appendix~}\ref{sec:bogolioubov_transfomation} provides technical details of the Bogolibubov transformation used in the multiple--Boson expansion.


\hyperref[sec:structure_factors_quantum]{Appendix~}\ref{sec:structure_factors_quantum} provides technical details of calculations of dynamical structure factors within a multiple--Boson expansion.


\hyperref[sec:Mermin_Wagner_Theorem]{Appendix~}\ref{sec:Mermin_Wagner_Theorem} provides details of analytic calculations of the ordered moment for finite--size clusters.


\hyperref[sec:table.of.integrals]{Appendix~}\ref{sec:table.of.integrals} lists integrals used in developing the analytic theory of FQ order.


\hyperref[section:anisotropy_FM]{Appendix~}\ref{section:anisotropy_FM} develops an analytic theory of the excitations of a 
spin--1 easy--plane ferromagnet, based on generators of $U(3)$.

\section{Description of a \mbox{spin--1} moment using a $u(3)$ algebra}
\label{section:maths.for.spin.1}

In this Section, we develop the mathematical tools needed to describe 
a \mbox{spin--1} moment, and explain how one naturally arrives at 
a general description in terms of operators satisfying a $u(3)$ algebra. 
These will form the basis for both the analytic calculations and the simulations 
described in the remaining parts of the Article.
Our analysis builds on the earlier work by Papanicolou \cite{Papanicolaou1988}, 
and will also make connection with the notation of ``{\bf d}--vectors'',  
used in \cite{Lauchli2006,Penc2011-Springer, Smerald2013,Ueda2016-PRA93}.


We start in \Autoref{sec:spin.half} by reviewing the familiar description of 
a spin--1/2 moment in terms of eigenstates of $S^z$, and describe how its  
classical limit, an $O(3)$ vector, can be used as a basis for numerical simulations.
In \Autoref{sec:spin1}, we show how the usual magnetic basis 
for a single \mbox{spin--1} moment (eigenstates of $S^z$), 
can be used to construct a new, non--magnetic, basis of states 
invariant under time--reversal.
This motivates a general description of a \mbox{spin--1} moment in terms 
of a complex vector ${\bf d}$, 
and of the introduction of the a set of $3 \times 3$ matrices $A^{\alpha}_{~\beta}$, 
which generate a representation of the algebra $u(3)$. 
%
%
Then, in \Autoref{sec:u3_HBBQ}, we show how the the BBQ model, \Autoref{eq:HamA1}, 
can be expressed in terms of the matrices $A^{\alpha}_{~\beta}$, providing a 
starting point for numerical simulation of thermodynamic properties.
Finally, in \Autoref{sec:u3.EoM}, we use this representation of the BBQ model 
to derive equations of motion for a spin--1 moment in terms of $A^{\alpha}_{~\beta}$, 
providing a starting point for numerical simulations of dynamics.


For compactness of notation, we set $\hbar = 1$.

\subsection{Mathematical description of spin--1/2 moments}
\label{sec:spin.half}

Before reviewing the mathematical description of a \mbox{spin--1} moment, 
it is helpful to have in mind the usual picture of a spin--1/2 moment, 
and its (semi--)classical ``large--S'' limit.
%
Any quantum spin can be completely described by the 
eigenstates of $\ops{}{z}$
\be
\ops{}{z}\ket{m}=m\ket{m}\; ,
\ee
where the $2s+1$ states $\ket{m}$ form a closed orthogonal basis 
for $m=-j,-j+1,\dots,j-1,j$ \cite{Landau1977}.
In the case of spin--1/2, $\ops{}{z}$ there are only two such eigenstates 
\be
\ops{}{z} \ket{\pm \textstyle{\frac{1}{2}}} = \pm \textstyle{\frac{1}{2}} \ket{\pm \textstyle{\frac{1}{2}}}\; .
\ee
These states form a Kramers pair, related by time--reversal symmetry
\begin{eqnarray}
\mathcal{T} \ket{\pm \textstyle{\frac{1}{2}}} =  \pm \ket{\mp \textstyle{\frac{1}{2}}}  \; .
\end{eqnarray}
As a consequence, individual spin--1/2 moments always break time--reversal 
symmetry, and always exhibit a finite spin---dipole moment.


From this starting point, it is possible to express any possible quantum state of a 
spin--1/2 moment in terms of two complex numbers 
\be
\ket{ \psi_{\sf 1/2} } = c_1 \ket{+\textstyle{\frac{1}{2}}} + c_2 \ket{-\textstyle{\frac{1}{2}}}
\ee
subject to the constraint 
\be
|c_1|^2 + |c_2|^2 = 1
\ee
Resolving this constraint reduces the number of real parameters to three.
And since the overall phase of $\textstyle{\ket{ \psi_{\sf 1/2} }}$ does not affect its 
physical properties, any state of a spin--1/2 moment can be specified using 
only two real numbers.
Geometrically, this is equivalent to specifying the two angles needed to define 
a point on a Bloch sphere \cite{Bloch1946}.
Formally, it is equivalent to working in the complex projective line $\mathbb{CP}^1$.


Any two states within this space can be connected by an $SU(2)$ rotation, 
for which the Pauli Matrices
\be
\sigma^x = 
\begin{pmatrix}
 0 & 1 \\
 1 & 0 
 \end{pmatrix} \; , \; 
 \sigma^y = 
\begin{pmatrix}
 0 & -i \\
 i & 0 
 \end{pmatrix}  \; , \; 
 \sigma^z = 
\begin{pmatrix}
 1 & 0 \\
 0 & -1 
 \end{pmatrix} \; . \label{eq:pauli.matrices}
\ee
provide a convenient basis, with commutation relations 
\be
[\sigma_\alpha, \sigma_\beta] = 2i \epsilon_{\alpha\beta\gamma}  \sigma_\gamma
\; .
\ee


The classical, ``large--S'' limit of a spin--1/2 can be taken through 
a coherent state representation \cite{Auerbach1994-Springer}, and is a
$O(3)$ vector 
\be
{\bf S}  = (S^x, S^y, S^z) \; .
\ee
Since, for a single spin--1/2, all higher--order spin moments vanish, 
this vector describes all possible magnetic degrees of freedom, and 
can form the starting point for Monte Carlo (MC) simulation of  
thermodynamic properties \cite{Landau2014-Cambridge}.

The representation of spin--1/2 moments in terms of $O(3)$ vectors 
also provides the starting point for (semi--)classical descriptions of their dynamics,  
as determined by the Heisenberg equation of motion (EoM)
\be
	\frac{d {\bf S}_i}{d t} 	= - i \big[  {\bf S}_i, {\cal H} \big]   
					= \frac{d {\cal H}}{d {\bf S}_i} \times {\bf S}_i		\; .
					\label{eq:Heisenberg.EoM.O3}
\ee
Numerical integration of these EoM, using spin configurations drawn from 
MC simulation, provides a (semi--)classical approach to spin dynamics 
which has been dubbed ``Molecular Dynamics'' (MD) 
simulation \cite{Moessner1998-PRL80,Zhang2019,Pohle2021}, and is closely 
analagous to simulations based on the (phenomenological) Landau--Lifshitz--Gilbert 
equations \cite{Gilbert2004}.

\subsection{Description of a single quantum spin-1}
\label{sec:spin1}

\subsubsection{Magnetic basis}

Several new features arise in the case of a spin--1.
Here, the eigenstates of $\ops{}{z}$ comprise the 3 states
\be
\mathcal{B}_1=\left\lbrace \ket{1},\ket{0},\ket{\overline{1}} \right\rbrace
\label{eq:B1}\; .
\ee
forming the ``magnetic'' basis illustrated in \Autoref{fig:magnetic.basis}. 
While the states $\ket{1}$ and $\ket{\overline{1}}$ are truly magnetic, in 
the sense of possesing a finite spin--dipole moment, the same is not 
true of $\ket{0}$, 
for which
 \be
 	\bra{0}\ops{}{x}\ket{0}=\bra{0}\ops{}{y}\ket{0}=\bra{0}\ops{}{z}\ket{0}=0 \; .
 \ee
This result follows straightforwardly from the fact that 
\be
	\ket{z} = -i \ket{0} \; ,	
\ee
is invariant under time--reversal symmetry \cite{Penc2011-Springer}
\be
	{\mathcal T} \ket{z} = \ket{z} \; .
\ee
It follows that $\ket{0} \propto \ket{z}$ is incapable of supporting a dipole moment 
since, such a moment  would, by definition, break time--reversal symmetry.


Instead, the state $\ket{0}$ posses a finite spin--quadrupole moment.
Spin quadrupoles are defined through the symmetric, traceless rank--2 tensor
 \be
 \opq{}{\alpha\beta} 
 	= \ops{}{\alpha}\ops{}{\beta} 
 	+ \ops{}{\beta}\ops{}{\alpha}
	- \frac{2}{3}\delta^{\alpha\beta}s(s+1)
	\label{eq:DefQ} \; ,
\ee
and so are invariant under time--reversal symmetry.
The state $\ket{0}$, exhibits two non--zero matrix elements
\begin{eqnarray}
 	 \bra{0}\ops{}{x}\ops{}{x}\ket{0}=\bra{0}\ops{}{y}\ops{}{y}\ket{0}=1 \; ,
\end{eqnarray}
implying that both $\rm{Q}^{xx} $ and $\rm{Q}^{yy}$ take on a finite value.
Moreover the fact that 
\begin{eqnarray}
\bra{0}\ops{}{z}\ops{}{z}\ket{0}=0
\end{eqnarray}
implies that $\ket{0}$ breaks spin--rotation invariance, even though it 
does not posses a finite dipole moment.



The possibility of finding a finite quadrupole moment on a single site 
sharply distinguishes \mbox{spin--1} moments from spin--1/2 moments.
%
And \mbox{spin--1} are special in the sense that they are the smallest spin able to 
support a quadrupole moment on a single site, making them a good candidate 
to illustrate both magnetism based on higher order--moments, 
and quantum effects. 


More generally, any state of a \mbox{spin--1} moment can be described through a 
linear superposition of the states $\mathcal{B}_1$
\be
\ket{ \psi_1 } = c_1 \ket{1} + c_2 \ket{0} + c_3 \ket{\overline{1}} \; ,
\label{eq:spin-1.magnetic.basis}
\ee
where the complex numbers $c_1$, $c_2$, $c_3$, are subject to the constraint
\be
|c_1|^2 + |c_2|^2 + |c_3|^2 = 1 \; .
\label{eq:normalisation}
\ee
Resolving this constraint immediately reduces the number of real parameters needed
to specify $\ket{ \psi_1 }$ to five.
Furthermore, no physical properties of the state depend on 
the overall phase of $\ket{ \psi_1 }$.
It follows that any state of a spin--1 moment can be fully characterised using 
a total of four real numbers.
Formally, this is equivalent to working in the complex projective plane $\mathbb{CP}^2$.


The algebra which connects states within this Hilbert space 
is $su(3)$, with eight generators, for which the Gell--Mann matrices 
 \begin{eqnarray}
\lambda_1 &=&
 \begin{pmatrix}
 	0 & 1 & 0 \\
 	1 & 0 & 0 \\
 	0 & 0 & 0
 \end{pmatrix} \; , \; 
\lambda_2 =
 \begin{pmatrix}
 	0 & -i & 0 \\
 	i & 0 & 0 \\
 	0 & 0 & 0
 \end{pmatrix} \; , \; 
\lambda_3 =
 \begin{pmatrix}
 	1 & 0 & 0 \\
 	0 & -1 & 0 \\
 	0 & 0 & 0
 \end{pmatrix} \; , \; \nonumber\\
\lambda_4 &=&
 \begin{pmatrix}
 	0 & 0 & 1 \\
 	0 & 0 & 0 \\
 	1 & 0 & 0
 \end{pmatrix} \; , \; 
\lambda_5 =
 \begin{pmatrix}
 	0 & 0 & -i \\
 	0 & 0 & 0 \\
 	i & 0 & 0
 \end{pmatrix} \; , \; 
\lambda_6 =
 \begin{pmatrix}
 	0 & 0 & 0 \\
 	0 & 0 & 1 \\
 	0 & 1 & 0
 \end{pmatrix} \; , \; \nonumber\\
\lambda_7 &=&
 \begin{pmatrix}
 	0 & 0 & 0 \\
 	0 & 0 & -i \\
 	0 & i & 0
 \end{pmatrix} \; , \; 
\lambda_8 = \frac{1}{\sqrt{3}}
 \begin{pmatrix}
 	1 & 0 & 0 \\
 	0 & 1 & 0 \\
 	0 & 0 & -2
 \end{pmatrix} \; , \; \label{eq:gellmann.matrices}
 \end{eqnarray}
provide a convenient representation, albeit one with complex
commutation relations \cite{Itzykson1980-McGraw-Hill}.


It is immediately apparent that algebra describing a \mbox{spin--1} moment 
is much richer than that describing a spin--1/2.
In fact the three $SU(2)$ rotations needed to describe a spin--1/2 
moment, \Autoref{eq:pauli.matrices},  correspond 
to the generators of rotations 
\be
{\bf \hat{S}} = ( \ops{}{x}, \ops{}{y}, \ops{}{z} )^{\rm{t}} \; ,
\label{eq:dipole.moments}
\ee
and form a closed sub--algebra of $su(3)$.
Meanwhile the five additional generators found in $su(3)$ can be identified 
with the quadrupole moments 
 \begin{align*}
\bf{\hat{Q}_{}}& 
=
 \begin{pmatrix}
 	\opq{}{x^2-y^2}\\
 	\opq{}{3z^2-s^2}\\
 	\opq{}{xy}\\
 	\opq{}{xz}\\
 	\opq{}{yz}\\
 \end{pmatrix}
= \begin{pmatrix}
 \frac{1}{2}(\opq{}{xx}-\opq{}{yy})\\
 \frac{1}{\sqrt{3}}(\opq{}{zz}- \frac{1}{2}(\opq{}{xx}+\opq{}{yy}))\\
 \opq{}{xy}\\
 \opq{}{xz}\\
 \opq{}{yz}\\
\end{pmatrix}\\
&= \begin{pmatrix}
(\ops{}{x})^2-(\ops{}{y})^2\\
\frac{1}{\sqrt{3}}(2(\ops{i}{z})^2-(\ops{}{x})^2-(\ops{}{y})^2)\\
\ops{}{x}\ops{}{y}+\ops{}{y}\ops{}{x}\\
\ops{}{x}\ops{}{z}+\ops{}{z}\ops{}{x}\\
\ops{}{y}\ops{}{z}+\ops{}{z}\ops{}{y}\\
\end{pmatrix}
\; . \numberthis
\label{eq:Def_Q_vec}
 \end{align*}
as previously listed in \Autoref{eq:def_Qvec}.
We note that the vector notation [\Autoref{eq:def_Qvec}], 
and tensor notation [\Autoref{eq:DefQ}], are linked by 
 \be
\bf{\hat{Q}_{}}\cdot \bf{\hat{Q}_{}}=\sum_{\alpha}{\opq{}{\alpha}}{\opq{}{\alpha}}=\frac{1}{2}\sum_{\alpha\beta}\opq{}{\alpha\beta}\opq{}{\alpha\beta}\label{eq:RelQvec_Qtens}\; ,
 \ee
 and in what follows we shall follow the Einstein convention of 
 assuming sums on repeated indices of tensors.


It is worth noting that, while the algebra $su(3)$ has eight generators, a general $SU(3)$ 
rotation can be constructed using a subset of four of these \cite{Nelson1967}.
It follows that (as argued above), only four real parameters are needed to parameterise 
any \mbox{spin--1} state.

 

\subsubsection{Time--reversal invariant basis}

\begin{figure}[t]
	\centering
	\includegraphics[width=0.55\textwidth]{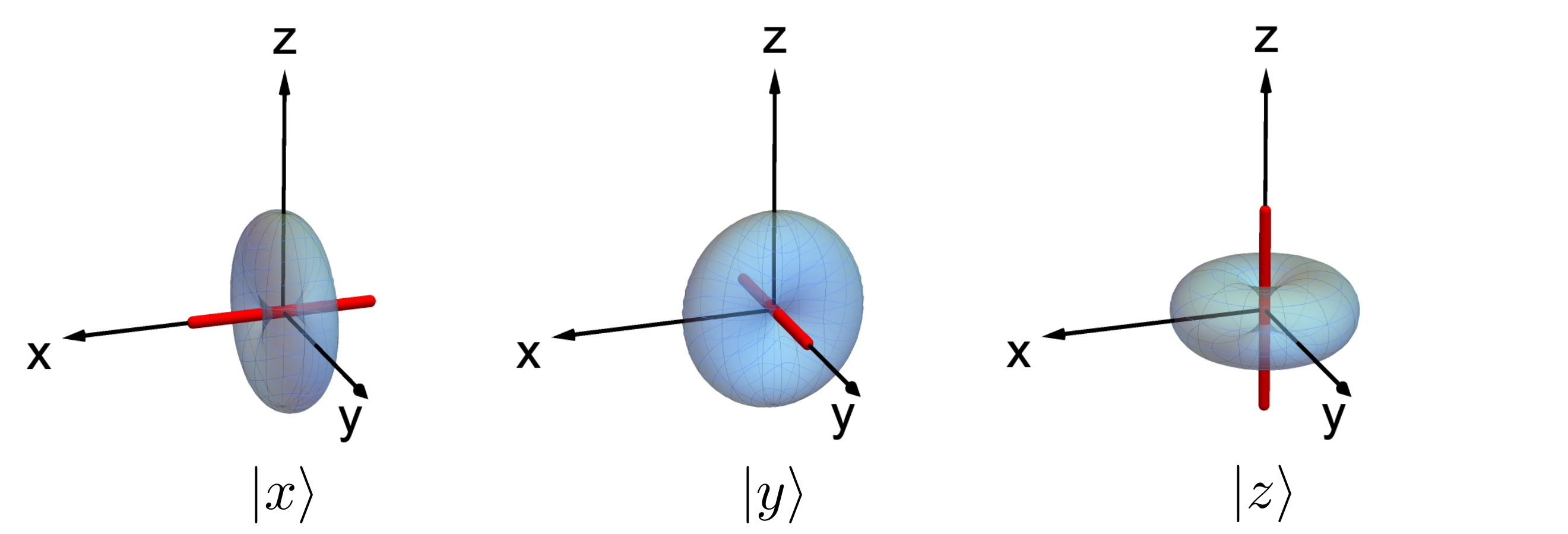}
	\caption{
	Time--reversal invariant basis for a \mbox{spin--1} moment.		
	The three states $| \alpha \rangle$, with $\alpha = x,y,z$, are invariant
	under time--reversal, and satisfy $\bra{\alpha}S^\mu \ket{\alpha} = 0$, 
	for $\mu= x,y,z$ referring to the usual spacial spin components .
	These states can be expressed in terms of the usual magnetic basis 
	[\protect\Autoref{fig:magnetic.basis}] through \protect\Autoref{eq:TRstates}, 
	and exhibit a characteristic ``doughnut--shaped'' profile of spin fluctuations 
	[\protect\hyperref[app:Spin_Fluct_Prob]{Appendix~}\ref{app:Spin_Fluct_Prob}].
	}
	\label{fig:TR.basis}
\end{figure}

The ``magnetic'' basis, \Autoref{eq:B1}, is the most commonly used description 
of a \mbox{spin--1} moment.
However this choice of basis is not unique, and any linear combination 
of \Autoref{eq:B1} which forms 3 orthogonal states can serve equally well.
For many purposes it is more convenient to describe \mbox{spin--1} moments 
in a basis of time--reversal invariant states, satisfying 
\be
{\mathcal T} \ket{\phi} = \ket{\phi} \; .
\ee
%
A suitable time--reversal invariant basis is given by 
 \be
 \mathcal{B}_2 = \left\lbrace \ket{x},\ket{y},\ket{{z}} \right\rbrace 
 \label{eq:TRB}
 \; ,
 \ee
 where 
 \be
 \begin{tabular}{ccc}
 	$\ket{x}=\frac{i}{\sqrt{2}}(\ket{1}-\ket{\overline{1}}) $,~ & $\ket{y}=\frac{1}{\sqrt{2}}(\ket{1}+\ket{\overline{1}})$,~ & $\ket{z}=-i\ket{0}$\\
 \end{tabular}\; .
 \label{eq:TRstates}
 \ee
This basis is illustrated in  \Autoref{fig:TR.basis}.
Within this basis, any state of a \mbox{spin--1} can be decomposed 
in terms of complex coefficients $d_{\alpha}^{\ast}$ 
 \be
 \ket{\V{d}} 
 	= \sum_{\alpha=x,y,z}d_{\alpha}^{\ast}\ket{\alpha},~~~d_{\alpha}^{\ast} \in \mathbb{C}
\label{eq:Defd} \; ,
 \ee
which we collect in a complex vector (director) $\V{d}$, of unit length
\be
\V{d}^{\ast} \cdot \V{d} = 1 \; .
\ee
We can further separate this 
into a real and imaginary parts
 \be
 \V{d}^{\ast}=\V{u}+i\V{v} \label{eq:Duv} \; ,
 \ee
providing a represent of a \mbox{spin--1} in terms of two real, three--dimensional vectors, 
subject to the constraint 
\be
\V{u} \cdot \V{u} + \V{v} \cdot \V{v} = 1 \; .
\ee
 

All of the operators needed to characterise a \mbox{spin--1} moment can also be written in 
terms of matrix elements of the time--reversal invariant basis [\Autoref{eq:TRstates}], 
with spin operators given by the antisymmetric contraction 
 \begin{align}
 \ops{}{\alpha}&=-i\epsilon^{\alpha\beta\gamma}\ket{\beta}\bra{\gamma}
 \label{eq:STRB} \; ,
 \end{align}
while quadrupoles by the symmetric contraction
 \begin{align}
 \opq{}{\alpha\beta}&=-\ket{\alpha}\bra{\beta}-\ket{\beta}\bra{\alpha}+\frac{2}{3}\delta^{\alpha\beta}\ket{\gamma}\bra{\gamma}
 \label{eq:QTRB} \; .
 \end{align}
We can use these results to express the expected dipole--moment
 \begin{align}
 \bra{\V{d}}\ops{}{\alpha}\ket{\V{d}}&=2\epsilon^{\alpha\beta\gamma}\rm{u}^{\beta}\rm{v}^{\gamma} \label{eq:S_uv} \; , 
\end{align}
and quadrupole--moment
 \begin{align}
 \bra{\V{d}}\opq{}{\alpha\beta}\ket{\V{d}}&=-2(\rm{u}^{\alpha}\rm{u}^{\beta}+\rm{v}^{\alpha}\rm{v}^{\beta})+\frac{2}{3}\delta^{\alpha\beta}(\rm{u}^{\gamma}\rm{u}^{\gamma}+\rm{v}^{\gamma}\rm{v}^{\gamma})\label{eq:Q_uv} \; ,
\end{align}
moments of a general state $ \ket{\V{d}}$ [\Autoref{eq:Defd}], 
in terms of  $\V{u}$ and $\V{v}$ [\Autoref{eq:Duv}].
 In vector form, the equation for the dipole moment [\Autoref{eq:S_uv}] 
 then becomes
 \be
 \langle \V{S} \rangle = 2\V{u}\times \V{v}\label{eq:Suv}\; ,
 \ee
 and we see that if the director $\V{d}$ is either purely real, or purely imaginary, 
 the associated dipole moments will be zero.
 
 \subsubsection{Description in terms of $u(3)$}

 The form of the expressions for spin-- [\Autoref{eq:STRB}] and quadrupole--moments [\Autoref{eq:QTRB}], motivates us to introduce an object 
 with matrix elements 
  \be
 \opa{}{\alpha}{\gamma} 
 = \ket{\alpha}\bra{\gamma} 
    \label{eq:defAop} \; ,
 \ee
 i.e. 
 \be
 \ophb{A} =
 \begin{pmatrix}
 	\ket{x}\bra{x} & \ket{x}\bra{y} & \ket{x}\bra{z}\\
 	\ket{y}\bra{x} & \ket{y}\bra{y} & \ket{y}\bra{z}\\
 	\ket{z}\bra{x} & \ket{z}\bra{y} & \ket{z}\bra{z}
 \end{pmatrix} \; , 
 \label{eq:def_A}
 \ee
colloquially referred to as the ``{\it A--matrix}'' \cite{Papanicolaou1988} . 
(More precisely formulated, $\ophb{A}$ is a tensor, as described in \Autoref{sec:u3_prop}).
The matrix $\ophb{A}$ acts on the basis of time--reversal invariant states, and is subject to the constraint
 \be
	 \text{Tr}\ \ophb{A} = 1 \; ,
	 \label{eq:traceA.on.A.matrix}
 \ee
following from the normalisation of the spin state, \Autoref{eq:normalisation}. 
 It is now straightforward to transcribe both spin-- 
 \begin{align}
 	\ops{}{\alpha}& = -i\epsilon^{\alpha~\gamma}_{~\beta}\opa{}{\beta}{\gamma} \; ,
 \label{eq:dipole.in.terms.of.A} 
 \end{align}
 and quadrupole--moments 
  \begin{align}
 	\opq{}{\alpha\beta}& 
	=-\opa{}{\alpha}{\beta}-\opa{}{\beta}{\alpha}
	+\frac{2}{3}\delta^{\alpha\beta}\opa{}{\gamma}{\gamma}
	\label{eq:quadrupole.in.terms.of.A} \; ,
 \end{align}
 in terms of matrix elements of $\ophb{A}$ [\Autoref{eq:defAop}].

 A convenient basis for $\ophb{A}$ is provided by a set 
 of matrices with a single non--zero element \cite{Papanicolaou1988},
 \begin{eqnarray}
 \opa{}{1}{1} &=&
 \begin{pmatrix}
 	1 & 0 & 0 \\
 	0 & 0 & 0 \\
 	0 & 0 & 0
 \end{pmatrix} \; , \; 
  \opa{}{1}{2}  =
 \begin{pmatrix}
 	0 & 1 & 0 \\
 	0 & 0 & 0 \\
 	0 & 0 & 0
 \end{pmatrix} \; , \; 
  \opa{}{1}{3}  =
 \begin{pmatrix}
 	0 & 0 & 1 \\
 	0 & 0 & 0 \\
 	0 & 0 & 0
 \end{pmatrix} \; , \; \nonumber\\
\opa{}{2}{1} &=&
 \begin{pmatrix}
 	0 & 0 & 0 \\
 	1 & 0 & 0 \\
 	0 & 0 & 0
 \end{pmatrix} \; , \; 
 \opa{}{2}{2}  =
 \begin{pmatrix}
 	0 & 0 & 0 \\
 	0 & 1 & 0 \\
 	0 & 0 & 0
 \end{pmatrix} \; , \; 
 \opa{}{2}{3}  =
 \begin{pmatrix}
 	0 & 0 & 0 \\
 	0 & 0 & 1 \\
 	0 & 0 & 0
 \end{pmatrix} \; , \; \nonumber\\
\opa{}{3}{1}   &=&
 \begin{pmatrix}
 	0 & 0 & 0 \\
 	0 & 0 & 0 \\
 	1 & 0 & 0
 \end{pmatrix} \; , \; 
 \opa{}{3}{2}   =
 \begin{pmatrix}
 	0 & 0 & 0 \\
 	0 & 0 & 0 \\
 	0 & 1 & 0
 \end{pmatrix} \; , \; 
  \opa{}{3}{3}   =
 \begin{pmatrix}
 	0 & 0 & 0 \\
 	0 & 0 & 0 \\
 	0 & 0 & 1
 \end{pmatrix} \; , \;  \nonumber\\
 \label{eq:u3.A.matrix.rep}
 \end{eqnarray}
 These matrices satisfy the closed algebra $u(3)$, with 
commutation relations
 \be
\begin{tabular}{rcl}
$\com{\opa{i}{\alpha}{\beta},\opa{i}{\gamma}{\eta}}
$&$=$&${\delta^{\gamma}}_{\beta}\opa{i}{\alpha}{\eta}-{\delta^{\alpha}}_{\eta}\opa{i}{\gamma}{\beta}$\; ,\\
& &\\
$\com{\opa{i}{\alpha}{\beta},\opa{j}{\gamma}{\eta}}$&=&0 \; ,
\end{tabular} \label{eq:u3.commutation.relations}
\ee
previously introduced in \Autoref{eq:ComRelAop}.


Alternative representations of $U(3)$ are possible, and 
have their own merits \cite{LieGroups}.
The specific advantage of the basis given in Eq.~(\ref{eq:u3.A.matrix.rep}) 
is its simplicity.
And, in conjunction with complex coefficients, this basis   
can be used to describe all possible states of a \mbox{spin--1} moment.
Once again, after constraints coming from the Hermitian nature of $\opa{}{\alpha}{\beta}$ [Eq.~(\ref{eq:def_A})], its trace [Eq.~(\ref{eq:traceA.on.A.matrix})], and the fact 
that it is proportional to a projection operator have been taken into account, this 
requires a total of four real coefficients \cite{Remund-unpub}.

 \subsubsection{Relationship between $U(3)$ and $SU(3)$ representations}
 \label{sec:u3}
 
The representation of a \mbox{spin--1} moment in terms of the nine 
matrices $\opa{}{\alpha}{\beta}$ [\Autoref{eq:u3.A.matrix.rep}], contains one 
additional operator, relative to the eight generators of 
$SU(3)$ [\Autoref{eq:gellmann.matrices}].
The resolution of this seeming paradox rests in enforcing the constraint 
that each site is occupied by a single spin--1 moment.


Relative to $U(3)$, the operator ``missing'' from $SU(3)$ is the spin--length $\opsb{}^2$.
Once this is included, there exists a specific transformation 
relating the representation of $SU(3)$ in terms of dipole [\Autoref{eq:dipole.moments}] 
and quadrupole [\Autoref{eq:Def_Q_vec}] moments, and the representation 
of $U(3)$ in terms of the nine generators $\ophb{A}^\alpha_\beta$ [\Autoref{eq:u3.A.matrix.rep}]
 \be
  \begin{pmatrix}
  	\opsb{}^2\\
  	\ops{}{x}\\
  	\ops{}{y}\\
  	\ops{}{z}\\ 
  	\opq{}{x^2-y^2}\\ 
  	\opq{}{3r^2-s^2}\\
  	\opq{}{xy}\\
  	\opq{}{xz}\\
  	\opq{}{yz}\\
  \end{pmatrix}=C
  \begin{pmatrix}
  	\opa{}{1}{1}\\
  	\opa{}{1}{2}\\
  	\opa{}{1}{3}\\
  	\opa{}{2}{1}\\
  	\opa{}{2}{2}\\
  	\opa{}{2}{3}\\
  	\opa{}{3}{1}\\
  	\opa{}{3}{2}\\
  	\opa{}{3}{3}\\
  \end{pmatrix}
  \label{eq:su3_u3}\; ,
\ee
where C is the $\rm{9\times 9}$ matrix.
\be
  C=
  \begin{pmatrix}
  	2 & 0 & 0 & 0 & 2 & 0 & 0 & 0 & 2  \\
  	0 & 0 & 0 & 0 & 0 & -i & 0 & i & 0 \\
  	0 & 0 & i & 0 & 0 & 0 & -i & 0 & 0 \\
  	0 & -i & 0 & i & 0 & 0 & 0 & 0 & 0 \\
    -1 & 0 & 0 & 0 & 1 & 0 & 0 & 0 & 0 \\
  	\frac{1}{\sqrt{3}} & 0 & 0 & 0 & \frac{1}{\sqrt{3}} & 0 & 0 & 0 & -\frac{2}{\sqrt{3}} \\
  	0 & -1 & 0 & -1 & 0 & 0 & 0 & 0 & 0 \\
  	0 & 0 & -1 & 0 & 0 & 0 & -1 & 0 & 0 \\
  	0 & 0 & 0 & 0 & 0 & -1 & 0 & -1 & 0 \\
  \end{pmatrix}\label{eq:C}\; ,
\ee
previously shown schematically as \Autoref{eq:u3}.

 
We can fix the spin sector, and thereby restrict fluctuations to the 
smaller group $SU(3)$, by imposing the constraint 
\be
\begin{matrix}
 	\text{Tr} \mathcal{A} 
 	= \sum_{\alpha}\opa{i}{\alpha}{\alpha} 
	= \sum_{\alpha}\frac{1}{2}\ops{i}{\alpha}\ops{i}{\alpha}
	= \frac{1}{2}s(s+1) 
	= 1 
 \end{matrix}
\label{eq:traceA} \; ,
\ee
where we have used the property 
\be
	\sum_{\alpha} \opq{i}{\alpha\alpha} = 0 \; .
\ee
It follows that, for purposes of simulation of \mbox{spin--1} moments, 
we can work directly with the matrices $\ophb{A}$, as long as 
these satisfy the constraint \Autoref{eq:traceA.on.A.matrix}. 
This constraint was previously introduced as \Autoref{eq:fix.spin.length}.

 \subsubsection{Mathmatical properties of A--matrices}
\label{sec:u3_prop}

While it is convenient to refer to the operators $\opa{}{\alpha}{\beta}$
as matrices, they are in fact tensors.
The tensor nature of these objects is explored in 
\hyperlink{sec:appA}{Appendix~}\ref{sec:appA}.  
Here we single out a property which will prove useful in subsequent 
derivations, namely the way in which  $\opa{}{\alpha}{\beta}$ transforms 
under a linear map.
  

The operator $\opa{}{\alpha}{\beta}$ is defined through matrix elements 
of the time--reversal invariant basis [\Autoref{eq:TRB}].  
In defining $ \opa{}{\alpha}{\gamma}$ [\Autoref{eq:defAop}], we introduced both a 
contravariant index ${\alpha}$, and a covariant index ${\gamma}$.
This distinction follows from the fact that the index ${\alpha}$ relates 
to a bra vector, while the index $\gamma$ related to a ket vector. 
Bra--vectors and ket--vectors (such as the states in the basis $\mathcal{B}_2$ [\Autoref{eq:TRB}]), 
inhabit mutually--dual vector spaces.
And for this reason, contravariant and covariant indexes will transform differently 
under a linear transformation of basis vectors.   
%


Let us consider a general linear transformation	
\be
\Lambda:V\rightarrow V \; , 
\ee
with
\be
\textrm{det}\ \Lambda \neq 0  \; ,
\ee
such that $\Lambda$ is invertible, and define 
\be
\tilde{\Lambda} = \left(\Lambda^{-1}\right)^T \; .
\ee
Under this transformation, the components of $\opa{}{\alpha}{\beta}$ will transform as
\be
(\opa{}{\alpha}{\beta})^{\mu}_{~\nu}
 = \Lambda^{\mu}_{~\gamma}\tilde{\Lambda}_{\nu}^{~\kappa}(\opa{}{\alpha}{\beta})^{\gamma}_{~\kappa}
 =\Lambda^{\mu}_{~\gamma}(\Lambda^{-1})_{~\nu}^{\kappa}(\opa{}{\alpha}{\beta})^{\gamma}_{~\kappa}  \; ,
 \label{eq:LTA}
\ee
where we once again assume the Einstein convention of summing on repeated indices.


This result can be interpreted as follows:  $\opa{}{\alpha}{\beta}$ is properly 
considered to be a $(1,1)$--tensor, implying that the linear map takes one element 
in the vector space $V$, and a second one in the dual vector space $V^{\ast}$, 
and assigns then a number in the field $F$, for which the multiplication of the 
vector space is defined.
And, crucially, the only non-zero component of 
$(\opa{}{\alpha}{\beta})^{\gamma}_{~\kappa}$ in the time--reversal invariant 
basis is 
\be
(\opa{}{\alpha}{\beta})^{\alpha}_{~\beta} \; .
\ee
This fact, and the mapping \Autoref{eq:LTA} will prove important where we use the 
generators $\opa{}{\alpha}{\beta}$ to derive a theory of small fluctuations 
about an ordered state, in \Autoref{section:small.fluctuations}.


We will briefly comment on two other mathematical properties of 
the operators $\ophb{A}$ which will prove useful in later calculations.
Firstly, it is possible to construct any state $\ket{\alpha = x,y,z}$ in the 
basis $\mathcal{B}_2$ [\Autoref{eq:TRB}] as
 \be
 \ket{\alpha} = \opdD{}{\alpha} \ket{\rm{vac}} \; ,
 \ee
 where $\op{}{\alpha}$ satisfies the Bosonic commutation relation
  \be
[ \op{}{\alpha} , \opdD{}{\alpha}] = \delta_{\alpha\beta} \; ,
 \ee
 and $\ket{\rm{vac}}$ is the vacuum.
 It follows that we can build the matrix $\ophb{A}$ as the exterior product 
 of the operators $\op{}{\alpha}$, 
  \be
 \opa{}{\alpha}{\beta} = \opdD{}{\alpha}\op{}{\beta} \; .
 \label{eq:A-matrix_Director_op}
 \ee
This particular representation will prove useful when is comes to construction a 
quantum theory of excitations in \Autoref{section:quantum.theory}.
And an interesting corollary of Eq.~(\ref{eq:A-matrix_Director_op}) is that 
the overall phase of the operator $\op{}{\alpha}$ plays no part in 
determining $\opa{}{\alpha}{\beta}$.


Secondly, it is helpful to note that 
\be
\sum_{\alpha,\beta}\opa{i}{\alpha}{\beta}\opa{j}{\beta}{\alpha}= \sum_{\alpha,\beta}\frac{1}{4}\opq{i}{\alpha\beta}\opq{j}{\beta\alpha}+\sum_{\alpha}\frac{1}{2}\ops{i}{\alpha}\ops{j}{\alpha}+\frac{1}{12}s^2(s+1)^2
\label{eq:sum.rule.AQS} \; .
\ee
This result will prove useful when considering the sum rules on structure 
factors in \Autoref{sec:classical.structure.factors} and \Autoref{sec:quantum.structure.factors}.

\subsection{Representation of the BBQ model within a $u(3)$ formalism}
\label{sec:u3_HBBQ}
  
\nic{
The most general form of nearest--neighbour Hamiltonian permitted 
by $SU(2)$ symmetry for a \mbox{spin--1} magnet is the 
bilinear--biquadratic (BBQ) model 
\be
	\Ham_{\sf BBQ} = \sum_{\nn{i,j}} 
		\left[ J_1\opsb{i} \cdot \opsb{j} + J_2( \opsb{i} \cdot \opsb{j} )^2 \right] \; ,
	\label{eq:H.BBQ.1}
\ee
previously introduced in \Autoref{eq:BBQ.model}.
This  model has been studied extensively, in the context of spin--1 magnets
\cite{Matveev1973,Andreev1984,Papanicolaou1988,Harada2002,Tsunetsugu2006,Lauchli2006,Kaul2012,Smerald2013,Voell2015,Penc2011-Springer}, 
systems of cold atoms \cite{Demler2002,Yip2003,Imambekov2003,Gorshkov2010,Rodriguez2011,DeChiara2011,Bauer2012,Kurn2013}, 
and as a toy model for nematic order in Fe--based superconductors \cite{Fernandes2014,Luo2016,Wang2016,Gong2017,Lai2017} 
and spin--1/2 magnets \cite{Smerald2015}.
}



\nic{
The physical nature of the interactions in the BBQ model is most obvious 
once it is recast in terms of generators of $SU(3)$, via \Autoref{eq:Def_Q_vec},  
%
\be
	\Ham_{\sf BBQ} = \sum_{\nn{i,j}} \left(J_1-\frac{J_2}{2}\right) \opsb{i} \cdot \opsb{j}
	+ \frac{J_2}{2} \left. \bf{\hat{Q}_{i}} \cdot \bf{\hat{Q}_{j}} \right. \\
	 + \frac{4 J_2}{3} 
	\label{eq:H.BBQ.2} \; ,
\ee
a form previously introduced in \Autoref{eq:BBQ.modelQ}.
Here biquadratic interactions are revealed as an interaction between on--site quadrupoles, 
which are explicitly forbidden for spin--1/2 moments.
}


\nic{
Biquadratic interactions can have a number of different microscopic origins.
In insulating magnets, high--spin moments involve electrons with more than 
one orbital, and 
biquadratic interactions follow from the exchange of electrons 
in different orbitals, on different sites \cite{Fazekas1999-WorldScientific}.
Similarly, in systems of cold atoms, biquadratic interactions follow 
from the structure of the underlying Mott physics, which may be 
Bosonic \cite{Jaksch1998,Demler2002,Yip2003,Imambekov2003,Rodriguez2011,DeChiara2011,Kurn2013,deForgesdeParny2014,Zibold2016}
or Fermionic \cite{Honerkamp2004,Gorelik2009,Gorshkov2010,Bauer2012} 
in character.    
More generally, biquadratic interactions can also arise as an effective interaction 
coming from spin--lattice coupling \cite{Kittel1960}, or as a consequence of integrating 
out quantum or thermal fluctuations \cite{Chandra1990}. 
}


\nic{
The $SU(2)$--invariance of $\Ham_{\sf BBQ}$ can be read directly from 
\Autoref{eq:H.BBQ.1} or \Autoref{eq:H.BBQ.2}.
The scalar contractions $\opsb{i} \cdot \opsb{j}$ and $\bf{\hat{Q}_{i}} \cdot \bf{\hat{Q}_{j}}$ 
are both unchanged by rotations belong to the group $O(3)$, 
which provides 2--fold cover for the group $SU(2)$.
However $SU(2)$ is not the highest symmetry which can be achieved, and for the specific 
choice of parameters $J_1 = J_2 = J$, the symmetry of the model is enlarged to $SU(3)$ \cite{Papanicolaou1988,Penc2011-Springer,Smerald2013}.
In this case, the BBQ model can be rewritten 
\be
	\Ham_{\sf BBQ} = \frac{J}{2} \sum_{\nn{i,j}} {\bf T}_i \cdot {\bf T}_j 
		+ \frac{4 J}{3}  \; ,
	\label{eq:H.BBQ.SU3}
\ee
where ${\bf T}_i$ is the eight--dimensional vector  
\be
{\bf T}_i = ( 
	\ops{i}{x}, 
	\ops{i}{y}, 
	\ops{i}{z} ,
	 \opq{i}{x^2-y^2}, 
 	\opq{i}{3r^2-s^2},
 	\opq{i}{xy},
 	\opq{i}{xz},
 	\opq{i}{yz}) \; ,
\ee
and it is possible to rotate dipole moments into quadrupoles 
(or vice versa) without any cost in energy \cite{Smerald2013}. 
Consistent with this, the high--symmetry $SU(3) $ points define the zero--temperature 
boundaries between phases with dipolar and quadrupolar character, as illustrated in \Autoref{fig:MF.phase.diagram}.
}


\nic{
It is also possible to transcribe $\Ham_{\sf BBQ}$ in terms of generators of $U(3)$.
Starting from \Autoref{eq:H.BBQ.2}, and using \Autoref{eq:dipole.in.terms.of.A} and \Autoref{eq:quadrupole.in.terms.of.A} 
--- or, equivalently, \Autoref{eq:su3_u3} --- we find} 
\kim{
\begin{align*}
\Ham_{\sf BBQ} = \sum_{\nn{i,j}}&\left[J_1\opa{i}{\alpha}{\beta}\opa{j}{\beta}{\alpha}+(J_2-J_1)\opa{i}{\alpha}{\beta}\opa{j}{\alpha}{\beta}\right.\\
&\left.+J_2\opa{i}{\alpha}{\alpha}\opa{j}{\beta}{\beta}\right] \numberthis\label{eq:eq:H.BBQ.3.0}
\end{align*}}
\nic{Imposing the constraint on the trace of the A--matrix [\Autoref{eq:traceA}], this simplifies to}
\be
	\Ham_{\sf BBQ} = \sum_{\nn{i,j}}
	\left[ 
	      J_1 \opa{i}{\alpha}{\beta} \opa{j}{\beta}{\alpha}
	     + (J_2 - J_1) \opa{i}{\alpha}{\beta} \opa{j}{\alpha}{\beta} 
	     + J_2 
	\right] \; ,
	\label{eq:H.BBQ.3}
\ee
where sums on repeated indices are assumed.
%
%
This result was previously introduced in \Autoref{eq:HamA1}.   
   

\nic{
The $U(3)$ formulation of the BBQ model, \Autoref{eq:H.BBQ.3}, 
contains terms which transform in two different ways under spin rotations.}
Using results of \Autoref{sec:u3_prop}, we can show that the second term, 
\be
\opa{i}{\alpha}{\beta} \opa{j}{\alpha}{\beta}   \nonumber 
\ee
is invariant under $O(3) \simeq SU(2)$ rotations.
Meanwhile the first term 
\be
\opa{i}{\alpha}{\beta} \opa{j}{\beta}{\alpha}  \nonumber 
\ee
has indices $\alpha$ and $\beta$ which transform contravariantly on one site,  
and covariantly on the other, and therefore possesses $U(3)$ symmetry. 
This is in turn broken down to $SU(3)$ by the constraint, \Autoref{eq:traceA}.
Thus, for general parameters, \Autoref{eq:H.BBQ.3} possesses $SU(2)$ symmetry, 
but for $J_1 = J_2$,  the second term vanishes, and the symmetry is enlarged to $SU(3)$.
Further details of this analysis can be found in \hyperref[sec:appA]{Appendix~}\ref{sec:prop_A}.


\nic{Crucially, once written in terms of generators of $U(3)$,  \Autoref{eq:H.BBQ.3}, 
the BBQ model takes on a form quadratic in $\opa{i}{\alpha}{\beta}$, which treats 
dipole and quadrupole moments on an equal footing.}
This quadratic form is well--suited to the development of  analytic, mean--field 
approaches, since it facilitates a straightforward decoupling of interactions \cite{Papanicolaou1988}.  
And in \Autoref{sec:MC} we show how it can also be used to develop 
classical Monte Carlo simulations of the thermodynamic properties 
of \mbox{spin--1} magnets, which respect the fact that the (semi--)classical limit of a \mbox{spin--1} moment is not an $O(3)$ vector.

\subsection{Heisenberg equations of motion within a $u(3)$ formalism}
\label{sec:u3.EoM}

The quadratic form of \Autoref{eq:H.BBQ.3} also makes it well--suited for the derivation 
of a Heisenberg EoM for a \mbox{spin--1} magnets, 
in analogy with the well--known result for $O(3)$--vectors, \Autoref{eq:Heisenberg.EoM.O3}.
By explicit calculation of commutators, using \Autoref{eq:u3.commutation.relations}, 
and setting $\hbar =1$, we find
\begin{eqnarray}
	\ddp_t \opa{i}{\gamma}{\eta} 
	&=& -i\com{\opa{{i}}{\gamma}{\eta},\Ham_{\sf BBQ}} \nonumber \\
	&=& -i\sum_{\delta}\left[\right.J_1(\opa{{i}}{\gamma}{\alpha}\opa{{i+\delta}}{\alpha}{\eta}-\opa{{i}}{\alpha}{\eta}\opa{{i+\delta}}{\gamma}{\alpha}) \nonumber \\
	 && +(J_2-J_1)(\opa{{i}}{\gamma}{\alpha}\opa{{i+\delta}}{\eta}{\alpha}-\opa{{i}}{\alpha}{\eta}\opa{{i+\delta}}{\alpha}{\gamma})\left.\right] \; ,
	\label{eq:EoM.u3}
\end{eqnarray}
a result previously introduced in \Autoref{eq:DA2}.


Like the Hamiltonian it descends from, the EoM, 
\Autoref{eq:EoM.u3}, treats dipole and quadrupole moments on a equal footing, 
and is ideally--suited to numerical integration, a subject we return 
to in \Autoref{sec:u3MD}.
But since this EoM is written in terms of a representation of the algebra $u(3)$, 
it also describes the dynamics of the operator for the total spin $\opsb{}^2$.
And to correctly describe the dynamics of a \mbox{spin--1} magnet, we 
require that 
\begin{eqnarray}
s = 1 \; \Rightarrow \; {\rm Tr}\ \hat{\mathcal{A}} = 1
\end{eqnarray}
throughout [cf. \Autoref{eq:traceA}].  
Happily, the EoM for \mbox{$\mathcal{A}$--matrices} conserves 
the trace of $\mathcal{A}$, 
a fact which follows straightforwardly from \Autoref{eq:EoM.u3} 
\begin{eqnarray}
	\ddp_t \left( \rm{Tr}\ \opa{i}{}{} \right)
	&=& -i\ \rm{Tr} \sum_{\delta} \left[ \right.J_1(\opa{{i}}{\gamma}{\alpha}\opa{{i+\delta}}{\alpha}{\eta}-\opa{{i}}{\alpha}{\eta}\opa{{i+\delta}}{\gamma}{\alpha}) \nonumber \\
	 && +(J_2-J_1)(\opa{{i}}{\gamma}{\alpha}\opa{{i+\delta}}{\eta}{\alpha}-\opa{{i}}{\alpha}{\eta}\opa{{i+\delta}}{\alpha}{\gamma}) \left.\right]  \nonumber \\
	 &\equiv& 0 \; ,
	\label{eq:conservation.of.trace}
\end{eqnarray}


The implication is that, as long as the EoM \Autoref{eq:EoM.u3} is applied to a valid 
$\mathcal{A}$--matrix configuration, with ${\rm Tr}\ \mathcal{A}_i \equiv 1$, 
the time--evolution of the operators $\opa{i}{\alpha}{\beta}$ will respect 
the constraint on spin length.
As we shall see in \Autoref{section:anisotropy}, this remains true for systems with  
interactions which are anisotropic in spin--space, making these EoM a powerful 
tool for the exploration of the dynamics of \mbox{spin--1} magnets.   

\section{Numerical simulation of \mbox{spin--1} magnets} 			
\label{section:numerical.method}

In \Autoref{section:maths.for.spin.1}, we introduced the technical framework 
needed to describe a spin--1 magnet in terms of a suitable representation of 
$u(3)$, $\opaA{}{\alpha}{\beta}$ [\Autoref{eq:def_A}].
This allowed us to write both the BBQ model, $\Ham_{\sf BBQ}$ [\Autoref{eq:H.BBQ.3}], 
and its associated equation of motion [\Autoref{eq:EoM.u3}], in a simple form, bilinear in 
$\opaA{}{\alpha}{\beta}$, without making any approximation as to its physical content.


	In what follows we develop these 
	results into a practical scheme for the numerical simulation of \mbox{spin--1} magnets,  
	providing technical details the updates needed for both classical Monte Carlo (MC) 
	and (semi-)classical Molecular Dynamics (MD) simulations, carried out in the space 
	of the  ``A-matrices'', $\opaA{}{\alpha}{\beta}$.
	We will refer to these approaches as ``u3MC'' and ``u3MD'', respectively.


	We demonstrate the validity of this approach by reproducing known results 
	for the thermodynamics of the \mbox{spin--1} BBQ model on the triangular lattice, 
	at the border of ferroquadrupolar (FQ) and antiferromagnetic (AFM) order 
	\cite{Stoudenmire2009}.
	We also obtain a complete finite--temperature phase diagram for this model, 
	previously exhibited in \Autoref{fig:MC.phase.diagram}.


	The detailed application of the method to the thermodynamics 
	and dynamics of the ferroquadrupolar (FQ) phase will be described 
	in \Autoref{section:numerics}.

\subsection{Monte Carlo simulations within $u(3)$ framework}  			 	
\label{sec:MC}

\subsubsection{Implementation of u3MC update}  			 	
\label{sec:u3.update}

The starting point for both MC and MD simulations of \mbox{spin--1} 
magnets, is a product wave function written in the space of $\mathcal{A}$--matrices,
\begin{eqnarray}
	\ket{ \Psi_\mathcal{A} }
	= \prod^N_{i=1} \ket{ \mathcal{A}_i }
	= \prod^N_{i=1} \nic{\sum_{\alpha\beta}} \hat{\mathcal{A}}^\alpha_{i, \beta} \ket{\beta}
	\equiv \prod^N_{i=1} \ket{ {\bf d_i} }
	\; ,
\label{eq:product.wf}
\end{eqnarray}
where $\mathcal{A}_i$ denotes the nine parameters $\mathcal{A}^\alpha_{i,\beta}$, 
$\ket{\beta}$ is the basis of time--reversal invariant states [\Autoref{eq:TRB}], 
and $ \ket{ {\bf d_i} }$ is defined through \Autoref{eq:Defd}.
From \Autoref{eq:H.BBQ.3}, this state has an associated  (classical) energy 
\begin{eqnarray}
	E [ \mathcal{A}_i ]  
	&=& \bra{  \Psi_\mathcal{A}  } \Ham_{\sf BBQ}  \ket{  \Psi_\mathcal{A}  } \nonumber\\
	&=& \sum_{\nn{i,j}} \sum_{\alpha\beta}
		\left[ J_1 \mathcal{A}^\alpha_{i,\beta} \mathcal{A}^\beta_{j,\alpha}
	     + (J_2 - J_1) \mathcal{A}^\alpha_{i,\beta} \mathcal{A}^\alpha_{j, \beta} 
	     + J_2 \right] \; .\nonumber\\
\label{eq:E.of.A}
\end{eqnarray}
By its nature, such a product wave function is unentangled, and cannot describe 
quantum effects extending beyond a single site.
However it remains a semi--classical approximation in the sense that the 
quantum mechanics of each \mbox{spin--1} moment is treated exactly at the 
level of a single site.


As can be seen from \Autoref{eq:product.wf}, the product wave function written 
in terms of $\mathcal{A}$--matrices is exactly equivalent to one be expressed 
in terms of ${\bf d}$--vectors.
It follows that MC simulations can equally well be carried out in the 
space of ${\bf d}$--vectors, with energy \cite{Penc2011-Springer,Smerald2013}
\begin{eqnarray}
	E [ {\bf d}_i ]  
	&=& \bra{  \Psi_{\bf d}  } \Ham_{\sf BBQ}  \ket{  \Psi_{\bf d}  } \nonumber\\
	&=& \sum_{\nn{i,j}} 
		\left[J_1 | {\bf d}_i \cdot \overline{\bf d}_j  |^2
	     + (J_2 - J_1)  | {\bf d}_i \cdot {\bf d}_j  |^2
	     + J_2 \right] \; .\nonumber\\
\label{eq:E.of.d}
\end{eqnarray}
This approach has been pursued elsewhere, under the name of 
``semiclassical $SU(3)$'' or ``s$SU(3)$'' simulation ~\cite{Stoudenmire2009}.
%
However for MD simulations, and many analytic calculations, 
$\mathcal{A}$--matrices offer a more convenient representation.
It is this line we pursue here.


\begin{figure}[t]
	\centering
	\includegraphics[width=0.45\textwidth]{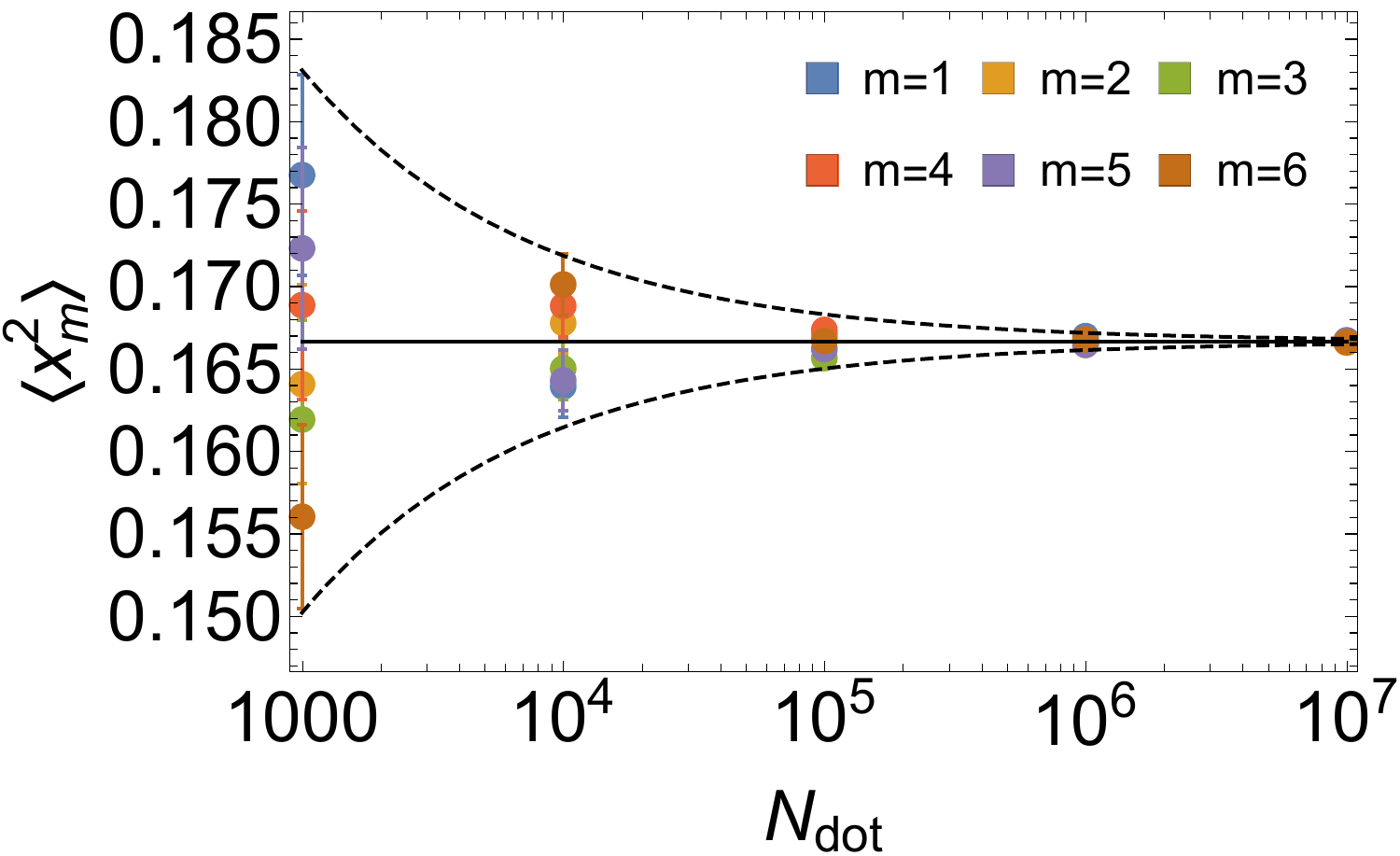}
	\caption{
			Statistical independence of points generated at random on a 5-dimensional sphere,  
			using \protect\Autoref{eq:sampling.d}.
			The second moment $\langle x_m^2 \rangle$, of variables $x_m$, $m = 1,...,6$ 
			is plotted as function of the number of points, $N_{\sf dot}$.
			In all cases $\langle x_m^2 \rangle \to 1/6$ (black line) as $N_{\sf dot} \to \infty$.
			%
			Statistical errors respect the central--limit theorem, and decrease 
			as $1/\sqrt{N_{\sf dot}}$ (dashed lines).
	}
	\label{fig:MC.update}
\end{figure}


The ingredient needed to convert \Autoref{eq:product.wf} 
and \Autoref{eq:E.of.A} into a practical MC scheme, is an update 
capable of generating a sequence of spin configurations 
$\{ \mathcal{A}_i \}$ corresponding to states drawn from a 
thermal ensemble.
We approach this by constructing a Metropolis--style  \cite{Metropolis1953}
update for a single \mbox{spin--1} moment, as represented by 
a matrix $\mathcal{A}_i$.  
More general cluster-- \cite{Landau2014-Cambridge} or worm-- 
\cite{Newman1999-ClarendonPress} updates could be built along 
similar lines, but will not be considered here.


\begin{figure*}[t]
	\centering
	\subfloat[Comparison between sSU(3) and u3MC approaches \label{fig:comp.Stoudenmire}]{
		\includegraphics[width=0.48\textwidth]{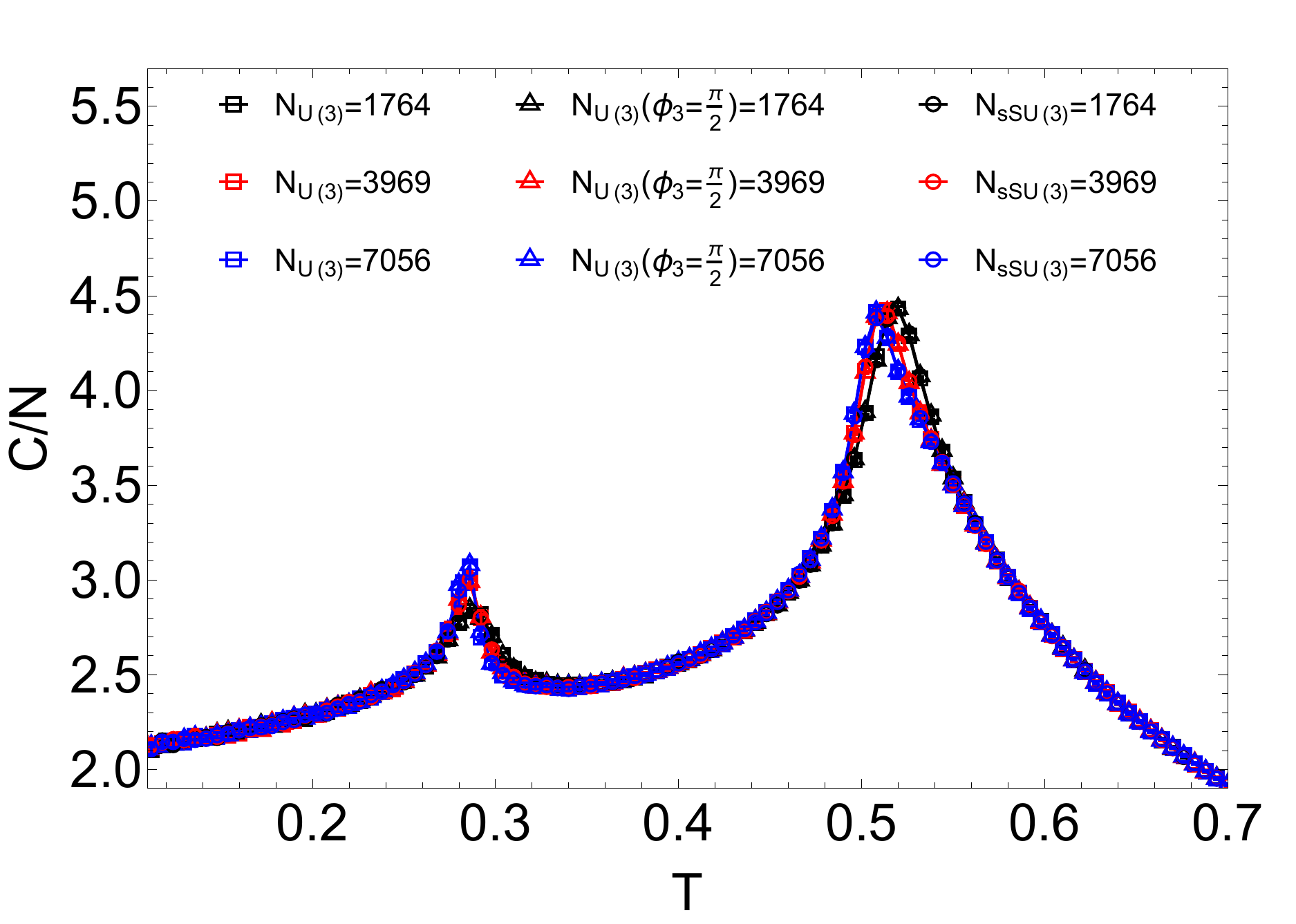}} 
	\subfloat[Detail of peak in $c(T)$ for $T \approx 0.285$ \label{fig:comp.Stoudenmire.detail}]{	
		\includegraphics[width=0.48\textwidth]{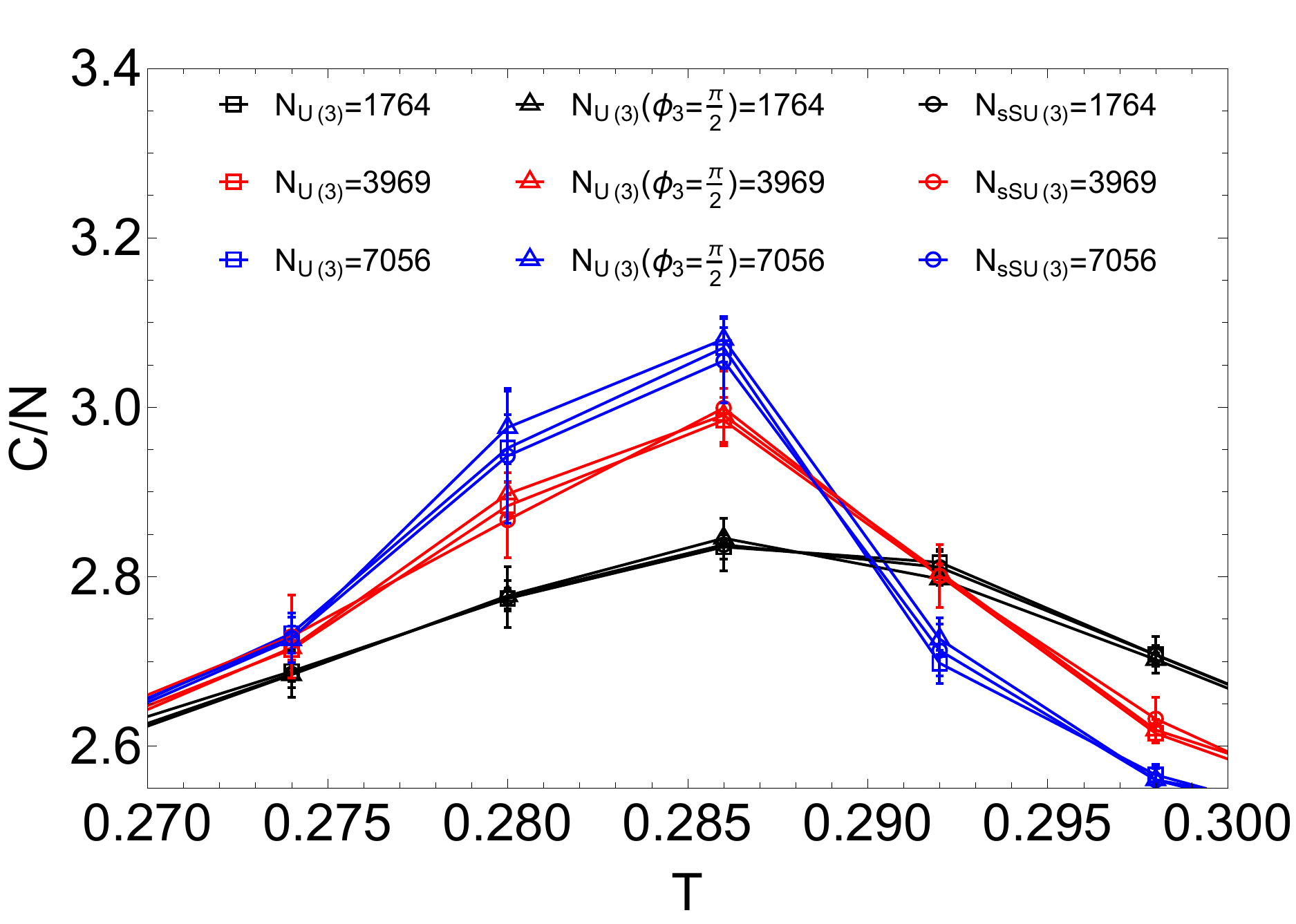}}\\
	\caption{
	\nic{
			Benchmark of the $U(3)$ Monte Carlo (u3MC) method against published results 
			for the \mbox{spin--1} bilinear---biquadratic (BBQ) model on the triangular lattice.
			(a) Specific heat C/N, for parameters $J_1 = 1$, $J_2 = -1.5$, showing 
			double--peak structure. 
			Results are shown for u3MC simulations using 
			both 5--dimensional updates [Eq.~\ref{eq:sampling.d}] 
			--- squares, and 4--dimensional updates with $\phi_3 = \pi/2$ 
			[Eq.~\ref{eq:alternative.sampling.d}] --- triangles, and for 
			``sSU(3)'' simulations in the space of the complex 
			vector ${\bf d}$, following Stoudenmire {\it et al.} 
			[\onlinecite{Stoudenmire2009}] --- circles.
			Simulations were carried out for clusters of size
			$N=1764$ spins (black symbols),
			$3969$ spins (red symbols),
			and $7056$ spins (blue symbols).  
			The results of the three different approaches 
			agree perfectly, within statistical errors.
			(b) Detail of peak in $C/N$ for $T \approx 0.285$.
			}
	}
	\label{fig:PD.Stoudenmire}
\end{figure*}


We start by revisiting the expression for an individual 
A--matrix in terms of the director ${\bf d}$ [\Autoref{eq:A-matrix_Director_op}]
\begin{equation}
		\opaA{}{\alpha}{\beta} = ({\rm d}^{\alpha})^* \  {\rm d}_{\beta} 	\, ,
		\label{eq:A-matrix_Director}
\end{equation}
where
\begin{equation}
		{\bf d} = \begin{pmatrix} 
			x_1 + i \ x_2		\\ 
			x_3 + i \ x_4		\\ 
			x_5 + i \ x_6		
		\end{pmatrix}  ; 		\hspace{1 cm}
		{\bf d}^* {\bf d}	= 
		|{\bf d}|^2 		= 1	\, .
		\label{eq:d-vector}   \\
\end{equation}
Written in this way, any matrix $\opaA{}{\alpha}{\beta}$ can be specified in terms of 
5 linearly--independent variables, coming from the six coefficients of ${\bf d}$, 
$x_1, x_2, \ldots x_6$,  and the constraint on its length.


Constructing a general update for a single \mbox{spin--1} moment 
therefore translates into sampling of statistically--independent, equally--distributed 
points on a 5--dimensional sphere, within a 6--dimensional space.
By direct analogy with the Marsaglia construction \cite{Marsaglia1972}, 
we write
\begin{subequations}
		\begin{align}
			x_1 &= \theta_2^{1/4} \ \theta_1^{1/2} \ \sin{\phi_1}		\ , \\ 
			x_2 &= \theta_2^{1/4} \ \theta_1^{1/2} \ \cos{\phi_1}		\ , \\ 
			x_3 &= \theta_2^{1/4} \ \sqrt{1-\theta_1} \ \sin{\phi_2}		\ , \\ 
			x_4 &= \theta_2^{1/4} \ \sqrt{1-\theta_1} \ \cos{\phi_2}	\ , \\ 
			x_5 &= \sqrt{1-\theta_2^{1/2}} \ \sin{\phi_3}			\ , \\ 
			x_6 &= \sqrt{1-\theta_2^{1/2}} \ \cos{\phi_3}			\ ,
		\end{align}
		\label{eq:sampling.d}
\end{subequations}
where $0 \leq \theta_1, \theta_2 \leq 1$ and $0 \leq \phi_1, \phi_2, \phi_3 < 2\pi$ 
are parameters chosen at random from a uniform distribution.
By construction, $| {\bf d} |^2 = 1$, and it follows from \Autoref{eq:A-matrix_Director} 
that $\text{Tr}\ \mathcal A = 1$. 
This ensures that all states generated remain within the Hilbert space for 
a \mbox{spin--1} moment [cf. \Autoref{eq:traceA.on.A.matrix}].


Evidence for the statistical validity of this generalised Marsaglia 
approach is shown in  \Autoref{fig:MC.update}.
The second moment $\langle x_m^2 \rangle$ of each variable $x_1, \cdots, x_6$ [\Autoref{eq:sampling.d}] converges to $1/6$ (black line) as the number of points 
$N_{dot} \to \infty$, implying that $x_m$ are uncorrelated.
Statistical errors respect the central--limit theorem and decrease as $1/\sqrt{N_{dot}}$, 
indicated with a dashed line. 


\nic{
Eq.~(\ref{eq:sampling.d}) provides a valid  
generalisaiton of Marsaglia construction 
from an $O(3)$ vector to a $U(3)$ matrix, and will form the basis for the majority 
of simulation results shown in this Article.
None the less, it is worth noting that this approach is redundant, 
in that the matrix $\opaA{}{\alpha}{\beta}$ 
can be fully characterised using only 4 parameters.
This fact is linked to the structure of representations of $SU(3)$ \cite{Nelson1967}, 
and can be understood directly from Eq.~(\ref{eq:A-matrix_Director}):   
By construction, $\opaA{}{\alpha}{\beta}$ is independent of the overall 
phase of ${\rm d}_\alpha$, leading to a gauge--like redundancy in the 
5--dimensional parameterisation, Eq.~(\ref{eq:sampling.d}).
%
It must therefore be possible to define a Monte Carlo update which acts within a 
4--dimensional subspace of the parameters in Eq.~(\ref{eq:sampling.d}), 
corresponding to the $\mathbb{CP}^2$ space of the spin--1 moment.
}


\nic{
There is no unique prescription for obtaining a \mbox{4--dimensional} 
update in the space of  $U(3)$ matrices.
But one very simple approach  \cite{Amari-private-communication} is to set 
$\phi_3 = \pi/2$ in Eq.~(\ref{eq:sampling.d}), so that the 
z--component of ${\bf d}$ is purely real, vis
\begin{subequations}
		\begin{align}
			x_1 &= \theta_2^{1/4} \ \theta_1^{1/2} \ \sin{\phi_1}		\ , \\ 
			x_2 &= \theta_2^{1/4} \ \theta_1^{1/2} \ \cos{\phi_1}		\ , \\ 
			x_3 &= \theta_2^{1/4} \ \sqrt{1-\theta_1} \ \sin{\phi_2}		\ , \\ 
			x_4 &= \theta_2^{1/4} \ \sqrt{1-\theta_1} \ \cos{\phi_2}	\ , \\ 
			x_5 &= \sqrt{1-\theta_2^{1/2}} 			\ , \\ 
			x_6 &= 0			\ ,
		\end{align}
		\label{eq:alternative.sampling.d}
\end{subequations}
We have confirmed that this alternative parameterisation of $\opaA{}{\alpha}{\beta}$ 
produces identical results in simulations of the BBQ model, a point which we return to below.
}


\nic{
Irrespective of whether the update is 4-- or 5--dimensional, our 
}
Monte Carlo scheme is defined by selecting a site within the lattice at random, 
and using \Autoref{eq:sampling.d} to generate a new configuration of the 
$\mathcal A$--matrix at that site. 
Following the standard Metropolis argument \cite{Metropolis1953}, 
the new state $\mu$ is accepted if 
\begin{equation}
	r_0 \leq e^{-\beta (E_{\mu} - E_{\nu})}		\ ,
\end{equation}
where $r_0$ is number chosen at random on the interval \mbox{$r_0 \in (0,1)$}, 
$\beta = \frac{1}{k_{\text{B}} T}$ (we set $k_{\text{B}} = 1$), and $E_{\nu}$ 
is the energy of the initial configuration.
A single MC step consists of $N$ such local updates, 
where N is the total number of sites in the system.  
In addition, we use the replica--exchange method (parallel tempering)  
to reduce auto--correlation within the resulting Markov 
chain \cite{Swendsen1986, Earl2005}.
An exchange of replicas is carried out every $100$ MC steps.


Simulations are initialized from a state with randomly chosen $\mathcal A$--matrices, 
	mimicking a high--temperature paramagnet.
	Thermalisation is accomplished by cooling the system adiabatically to the target 
	temperature over $10^6$ MC steps (simulated annealing), followed by a further 
	$10^6$ MC steps of thermalisation at that target temperature.
	Thermodynamic quantities were calculated using averages over $5 \times 10^5$ 
	statistically--independent samples.
%


Further insight into correlations can be gained by calculating the 
equal--time structure factors 
\begin{equation}
	S_{\rm{\lambda}}({\bf q}) 
		= \left\langle \sum_{\alpha\beta}
		 	| {m_{\lambda}}^{\alpha}_{~\beta}({\bf q}) |^2 \
		\right\rangle \, 
\label{eq:Sq.MC}
\end{equation}
where $\langle \ldots  \rangle$ represents an average over statistically--independent 
states, and we consider structure factors associated with 
dipole moments, $\lambda = \mathcal{S}$;
quadrupole moments, $\lambda = \mathcal{Q}$; 
and A--matrices, $\lambda = \mathcal{A}$.
Numerically, it is convenient to work with the lattice Fourier transform of 
$\opaA{i}{\alpha}{\beta}$, 
\begin{equation}
	{m_{A}}^{\alpha}_{~\beta}({\bf q}) 
		= \frac{1}{\sqrt{N}} \sum_i^N  e^{i {\bf r}_i {\bf q}} \opaA{i}{\alpha}{\beta}  \, ,
\label{eq:defn.m.A}
\end{equation}
which can be found by fast Fourier transform (FFT).  
From this we can obtain structure factors for both dipole moments [\Autoref{eq:dipole.in.terms.of.A}], 
\begin{align}
	{m_{S}}^{\alpha}_{~\alpha}({\bf q}) 
		&= -i \sum_{\beta,\gamma} \epsilon^{\alpha~\gamma}_{~\beta} {m_{A}}^{\beta}_{~\gamma}({\bf q})  \, , 
\label{eq:defn.m.S}
\end{align}
and quadrupole moments [\Autoref{eq:quadrupole.in.terms.of.A}], 
\begin{align}
	{m_{Q}}^{\alpha}_{~\beta}({\bf q}) 
		&= -{m_{A}}^{\alpha}_{~\beta}({\bf q}) - {m_{A}}^{\beta}_{~\alpha}({\bf q})  
			+\frac{2}{3} \delta^{\alpha\beta} 
			\sum_\gamma {m_{A}}^{\gamma}_{~\gamma}({\bf q}) \; ,	
\label{eq:defn.m.Q}
\end{align}
by direct substitution in \Autoref{eq:Sq.MC}.

\begin{figure*}[t]
	\centering
	\subfloat[Dipolar structure factor $S_{\rm{S}}(\q)$ \label{fig:PD.orderparameter.dipole}]{
		\includegraphics[width=0.48\textwidth]{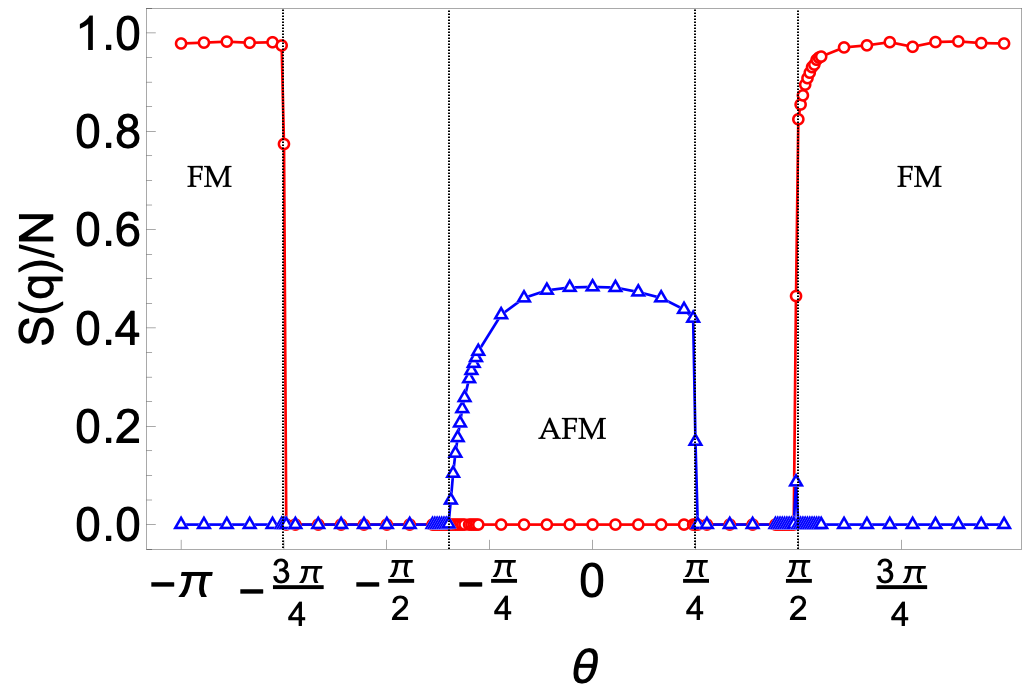}} 
	\subfloat[Quadrupolar structure factor $S_{\rm{Q}}(\q)$  \label{fig:PD.orderparameter.quadrupole}]{
		\includegraphics[width=0.48\textwidth]{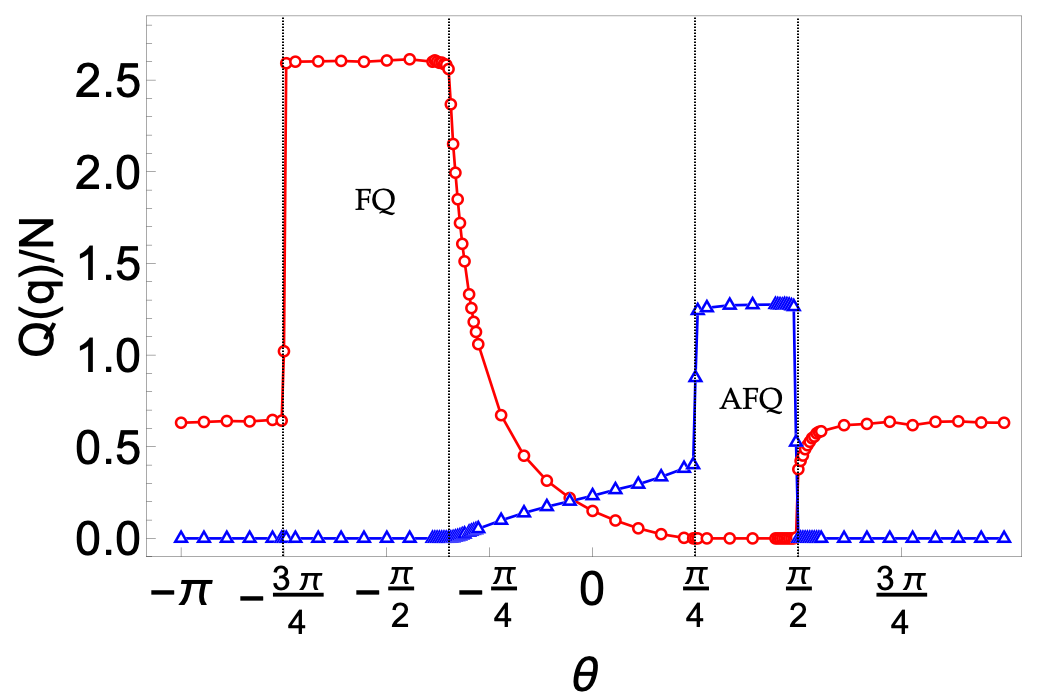}} \\
	\caption{
			Phases occurring in the \mbox{spin--1} BBQ model on a triangular lattice 
			at finite temperature, as found in classical Monte Carlo simulation 
			in the space of $u(3)$ matrices (u3MC) .
			\protect\subref{fig:PD.orderparameter.dipole} Dipolar structure factor $S_{\rm{S}}({\bf q})$ [\protect\Autoref{eq:Sq.MC}], 
			showing ferromagnetic (FM) correlations for ${\bf q} = \Gamma$ (red circles)
			and 3--sublattice antiferromagnetic (AFM) correlations for ${\bf q} = K$ (blue triangles).
			\protect\subref{fig:PD.orderparameter.quadrupole} Corresponding results for the quadrupolar structure factor $S_{\rm{Q}}({\bf q})$ 
			[\protect\Autoref{eq:Sq.MC}], showing ferroquadrupolar (FQ) correlations for
			${\bf q} = \Gamma$ (red circles) and 3--sublattice antiferroquadrupolar (AFQ) 
			correlations for ${\bf q} = K$ (blue triangles).	
			Simulations of \protect\Autoref{eq:BBQ.model} were carried out using the u3MC method 
			described in \protect\Autoref{sec:u3.update}, at a temperature $T = 0.01\ J$, for cluster of 
			linear dimension $L = 48$ ($N=2304$ spins), with parameters given by \protect\Autoref{eq:J.cos.theta}. 
			The phases found are in direct correspondence with known results 
			for the mean--field ground state \cite{Lauchli2006, Smerald2013,Penc2011-Springer}, 
			summarised in \protect\Autoref{fig:MF.phase.diagram}.
			In each case the temperature associated with the onset of fluctuations corresponds 
			to the peak found in specific heat, cf. \protect\Autoref{fig:MC.phase.diagram}.
	}
	\label{fig:PD.StructureFactor}
\end{figure*}

\subsubsection{Phase diagram and comparison with published results}
\label{sec:phase.diagram}

As a first check on the method, we have carried out u3MC 
simulations of the thermodynamic properties of \mbox{spin--1} 
BBQ model \Autoref{eq:H.BBQ.1} on a triangular lattice for comparison with 
published results \cite{Lauchli2006,Stoudenmire2009,Smerald2013}.
Typical results for the heat capacity are shown in \Autoref{fig:PD.Stoudenmire}, 
for parameters $J_1=1$, $J_2=-1.5$, chosen to facilitate comparison with earlier 
work \cite{Stoudenmire2009}.
For these parameters, mean--field calculations find a ground state 
with 3--sublattice antiferromagnetic (AFM) order, close to a phase boundary 
with ferroquadrupolar (FQ) order \cite{Lauchli2006,Smerald2013}.


Simulating in the space of $\mathcal{A}$--matrices [cf. \Autoref{sec:u3.update}], 
we find two peaks in heat capacity, one at $T \approx 0.5\ J_1$, 
corresponding to the onset of FQ fluctuations, and one at $T \approx 0.3\ J_1$ 
corresponding to the onset of AFM fluctuations.
\nic{
In \Autoref{fig:PD.Stoudenmire} we show results obtained using both 
5--dimensional [Eq.~(\ref{eq:sampling.d})] and 4--dimensional [Eq.~(\ref{eq:alternative.sampling.d})] u3MC updates.   
For comparison, we have also} carried out equivalent simulations in the space of  
${\bf d}$--vectors, following the 
sSU(3) approach of Stoudenmire {\it et al.} \cite{Stoudenmire2009}.
\nic{Within statistical errors, we find quantitative agreement between the 
three different methods.}


We have extended this analysis to obtain a complete finite--temperature phase diagram 
for the BBQ model, previously shown in \Autoref{fig:MC.phase.diagram}.
Results are shown for a cluster of linear dimension $L = 48$ [N=2304 spins].  
Phase boundaries were obtained by tracking the evolution of peaks in heat capacity 
as a function of 
\be
	J_1 = J \cos \theta \; , \; J_2 = J \cos \theta
\label{eq:J.cos.theta}
\ee
and using the equal--time structure factors $S_{\mathcal{S}} ({\bf q})$ and 
$S_{\mathcal{Q}} ({\bf q})$ [\Autoref{eq:Sq.MC}] to determine the nature of 
each phase.
Typical results for structure factors evaluated at known ordering vectors,  
for a temperature $T/J = 0.01$, are shown in \Autoref{fig:PD.StructureFactor}.


The correlations found at low temperature exactly correspond to the four 
known mean--field ground states \cite{Lauchli2006,Smerald2013}, 
having ferromagnetic (FM), antiferromagnetic (AFM), 
ferroquadrupolar (FQ) and antiferroquadrupolar (AFQ) order, 
as illustrated in \Autoref{fig:MF.phase.diagram}.
As previously noted by Stoudenmire {\it et al.} \cite{Stoudenmire2009}, 
FQ order occurs as a secondary order parameter within the coplanar AFM 
ground state.
Consistent with this, for $\theta \lesssim - \pi/4$ the onset of FQ fluctuations 
occurs at a higher temperature than the onset of AFM fluctuations 
(cf. results for $\theta \approx -0.313\ \pi$ in \Autoref{fig:PD.Stoudenmire}).


We also find that there is a range of parameters $\theta \sim \pi/2$, near the border 
between FM and AFQ phases for which the onset of FM fluctuations occurs at 
a higher temperature than the onset of AFQ fluctuations.
Here no interpretation in terms of a secondary order--parameter is possible, 
but once again it is the single--sublattice phase which dominates at higher temperatures.
We infer that the entropy of fluctuations about the FM ground state 
is higher than the entropy of fluctuations about the AFQ ground state, 
presumably because of the $k^2$ dispersion of its excitations.


Perhaps the most striking feature of the phase \kim{diagram} in \Autoref{fig:MC.phase.diagram} 
are the ``vertical'' phase boundaries between dipolar and quadrupolar phases 
at the two $SU(3)$ points, shown as solid red lines.
These are consistent with the $SU(3)$ symmetry of the ground--state manifolds being 
preserved up to temperature associated with the onset of correlations, $T^*$.
And this in turn raises the possibility of finding exotic topological phase 
transitions at $T^*$, mediated by topological defects specific to the 
$SU(3)$ points \cite{Ivanov2008,Ueda2016-PRA93}.
We leave this interesting topic for future studies.


In conclusion, our survey of correlations at finite temperature, summarised 
in \Autoref{fig:MC.phase.diagram} and \Autoref{fig:PD.StructureFactor}, 
provides strong {\it prima facie} evidence that the u3MC approach introduced 
in \mbox{\Autoref{sec:u3.update}} can describe the thermodynamic properties 
of \mbox{spin--1} magnets.
In \Autoref{section:numerics} we present a more rigorous test, 
in the form of a detailed study of the thermodynamic properties of 
the FQ phase at low temperatures, where we are able to make 
quantitative comparison with analytic predictions.

\subsection{Molecular Dynamics simulations within $u(3)$ framework}			
\label{sec:u3MD}

	Numerical integration of equations of motion provides a powerful approach to describing the 
	(semi--)classical dynamics of quantum magnets, which can readily be combined with classical 
	Monte Carlo simulation, an approach which has been referred to as ``molecular dynamics'' (MD) 
	simulation \cite{Moessner1998-PRL80,Moessner1998-PRB58,Conlon2009,Taillefumier2014,Zhang2019}.
	Microscopic approaches typically start from the Heisenberg equation of motion 
	for an $O(3)$ spin, \Autoref{eq:Heisenberg.EoM.O3}, and have proved surprisingly 
	effective in describing the dynamics of quantum magnets \cite{Samarakoon2017,Chern2018,Zhang2019,Pohle2021}.  


\begin{figure*}[t]
	\centering
	\subfloat[time-dependence of  $\text{Tr}\ \mathcal{A}$  \label{fig:RK4.TrA}]{
		\includegraphics[width=0.47\textwidth]{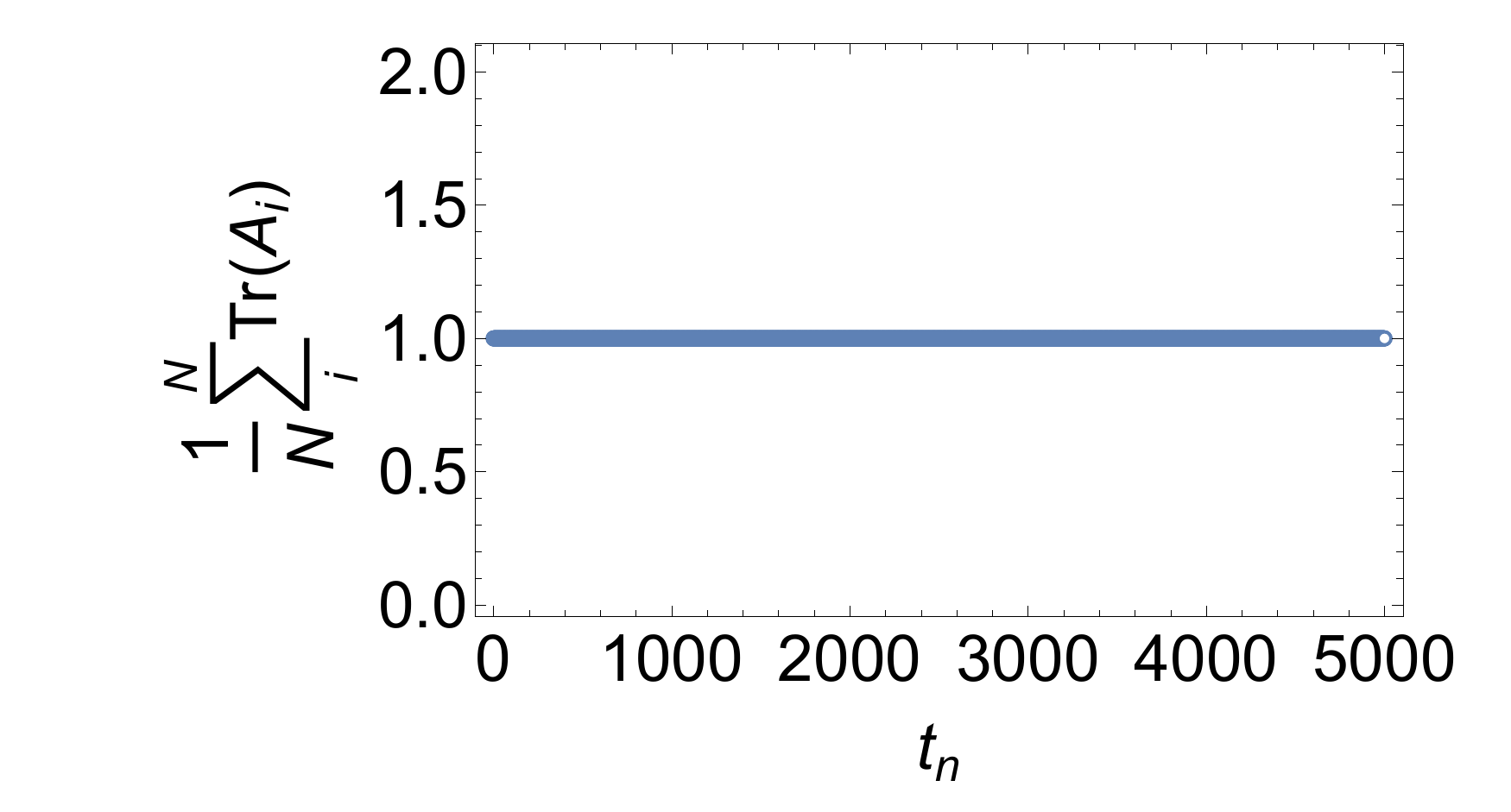}} 
	\hspace{0.5cm}
	\subfloat[time-dependence of total energy \label{fig:RK4.energy}]{
		\includegraphics[width=0.47\textwidth]{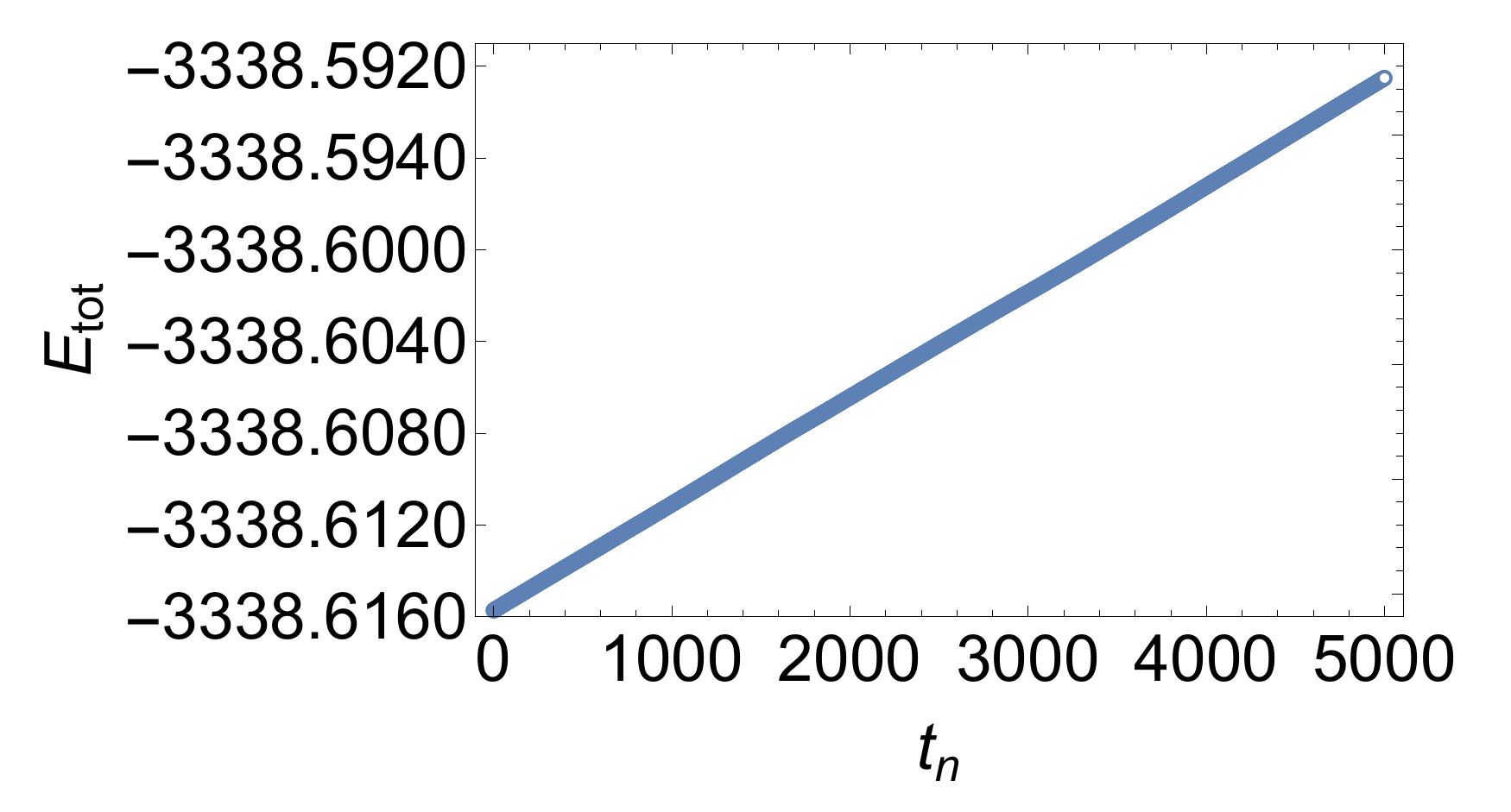}} 
\caption{
	Evidence of stability of numerical integration of equations of motion.
	\protect\subref{fig:RK4.TrA} Time--dependence of $\text{Tr}\ \mathcal{A}$ [\protect\Autoref{eq:traceA}], 
	showing conservation of spin to numerical precision.
	\protect\subref{fig:RK4.energy} Time--dependence of energy $E = \langle \mathcal{H}_{\sf BBQ} \rangle$, 
	showing conservation to the level expected for a $4^{th}$--order 
	Runge--Kutta (RK--4) algorithm.
	Simulations were carried out for the \mbox{spin--1} bilinear--biquadratic 
	model [\protect\Autoref{eq:H.BBQ.3}], on a triangular--lattice cluster 
	with linear dimension $\text{L} = 24$ (\mbox{$\text{N}=2304$} spins), 
	for parameters $J_1 = 0$, $J_2 = -1$, 
	at a temperature $T = 0.1\ J $, using the equation of motion 
	\protect\Autoref{eq:EoM.u3}, 
	with a time--step $\delta t = 0.4\ J^{-1}$.
	}
	\label{fig:conservation.E.TrA}
\end{figure*}


The success of the MD approach in these cases rests on the fact that an $O(3)$ vector 
provides an appropriate \mbox{(semi--)classical} description of a spin--1/2 moment.    
However, in the case of \mbox{spin--1} magnets, $O(3)$ vectors fail to provide an adequate 
description, since they do not properly account for quadrupole degrees of freedom 
[cf. \Autoref{section:maths.for.spin.1}].
This problem has long been understood in the context of the analytic theory 
of nematic phases \cite{Matveev1973,Papanicolaou1988}.
And in general the band--like excitations of \mbox{spin--1} magnets include both spin waves 
and quadrupole waves \cite{Akaki2017}.
These can be addressed analytically through a multiple--Boson expansion, also known as 
``flavour--wave'' theory \cite{Penc2011-Springer}.


Equation of motion approaches to the dynamics of \mbox{spin--1} magnets have also 
been developed in terms of spin-- and quadrupole--operators 
\cite{Balla2014-Thesis,Remund2015-Thesis,Zhang2021}.
However these approaches are complicated by the convoluted nature of the structure 
constants of the algebra $su(3)$.
In contrast, the $u(3)$ framework established in \Autoref{section:maths.for.spin.1} 
leads to a very compact EoM for $\mathcal{A}$--matrices,  
\Autoref{eq:EoM.u3}, ideally--suited to numerical integration.
And the power of these EoM are greatly enhanced by the fact that they can 
be combined with the MC methods developed in \Autoref{sec:MC}, providing 
an ``u3MD'' approach to \mbox{spin--1} magnets, on the same footing as the $O(3)$ 
methods applied to spin--1/2 magnets.

\subsubsection{Implementation of u3MD update}

Our MD simulations, like the MC simulations described in \Autoref{sec:u3.update},  
are carried out in the basis of states defined by products of 
${\mathcal A}$--matrices [\Autoref{eq:product.wf}].
We implement simulations by using a 4$^{th}$ order Runge-Kutta (RK--4) 
algorithm \cite{NumericalRecipes2007, OrdinaryDiffEquations1} to 
numerically integrate \Autoref{eq:EoM.u3} for each component 
of ${\mathcal A}^\alpha_{i, \beta}$, using a fixed timestep $\delta t_{\sf RK}$.  
%
Iterative application of RK--integration 
\begin{eqnarray}
	\{ {\mathcal A}^\alpha_{i, \beta} (t) \}
		\mapsto  \{ {\mathcal A}^\alpha_{i, \beta} (t + \delta t_{\sf RK}) \}
		+ \mathcal{O} (\delta t_{\sf RK}^5)
\end{eqnarray}
generates a time series for $\{ {\mathcal A}^\alpha_{i, \beta} (t) \}$ with errors 
which are controlled by the size of $\delta t_{\sf RK}$.
A single RK update is defined through numerical integration of \Autoref{eq:EoM.u3}
for every spin in the lattice.
In order to work with a manageable set of data, while retaining sufficient precision in 
numerical integration, we store only the result of every 20$^{th}$ global update.


The stored data defines a time series 
\begin{eqnarray}
\{ {\mathcal A}^\alpha_\beta (i, t_n) \} \; , \quad
	t_n = n\ \delta t  \; , 
	\quad n = 1 \ldots N_{\sf t} \; , 
\end{eqnarray}
%
where
the size of the effective time step, $\delta t$, determines the 
highest frequency 
we are able to resolve
\begin{eqnarray}
	\delta t = \frac{2 \pi}{\omega_{\sf max}}   \; .
\end{eqnarray}
Meanwhile the duration of the simulation 
\begin{eqnarray}
	\Delta t = N_t\ \delta t \; ,
\end{eqnarray}
determines the energy--resolution of results 
\begin{eqnarray}
	\delta \omega = \frac{2\pi}{\Delta t} \; ,
\label{eq:freq.resolution}
\end{eqnarray}
where we work in units such that $\hbar = 1$.


In practice, we typically work with a time--series of length 
\begin{eqnarray}
	N_t = 1600 \; , 
\end{eqnarray}
with effective time--step 
\begin{eqnarray}
	\delta t \approx 0.4\ J^{-1} \; .
\label{eq:MD.parameter.delta.t}
\end{eqnarray}
It follows that the time--interval used in RK integration for 
an individual spin is 
\begin{eqnarray}
	\delta t_{\sf RK} \equiv \delta t/20 \approx 0.02\ J^{-1} \; .
\end{eqnarray}
This choice of parameters is adequate to resolve excitations 
with energy up to 
\begin{eqnarray}
	\omega_{\sf max} \approx 16\ J \; ,
\end{eqnarray}
twice what is needed for individual excitations of the 
FQ state with parameters 
\begin{eqnarray}
(J=1, \theta = -\frac{\pi}{2}) \Rightarrow (J_1 = 0.0,  J_2 = -1.0)   \; ,
\label{eq:model.parameters}
\end{eqnarray}
cf.~\Autoref{fig:illustration.of.results}.
The corresponding energy resolution 
\begin{eqnarray}
\delta \omega \approx  10^{-2}\ J \; ,
\end{eqnarray}
is sufficient to resolve fine--structure in dynamical 
structure factors, described below.


The validity of this MD approach
depends on the satisfaction of both the constraint on spin--length 
[\Autoref{eq:traceA}], and on the conservation 
of the total energy of the system, $E[ \mathcal{A}_i ]$ [\Autoref{eq:E.of.A}].
In \Autoref{fig:conservation.E.TrA} we show evidence that both are 
satisfied, within controlled errors, for simulations of a triangular--lattice 
cluster of linear dimension $\text{L} = 24$ ($\text{N}=2304$ spins), 
with model parameters \Autoref{eq:model.parameters}, 
and time--step \Autoref{eq:MD.parameter.delta.t}, 
at a temperature $T = 0.1\ J$.    


We consider first the constraint on spin--length.
As discussed in \Autoref{sec:u3.EoM}, as long as the initial 
configuration $\{ \opa{i}{\alpha}{\beta} (t=0) \}$ satisfies the spin--length 
constraint
\begin{eqnarray}
\text{Tr}\ \opa{i}{}{} \equiv 1 \; \forall \; i \in  ( 1 \ldots \text{N} )  \; ,  
\end{eqnarray}
[\Autoref{eq:traceA}], its continued satisfaction 
is guaranteed by the structure of the EoM, \Autoref{eq:EoM.u3}.
%
From \Autoref{fig:conservation.E.TrA}~\subref{fig:RK4.TrA}, we see that 
the trace of ${\mathcal A}_i$ is conserved, 
up to numerical precision, for simulations 
of duration $N_t = 5000$.  
This confirms that simulations of any feasible duration, 
continue to describe \mbox{spin--1} moments.


We now turn to the conservation of energy.
RK--integration is not a symplectic (energy--conserving) method.
However the rate at which error accumulates depends on 
the size of the RK time step, $\delta t_{\sf RK}$.
And, by making $\delta t_{\sf RK}$ sufficiently small, errors in 
energy can be kept bounded. 
%
From \Autoref{fig:conservation.E.TrA}~\subref{fig:RK4.energy}, we see that the error in energy 
which accumulates over simulations of duration $N_t = 5000$ 
is $\approx 0.03\ J$.
This implies that one ``MD step'', i.e. a single sweep of the entire lattice 
using an RK--4 algorithm, introduces an error in total energy 
of order $\sim 10^{-6}\ J$.
This is sufficiently small to ensure adequate conservation of energy 
for simulations of practical duration, i.e. $N_t = 1600$, 
$\Delta t \approx 1600\times \delta t = 640\ J^{-1}$.

\subsubsection{Calculation of dynamical structure factors}
\label{section:calculation.of.dynamical.structure.factors}

We can analyse the time series $\{ {\mathcal A}^\alpha_\beta (i, t_n) \}$ 
by directly animating the evolution of spin configurations \cite{Pohle-in-preparation}, 
or by calculating dynamical structure factors of the form 
\begin{eqnarray}
	S_{\rm{\lambda}}({\bf q}, t_n) 
		= \left \langle\sum_{\alpha\beta} 
		 	\left( {m_{\lambda}}^{\alpha}_{~\beta}({\bf q}, t_n) \right)^*  
			{m_{\lambda}}^{\alpha}_{~\beta}({\bf q}, 0) 
		\right \rangle \; ,
\label{eq:Sq.MD}
\end{eqnarray}
with $\lambda = \mathcal{S},  \mathcal{Q}, \mathcal{A}$ 
[cf. \Autoref{eq:Sq.MC}].  
The dynamical structure factor 
for $\mathcal{A}$--matrices is defined through
\begin{eqnarray}
	{m_{A}}^{\alpha}_{~\beta}({\bf q}, t_n) 
		= \frac{1}{\sqrt{N}} \sum_{i=1}^N  e^{i {\bf r}_i {\bf q}} \opaA{i}{\alpha}{\beta} (t_n)  \, ,
\end{eqnarray}
[cf. \Autoref{eq:defn.m.A}], and equivalent structure factors for 
dipole-- ${m_{S}}^{\alpha}_{~\alpha}({\bf q}, t)$ 
and quadrupole--moments ${m_{Q}}^{\alpha}_{~\beta}({\bf q}, t)$, can be  
defined by extension of \Autoref{eq:defn.m.S} and \Autoref{eq:defn.m.Q}.


For purposes of comparison with experiment, it is 
usually more convenient to work with the Fourier transform
\begin{eqnarray}
	S_{\rm{\lambda}} ({\bf q}, \omega_m) 
		= \frac{1}{\sqrt{N_t}} \sum^{N_t}_{n=1}  
		e^{i \omega_n t_n}\ 
		S_{\rm{\lambda}} ({\bf q}, t_n)  \; ,
\label{eq:Sq.MD.FT}
\end{eqnarray}
where $\omega_m$, like ${\bf q}$, 
takes on discrete values
\begin{eqnarray}
	\omega_m = m\ \delta \omega  \; , 
	\quad m = 0 \ldots N_{\sf t} - 1 \; .
\end{eqnarray}
To avoid numerical artefacts (Gibbs phenomenon) coming 
from discontinuities at \mbox{$t = 0$} and \mbox{$t =  \Delta t$} \cite{MathematicalPhyics}, 
we multiply the time--series entering \Autoref{eq:Sq.MD.FT}
by a Gaussian envelope centred on $t_n = \Delta t/2$. 
In practice, we evaluate the dynamical structure factor as
\begin{eqnarray}
	S_{\rm{\lambda}} ({\bf q}, \omega_m) = \left\langle \sum_{\alpha\beta} 
		| {\overline{m}_{\lambda}}^{\alpha}_{~\beta}({\bf q}, \omega_m)  |^2 
		\right\rangle \; ,
\label{eq:dynamical.structure.factor.in.simulation}
\end{eqnarray}
where 
\begin{eqnarray}
	{\overline{m}_{\lambda}}^{\alpha}_{~\beta}({\bf q}, \omega_m) 
		&=& \frac{1}{\sqrt{N_t}} \sum_{n=1}^{N_t} e^{i \omega_m t_n}  
			\sqrt{\overline{g}(t_n)} 
		\  {m_\lambda}^{\alpha}_{~\beta}({\bf q}, t_n)  \, , \nonumber \\
\label{eq:mAqw}
\end{eqnarray}
is found by Fast Fourier transform (FFT) \cite{NumericalRecipes2007}.
The Gaussian envelope is implemented through the function 
 \begin{eqnarray}
	\overline{g} (t_n) = \frac{\delta t}{\delta \omega_n} \frac{\sigma }{\sqrt{2 \pi}} 
			e^{-\frac{\sigma^2}{2} (t_n - \Delta t/2)^2}  \, ,
 \end{eqnarray}
and absorbs a dimensional factor  $\delta t/\delta \omega$ 
associated with integrals.
The value of $\sigma$ is chosen such that the full--width half maximum 
(FWHM) of $\overline{g} (t_n)$ is $\approx \Delta t$.
Introducing this envelope in time is equivalent to convoluting 
$S_{\rm{\lambda}} ({\bf q}, \omega)$ with a Gaussian in frequency space, 
with
\begin{eqnarray}
	\text{FWHM} = 2 \sqrt{2 \ln{2}} \times \sigma  \; ,
\label{eq:FWHM}
\end{eqnarray}
approximately equal to $\delta \omega$.
This determines the ultimate energy resolution of results.
Structure factors calculated in this way are averaged over 
$500$ independent time--series, each determined by a separate 
initial state drawn from classical MC simulation.

 
An example of a dynamical structure factor calculated using the u3MD approach 
has been presented in \Autoref{fig:illustration.of.results}, where results 
are shown for the FQ phase of the BBQ model on the triangular lattice.

 
\nic{
It is important to note that the EoM, Eq.~(\ref{eq:EoM.u3}), are invariant under 
time--reversal symmetry.   
Solutions to these equations therefore occur in pairs, with positive and negative 
eigenvalues 
\begin{eqnarray}
\omega = \pm \epsilon_{\bf k} \; .
\end{eqnarray}
Both positive and negative energy solutions play a role in experimental 
response functions, reflecting the absorption and emission of energy by 
the system.
And numerical integration of EoM will generally recover both solutions with equal weight, 
leading to structure factors which are even functions of frequency
\begin{eqnarray}
	S^{\sf MD}_{\rm{\lambda}} ({\bf q}, \omega) = S^{\sf MD}_{\rm{\lambda}} ({\bf q}, -\omega) 
\label{eq:symmetry.of.Sqomega} 
\end{eqnarray}
%
%
None the less, in \Autoref{fig:illustration.of.results}, and elsewhere in this Article, 
we concentrate on solutions at positive energy, $\omega > 0$, since these are the most 
relevant for the low--temperature properties of quantum magnets.
}

 
\nic{
In \Autoref{section:numerics} and \Autoref{section:quantum.vs.classical} 
we delve deeper into u3MD results, and their connection with the analytic
theory of the excitations about a FQ ground state.
But before doing so, we first develop the analytic theory necessary to 
understand simulation.
}


\begin{figure}[t]
	\centering
	\includegraphics[width=0.45\textwidth]{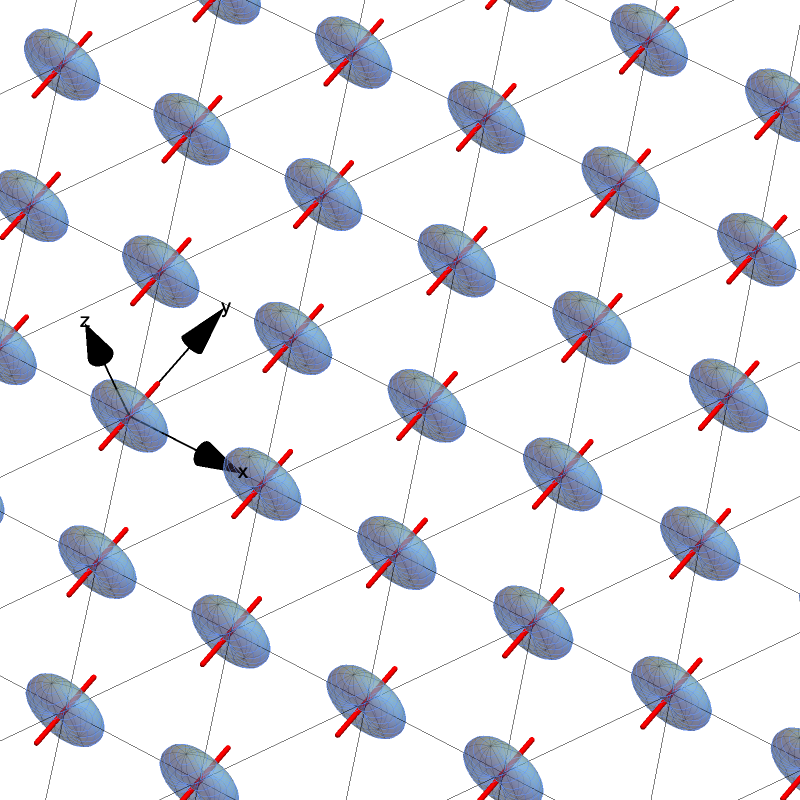}
	\caption{
		Ferroquadrupolar (FQ) ground state of a spin--1 magnet on a triangular lattice. 
		Each magnetic moment has been plotted in the state $\ket{y}$, or 
		equivalently in director representation  \mbox{$d^{x}=0$}, \mbox{$d^{y}=1$}, 
		\mbox{$d^{z}=0,$} in \protect\Autoref{eq:Defd}.
		The corresponding spin probability distribution [\protect\Autoref{eq:FluctProb}] 
		is shown in grayish blue, and the associated director in red.
	}
	\label{fig:FQ.order}
\end{figure}

\section{Classical theory of fluctuations about a ferroquadrupolar ground state} 		
\label{section:classical.theory}

In this Section we use the $u(3)$ formalism introduced in \Autoref{section:maths.for.spin.1}
to develop a classical theory of fluctuations about a ferroquadrupolar (FQ) ground state.
This will serve as a benchmark for the classical MC simulations presented 
in \Autoref{section:numerics}, and as the starting point for an analysis of 
quantum--classical correspondence in \Autoref{section:quantum.vs.classical}. 


We chose to work with FQ order, since this is simplest of the non--trivial phases found 
in the BBQ model.
Mean--field (MF) calculations for the \mbox{spin--1} BBQ model on a triangular lattice 
\cite{Lauchli2006,Smerald2013} predict a FQ ground state for a broad range of parameters 
[\Autoref{fig:MF.phase.diagram}], and its existence has since been confirmed using 
exact--diagonalisation \cite{Lauchli2006},  
QMC \cite{Kaul2012,Voell2015}
and tensor--network approaches \cite{Niesen2018}.
The dynamics of this state have also been explored through both
``flavour wave'' theory \cite{Lauchli2006,Penc2011-Springer,Matveev1973,Onufrieva1985} 
and QMC simulation \cite{Voell2015}.
This makes FQ order a convenient point of reference, with many published results available 
for comparison.
A classical theory of its low--temperature properties, however, is lacking.


We first show how small fluctuations about FQ order can be described using 
four of the nine generators of $U(3)$ [\Autoref{section:small.fluctuations}]. 
This leads naturally to a low--temperature expansion scheme for the classical 
thermodynamic properties of FQ order [\Autoref{section:low.T}]. 
This theory is used to make explicit predictions for the classical 
thermodynamic properties [\Autoref{section:low.T}] of FQ order, 
for later comparison with simulation.

\subsection{Expansion of small fluctuations} 
\label{section:small.fluctuations}

	Our starting point is the FQ ground state found in mean--field calculations; 
	a product wave function of on--site quadrupolar moments with 
	a common orientation
	\be
	\ket{\Psi_0}^{\sf MF}_{\sf FQ} = \prod_{i=1}^{N} \ket{\bf{d}^{\sf FQ}_i} \; ,
	\label{eq:FQ.GS.d}
	\ee
	where $\ket{\bf{d}_i}$ is defined through \Autoref{eq:Defd}.
	For concreteness, we assume the director to be along the y-axis for all lattice sites, i.e. 
	\be
	\begin{matrix}
		\ket{\V{d}_{i}^{\sf FQ}}=\ket{y}&\textrm{or equivalently}& \bf{d}^{\sf FQ}=\begin{pmatrix}
			0\\
			1\\
			0
		\end{pmatrix}
	\end{matrix}\; .
	\label{eq:GS_FQ}
	\ee
	%
	%
	This state is illustrated in \Autoref{fig:FQ.order}.    
	Once corrections to mean--field theory are taken into account, this product state will 
	be dressed with thermal and/or quantum fluctuations, reducing the expectation 
	value of the quadrupolar order parameter.
	We now derive a framework for describing these fluctuations in 
	terms of generators belonging to the Lie algebra $u(3)$.


\begin{figure}[t]
	\centering
	\includegraphics[width=0.50\textwidth]{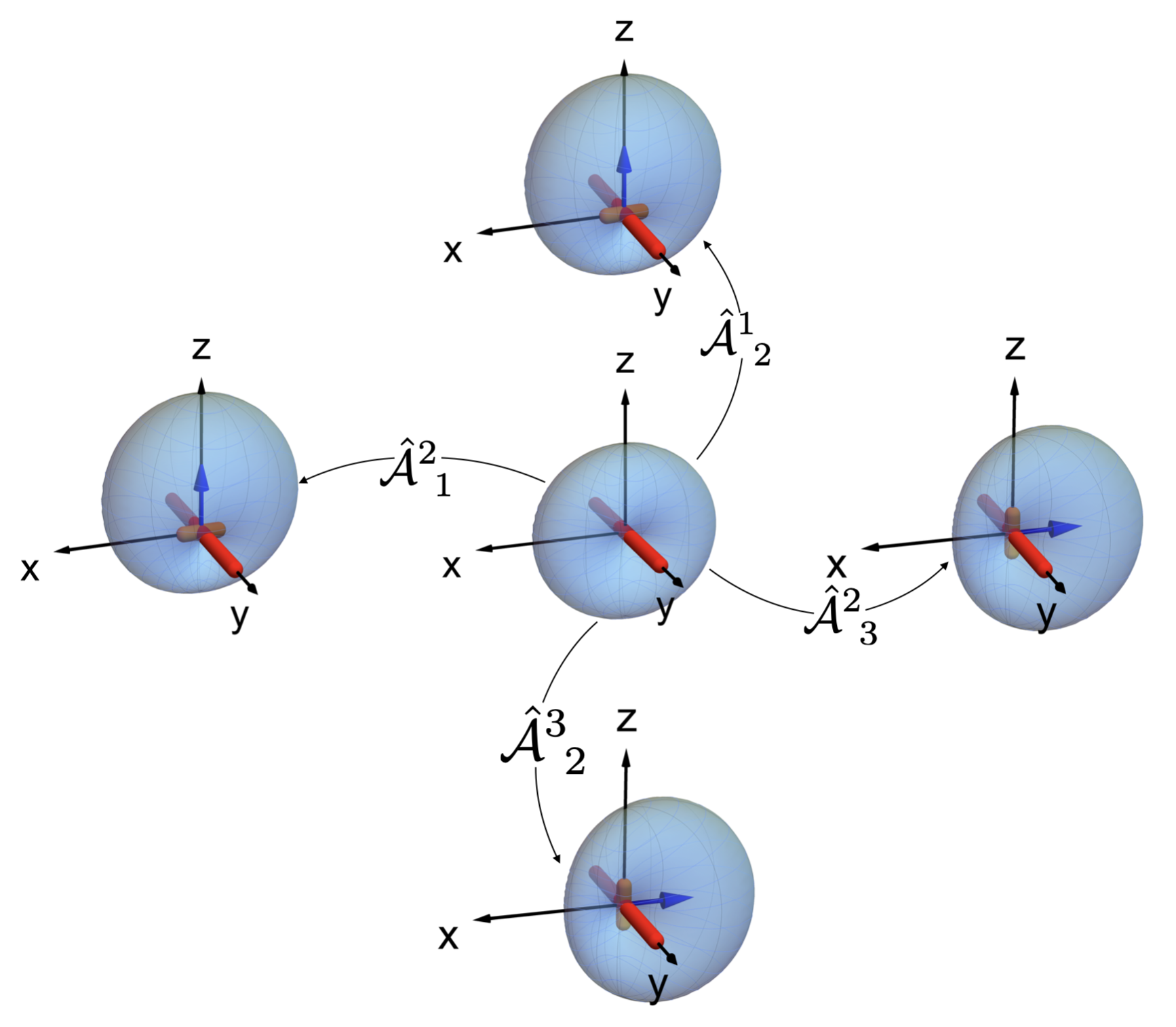}
	\caption{
			Effect of the four operators generating fluctuations about the 
			ferroquadrupolar (FQ) ground state $\ket{y}$.  
			The generators $\opa{}{1}{2}$ (acting on the right) and $\opa{}{2}{1}$ 
			(acting on the left) introduce a complex component of the director 
			${\bf d}$, parallel to the x--axis.
			This has the effect of rotating the quadrupole moment about the z--axis, 
			while simultaneously inducing a dipole moment along the z--axis 
			[\protect\Autoref{eq:Suv}].
			Meanwhile the generators $\opa{}{2}{3}$ and $\opa{}{3}{2}$ rotate the 
			quadrupole about the x--axis, and induce a dipole moment along 
			the same axis. 
			The effect of each generator is computed according to \protect\Autoref{eq:NewPsi}, 
			for a rotation through an angle of $\phi=\frac{\pi}{8}$.
		The red bar represents the real part $\bf{u}$ of the coefficients $d^{\alpha}$ 
		[\protect\Autoref{eq:Duv}] in \protect\Autoref{eq:Defd}, while the orange bar 
		represents the imaginary part  $\bf{v}$.}
	\label{fig:gen_fluct}
\end{figure}


	The first step of our analysis is to transcribe the MF ground state, \Autoref{eq:FQ.GS.d},  
	in terms of ${\mathcal A}$--matrices.
	Using \Autoref{eq:defAop}, we represent the ground state as
	\be
	\bra{\V{d}_{i}^{\sf FQ}} {\mathcal A} \ket{\V{d}_{i}^{\sf FQ}}=\bf{A}_0=
	\begin{pmatrix}
		0& 0 & 0 \\
		0 & 1 & 0 \\
		0& 0 & 0 \\
	\end{pmatrix}\; .
	\label{eq:GSA_FQ}
	\ee
	This in turn forms the basis for a product state 
	\be
	\ket{\Psi_0}^{\sf MF}_{\sf FQ} = \prod_{i=1}^{N} \ket{ \bf{A}_0 } \; ,
	\label{eq:FQyGS}
	\ee
	in the space of ${\mathcal A}$--matrices.


	Within a Lie algebra, local fluctuations about any state $\ket{\psi_0}$ 
	can be written  
	%
	\be
	\ket{\psi} = \hat{R} ( \vec\phi ) \ket{\psi_0} \; ,
	\label{eq:NewPsi}
	\ee
	where the operator  $\hat{R}(\vec\phi)$ has the form
	\be
	\hat{R}(\vec\phi) = \exp{-i \sum_p \phi_p \hat{G}_p} \; ,
	\ee
	and $G_p$ are elements of the algebra, with $\phi_p \in \mathbb{R}$.  
	In the case of $u(3)$, a suitable set of generators 
	$\hat{G}_p ~(p=1,2,\dots,9)$ are the matrices $\opa{}{\alpha}{\beta}$ 
	[\Autoref{eq:u3.A.matrix.rep}], and we can write 
	\be
	\hat{R}(\vec\phi) = \exp{-i \sum_{\alpha\beta} \phi_{\alpha,\beta} \opa{}{\alpha}{\beta}}  \; ,
	\label{eq:defR}
	\ee
	where $\alpha,\beta=1,2,3$.
	Under this operation, ${\mathcal A}$--matrices transform as
	\be
	\textbf{A}(\vec\phi) = 
	\hat{R}(\vec\phi) \textbf{A} \hat{R}(\vec\phi)^{\dagger} \; .
	\label{eq:NewA}
	\ee
	as determined by \Autoref{eq:LTA} [cf. \hyperref[sec:appA]{Appendix~}\ref{sec:appA}]. 


	If we assume fluctuations to be small, i.e. $\phi_{\alpha,\beta} \ll 1$, we can expand 
	the exponential in \Autoref{eq:defR} 
	\be
	\hat{R}(\vec\phi) 
	\simeq \mathbb{I} + i \sum_{\alpha\beta} \phi_{\alpha,\beta} \opa{}{\alpha}{\beta} + \cdots \; .
	\label{eq:expanding.fluctuation}  
	\ee
	Considering the action of this operator on the FQ ground state, as characterised by 
	the matrix $\bf{A}_0$ [\Autoref{eq:FQyGS}], the only $\opa{}{\alpha}{\beta}$ that will give a 
	non--zero result are $\opa{}{1}{2},\opa{}{3}{2}$ on the left, $\opa{}{2}{1},\opa{}{2}{3}$ 
	on the right, and $\opa{}{2}{2}$, which preserves the ground state.
	Consequently, if we wish to describe fluctuations about the ground state, we need only keep 
	four generators $\opa{}{1}{2},\opa{}{3}{2},\opa{}{2}{1},\opa{}{2}{3}$.
	We can think of these operators 
	as performing rotations in the space of ${\mathcal A}$--matrices or, 
	equivalently, of ${\bf d}$--vectors \cite{Smerald2013}.   
	Their effect is 
	illustrated in \Autoref{fig:gen_fluct}.   


	Restricting \Autoref{eq:expanding.fluctuation} to the four relevant generators, 
	and retaining all terms to order $\phi^2$, we arrive at a general expression for 
	infinitesimal fluctuations about the FQ ground state
	\be
	\hat{R}(\vec\phi)=
	\begin{pmatrix}
		1 & i \phi ^{1,2} & 0 \\
		0 & 1-\frac{1}{2}\phi^{1,2}\phi^{2,1}-\frac{1}{2}\phi^{2,3}\phi^{3,2} & 0 \\
		0 & i \phi _{3,2} & 1 \\
	\end{pmatrix}\; ,\label{eq:Rphi2} 
	\ee
	where $\hat{R}(\vec\phi)^{\dagger}$ is the transpose of \Autoref{eq:Rphi2}, 
	with the small subtlety that ${\phi ^{1,2}}^\dagger=\phi ^{2,1}$.

	Substituting \Autoref{eq:Rphi2} into \Autoref{eq:NewA}, we find 
	\begin{align*}
		\ophb{A}(\vec\phi) &=\hat{R}(\vec\phi) \textbf{A}_0 \hat{R}(\vec\phi)^{\dagger}\\
		&=\begin{pmatrix}
			\phi^{1,2} \phi^{2,1} & i \phi^{1,2} & \phi^{1,2}\phi^{2,3}\\
			-i \phi^{2,1}& 1-\phi^{1,2}\phi^{2,1}-\phi^{2,3}\phi^{3,2}  &-i  \phi^{2,3}\\
			\phi^{2,1}\phi^{3,2} & i \phi^{3,2} &\phi^{2,3}\phi^{3,2}\\
		\end{pmatrix} \; . 
		\numberthis
		\label{eq:A_phi}
	\end{align*}
	Crucially,  \Autoref{eq:A_phi} satisfies the contstaint \mbox{$\text{Tr}\ \mathcal{A} = 1$} 
	[\Autoref{eq:traceA.on.A.matrix}], implying that the spin length is conserved.
	%

	The effect of the four generators $\opa{}{1}{2},\opa{}{3}{2},\opa{}{2}{1},\opa{}{2}{3}$ 
	on the state $\ket{y}$ can now be quantified directly.
	Substituting \Autoref{eq:A_phi} in \Autoref{eq:quadrupole.in.terms.of.A}, and keeping 
	terms to $\mathcal{O}(\phi^2)$, we find
\begin{widetext}
	\begin{eqnarray}
			\opqb{}(\vec\phi)
			&=& \begin{pmatrix}
				\frac{2}{3}-2\phi^{1,2}\phi^{2,1} & i(\phi^{2,1}-\phi^{1,2})&  -\phi^{1,2}\phi^{2,3}-\phi^{2,1}\phi^{3,2} \\
				i(\phi^{2,1}-\phi^{1,2})& -\frac{4}{3}+2\phi^{1,2}\phi^{2,1} +2\phi^{2,3}\phi^{3,2}&  i(\phi^{2,3}-\phi^{3,2})\\
				-\phi^{1,2}\phi^{2,3}-\phi^{2,1}\phi^{3,2}  & i(\phi^{2,3}-\phi^{3,2})& \frac{2}{3}-2\phi^{2,3}\phi^{3,2}
			\end{pmatrix} \; .
			\label{eq:Q.of.phi}
	\end{eqnarray}
\end{widetext}
Similarly, from \Autoref{eq:dipole.in.terms.of.A}, to $\mathcal{O}(\phi^2)$
we find
\begin{eqnarray}
	\opsb{}(\vec\phi)&=& \begin{pmatrix}
			-\phi^{2,3}-\phi^{3,2}\\
			i(\phi^{1,2}\phi^{2,3}-\phi^{2,1}\phi^{3,2})\\
			\phi^{1,2}+\phi^{2,1}
		\end{pmatrix}
		\; .
\label{eq:S.of.phi}
\end{eqnarray}


From these results we understand that fluctuations introduce a small 
imaginary part to the director ${\bf d}$, parallel to either the x-- or the the z--axis.
This leads to a rotation of the quadrupole moment about either 
the z-- or the the x--axis, and simultaneously introduces a dipole 
moment along that axis of rotation.
These changes are clearly visible in \Autoref{fig:gen_fluct}, 
where the orientation of the quadrupole moment is shown as a red bar.
Meanwhile, the dipole moment induced by each fluctuation is indicated  
with a blue arrow, and is also visible as a (small) distortion of the 
spin--probability distribution.


We are now in a position to derive a Hamiltonian describing fluctuations   
about FQ order.  
Substituting \Autoref{eq:A_phi} in \Autoref{eq:H.BBQ.3}, we find
\begin{align}
		\Ham^\prime_{\sf BBQ} &= E_0 + \frac{1}{2} \sum_{\k} 
		\left[ \vec{\phi}_{\k}^T M_{\k} \vec{\phi}_{-\k} \right]
		+	\O(\phi^4) \; , 
		\label{eq:H.prime}
\end{align}
	where the energy of the MF ground state is 
	\begin{eqnarray}
		E_0 &=& N z J_2  \; ,
		\label{eq:E0}
	\end{eqnarray}
	and fluctuations are described by 
	\begin{eqnarray}
		\vec{\phi}_{\k}  =
		\begin{pmatrix}
			\phi_{\k}^{2,1}\\
			\phi_{\k}^{1,2}\\
			\phi_{\k}^{3,2}\\
			\phi_{\k}^{2,3}\\
		\end{pmatrix}
		= \frac{1}{\sqrt{N}}\sum_{i} 
		\begin{pmatrix}
			e^{i \k \cdot \r_i} \phi^{2,1}_{\r_i} \\
			e^{i \k \cdot \r_i} \phi^{1,2}_{\r_i} \\
			e^{i \k \cdot \r_i} \phi^{3,2}_{\r_i} \\
			e^{i \k \cdot \r_i} \phi^{2,3}_{\r_i} 
		\end{pmatrix} \; ,
		\label{eq:abFT} 
	\end{eqnarray}
	with energy determined by a matrix 
	\begin{eqnarray}
		M_{\k} &=& 
		\begin{pmatrix}
			A_{\k} & -B_{\k} &  0& 0 \\
			-B_{\k} &A_{\k} &  0& 0 \\
			0 &  0 & A_{\k} &- B_{\k}  \\
			0 & 0 & -B_{\k} & A_{\k} \\
		\end{pmatrix} \; , 
		\label{eq:M}
	\end{eqnarray}
	for which
	\begin{subequations}
		\begin{eqnarray}
			A_\k &=& z(J_1\gamma(\k)-J_2)  \; , \\
			B_\k &=& z\gamma(\k)(J_2-J_1 ) \; ,
		\end{eqnarray} \label{eq:DefAB}
	\end{subequations}
	with lattice structure factor 
	\begin{eqnarray}
		\gamma(\k) &=& \frac{1}{z} \sum_{\boldsymbol\delta} e^{-i\k\cdot \boldsymbol\delta}  \; .
		\label{eq:def_gamma}
	\end{eqnarray}
	For the triangular lattice, the lattice coordination number \mbox{$z=6$}, 
	and the vectors  which connect neighbouring lattice 
	sites, $\{ \boldsymbol\delta \}$, are listed in  \hyperref[sec:conv_trig_lat]{Appendix~}\ref{sec:conv_trig_lat}.
We note also that the transpose vector for the fluctuations has the property 
	\begin{align}
		\phi_{\k}^{T\mu,\nu}&=\phi_{\k}^{\nu,\mu} \; .
	\end{align}
	This implies
\begin{eqnarray}
		\vec{\phi}_{\k}^T =
		\begin{pmatrix}
			\phi_{\k}^{1,2},&
			\phi_{\k}^{2,1},&
			\phi_{\k}^{2,3},&
			\phi_{\k}^{3,2}&
		\end{pmatrix} \; .
		\label{eq:phiT}
\end{eqnarray}

	The Hamiltonian $\Ham^\prime_{\sf BBQ}$ [\Autoref{eq:H.prime}] describes all 
	possible fluctuations about FQ order at a Gaussian (i.e. non--interacting) level, 
	and can be used as a starting point for both classical and quantum theories 
	of its excitations.
	The absence of terms linear in $\phi$ in  \Autoref{eq:H.prime}  
	confirms that the MF ground state, \Autoref{eq:FQyGS}, minimises energy, 
	and is therefore a valid starting point for 
	describing FQ order.

\subsection{Classical low--temperature expansion} 
\label{section:low.T}

	We now use the results of \Autoref{section:small.fluctuations} to develop 
	a classical theory of thermal fluctuations about FQ order at low temperature.
	From this we can calculate thermodynamic quantities in a form suitable 
	for comparison with classical Monte Carlo simulation.
	Results will be quoted to linear order in $T$ (i.e. quadratic in fluctuations).

\subsubsection{Expression for free energy}
\label{sec:Class_fluct}

	Within the framework of \Autoref{section:small.fluctuations}, fluctuations 
	about FQ order can be described by the partition function  
	\begin{eqnarray}
		Z_0  &=& \int \dd \vec{\phi}_{\k}\ 
		\exp{ -\beta \mathcal{H}^\prime_{\sf BBQ} [\vec{\phi_{\k}}] }  \; ,
		\label{eq:defZ} 
	\end{eqnarray}
	where the measure of integration is 
	\be
	\dd \vec{\phi}_{\k}=\dd \phi^{1,2}_{\k}\dd \phi^{2,1}_{\k}\dd \phi^{2,3}_{\k} \dd \phi^{3,2}_{\k}\; ,
	\ee
	the inverse temperature 
	\be
	\begin{matrix}
		\beta=\frac{1}{k_B T} \; ,
	\end{matrix}
	\label{eq:def_beta}
	\ee
	and $\mathcal{H}^\prime_{\sf BBQ} [\vec{\phi_{\k}}] $ is defined through 
	\Autoref{eq:H.prime}.  
	Neglecting ${\mathcal O}(\phi^4)$ terms, we find
	\begin{subequations}
		\begin{eqnarray}
			Z_0 &=&  \prod_{\k}^{N}\int \exp{-\beta\frac{1}{2}
				\vec{\phi}_{\k}^T M_{\k} \vec{\phi}_{-\k}} \exp{-\beta \frac{E_0}{N}} \dd \vec{\phi}_{\k}  \\
			&=&  \exp{-\beta E_0} \prod_{\k}^{N}\left[\sqrt{\frac{(2\pi)^n}{\beta^n \det M_{\k}}}\right]  \; , 
			\label{eq:Z0}
		\end{eqnarray}
	\end{subequations}
	where $E_0$ is defined through \Autoref{eq:E0}, the $4 \times 4$ matrix $M_{\k}$ through \Autoref{eq:M}, 
	$N$ is the number of lattice sites and $n$ is the dimension of  $M_{\k}$ (in this case, \mbox{$n=4$}).
	It follows that the free energy per site is 
	\begin{align}
		f_0 &=-\frac{\log(Z_0)}{\beta N} \nonumber\\
		&=\frac{E_0}{N}+\frac{k_B T}{2 N}\sum_{\k}\sum_{\lambda=1}^{N_{\lambda}}	
		\log(\frac{\omega_{\k,\lambda}}{2\pi k_{\sf B} T}) + {\mathcal O}(T^2) \; ,
		\label{eq:f0}
	\end{align}
	where $\omega_{\k,\lambda}$ are the eigenvalues of $M_{\k}$, 
	and we have used the fact that 
	\be
	\log [\det M_{\k}] = \text{Tr} \log M_{\k} = \sum_{\lambda=1}^4 \log \omega_{\k,\lambda} \; .\label{eq:log.det.M}
	\ee


	The free energy, \Autoref{eq:f0}, represents the first term in a  
	classical low--temperature expansion of the thermodynamic properties
	of the BBQ model. 
	To ${\mathcal O} (T)$, these are completely conditioned by the 
	solutions of the eigensystem 
	\be
	M_{\k} \bf{v}_{\k,\lambda} = \omega_{\k,\lambda} \bf{v}_{\k, \lambda} \; .
	\ee
	Working in the basis 
	\be 
	\{\phi_{\k}^{2,1},~\phi_{\k}^{1,2},~\phi_{\k}^{3,2},~\phi_{\k}^{2,3}\} \; , 
	\ee
	we find the eigenvalues 
	\begin{subequations}
		\begin{eqnarray}
			\omega^+_\k &=& \omega_{\k,1}^{} = \omega_{\k,3}^{} = A_\k + B_\k \; ,\label{eq:omega_vec_+} \\	
			\omega^-_\k &=& \omega_{\k,2}^{} = \omega_{\k,4}^{} =  A_\k - B_\k \; , \label{eq:omega_vec_-}
		\end{eqnarray}\label{eq:omega_vec_1}
	\end{subequations}
	with associated eigenvectors 
	\be
	\begin{matrix}
		v_{1}^{}=	\frac{1}{\sqrt{2}}\begin{pmatrix}
			-1 \\ 1\\ 0\\ 0
		\end{pmatrix}&,~&v_{2}^{}=	\frac{1}{\sqrt{2}}\begin{pmatrix}
			1 \\ 1 \\ 0 \\ 0
		\end{pmatrix}&,\\
		v_{3}^{}=	\frac{1}{\sqrt{2}}\begin{pmatrix}
			0 \\ 0 \\-1 \\ 1
		\end{pmatrix}&,~&v_{4}^{}=	\frac{1}{\sqrt{2}}\begin{pmatrix}
			0 \\ 0 \\ 1 \\ 1
		\end{pmatrix}&\; . \end{matrix}
	\label{eq:omega_vec_2} 
	\ee
	For quadrupolar order $\sim \ket{y}$, $v_{1}^{}$ and  $v_{2}^{}$ are associated 
	with rotations of quadrupole moments about the $z$--axis, while  $v_{3}^{}$ 
	and $v_{4}^{}$ are associated with rotations about the $x$--axis 
	[cf. \Autoref{fig:gen_fluct}].


	By construction, to ${\mathcal O}(\phi^2)$, the Hamiltonian 
	$\mathcal{H}^\prime_{\sf BBQ}$ is diagonal in the basis
	\begin{align}
		\vec{v}^{T}_{\k}&= (
		v_{\k,1},
		v_{\k,2},
		v_{\k,3},
		v_{\k,4},)
		\label{eq:diagonal.basis}\; ,
	\end{align}
	and can be written 
	\begin{align*}
		\Ham^\prime_{\sf BBQ} 
		&= E_0 + \frac{1}{2} \sum_{\k} \vec{v}_{\k}^{T} \tilde{M}_{\k} \vec{v}_{-\k} + \O(\vec{v}_{}^4)\\
		& = E_0 + \frac{1}{2} \sum_{\k} \sum_{\lambda=1}^4 
		\omega_{\k,\lambda} v_{\k,\lambda}^{T} v_{-\k,\lambda} 
		+ \O(\vec{v}_{}^4) \; , 
		\numberthis\label{eq:HBBQ_diag}
	\end{align*}
	where
	\begin{align}
		\tilde{M}_{\k}=O^TM_{\k}O &= 
		\begin{pmatrix}
			\omega_{\k,1}& 0&0 &0\\
			0& \omega_{\k,2}&0 &0\\
			0&0& \omega_{\k,3}&0\\
			0&0& 0 &\omega_{\k,4}
		\end{pmatrix}\; . 
		\label{eq:orthogonal.transformation2_M}
	\end{align}
	and the orthogonal transformation $O$ is defined by 
	\be
	\begin{matrix}
		\begin{pmatrix}
			\phi_{}^{2,1}\\
			\phi_{}^{1,2}\\
			\phi_{}^{3,2}\\
			\phi_{}^{2,3}\\
		\end{pmatrix}&=O\begin{pmatrix}
			v_{1}\\
			v_{2}\\
			v_{3}\\
			v_{4}\\
		\end{pmatrix}
		&\rm{where} & 
		O=	\frac{1}{\sqrt{2}}\begin{pmatrix}
			-1 & 1 & 0 & 0 \\
			1 & 1 & 0 & 0\\
			0 & 0 &-1 & 1 \\
			0 & 0 & 1 & 1 
		\end{pmatrix}
		\label{eq:orthogonal.transformation}
	\end{matrix} \; . 
	\ee
	%
	%
Of necessity, eigenmodes form an orthonormal set
\be
	v_{\k,\lambda}^{T} v_{-\k,\lambda'}  = \delta_{\lambda\lambda'} \; .
\label{eq:orthogonality.of.vk}
\ee
This coordinate system will prove useful in the subsequent calculation of 
correlation functions and ordered moments, described below.

\subsection{Calculation of thermodynamic quantities}
\label{section:theory.classical.thermodynamics}

Starting from the free energy $f_0$ [\Autoref{eq:f0}], it is possible to calculate all 
thermodynamic properties of the FQ state 
as the leading term in a perturbative expansion about $T=0$.
This can be accomplished by taking appropriate (functional) 
derivatives of the free energy.

\subsubsection{Heat capacity}
\label{sec:classal.heat.capacity}

The simplest thermodynamic property we can consider is the specific heat
	\be
	c_v = \frac{C_v}{N} = -T\left(\frac{\ddp^2 f_0}{\ddp T^2}\right)_{V} \; .
	\label{eq:Defcv}
	\ee
\nic{	
In the limit $T\to 0$, this is controlled by the classical limit 
of the equipartition theorem, which implies that each quadratic 
mode contributes $k_B/2$ to 
$c_v(T \to 0)$ \cite{Moessner1998-PRB58,Zhitomirsky2008,Shannon2010}}.
In the present case
\be
	c_v = -T \frac{-k_BN_{\lambda}}{2T} = k_B \frac{N_\lambda}{2}
\label{eq:cv}\; ,
\ee
where 
\be
	N_{\lambda} = 4\; ,
\label{eq:N_lambda}
\ee 
counts the number of normal modes accessible per \mbox{spin--1} 
moment [cf. \Autoref{fig:gen_fluct}].
It follows that, in the limit $T \to 0$, 
	\be
	c_v \to 2  \qquad  \text{[$u(3)$ matrix]}
	\; ,
	\label{eq:cv.U3} 	
	\ee
where, for simplicity, we  set
	\be
	k_B = 1 \; .
	\label{eq:kB}
	\ee
This should be contrasted with the usual result for classical fluctuations 
about an ordered state composed of $O(3)$ vectors: 
Here, at the level of a single spin, only two orthogonal fluctuations  
are possible, and so $N_{\lambda} = 2$ \cite{Moessner1998-PRB58,Zhitomirsky2008,Shannon2010}:
	\be
	c_v \to 1  \qquad  \text{[$O(3)$ vector]}
	\; .
	\label{eq:cv.O3} 	
	\ee
The zero--temperature limit of specific heat will prove important in the 
interpretation of the results of Monte Carlo simulation, as 
discussed in \Autoref{section:numerics}.

\subsubsection{Structure factors: general considerations}
\label{sec:classical.structure.factors}

We now turn to the calculation of the structure factors associated with 
dipole moments, quadrupole moments, and A--matrices.   
To facilitate this, it will prove useful to introduce source terms 
\begin{eqnarray}
	{\mathcal H} = {\mathcal H}_{\bf BBQ} + \Delta \Ham [{h}^\alpha_{i,\beta}] \; ,
\label{eq:Hfield}
\end{eqnarray}
where 
\begin{eqnarray}
		\Delta {\mathcal H} [{h}^\alpha_{i,\beta}] 
		&=& -\sum_{i,\alpha} {h}^\alpha_{i,\beta} \hat{O}^{\alpha}_{i,\beta} \; , 
\label{eq:DeltaH}
\end{eqnarray}
describes the coupling of a ficticious field ${h}^\alpha_{i,\beta}$
to the observable  
\begin{eqnarray}
		\hat{O}^{\alpha}_{i,\beta} 
		&\rightarrow& \ops{i}{\alpha}\ \delta_{\alpha\beta} 
		\; , \; \opq{i}{\alpha\beta} 
		\; , \; \opa{i}{\alpha}{\beta} \; .
\label{eq:O.definition}
\end{eqnarray}

%
Calculations proceed by expanding the observable $\hat{O}^{\alpha}_{i,\beta}$ 
in terms of the orthogonal eigenmodes $v_{\k,\lambda}$ [\Autoref{eq:orthogonal.transformation}], 
and calculating thermodynamic averages through functional derivatives of the 
free energy [\Autoref{eq:f0}] with respect to ${h}^\alpha_{i,\beta}$.


Details of these calculations, which involve contributions from both the 
ground state and thermal excitations, are given in \hyperref[sec:structure_factors_classical]{Appendix~}\ref{sec:structure_factors_classical}.
Where we come to compare with a quantum theory in \Autoref{section:quantum.theory}, 
it will also prove useful to introduce a spectral decomposition of the structure factors,  
which resolves contributions from eigenmodes at different energies.
These are defined in \Autoref{eq:SOF_class_q_w3}.


In what follows, we list the results needed for subsequent comparison with 
numerics in \Autoref{section:numerics}.



\begin{figure*}[]
	\begin{center}
		\begin{minipage}[b]{\textwidth}
			\subfloat[	$S_{\rm{S}}^{\sf CL}(\q, \omega)$	\label{fig:Ana_Cl_Sqw}]{
				\includegraphics[width=0.3\linewidth]{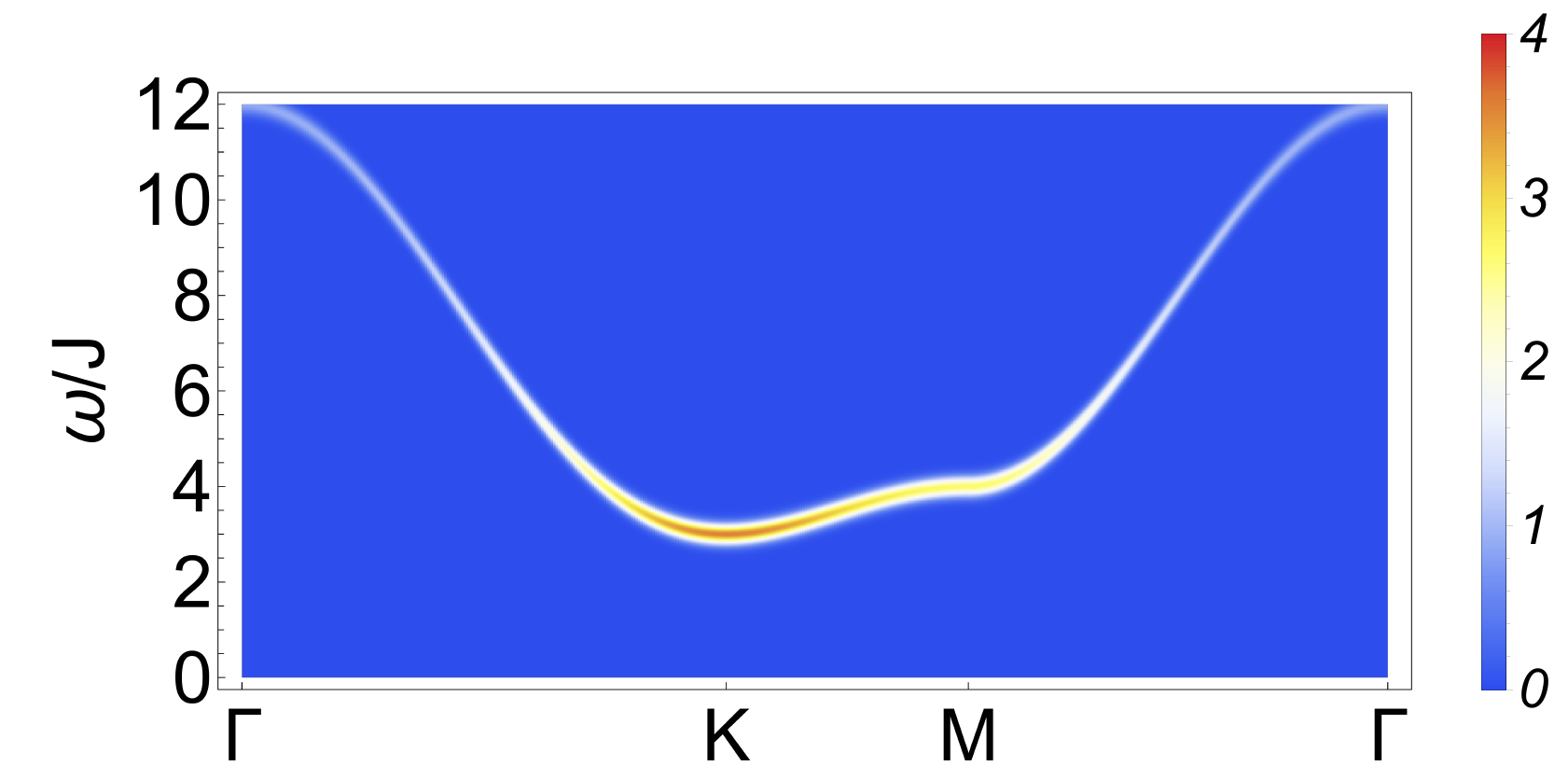}
			}
			\subfloat[$S_{\rm{Q}}^{\sf CL}(\q, \omega)$		\label{fig:Ana_Cl_Qqw}]{
				\includegraphics[width=0.3\linewidth]{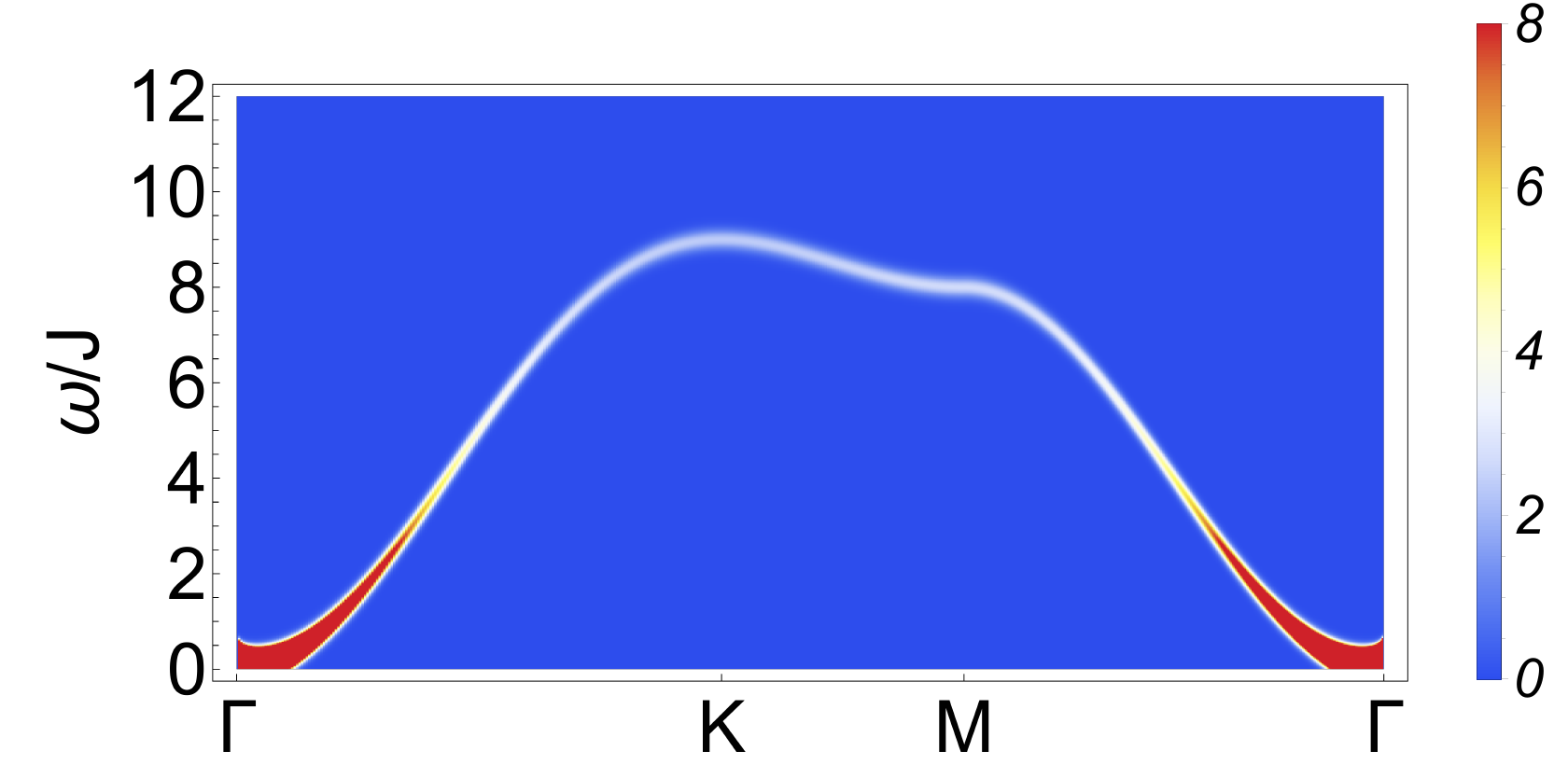}
			}
			\subfloat[$S_{\rm{A}}^{\sf CL}(\q, \omega)$		\label{fig:Ana_Cl_Aqw}]{
				\includegraphics[width=0.3\linewidth]{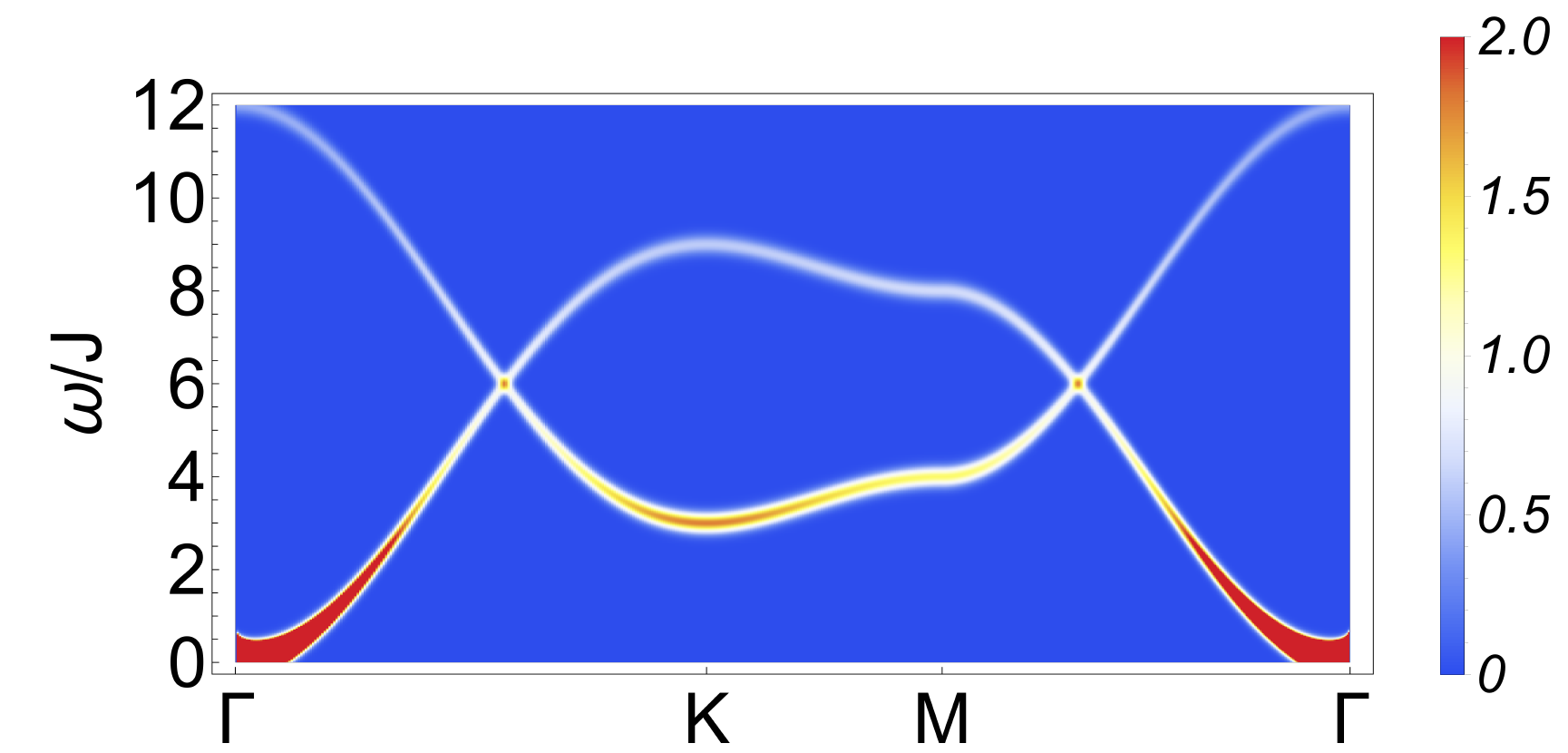}`
			}\\
			\subfloat[$S_{\rm{S}}^{\sf QM}(\q, \omega)$		\label{fig:Ana_FW_Sqw}]{
				\includegraphics[width=0.3\linewidth]{Ana_FW_Sqw}
			}
			\subfloat[ $S_{\rm{Q}}^{\sf QM}(\q, \omega)$		\label{fig:Ana_FW_Qqw}]{
				\includegraphics[width=0.3\linewidth]{Ana_FW_Qqw}
			}
			\subfloat[$S_{\rm{A}}^{\sf QM}(\q, \omega)$	\label{fig:Ana_FW_Aqw}]{
				\includegraphics[width=0.3\linewidth]{Ana_FW_Aqw}
			}
		\end{minipage}
	\end{center}
\caption{
Comparison of band--like excitations found in classical and quantum theories 
of fluctuations about a Ferroquadrupolar (FQ) ground state.
\protect\subref{fig:Ana_Cl_Sqw} Spectral representation of structure factor associated with dipole moments,
$S^{\sf CL}_{\rm{S}} (\q, \omega)$ [\protect\Autoref{eq:SF_class}], 
within classical low--temperature expansion of \protect\Autoref{section:classical.theory}, 
at a notional temperature $T = J$. 
\protect\subref{fig:Ana_Cl_Qqw} Equivalent results for quadrupole moments,
$S^{\sf CL}_{\rm{Q}}(\q,\omega)$ [\protect\Autoref{eq:QF_class}].
\protect\subref{fig:Ana_Cl_Aqw} Equivalent results for A--matrices, 
$S^{\sf CL}_{\rm{A}}(\q,\omega)$ [\protect\Autoref{eq:AF_class}].
\protect\subref{fig:Ana_FW_Sqw} Dynamcial structure factor associated with dipole moments,
$S^{\sf QM}_{\rm{S}} (\q, \omega)$ [\protect\Autoref{eq:Sqw_qm}], 
within \mbox{$T=0$} quantum theory.
\protect\subref{fig:Ana_FW_Qqw} Equivalent results for 
quadrupole moments, $S^{\sf QM}_{\rm{Q}}(\q, \omega)$ [\protect\Autoref{eq:Qqw_qm}].
\protect\subref{fig:Ana_FW_Aqw} Equivalent results for 
A--matrices, $S_{\rm{A}}^{\sf QM}(\q, \omega)$ [\protect\Autoref{eq:Aqw_qm}].
Comparing classical and quantum results, we see that the dispersion 
found in the quantum case is the geometric mean of the two different 
dispersions found for dipolar and quadrupolar excitations in the classical case.
All results are shown for parameters \protect\Autoref{eq.band.plot.parameters}, 
and have been convoluted with Gaussian of FWHM~$0.35$~J.
Bragg peaks have been omitted for simplicity.
Details of the quantum theory are given in \protect\Autoref{section:quantum.theory}.
}
\label{fig:comparison.classical.quantum.dispersion}
\end{figure*}

\subsubsection{Structure factor for dipole moments}
\label{sec:SDSF}

We first consider the structure factor for dipole moments of spin
\begin{eqnarray}
		S^{\sf CL}_{\rm{S}}(\q) 
		&=& \sum_{\alpha} \nn{\ops{\q}{\alpha} \ops{-\q}{\alpha}}  \; .
\end{eqnarray}
Within the classical low--temperature expansion, we find 
\begin{eqnarray}
	S^{\sf CL}_{\rm{S}} (\q) 
	&=& 	\frac{4}{\beta} \frac{1}{\omega^-_\q} + {\mathcal O}(T^2)\; ,
\label{eq:classical.structure.factor.S}
\end{eqnarray}
where $\omega^\pm_\q$ are defined through \Autoref{eq:omega_vec_1}.
Because the FQ phase does not break time--reversal symmetry, all ground--state 
averages of dipole moments vanish.
None the less, fluctuations restore a finite value of $S^{\sf CL}_{\rm{S}} (\q)$ 
at finite temperature.
The absence of terms in $\omega^+_\q$ reflects the fact that only 
the ``odd'' modes $\lambda =2,4$ contribute to dipolar fluctuations.


The spectral decomposition of the structure factor, \protect\Autoref{eq:classical.structure.factor.S}, 
is given by 
\begin{eqnarray}
S^{\sf CL}_{\rm{S}} (\q, \omega) 
	&=& 	\frac{4}{\beta} \frac{1}{\omega^-_\q} \delta( \omega - \omega^-_\q )
	+ {\mathcal O}(T^2) \; .
\label{eq:SF_class}
\end{eqnarray}
This is plotted in \Autoref{fig:comparison.classical.quantum.dispersion}~\subref{fig:Ana_Cl_Sqw}, for parameters
\begin{eqnarray}
	J_1 = 0.0 \; , \; J_2 = -1.0 \; ,
\label{eq.band.plot.parameters}
\end{eqnarray}
consistent with a FQ ground state, at a notional temperature $T/J =1$.
Within the classical theory, excitations with a dipolar character form a gapped, 
dispersing band, with spectral weight concentrated at $\q = {\sf K}$.


Further details of these calculations can be found in \mbox{\hyperref[sec:structure_factors_classical_S]{Appendix~}\ref{sec:structure_factors_classical_S}} [$\q\neq0$] and \mbox{\hyperref[sec:structure_factors_classical_Sq0]{Appendix~}\ref{sec:structure_factors_classical_Sq0}} [$\q=0$].

\subsubsection{Structure factor for quadrupole moments}
\label{sec:SQSF}

Next we consider the structure factor for quadrupole moments 
	\begin{eqnarray}
		S^{\sf CL}_{\rm{Q}}(\q) 
		&=& \sum_{\alpha\beta} 
		\nn{\opq{\q}{\alpha\beta} \opq{-\q}{\beta\alpha}} \; ,
	\end{eqnarray}
where the scalar contraction accomplished by the sum on $\alpha$, $\beta$ respects 
$SU(2)$ symmetry.
%
%
We obtain 
\begin{eqnarray}
	S^{\sf CL}_{\rm{Q}}(\q) 
	&=& \frac{8}{\beta} \frac{1}{\omega^+_{\q}}  (1 - \delta_{\q,0}) \nonumber\\ 
		 && \;  + \frac{8}{3} \left[ N - \frac{\Delta}{\beta} \right] \delta_{\q,0} 
		 + {\mathcal O}(T^2) \; ,
\label{eq:classical.structure.factor.Q}
\end{eqnarray}
where 
\begin{eqnarray}
	\Delta &=& 3 \sum_{\k\neq 0} \left[ \frac{1}{\omega^+_\k}  
	+ \frac{1}{\omega^-_\k} \right]+\frac{1}{\omega^-_{\q}}  \; .
\label{eq:delta}
\end{eqnarray}	
The structure factor for quadrupole moments is sensitive to the FQ ground state, 
and the  term $\Delta/\beta$ describes corrections to ground--state averages for $T>0$.
The absence of terms in $\omega^-_\q$ for $q\neq 0$ reflects the fact that only 
the ``even'' modes $\lambda =1,3$ contribute to quadrupolar fluctuations.
However all four modes, $\lambda =1,2,3,4$, contribute to the reduction of the 
ordered moment, through \Autoref{eq:delta}.

The spectral decomposition of the structure factor, \protect\Autoref{eq:classical.structure.factor.Q}, 
is given by
\begin{eqnarray}
	S^{\sf CL}_{\rm{Q}}(\q,\omega) 
		&=& \frac{8}{\beta} \frac{1}{\omega^+_{\q}}  
		(1 - \delta_{\q,0}) \delta(\omega - \omega^+_\q)	\nonumber\\
		&+& \frac{8}{3} \left[ N - \frac{\Delta}{\beta} \right]  
		\delta_{\q,0} \delta(\omega)+{\mathcal O}(T^2) \; .
\label{eq:QF_class}
\end{eqnarray}
This is illustrated in \Autoref{fig:comparison.classical.quantum.dispersion}~\subref{fig:Ana_Cl_Qqw}, where the Bragg peak 
at $\q = \Gamma$ has been suppressed for simplicity.
Within a classical theory, excitations with a quadrupolar character form a gapless  
dispersing band, with spectral weight concentrated at $\q = 0$.


Further details of these calculations are given in 
\mbox{\hyperref[sec:structure_factors_classical_Q]{Appendix~}\ref{sec:structure_factors_classical_Q}} [$\q\neq0$]
and 
\mbox{\hyperref[sec:structure_factors_classical_Qq0]{Appendix~}\ref{sec:structure_factors_classical_Qq0}} [$\q=0$]. 

\subsubsection{Structure factor for A--matrix}
\label{sec:SASF}

We now turn to the structure factor for the most fundamental object describing 
the spins, the matrix $\opa{}{\alpha}{\beta}$.   
This is defined by 
\begin{eqnarray}
	S^{\sf CL}_{\rm{A}}(\q) 
		&=& \sum_{\alpha\beta} \nn{\opa{\q}{\alpha}{\beta} \opa{-\q}{\beta}{\alpha}}  \; ,
\end{eqnarray}
where the scalar contraction accomplished through the sum on $\alpha$, $\beta$ 
preserves the full $U(3)$ symmetry of the representation. 
To leading order in $T$, we find 
\begin{eqnarray}
S^{\sf CL}_{\rm{A}}(\q) 
	&=&   \frac{2}{\beta} \frac{1}{\omega^-_\q}
	+ \frac{2}{\beta} \frac{1}{\omega^+_\q} (1-\delta_{\q,0}) 
	\nonumber\\
	&& + \left[N - \frac{2}{3} \frac{\Delta}{\beta} \right] \delta_{\q,0} + {\mathcal O}(T^2) \; .
	\nonumber\\
\label{eq:classical.structure.factor.A}
\end{eqnarray}
This structure factor encompasses both dipoles and quadrupoles, 
and so is sensitive to FQ ground state order.
All four modes, $\omega^\pm_\q$, contribute to fluctuation terms for $\q \ne 0$.


The spectral decomposition of the structure factor, \protect\Autoref{eq:classical.structure.factor.A}, 
is given by
\begin{eqnarray}
		S^{\sf CL}_{\rm{A}}(\q,\omega) 
		&=&  \frac{2}{\beta}\frac{1}{\omega^-_\q} \delta(\omega - \omega^-_\q) 
		\nonumber\\
		&& + \frac{2}{\beta}\frac{1}{\omega^+_{\q}} (1-\delta_{\q,0}) \delta(\omega - \omega^+_\q)
		\nonumber\\
		&& +  \left[ 1 - \frac{2}{3} \frac{\Delta}{\beta} \right]\ \delta_{\q,0}\ \delta(\omega) +{\mathcal O}(T^2)\; .
		\label{eq:AF_class}
\end{eqnarray}
This is illustrated in \Autoref{fig:comparison.classical.quantum.dispersion}~\subref{fig:Ana_Cl_Aqw}, where the Bragg peak 
at $\q = \Gamma$ has been suppressed for simplicity.
Both dipolar and quadrupolar fluctuations are visible as independent, 
dispersing, bands in $S^{\sf CL}_{\rm{A}}(\q,\omega)$.
	
	
Further details of these calculations can be found in 
\mbox{\hyperref[sec:structure_factors_classical_A]{Appendix~}\ref{sec:structure_factors_classical_A}} and [$\q\neq0$] \mbox{\hyperref[sec:structure_factors_classical_Aq0]{Appendix~}\ref{sec:structure_factors_classical_Aq0}} [$\q=0$].

\subsubsection{Sum rule for structure factors}
\label{sec:sum.rule}

The sum rule associated with A--matrices,  \Autoref{eq:sum.rule.AQS}, implies that the structure
factors $S^{\sf CL}_{\rm{S}}(\q)$, $S^{\sf CL}_{\rm{Q}}(\q)$ and $S^{\sf CL}_{\rm{A}}(\q)$, 
must also satisfy a sum rule.
By Fourier transform of \Autoref{eq:sum.rule.AQS}, we find
\begin{align*}
		\opa{\k}{\alpha}{\beta}\opa{-\k}{\beta}{\alpha}= &\frac{1}{4} \opq{\k}{\alpha\beta}\opq{-\k}{\beta\alpha}+\sum_{\alpha}\frac{1}{2} \ops{\k}{\alpha}\ops{-\k}{\alpha}\\
		&+\frac{1}{12}s^2(s+1)^2N \delta_{\k,0}\numberthis
\label{eq:sum.rule.AQS_k} \; .
\end{align*}
It follows that 
\begin{align}
S^{\sf CL}_{\rm{A}}(\q) 
		&= \frac{1}{4} S^{\sf CL}_{\rm{Q}}(\q) 
		+  \frac{1}{2} S^{\sf CL}_{\rm{S}}(\q) 
		+  \frac{1}{3}N \delta_{\q,0} \; .
\label{eq:sum.rule.classical.structure.factorq0}
\end{align}
By direct substitution of \Autoref{eq:classical.structure.factor.A}, 
\Autoref{eq:classical.structure.factor.S} and \Autoref{eq:classical.structure.factor.Q}, 
it is easy to see that the results of the low--temperature expansion satisfy the 
sum rule \Autoref{eq:sum.rule.classical.structure.factorq0}.

\subsubsection{Ordered Moments}
\label{sec:ordered.moment}

Finally, we consider the quadrupole moment which characterises the FQ state, 
$\langle {\bf Q} \rangle$.
This is most easily calculated through the associated equal--time structure factor 
\be
	\langle {\bf Q} \rangle^2 = \frac{S^{\sf CL}_{\rm{Q}}(\q=\Gamma)}{N} 	\; ,
\ee
where $N$ denotes the number of lattice sites.
%
%
From \Autoref{eq:classical.structure.factor.Q}, we find 
\be
	\langle {\bf Q} \rangle^2
	\simeq \frac{8}{3}  - \frac{8}{N\beta}\sum_{\k\neq 0} 
		\left[ \frac{1}{\omega^+_\k} + \frac{1}{\omega^-_\k} \right] 
		+ {\mathcal O}(T^2) \; .
\label{eq:OM}
\ee
This is the result used for comparison with MC simulation in \Autoref{section:ordered.moment}.

\section{Quantum theory of fluctuations about a ferroquadrupolar ground state}
\label{section:quantum.theory}

We now construct a quantum theory of fluctuations about FQ order, starting 
from the $u(3)$ formalism introduced in \Autoref{section:maths.for.spin.1}.
First we show  how quantization of the fluctuations introduced in 
\Autoref{section:small.fluctuations} leads to a multiple--Boson expansion 
exactly equivalent to published ``flavour--wave'' theory \cite{Lauchli2006}. 
\nic{
As with the classical theory of \Autoref{section:low.T}, we treat 
fluctuations at a Gaussian level (i.e. one quadratic in bosons). }
%
%
In \Autoref{sec:quantum.structure.factors} we go on to provide 
explicit results for the dynamical structure factors associated with 
spin--dipole and quadrupole moments, and with the A--matrix describing 
fluctuations.

\subsection{Quantization of fluctuations}
\label{section:quantization.of.fluctuations}

	In \Autoref{section:small.fluctuations} we have shown that fluctuations about FQ order 
	can be fully described using four generators $\opa{}{1}{2}$, $\opa{}{3}{2}$, 
	$\opa{}{2}{1}$, $\opa{}{2}{3}$, which naturally form conjugate pairs [\Autoref{fig:gen_fluct}].
	It follows that fluctuations can be parameterized through two pairs of real fields 
	$(\phi^{1,2}, \phi^{2,1})$, and $(\phi^{2,3}, \phi^{3,2})$.   
	Once quantum dynamics are taken into account, low--energy fluctuations 
	must take the form of ``quadrupole waves'', which are the Goldstone modes of FQ order.
	These carry integer spin, and will be Bosons.
	And since each Boson must be described by a complex field, we anticipate that 
	the pairs of fields $(\phi^{1,2}, \phi^{2,1})$, and $(\phi^{2,3}, \phi^{3,2})$ will 
	combine to give a total of two Bosonic degrees of freedom per site.


	With these expectations in mind, we quantize fluctuations of each pair of 
	fields through the Bosonic commutation relations 
	\begin{subequations}
		\begin{eqnarray}
			[ \phi^{2,1}_i, \phi^{1,2}_j ] &=& \delta_{ij} \; ,
		\end{eqnarray}	
		\begin{eqnarray}
			[ \phi^{2,3}_i, \phi^{3,2}_j ] &=& \delta_{ij} \; .
		\end{eqnarray}	
	\end{subequations}
	We can now associate each field $\phi^{\alpha,\beta}$ 
	with a creation or annihilation operator
	\begin{subequations}
		\begin{eqnarray}
			\phi_i^{1,2} &=& (\phi_i^{2,1})^{\dagger} = -i \bad{i}  \; ,
		\end{eqnarray}	
		\begin{eqnarray}	
			\phi_i^{2,3} &=& (\phi_i^{3,2})^{\dagger} = ~i \bb{i} \; .
			\label{eq:fluct_qm}
		\end{eqnarray}	
		\label{eq:quantization.as.bosons}
	\end{subequations}
	In this basis, a general fluctuation about a state $\ket{y_i}$ 
	[\Autoref{eq:FQyGS}], can be written 
	\be
	\ophb{A}_i =
	\begin{pmatrix}
		\bad{i}\ba{i} & \bad{i} & \bad{i}\bb{i}\\
		\ba{i} & 1- \bad{i}\ba{i} - \bbd{i}\bb{i} & \bb{i}\\
		\bbd{i}\ba{i} & \bbd{i} & \bbd{i}\bb{i}
	\end{pmatrix} \; ,
	\label{eq:Acond2}
	\ee
	cf. \Autoref{eq:A_phi}.
	To quadratic order in Bosons, the BBQ model, \Autoref{eq:H.BBQ.3}, 
	then reads
	\be
	\Ham^\prime_{\sf BBQ} 
	= E_0 + \frac{1}{2} \sum_{\k} \left[ \opdx{w}{\k}{M}_{\k} \opx{w}{\k} \right] 
	+ \O(\vec{w}_{}^4) \; ,
	\label{eq:HBBQ_qm}
	\ee
	where 
	\begin{align}
		\opdx{w}{\k}=
		\begin{pmatrix}
			\bad{\k}, & \ba{-\k}, & \bbd{\k}, &\bb{-\k}
		\end{pmatrix}&\; ,
		\opx{w}{\k}=
		\begin{pmatrix}
			\ba{\k} \\
			\bad{-\k} \\
			\bb{\k} \\
			\bbd{-\k} \\
		\end{pmatrix} \; ,
		\label{eq:def_wboson}
	\end{align}
	with ground--state energy $E_0$ [\Autoref{eq:E0}], 
	and fluctuations conditioned by the same 
	matrix ${M}_{\k}$ [\Autoref{eq:M}] as appears in the 
	classical theory [\Autoref{eq:H.prime}].


	From the Bosonic commutation relations [\Autoref{eq:quantization.as.bosons}], 
	it follows that 
	\be
	\com{ \opw{w}{\k}{\alpha}, \opdw{w}{\q}{\beta} } 
	= {\gamma_0}_{\alpha}^{~\beta} \delta_{\k,\q} \; ,
	\label{eq:comrel}
	\ee
	where
	\begin{align}
		\gamma_0=\begin{pmatrix}
			1 & 0 & 0 & 0 \\
			0  & -1 & 0 & 0 \\
			0 & 0 & 1 & 0 \\
			0 & 0 &0  & -1 
		\end{pmatrix}\; .\label{eq:gamma_0}
	\end{align}
	%
	%
	The excitations described by these operators form bands, whose 
	dispersion can be found by solving the eigensystem
	\be
	\begin{matrix}
		\gamma_0 M_{\k} u_{\k, \lambda} = \epsilon_{\k,\lambda} u_{\k,\lambda} & ~&\lambda=1,2,3,4 
	\end{matrix} \; ,
	\label{eq:quantum.eigensystem}
	\ee
	with eigenvectors $u_{\lambda,\k}$, and associated eigenvalues $\epsilon_{\k,\lambda}$.
	This is equivalent to diagonalising the matrix
	\be
	\gamma_0 M_{\k}=
	\begin{pmatrix}
		A_{\k}& -B_{\k} &  0& 0 \\
		B_{\k} &-A_{\k} &  0& 0 \\
		0 &  0 & A_{\k} &- B_{\k}  \\
		0 & 0 &  B_{\k} & -A_{\k}\\
	\end{pmatrix} \; ,
	\label{eq:quantum.eigensystem2}
	\ee
	where $A_{\k}$, $B_{\k}$ are defined through \Autoref{eq:DefAB}.
	This is a task which can, if necessary, be performed numerically.
	But in the present case, closed--form analytic solution is possible, 
	and we find
	%
	\be
	\epsilon_{\k,1} = -\epsilon_{\k,2} = \epsilon_{\k,3} 
	= -\epsilon_{\k,4} = +\sqrt{A_{\k}^2-B_{\k}^2} \; .
	\label{eq:omega_qm}
	\ee
	Of these, only the two solutions with positive energy, $\epsilon_{\k,1}$ 
	and $\epsilon_{\k,3}$, correspond to physical modes of the system, and 
	so we have a total of two Bosonic modes per site, as anticipated.
	Further details of this calculation are given in \hyperref[sec:bogolioubov_transfomation]{Appendix~}\ref{sec:bogolioubov_transfomation}.


	The solution of the quantum eigensystem, \Autoref{eq:quantum.eigensystem}, 
	is equivalent to performing generalised Bogoliubov transformation between 
	the original set of Bosons, \Autoref{eq:quantization.as.bosons}], 
	and a new set of Bosonic operators 
	\be
	[\balpha{\k}{}, \balphad{\k^\prime}{}]  = [\bbeta{\k}{}, \bbetad{\k^\prime}{}] = \delta_{\k{\k^\prime}} \; ,
	\ee
	which diagonalize the Hamiltonian.
	These are defined through 
	\be
	\opdx{w}{\k} = \frac{1}{\sqrt{\Delta_{\k}^2- B_{\k}^2}}
	   \begin{pmatrix}
		\Delta_{\k}&-B_{\k}& 0& 0\\
		-B_{\k} &\Delta_{\k}& 0& 0\\
		0& 0 &	\Delta_{\k}&-B_{\k}\\
		0& 0 &-B_{\k} &\Delta_{\k}\\
	\end{pmatrix}\opdx{u}{\k}\; ,
	\label{eq:v_e}
	\ee
	where
	\be
	\begin{matrix}
		\opdx{w}{\k}=
		\begin{pmatrix}
			\ba{\k}{}\\
			\bad{-\k}{}\\
			\bb{\k}{}\\
			\bbd{-\k}{}\\
		\end{pmatrix}
		&; &\opdx{u}{\k}=
		\begin{pmatrix}
			\balpha{\k}{}\\
			\balphad{-\k}{}\\
			\bbeta{\k}{}\\
			\bbetad{-\k}{}\\
		\end{pmatrix}&;& \Delta_{\k}=A_{\k} +\sqrt{A_{\k}^2-B_{\k}^2} \label{eq:Delta}
	\end{matrix}\; .
	\ee


In this new basis, the Hamiltonian can be written 
\begin{eqnarray}
	\Ham^\prime_{\sf BBQ} 
		&=& E_0 + \Delta E_0  
		+\sum_{\k} \epsilon(\k)  \left[\balphad{\k}\balpha{\k} +  \bbetad{\k}\bbeta{\k} \right]
		\nonumber\\
		&& + \text{[higher order terms]} \; ,
\label{eq:H.quantum}
\end{eqnarray}
where
\begin{eqnarray}
	\epsilon(\k) = \sqrt{A_{\k}^2-B_{\k}^2}\; ,
\label{eq:omega.k}
\end{eqnarray}
and 
\begin{eqnarray}
	\Delta E_0 &=& \sum_\k A_{\k} + \epsilon(\k) \; ,
\end{eqnarray}
represents the contribution to the ground state energy 
coming from the zero--point fluctuations, and $E_0$ is the ground state energy 
given in \Autoref{eq:E0}.
Written in this form, the result is exactly equivalent to that found in an earlier, linear 
``flavour wave'' treatment of FQ order \cite{Lauchli2006}, obtained through 
condensation of Schwinger Bosons \cite{Penc2011-Springer}.

\subsection{Dynamical structure factors within zero--temperature quantum theory} 		
\label{sec:quantum.structure.factors}

From this starting point, it is a straightforward, if involved, exercise 
to calculate the dynamical structure factors which characterize the excitations of FQ order.
These have the form 
\begin{eqnarray}
	S^{\sf QM}_{\rm{O}} (\q, \omega) 
		= \int_{-\infty}^{\infty} \frac{\dd t}{2\pi} \exp{i\omega t} 
		\sum_{\alpha,\beta} \Av{ \hat{O}_{\q,\beta}^{\alpha}(t) \hat{O}_{-\q,\alpha}^{\beta}(0) }\; .
	\label{eq:SqOF}
\end{eqnarray}
where 
\begin{eqnarray}
	\hat{O}^{\alpha}_{\q,\beta} = \frac{1}{\sqrt{N}} \hat{O}^{\alpha}_{i,\beta} \exp{i \q {\bf r}_i} \; ,
\end{eqnarray}
and the operator $\hat{O}^{\alpha}_{~\beta}$ can reflect fluctuations of 
dipole moments, $\ops{}{\mu}$; quadrupole moments, $\opq{}{\mu\nu}$;
or the underlying representation of $u(3)$, $\opa{}{\mu}{\nu}$.
	
	
We evaluate dynamical structure factors at finite energy ($\omega > 0$) through the 
explicit calculation of matrix elements within a multiple--Boson expansion.
The structure of these calculations is described in  
in \hyperref[sec:structure_factors_quantum_qn0]{Appendix~}\ref{sec:structure_factors_quantum_qn0}.
Static structure factors ($\omega = 0$) can also be calculated through functional 
derivatives of the ground--state energy, in analogy with \Autoref{section:theory.classical.thermodynamics}.
Details of this approach are given in \hyperref[sec:structure_factors_quantum_q0]{Appendix~}\ref{sec:structure_factors_quantum_q0}.
	
	
Below, we sketch key results at \mbox{$T=0$} which are needed for 
subsequent comparison with numerics [\Autoref{section:numerics}],
and the exploration of the relationship between quantum and 
classical results [\Autoref{section:quantum.vs.classical}].

\subsubsection{Dynamical spin structure factor}

We consider first the dynamical spin structure factor 
	\be
	S^{\sf QM}_{\rm{S}} (\q, \omega) 
	= \int_{-\infty}^{\infty} \frac{\dd t}{2\pi} \exp{i\omega t} 
	\sum_\mu \Av{ \ops{\q}{\mu}(t) \ops{-\q}{\mu}(0) } \; .
	\label{eq:SqSF}
	\ee
Substituting \Autoref{eq:Acond2} in the expression for spin 
operators, \Autoref{eq:dipole.in.terms.of.A}, and keeping terms to linear order, we find 
	\begin{subequations}
		\begin{align}
			\ops{i}{x}&\simeq i(\bbd{i}-\bb{i})\; ,\\
			\ops{i}{y}&\simeq 0\; ,\\
			\ops{i}{z}&\simeq- i(\bad{i}-\ba{i})\; .
		\end{align}
	\end{subequations}
Performing a Fourier transform and using the Bogoliubov transformation \Autoref{eq:v_e}, we can express these as 
	\begin{subequations}
		\begin{align}
			\ops{\q}{x}& 
			\simeq i \xi_{\sf S}(\q) (\bbetad{-\q}-\bbeta{\q}) \label{eq:Sqm_alpha_x} \; , \\
			\ops{\q}{y}&\simeq 0\; ,\\
			\ops{\q}{z}& 
			\simeq -i \xi_{\sf S}(\q) (\balphad{-\q}-\balpha{\q})\label{eq:Sqm_alpha_z} \; , 
		\end{align}
	\end{subequations}
where $\xi_{\sf S}(\q)$ is the coherence factor
\be
	\xi_{\sf S} (\q) = \frac{\Delta_{\q}+B_{\q}}{\sqrt{\Delta_{\q}^2- B_{\q}^2}}\; .
\label{eq:coherence.factor}
\ee
From this starting point we can connect $S^{\sf QM}_{\rm{S}} (\q, \omega)$ 
directly with the multiple--Boson expansion of \Autoref{section:quantization.of.fluctuations}.


Since FQ order does not break time--reversal symmetry, static averages of dipole moments 
vanish, and all contributions to $S^{\sf QM}_{\rm{S}} (\q, \omega)$ come from excitations.
Evaluating these, we find 
\be
	S^{\sf QM}_{\rm{S}} (\q, \omega) 
	= 2\frac{\sqrt{A_{\q}+B_{\q}}}{\sqrt{A_{\q}-B_{\q}}} \delta(\omega-\omega_{\q}) \;, 
\label{eq:Sqw_qm}
\ee
leading to an equal--time structure factor 
\be
	S^{\sf QM}_{\rm{S}} (\q) 
	= \int d\omega\ S^{\sf QM}_{\rm{S}} (\q, \omega) 
	= 2 \frac{\sqrt{A_{\q}+B_{\q}}}{\sqrt{A_{\q}-B_{\q}}} \; .
\label{eq:S.equal.time.structure.factor}
\ee
\nic{
Using \Autoref{eq:omega.k}, \Autoref{eq:Delta} 
and \Autoref{eq:coherence.factor} we can show that 
\begin{eqnarray}
	\xi_{\sf S}^2(\q) = \frac{ \sqrt{A_\q + B_\q } }{ \sqrt{ A_\q - B_\q } } 
	\; ,
\label{eq:simplification}
\end{eqnarray}
and write \Autoref{eq:S.equal.time.structure.factor} as
\begin{eqnarray}
	S^{\sf QM}_{\rm{S}} (\q) 
	= 2 \xi_{\sf S}^2 (\q)
	\; .
\label{eq:Sqomega.simplified}
\end{eqnarray}
This is a fact we will return to in \Autoref{section:MD.deconstructed}.
}


\nic{
Further details of calculation of $S^{\sf QM}_{\rm{S}} (\q, \omega)$ 
are given in \hyperref[sec:structure_factors_quantum_Sq]{Appendix~}\ref{sec:structure_factors_quantum_Sq} for $\q \ne 0$, and in \hyperref[sec:structure_factors_quantum_Sq0]{Appendix~}\ref{sec:structure_factors_quantum_Sq0} for $\q=0$.   
}

\subsubsection{Dynamical quadrupole structure factor}

We now consider the dynamical structure factor associated with 
quadrupole moments 
\be
S^{\sf QM}_{\rm{Q}} (\q, \omega) 
	= \int_{-\infty}^{\infty} \frac{\dd t}{2\pi} \exp{i\omega t} 
	\sum_{\mu\nu}  \Av{ \opq{\q}{\mu\nu}(t) \opq{-\q}{\mu\nu}(0) } \; .
\label{eq:dynamical.quadrupole.structure.factor}
\ee
Following the same steps as for the spin--structure factor,  starting from \Autoref{eq:quadrupole.in.terms.of.A}, we find
\begin{align}
\ophb{Q}_i & \cong
		\begin{pmatrix}
			\frac{2}{3} & -\bad{i}-\ba{i} & 0\\
			-\bad{i}-\ba{i} & -\frac{4}{3} & -\bbd{i}-\bb{i} \\
			0 &  -\bbd{i}-\bb{i}  & \frac{2}{3}
		\end{pmatrix} \; .
\label{eq:Q_bosons}
\end{align}
After Fourier transform, and transcription into the Bogoliubov basis, this 
yields
\begin{align}
	\ophb{Q}_{\q}&\cong\nonumber\\
	&	\small{\begin{pmatrix}
		\frac{2}{3}\sqrt{N}\delta(\q)  &	\xi_{\sf{Q}}(\q) (\balphad{-\q}+\balpha{\q}) & 0\\
		\xi_{\sf{Q}}(\q) (\balphad{-\q}+\balpha{\q})& 	-\frac{4}{3}\sqrt{N}\delta(\q)& 		\xi_{\sf{Q}}(\q)(\bbetad{-\q}+\bbeta{\q})\\
		0 &  	\xi_{\sf{Q}}(\q)(\bbetad{-\q}+\bbeta{\q}) & 	\frac{2}{3}\sqrt{N}\delta(\q)
		\end{pmatrix}}\; ,
\end{align}
where $N$ is the number of sites, and the relevant coherence factor is given by 
\be
	\xi_{\sf{Q}}(\q) = \frac{B_{\q}-\Delta_{\q}}{\sqrt{\Delta_{\q}^2- B_{\q}^2}}\; .\label{eq:coherence.factor.quadrupole}
\ee
%


Quadrupole moments at $\q = 0$ take on a finite value in a FQ state, 
and both the ground state and excitations contribute to 
the structure factor $S^{\sf QM}_{\rm{Q}} (\q, \omega)$.   
Evaluating both, we find 
\begin{align}
S^{\sf QM}_{\rm{Q}}(\q, \omega)
		=& \frac{8}{3}N(1-\Delta^{\sf{QM}}) \delta(\q) \delta(\omega)\nonumber\\  
		&+ 4 \frac{\sqrt{A_{\q}-B_{\q}}}{\sqrt{A_{\q}+B_{\q}}} \delta(\omega-\omega_{\q}) \; ,
\label{eq:Qqw_qm}
\end{align}
where $\Delta^{\sf{QM}}$ is given by as
\be
\Delta^{\sf{QM}}
	=\frac{3}{N}\sum_{\k}\frac{A_{\k}}{\sqrt{A_{\k}^2-B_{\k}^2}}  \; . 
\label{eq:Delta_QM}
\ee
The corresponding equal--time structure factor is given by
\begin{align}
S^{\sf QM}_{\rm{Q}}(\q)
	= \frac{8}{3}N(1-\Delta^{\sf{QM}}) \delta(\q) 
	+ 4\frac{\sqrt{A_{\q}-B_{\q}}}{\sqrt{A_{\q}+B_{\q}}}  \; .
\label{eq:Q.equal.time.structure.factor}
\end{align}
%


Details of these calculations can be found in \hyperref[sec:structure_factors_quantum_Qq]{Appendix~}\ref{sec:structure_factors_quantum_Qq} and in \hyperref[sec:structure_factors_quantum_Qq0]{Appendix~}\ref{sec:structure_factors_quantum_Qq0} for $\q=0$.

\subsubsection{Structure factor for A matrices }

	The most fundamental objects in our theory are not dipoles or quadrupoles, 
	but the A--matrices which describe the quantum state of the \mbox{spin--1} moment.
	It is therefore useful to introduce a dynamical structure factor 
\be
S^{\sf QM}_{\rm{A}} (\q, \omega) 
	= \int_{-\infty}^{\infty} \frac{\dd t}{2\pi} \exp{i\omega t} 
	\sum_{\mu\nu} \nn{\opa{}{\mu}{\nu}(t) \opa{}{\nu}{\mu}(0)}  \; .
\label{eq:dynamical.A.matrix.structure.factor}
\ee
This structure factors captures all dynamics that can be resolved at the level 
of a two--point correlation function, regardless of how that dynamics is 
expressed in spin correlations.
Neglecting 2\nd order and higher terms, \Autoref{eq:Acond2} implies
\be
	\ophb{A}_i\cong
	\begin{pmatrix}
		0& \bad{i} & 0\\
		\ba{i} & 1 & \bb{i}\\
		0 & \bbd{i} & 0
	\end{pmatrix}\; .
\ee                 
Fourier transforming, and resolving non--zero matrix elements in 
terms of the Bogolibov basis \Autoref{eq:dynamical.A.matrix.structure.factor}, 
we find
\be
\ophb{A}_{\q}\simeq
\begin{pmatrix}
	\begin{array}{c c c c c c}
		\multicolumn{2}{c}{\multirow{2}{*}[-1.5ex]{0}}& \multirow{2}{*}{{\small$\xi_{\sf{A}}^{-}(\q)\balphad{-\q}$}}& \multicolumn{2}{c}{\multirow{2}{*}[-1.5ex]{0}}\\
		& & \multirow{2}{*}{{\small$-\xi_{\sf{A}}^{+}(\q)\balpha{\q} $}}& & \\
		\multirow{2}{*}{{\small$-\xi_{\sf{A}}^{+}(\q)\balphad{-\q}$ }}& \multicolumn{2}{c}{\multirow{2}{*}[-1.5ex]{$\sqrt{N}\delta_{\q, 0}$ }}&	\multirow{2}{*}{{\small$-\xi_{\sf{A}}^{+}(\q)\bbetad{-\q}$}}\\
		\multirow{2}{*}{{\small$+\xi_{\sf{A}}^{-}(\q)\balpha{\q}$}} & & &\multirow{2}{*}{{\small$+\xi_{\sf{A}}^{-}(\q)\bbeta{\q} $}}\\
		\multicolumn{2}{c}{\multirow{2}{*}[-1.5ex]{0}}& \multirow{2}{*}{{\small$\xi_{\sf{A}}^{-}(\q)\bbetad{-\q}$}}& \multicolumn{2}{c}{\multirow{2}{*}[-1.5ex]{0}}\\
		& & \multirow{2}{*}{{\small$-\xi_{\sf{A}}^{+}(\q)\bbeta{\q}$}}& & 
	\end{array}
\end{pmatrix}
\ee
where $N$ is the number of sites and $\xi_{\sf{A}}^{+}(\q)$ and $\xi_{\sf{A}}^{-}(\q)$ are the coherence factors for A--matrices defined as
\begin{subequations}
	\begin{align}
		\xi_{\sf{A}}^{+}(\q)= \frac{\xi_{\sf{S}}(\q)+\xi_{\sf{Q}}(\q)}{2}\; ,\\
		\xi_{\sf{A}}^{-}(\q)=\frac{\xi_{\sf{S}}(\q)-\xi_{\sf{Q}}(\q)}{2}\; ,
	\end{align}\label{eq:coherence.factor.amatrix}
\end{subequations}
where $\xi_{\sf{S}}(\q)$ and $\xi_{\sf{Q}}(\q)$ are defined in \Autoref{eq:coherence.factor} and \Autoref{eq:coherence.factor.quadrupole} respectively.


From this starting point, we can calculate all of the quantum averages
which enter into $S^{\sf QM}_{\rm{A}} (\q, \omega)$.
Like the structure factor for quadrupole moments, this 
entails contributions from both ground state and excitations.
Evaluating these, we find 
\begin{align}
S_{\rm{A}}^{\sf QM}(\q, \omega) 
	=& N(1-\frac{2}{3}\Delta^{\sf{QM}}) \delta(\q) \delta(\omega) \nonumber\\
	&+ 2\frac{A_{\q}}{\sqrt{A_{\q}^2-B_{\q}^2}} 
		\delta(\omega-\omega_{\q}) \; ,
\label{eq:Aqw_qm}
\end{align}
where $\Delta^{\sf{QM}}$ is defined in \Autoref{eq:Delta_QM}.
It follows that the equivalent equal--time structure factor given by
\begin{align}
S_{\rm{A}}^{\sf QM}(\q) =& N(1-\frac{2}{3}\Delta^{\sf{QM}}) \delta(\q) \nonumber\\
		&+ 2\frac{A_{\q}}{\sqrt{A_{\q}^2-B_{\q}^2}} \; .
\label{eq:A.equal.time.structure.factor}
\end{align}


Details of these calculations can be found in  and  \hyperref[sec:structure_factors_quantum_Aq]{Appendix~}\ref{sec:structure_factors_quantum_Aq} and in \hyperref[sec:structure_factors_quantum_Aq0]{Appendix~}\ref{sec:structure_factors_quantum_Aq0} for $\q=0$.

\subsubsection{Sum rule on structure factors}

The sum rule on moments, \Autoref{eq:sum.rule.AQS}, implies that 
dynamical structure factors must satisfy a sum rule 
\begin{eqnarray}
S_{\rm{A}} (\q, \omega) 
	= \frac{1}{4} S_{\rm{Q}} (\q,\omega) 
	+ \frac{1}{2} S_{\rm{S}} (\q,\omega) 
	+ \frac{1}{3} N \delta(\omega) 	\; , \nonumber\\
\label{eq:quantum.sum.rule}
\end{eqnarray}
of the same form as the sum rule for equal--time structure factors, 
\Autoref{eq:sum.rule.classical.structure.factorq0}.


It is easy to confirm, by direct substitution in \Autoref{eq:quantum.sum.rule}, 
that the quantum results at $T=0$ for 
$S_{\rm{S}}^{\sf QM}(\q,\omega)$ [\Autoref{eq:S.equal.time.structure.factor}], 
$S_{\rm{Q}}^{\sf QM}(\q,\omega)$ [\Autoref{eq:Q.equal.time.structure.factor}]
and $S_{\rm{A}}^{\sf QM}(\q, \omega)$ [\Autoref{eq:A.equal.time.structure.factor}], 
satisfy this sum rule.
It is also informative to verify the sum rule visually, by examining how the intensities
in the dipole channel [\Autoref{fig:comparison.classical.quantum.dispersion}~\subref{fig:Ana_FW_Sqw}] 
and quadrupole channel [\Autoref{fig:comparison.classical.quantum.dispersion}~\subref{fig:Ana_FW_Qqw}] 
``add up'' to give the intensity for A--matrices [\Autoref{fig:comparison.classical.quantum.dispersion}~\subref{fig:Ana_FW_Aqw}].

\section{Low--temperature properties of ferroquadrupolar order from numerical simulation} 	
\label{section:numerics}

In this Section, we use the $U(3)$ Monte Carlo (u3MC) and 
Molecular Dynamics (u3MD) simulation schemes  
developed in \Autoref{section:numerical.method} to explore  
thermodynamic and dynamic properties of ferroquadrupolar (FQ) order at low
temperatures.
Simulation results are compared directly with the analytic theory developed in 
\Autoref{section:classical.theory}. 
We start by analysing the heat capacity, which is shown to satisfy the 
correct classical limit $c(T \to 0) \to 2$ [\Autoref{section:heat.capacity}].
Next we consider the low--temperature properties of the ordered moment 
$\langle Q \rangle$.
This takes on a finite value in simulation, but is shown to exhibit  
finite--size scaling consistent with the Mermin--Wagner theorem 
[\Autoref{section:ordered.moment}].


We then turn to the equal--time structure factors associated 
with dipole and quadrupole moments.
At low--temperatures, these conform to the predictions of 
\Autoref{section:classical.theory}, confirming that simulations 
accurately describe correlations within the FQ state [\Autoref{section:S.of.q}].


Finally, we present ``raw'' simulation results for dynamical structure factors 
[\Autoref{section:numerics.dynamics}].
These reproduce the dispersion predicted by the zero--temperature quantum 
theory [\Autoref{section:quantum.theory}], but with a mismatch in intensities
The way in which this mismatch can be corrected to achieve agreement with 
quantum theory in the limit \mbox{$T \to 0$}  will be analysed in \Autoref{section:quantum.vs.classical}.  


\begin{figure}[t]
\centering
\includegraphics[width=0.45\textwidth]{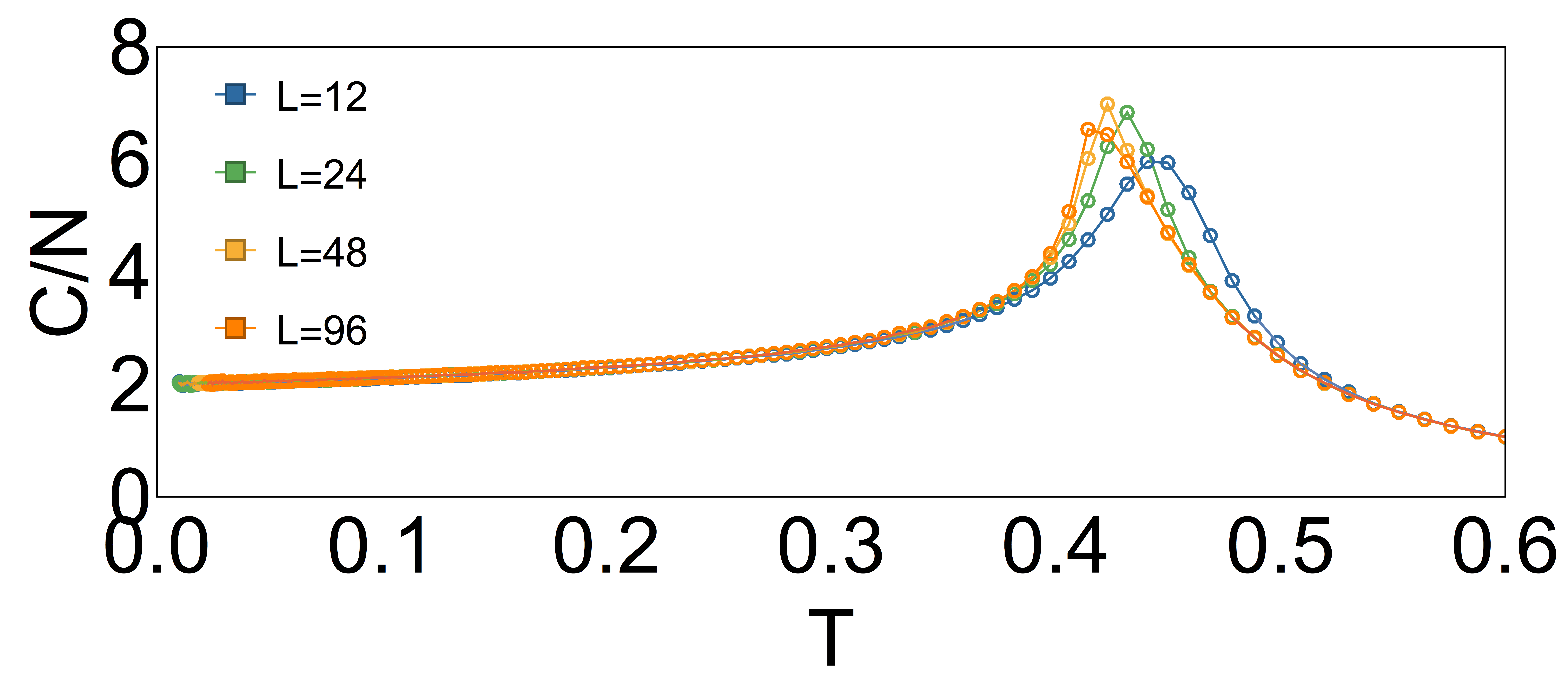}
\caption{
Temperature dependence of the specific heat per spin $c(T)$, 
found in $U(3)$ Monte Carlo (u3MC) simulations of 
$\Ham_{\sf BBQ}$ [\protect\Autoref{eq:H.BBQ.3}],  
for parameters consistent with a ferroquadrupolar (FQ) ground state.
Results are shown for a series of clusters of increasing linear dimension $L$.
The peak at $c(T)$ at $T^* \sim 0.43$ corresponds to the onset of 
fluctuations of FQ order, as shown in \protect\Autoref{fig:ordered.moment}~\protect\subref{fig:MW_theorem1}. 
The low temperature asymptote $c(T \to 0) \to 2$ is consistent 
with the existence of four independent excitations about the the FQ 
ground state, as discussed in \protect\Autoref{section:theory.classical.thermodynamics}.
All simulations were carried out 
with parameters \protect\Autoref{eq:simulation.parameters}, 
using the MC scheme described in \protect\Autoref{sec:MC}.
}
\label{fig:heat.capacity}
\end{figure}


\begin{figure*}[t]
	\centering
	
	\subfloat[Temperature dependence of ordered moment \label{fig:MW_theorem1}]{
		\includegraphics[width=0.45\textwidth]{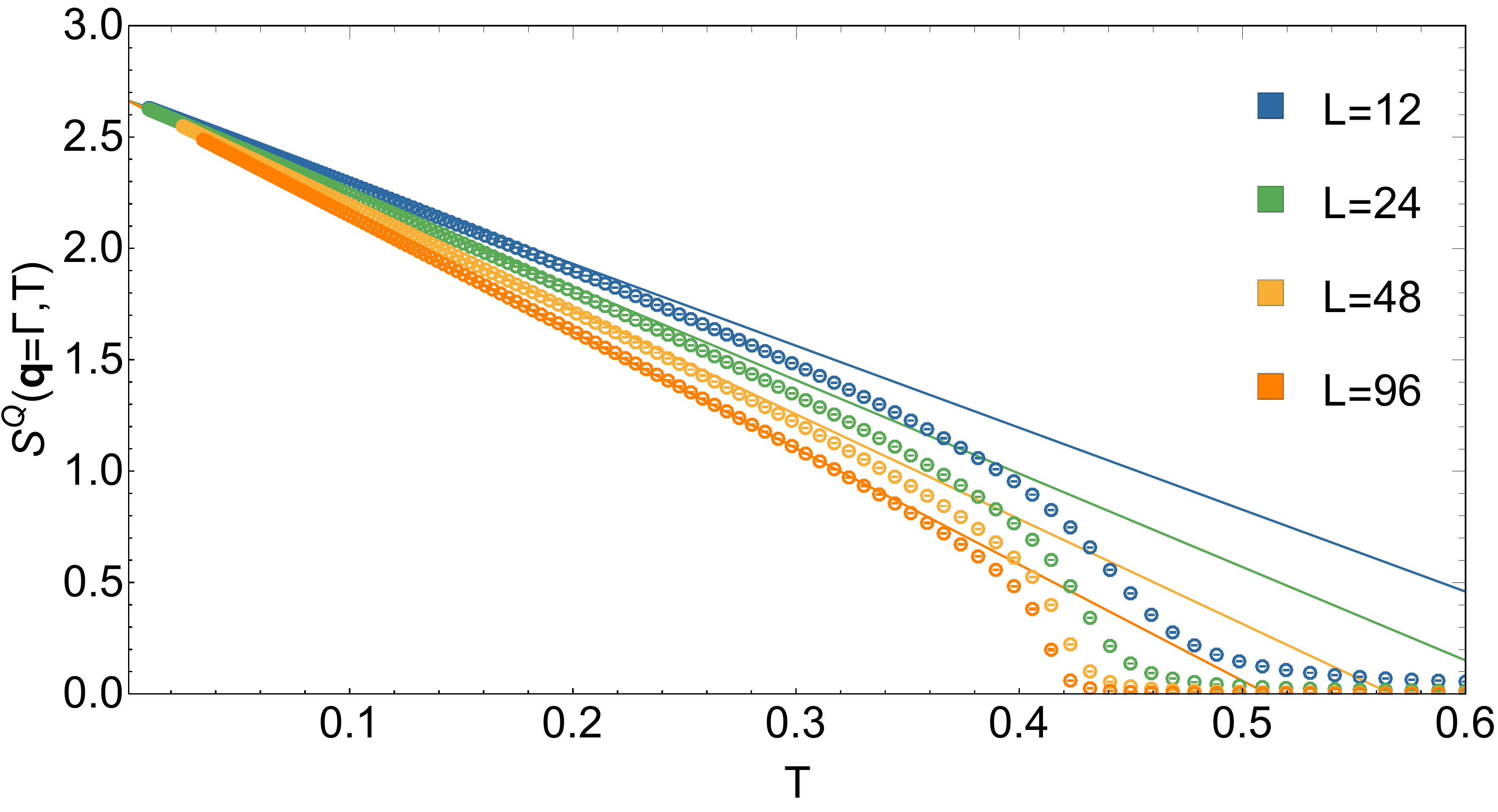}
	} 
	%
	\subfloat[Coefficient of leading temperature correction \label{fig:MW_theorem2}]{
		\includegraphics[width=0.45\textwidth]{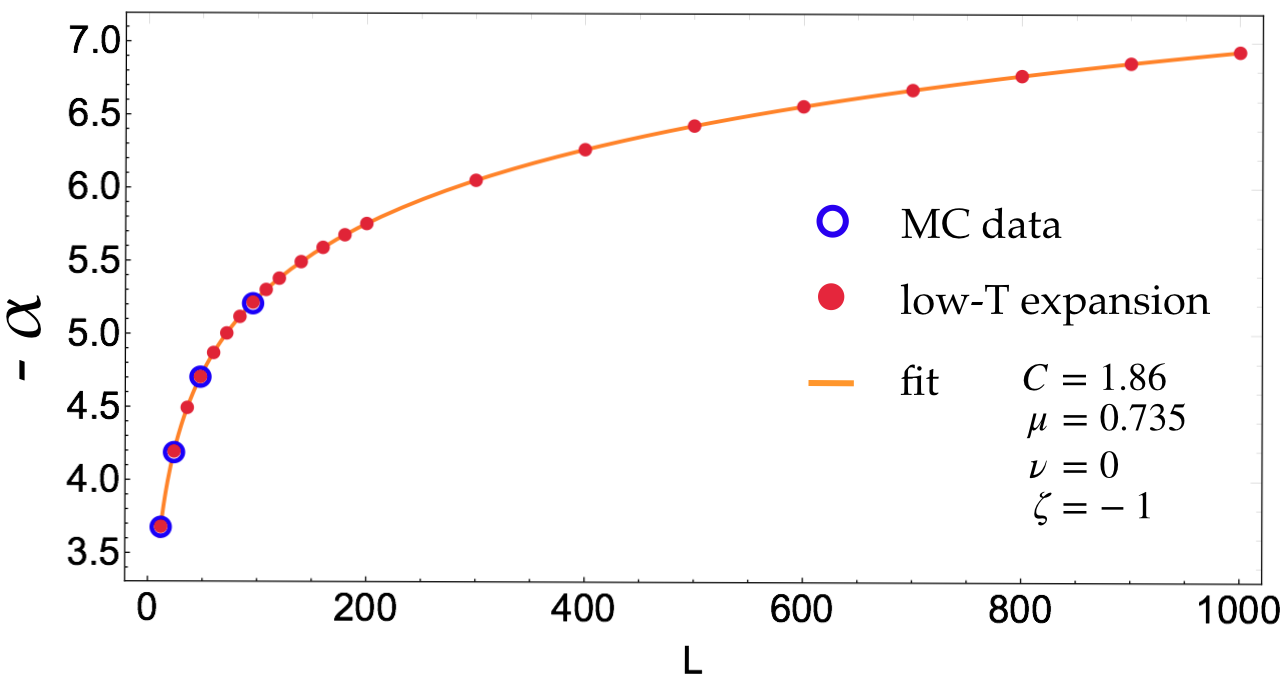}
	}
\caption{	
Temperature dependence of the quadrupole moment $\bf Q$ 
found in $U(3)$ Monte Carlo (u3MC) simulations of 
$\Ham_{\sf BBQ}$ [\protect\Autoref{eq:H.BBQ.3}],  
for parameters consistent with a ferroquadrupolar (FQ) ground state.
\protect\subref{fig:MW_theorem1}~Results for ${\bf Q}^2$ in a series of clusters of increasing linear 
dimension, $L$.
The onset of fluctuations of FQ order at ${\bf Q}^2$ at $T^* \sim 0.43$, 
corresponds to the peak in heat capacity, shown in \protect\Autoref{fig:heat.capacity}. 
At low temperatures, ${\bf Q}$ tends to the ordered moment of the 
FQ ground state, $Q_0$ [\protect\Autoref{eq:Q0}].
\protect\subref{fig:MW_theorem2}~Finite--size scaling of the coefficient $\alpha(L)$ [\protect\Autoref{eq:lowTfit1}], 
showing a logarithmic divergence in temperature--corrections to the ordered 
moment 
[\protect\Autoref{eq:lowTfit2}], consistent with the 
Mermin--Wagner Theorem.
Results are shown for both u3MC simulations, and the analytic theory 
developed in \protect\Autoref{section:low.T}.
All simulations were carried out 
with parameters \protect\Autoref{eq:simulation.parameters}, 
using the u3MC scheme described in \protect\Autoref{sec:MC}.	
}
\label{fig:ordered.moment}
\end{figure*}

\subsection{Heat capacity} 							
\label{section:heat.capacity}

In \Autoref{fig:heat.capacity} we present results for the heat capacity per spin 
\begin{eqnarray}
c(T) 
= C(T)/N 
= \frac{1}{N} \frac{1}{T^2} \big[ \langle E(T)^2 \rangle - \langle E(T) \rangle^2 \big] \; ,\nonumber\\
\end{eqnarray}
obtained in simulations of $\Ham_{\sf BBQ}$ [\Autoref{eq:H.BBQ.3}], 
for the same parameter set used in \Autoref{fig:comparison.classical.quantum.dispersion}
\begin{eqnarray}
	J_1 = 0.0 \; , \; J_2 = -1.0 \; .
\label{eq:simulation.parameters}
\end{eqnarray}
Results were obtained using the $U(3)$ Monte Carlo (u3MC)  formalism developed 
in \Autoref{sec:MC}, for clusters of linear dimension up to $L = 96$ ($N = 9216$ spins).


At low temperature, we find 
\begin{eqnarray}
c(T \to 0) \to 2 \; .
\end{eqnarray}
This is the result anticipated from the classical theory  
developed in \Autoref{section:low.T} [cf. \Autoref{eq:cv.U3}], 
and reflects the fact that the $u(3)$ formalism correctly 
describes the 4 orthogonal generators of fluctuations about the FQ ground state.
Each of these contribute $1/2$ to $c(T)$ in the limit $T \to 0$, as 
discussed in \Autoref{section:theory.classical.thermodynamics}.
This should be contrasted with classical MC simulations in an $O(3)$ basis, 
where at most two generators per spin are accessible and 
$c(T \to 0) \leq 1$ [\Autoref{eq:cv.O3}].


Meanwhile, the onset of fluctuations of FQ order is signaled by a pronounced peak 
at  $T^* \sim 0.43$, which gradually sharpens and moves to lower temperatures 
with increasing system size.
The scaling of this peak is not consistent with a conventional phase transition, and 
long range FQ order is not expected to occur in the two--dimensional BBQ model 
at finite temperature, because of the Mermin--Wagner theorem \cite{Mermin1966}.  


None the less, a BKT--like topological phase transition into \kim{a} phase with algebraic 
correlations of FQ order is permitted, and would also give rise to a peak in heat capacity.
Such a phase transition can be mediated by point--like, 
\begin{eqnarray}
	\pi_1(RP_2)={\mathbb Z}_2 \; ,
\end{eqnarray}
topological defects of FQ order, and has been observed in previous MC simulations 
of the $O(3)$ BBQ model on the triangular lattice \cite{Kawamura2007}.
A detailed analysis of topological phase transitions in the \mbox{spin--1} BBQ model 
lies outside the scope of this paper, but contains many interesting features, 
which will be discussed elsewhere \cite{Pohle-in-preparation}.


\subsection{Ordered moment} 							\label{section:ordered.moment}

We now consider the behaviour of the quadrupole--moment ${\bf Q}$,  
which acts as an order parameter for the FQ state.
In \Autoref{fig:ordered.moment} we show simulation results, obtained 
for the same parameter set, \Autoref{eq:simulation.parameters}.
The ordered moment was calculated through the equal--time structure factor
\begin{eqnarray}
	{\bf Q}^2 = \frac{S^{\sf CL}_{\rm{Q}}(\q=\Gamma)}{N} \; .
\end{eqnarray}
and takes on a finite value in finite--size clusters, as shown 
in \Autoref{fig:ordered.moment}~\subref{fig:MW_theorem1}.   
At low temperature, these results extrapolate to the expected 
ground--state value [\Autoref{eq:OM}]
\be
	{\bf Q}^2 \big|_{T \to 0}  = Q^2_0 = \frac{8}{3} \; ,
\label{eq:Q0}
\ee
and are well--described by the function 
\begin{equation}
	{\bf Q}^2 = Q^2_0 + \alpha(L) T + \beta(L) T^2 + \cdots		\, ,
	\label{eq:lowTfit1}
\end{equation}
where the coefficients $\alpha(L)$ and $\beta(L)$ are determined 
by fits to simulation results.
At a temperature corresponding to the peak in heat capacity, 
$T \approx T^* \sim 0.43$ [\Autoref{fig:heat.capacity}], 
the value of ${\bf Q}^2$ collapses rapidly.
Above this temperature, ${\bf Q}^2$ tends rapidly to zero 
with increasing system size.


For the Mermin--Wagner theorem to hold, we must find ${\bf Q}^2 \equiv 0$ 
in the thermodynamic limit, at any finite temperature \cite{Mermin1966}.
It follows that the coefficient $\alpha(L)$ in \Autoref{eq:lowTfit1} must  
diverge as $L \to \infty$. 
The trend in $\alpha(L)$ with increasing $L$ is immediately apparent 
from \Autoref{fig:ordered.moment}~\subref{fig:MW_theorem1}: 
the rate at which thermal fluctuations reduce the ordered moment 
is a monotonically increasing function of $L$.
However, for all system sizes accessible to simulation, the ordered moment
still takes on a substantial value at low temperatures.


This seeming--paradox can be resolved by turning to the analytic theory 
developed in \Autoref{section:low.T}.
In \Autoref{fig:ordered.moment}~\subref{fig:MW_theorem2} we plot the values of $\alpha(L)$ obtained 
in simulation, together with analytic results for systems of size up to 
$L = 1000$ ($N = 10^6$ spins).
Analytic estimates of $\alpha(L)$ were found by evaluating the sum on 
$\bf k$ in \protect\Autoref{eq:OM} numerically, for the specific set of wave vectors 
allowed by the geometry of the clusters.
Evaluating the leading contribution to this sum as an integral, 
we can identify a logarithmic divergence in $\alpha(L)$ for large $L$.
And consistent with this, both analytic and numerical results are well described by the function 
\begin{equation}
	-\alpha(L) =  \alpha_0 + \mu \log{L} + \nu \frac{1}{L} + \xi \frac{1}{L^2}   \, ,
\label{eq:lowTfit2}
\end{equation}
with fit parameters
\begin{eqnarray}
\alpha_0 = 1.86 \; , \; 
\mu = 0.735 \; , \; 
\nu = 0 \; , \; 
\xi = -1 \; .
\end{eqnarray}
It follows that $-\alpha(L \to \infty) \to \infty$, and the Mermin--Wagner Theorem is respected.
Further details of this analysis can be found in 
\hyperref[sec:Mermin_Wagner_Theorem]{Appendix~}\ref{sec:Mermin_Wagner_Theorem}	


\begin{figure}[t]
	\centering
	\subfloat[Comparison of $S^{\sf MC}_{\sf S} (\q)$ and $S^{\sf CL}_{\sf S} (\q)$ \label{fig:structure.factor.S}]{
		\includegraphics[width=0.45\textwidth]{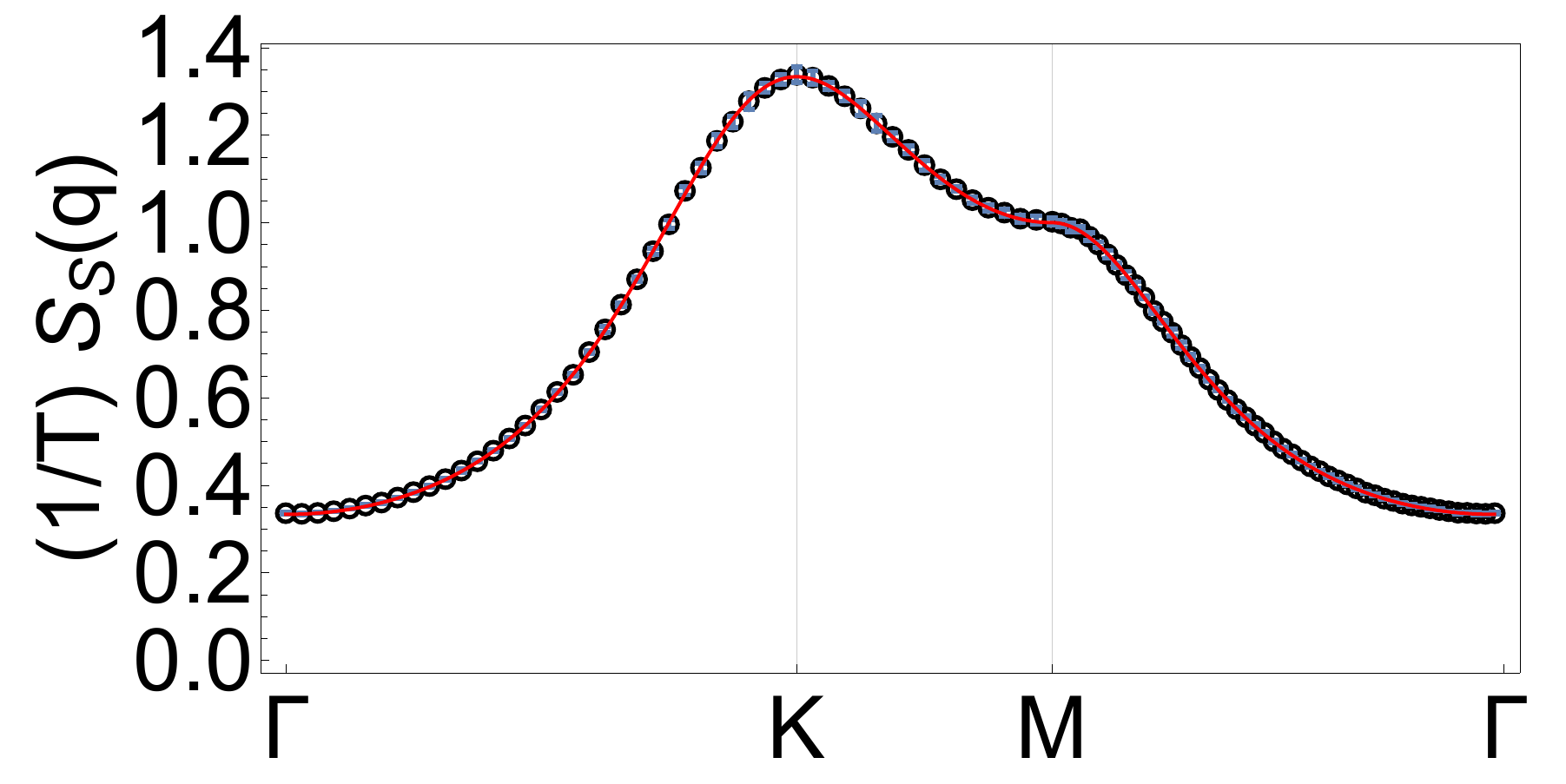}} \\
	\subfloat[Comparison of $S^{\sf MC}_{\sf Q} (\q)$ and $S^{\sf CL}_{\sf Q} (\q)$ \label{fig:structure.factor.Q}]{
		\includegraphics[width=0.45\textwidth]{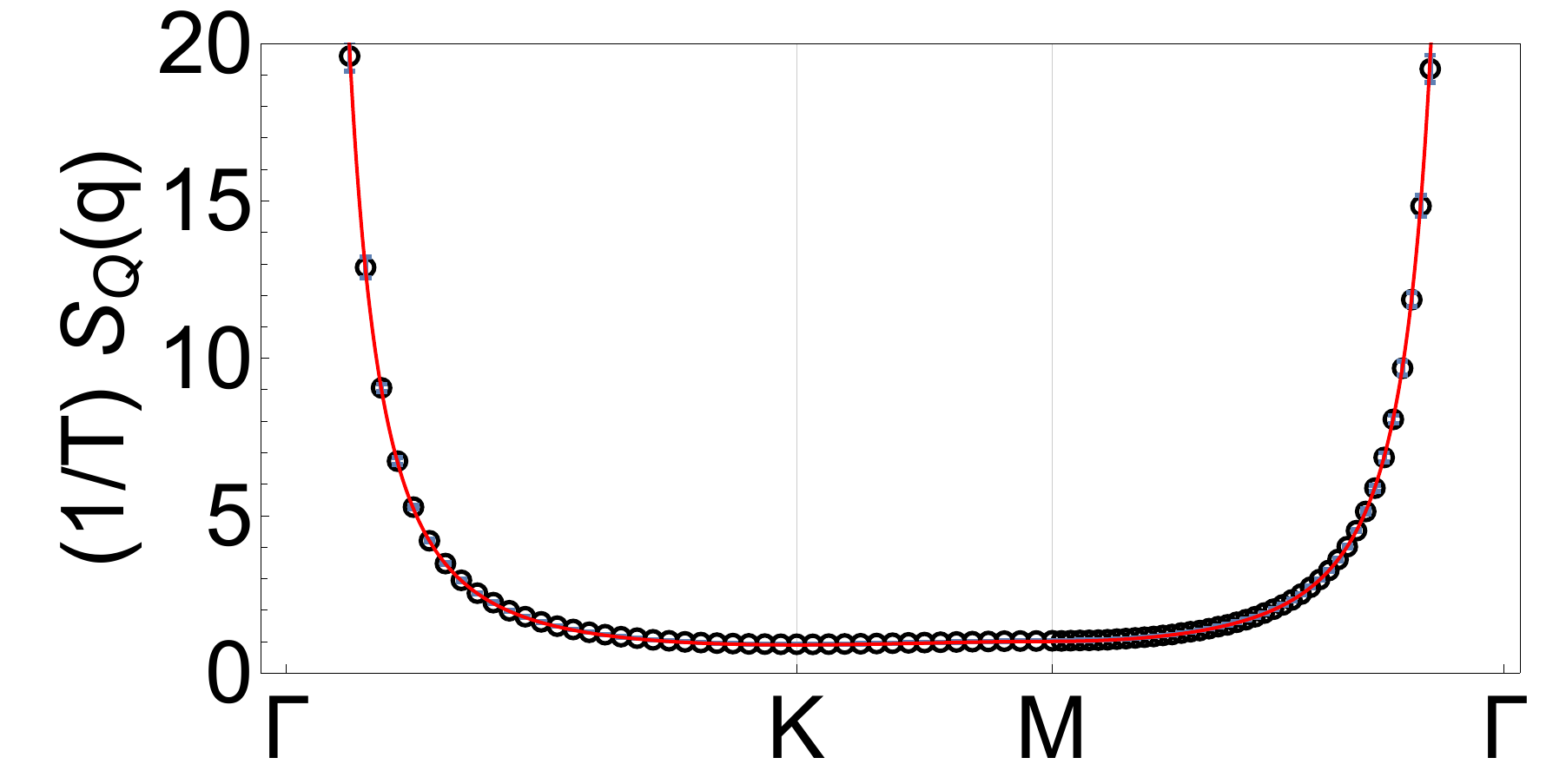}} \\
	\subfloat[Comparison of $S^{\sf MC}_{\sf A} (\q)$ and $S^{\sf CL}_{\sf A} (\q)$ \label{fig:structure.factor.A}]{
		\includegraphics[width=0.45\textwidth]{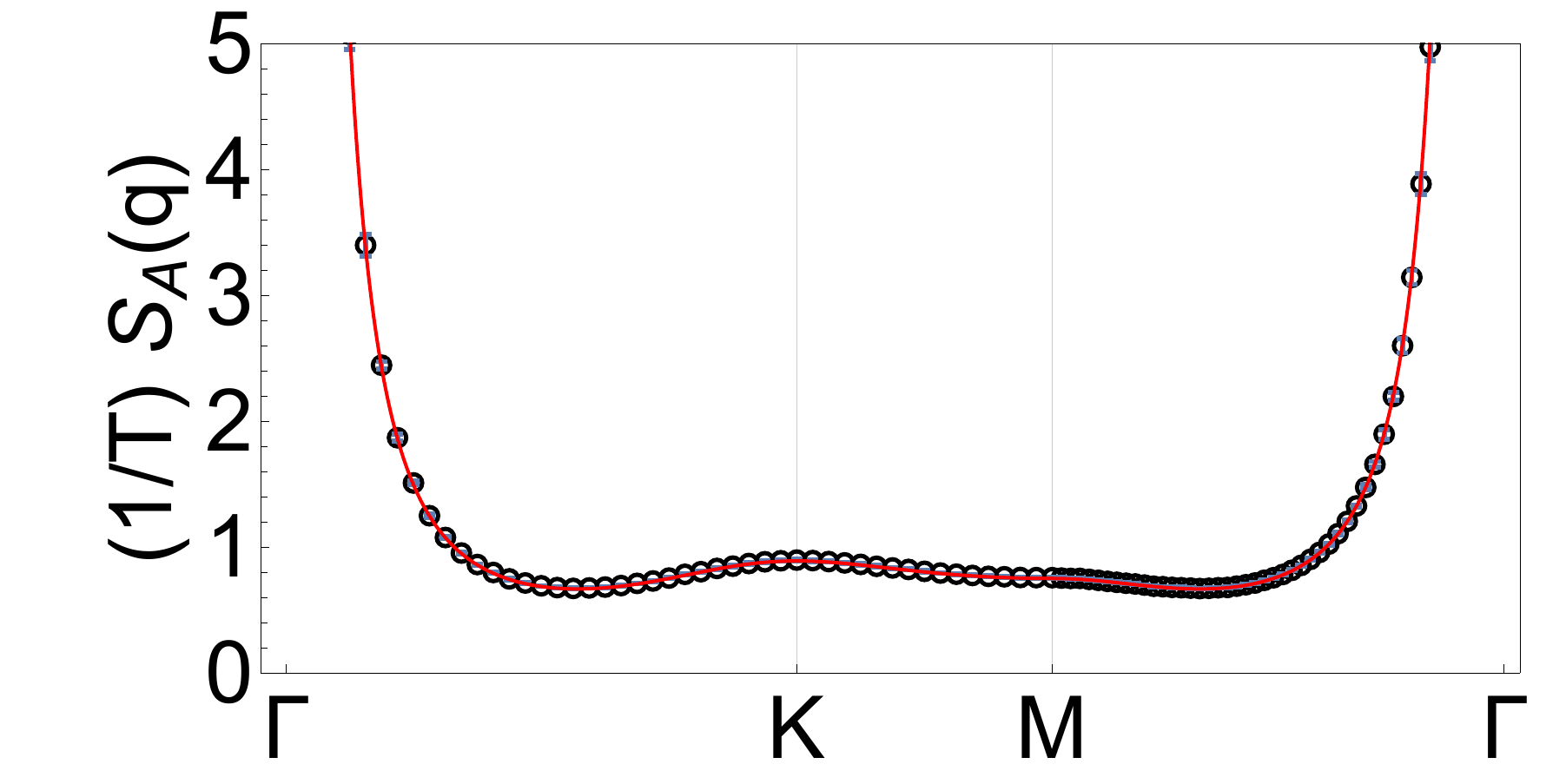}} \;
\caption{
Results for equal--time structure factors $S^{\sf CL}_\lambda (\q)$ [\protect\Autoref{eq:Sq.MC}] 
found in $U(3)$ Monte Carlo (u3MC) simulations of 
$\Ham_{\sf BBQ}$ [\protect\Autoref{eq:H.BBQ.3}], for parameters consistent 
with a ferroquadrupolar (FQ) ground state.
\protect\subref{fig:structure.factor.S}~Structure factor associated with dipole moments, $S^{\sf CL}_{\sf S} (\q)$, showing 
correlations at the 3--sublattice ordering vector, ${\sf K}$.  
\protect\subref{fig:structure.factor.Q}~Structure factor associated with quadrupole moments, $S^{\sf CL}_{\sf S} (\q)$, 
showing divergence associated with fluctuations of FQ order for $\q~\to~\Gamma$.  
\protect\subref{fig:structure.factor.A}~Structure factor associated with A--matrices, $S^{\sf CL}_{\sf A} (\q)$, 
sensitive to both dipolar and quadrupolar fluctuations.   
In all cases, simulation results (points) have been divided by temperature $T$, and agree 
perfectly with the predictions of low--temperature analytic theory (line).   
All simulations were carried out with parameters \protect\Autoref{eq:simulation.parameters}, 
for a cluster with linear dimension $L = 96$ (\mbox{$N=9216$} spins), 
at ${\rm{T \approx 0.03}}$, using the u3MC scheme described in \protect\Autoref{sec:MC}.	
}
	\label{fig:S.of.q}
\end{figure}


\begin{figure*}[t]
	\begin{center}
		\begin{minipage}[b]{\textwidth}
			\subfloat[$S^{\sf MD}_{\rm{S}}(\q, \omega), T = 0.05$	\label{fig:Sqw_MC}]{
				\includegraphics[width=0.32\linewidth]{Fig16_a.pdf}
			}
			\subfloat[$S^{\sf MD}_{\rm{Q}}(\q, \omega), T = 0.05$	\label{fig:Qqw_MC}]{
				\includegraphics[width=0.32\linewidth]{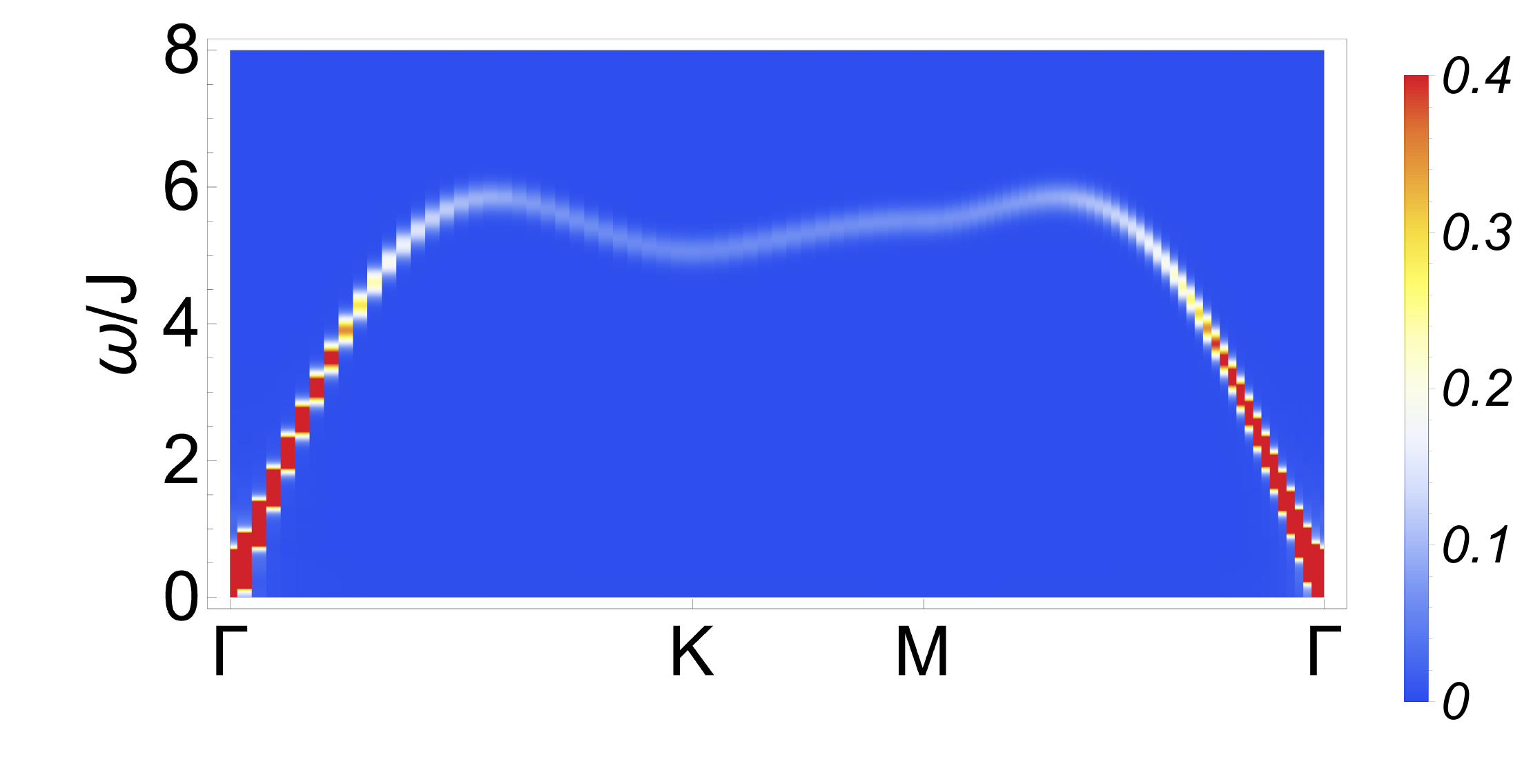}
			}
			\subfloat[$S^{\sf MD}_{\rm{A}}(\q, \omega), T = 0.05$	\label{fig:Aqw_MC}]{
				\includegraphics[width=0.32\linewidth]{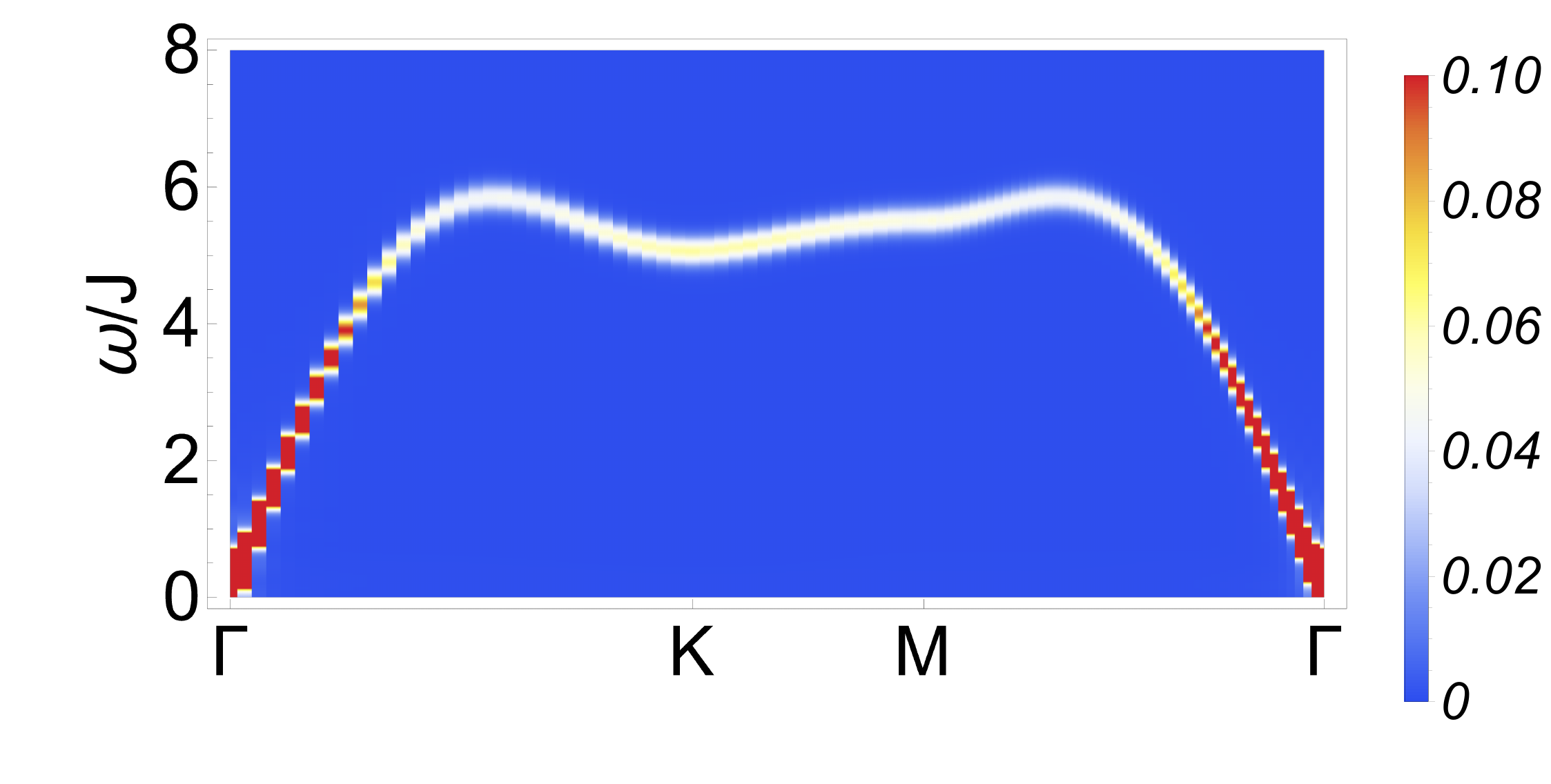}
			}
		\end{minipage}
	\end{center}
\caption{
``Raw'' results of $U(3)$ Molecular Dynamics (u3MD) 
simulations for parameters consistent with a ferroquadrupolar (FQ) ground state.
\protect\subref{fig:Sqw_MC}~Dynamical structure factor associated with dipole moments, $S^{\sf MD}_{\rm{S}}(\q, \omega)$.
%
%
\protect\subref{fig:Qqw_MC}~Dynamical structure factor associated with quadrupole moments, $S^{\sf MD}_{\rm{Q}}(\q, \omega)$.
%
\protect\subref{fig:Aqw_MC}~Dynamical structure factor associated with A--matrices, $S^{\sf MD}_{\rm{A}}(\q, \omega)$.
%
Comparison with the predictions of a quantum theory, 
\protect\Autoref{fig:comparison.classical.quantum.dispersion}~\protect\subref{fig:Ana_FW_Sqw}--\protect\subref{fig:Ana_FW_Aqw}, 
suggests that ``raw'' simulation results 
accurately describe the dispersion of excitations,  
but with an incorrect distribution of intensity.  
Simulations were carried out using the u3MD simulation scheme described 
in \protect\Autoref{sec:u3MD}, for $\Ham_{\sf BBQ}$ [\protect\Autoref{eq:H.BBQ.3}] with 
parameters \protect\Autoref{eq:simulation.parameters}, at a temperature $T = 0.05\ J$, 
in a cluster of linear dimension $L = 96$ ($N = 9216$ spins).
\nic{
Results are shown only for positive energy, $\omega > 0$, and have been convoluted 
with a Gaussian envelope of \mbox{$\text{FWHM} =  0.35~\text{J}$}.
}
}
\label{fig:dispersion_rawMD}
\end{figure*}

\subsection{Equal--time structure factor} 					
\label{section:S.of.q}

We now turn to correlations between magnetic moments, as described 
by the equal--time structure factors $S^{\sf CL}_\lambda (\q)$ [\Autoref{eq:Sq.MC}], 
found in u3MC simulations of $\Ham_{\sf BBQ}$ [\Autoref{eq:H.BBQ.3}].
In \Autoref{fig:S.of.q}, results are shown for the structure factors 
associated with dipole moments, $S_{\rm{S}}(\q)$, 
quadrupole moments $S_{\rm{Q}}(\q)$, 
and A--matrices, $S_{\rm{A}}(\q)$.
Simulations were carried out for parameters consistent with a FQ ground state [\Autoref{eq:simulation.parameters}], at a temperature ${\rm{T \approx 0.03}}$, 
in a cluster of linear dimension $L = 96$ (\mbox{$N=9216$} spins).
All results are plotted on an irreducible wedge 
$\Gamma$--${\sf K}$--${\sf M}$--$\Gamma$ [cf. \hyperref[sec:conv_trig_lat]{Appendix~}\ref{sec:conv_trig_lat}], 
and have been been divided by temperature, $T$, to extract their leading 
temperature dependence.


Fluctuations of dipole moments vanish in the FQ ground state, but take 
on a finite value at finite temperature, as shown in \Autoref{fig:S.of.q}~\subref{fig:structure.factor.S}.
Simulation results for $S_{\rm{S}}(\q)/T$ at low temperatures (points) 
are perfectly described by the low--temperature analytic prediction, 
\Autoref{eq:classical.structure.factor.S}, (solid line).
A broad peak in $S_{\rm{S}}(\q)$ for $\q = \sf{K}$ reflects the proximity of 3--sublattice 
antiferromagnetic order (AFM), as discussed in \Autoref{sec:phase.diagram}.


Meanwhile, the quadrupolar structure factor $S_{\rm{Q}}(\q)$ is sensitive 
to fluctuations of FQ order, and exhibits 
a $\q$--dependent contribution that diverges for $\q \to 0$, as shown in 
\Autoref{fig:S.of.q}~\subref{fig:structure.factor.Q}.
Once again, the agreement between simulation results for $S_{\rm{Q}}(\q)/T$
at low temperatures (points) and the low--temperature analytic prediction, 
\Autoref{eq:classical.structure.factor.Q}, (line), is perfect.


Finally, the structure factor for A--matrices $S_{\rm{A}}(\q)$, shown in 
\Autoref{fig:S.of.q}~\subref{fig:structure.factor.A}, is sensitive to both quadrupolar and dipolar 
fluctuations.
In keeping with this, it exhibits both a diverging contribution for $\q \to 0$, 
and a small peak at $\q = \sf{K}$.
Perfect agreement is found between simulation results for $S_{\rm{A}}(\q)/T$
at low temperatures (points) and the low--temperature analytic prediction, 
\Autoref{eq:classical.structure.factor.A}, (line).


Taken together, these results for $S^{\sf CL}_\lambda (\q)$ 
confirm the ability of the u3MC scheme developed in \Autoref{sec:MC}, 
to describe classical correlations of \mbox{spin--1} magnets at low temperature.
They will also play an important role in determining the quantum--classical 
correspondence discussed in \Autoref{section:quantum.vs.classical}.

\subsection{Dynamics} 				
\label{section:numerics.dynamics}

We complete our survey of simulation results for the FQ phase of the \mbox{spin--1} BBQ model 
by exploring the dynamics found in numerical integration of the equations of motion, \Autoref{eq:EoM.u3}, following the $U(3)$ Molecular Dynamics (u3MD) scheme 
introduced \Autoref{sec:u3MD}. 


In \Autoref{fig:dispersion_rawMD} we present ``raw'' results for the 
dynamical structure factors $S^{\sf MD}_\lambda(\q, \omega)$ [\Autoref{eq:Sq.MD.FT}] associated with dipole moments ($\lambda = {\sf S}$), 
quadrupole moments ($\lambda = {\sf Q}$), 
and A--matrices ($\lambda = {\sf A}$).
\nic{
Results are plotted for the same path in reciprocal space 
as was used for $S^{\sf MC}_\lambda(\q)$ in \Autoref{fig:S.of.q}, 
for positive frequency $\omega > 0$.   
MD solutions at negative energy will contribute with equal weight [\Autoref{eq:symmetry.of.Sqomega}].
For convenience of visualization, all results 
have been convoluted with a Gaussian of \mbox{$\text{FWHM} = 0.35\ J$}.
}


Comparing with the predictions of the zero--temperature quantum theory,   
$S^{\sf QM}_\lambda(\q, \omega)$ [\Autoref{fig:comparison.classical.quantum.dispersion}], 
we see that u3MD correctly reproduces a dispersing band of excitations, with predominantly quadrupolar character for $\omega \to 0$, and predominantly dipolar character 
at the top of the band.
Closer examination, however, reveals small differences in the energy of excitations, 
and dramatic differences in the distribution of spectral weight across the band.
In particular, while analytic results for the dipolar fluctuations 
[\Autoref{fig:comparison.classical.quantum.dispersion}~\subref{fig:Ana_FW_Sqw}, \Autoref{eq:Sqw_qm})],  
exhibit a characteristic linear loss of spectral weight at 
low energies \cite{Smerald2013}
\begin{eqnarray}
	S^{\sf QM}_{\sf S}(\q \to \bf{0}, \omega) 
		\propto \omega\ \delta(\omega - v |\q|) \; ,
\end{eqnarray}
numerical results for $S^{\sf MD}_{\sf S}(\q , \omega)$ [\Autoref{fig:dispersion_rawMD}~\subref{fig:Sqw_MC}] 
show a roughly constant spectral weight for $\omega \to 0$.
The distribution of spectral weight in the quadrupolar channel 
$S^{\sf MD}_{\sf Q}(\q , \omega)$ [\Autoref{fig:dispersion_rawMD}~\subref{fig:Qqw_MC}], 
is also visibly different from analytic predictions 
[\Autoref{fig:comparison.classical.quantum.dispersion}~\subref{fig:Ana_FW_Qqw}, \Autoref{eq:Qqw_qm}].

\begin{figure}[t]
	\centering
	\includegraphics[width=0.48\textwidth]{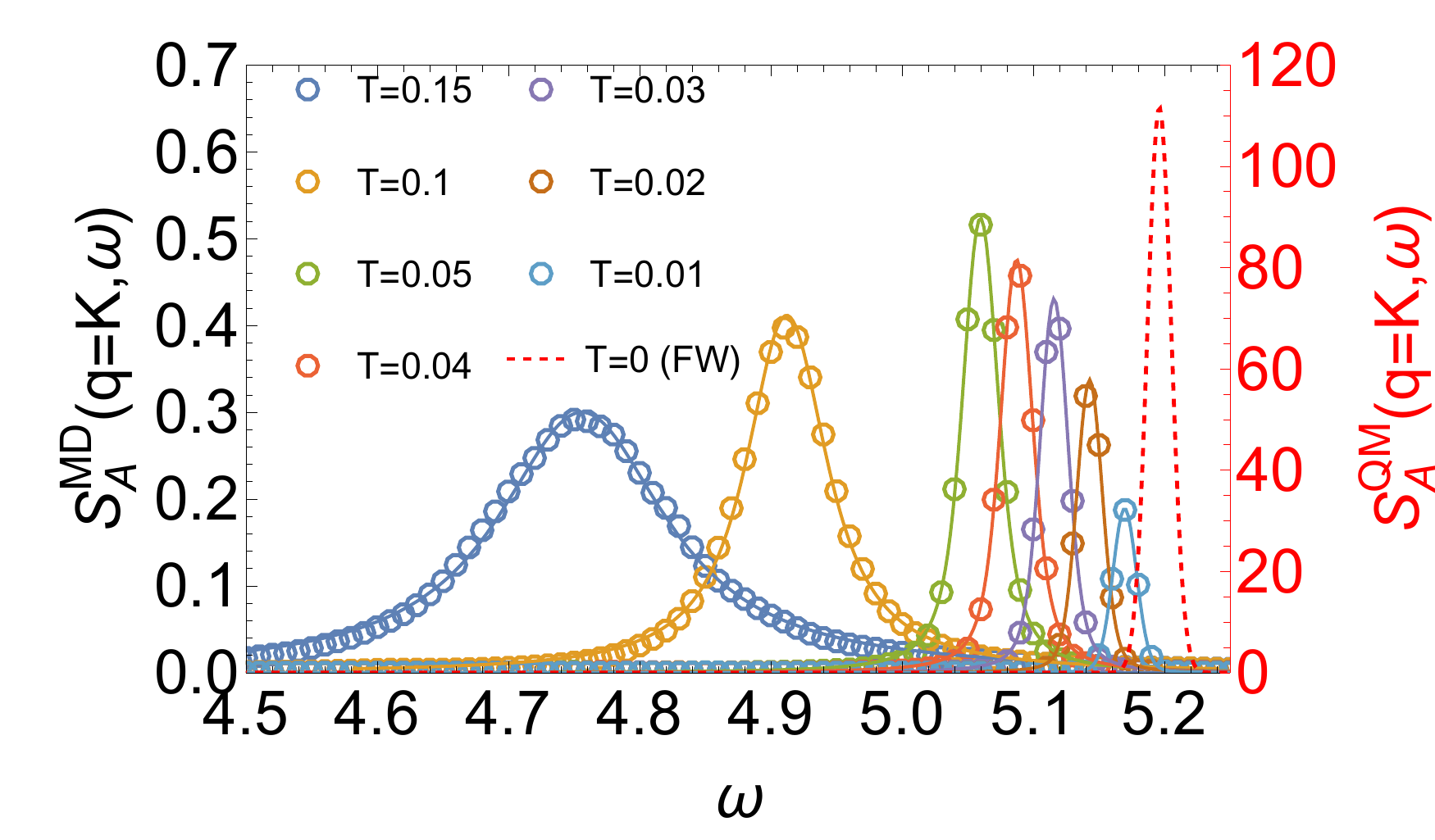}
\caption{
Temperature dependence of ``raw'' results of $U(3)$ Molecular Dynamics (u3MD) 
simulation, showing failure to converge to the predictions of a \mbox{$T=0$} quantum theory 
for \mbox{$T \to 0$}. 
Results are shown for the dynamical structure factor associated with A--matrices, 
$S^{\sf MD}_{\rm{A}}(\q, \omega)$ [\protect\Autoref{eq:dynamical.structure.factor.in.simulation}], 
for wave vector $\q=\K$, and temperatures ranging from $T=0.01\ J$ to $T=0.15\ J$.
The \mbox{$T=0$} prediction of a quantum theory, 
$S^{\sf QM}_{\rm{A}}(\q, \omega)$ [\protect\Autoref{eq:Aqw_qm}], 
is shown with a red dashed line.
Solid lines represent fits to u3MD data using a Voigt profile [\protect\Autoref{eq:voigt}].
 ``Raw'' simulation results converge on energy predicted by the quantum 
theory for $T\to 0$, but suffer a dramatic loss of intensity.
Simulations of $\Ham_{\sf BBQ}$ [\protect\Autoref{eq:H.BBQ.3}] were carried 
out using the MD simulation scheme described in \protect\Autoref{sec:u3MD}, 
for parameters \protect\Autoref{eq:simulation.parameters}, in a cluster of linear 
dimension $L = 96$ ($N = 9216$ spins).
Both simulation results and analytic prediction have been convoluted 
with a Gaussian of $\text{FWHM} = 0.02\ J$.
}
\label{fig:T-scaling_Sqw_no_wT}
\end{figure}

A more precise portrait of the ``raw'' u3MD results can be found by examining 
the temperature dependence of dynamical structure factors 
at fixed wavevector $\q$.
In \Autoref{fig:T-scaling_Sqw_no_wT} we present results 
for $S^{\sf MD}_{\rm{A}}(\q, \omega > 0)$ (symbols), at wave vector $\q=\K$, 
with temperatures ranging from $T=0.01\ J$ to $T=0.15\ J$.
The prediction of a zero--temperature quantum theory, 
\mbox{$S^{\sf QM}_{\rm{A}}(\q=\K, \omega)$} [\Autoref{eq:Aqw_qm}], 
is shown for comparison (dashed line).
Both analytic and simulation results (symbols) have been convoluted 
with a Gaussian of \mbox{$\text{FWHM} = 0.02\ J$}.


``Raw'' simulation results show a single peak in 
\mbox{$S^{\sf MD}_{\rm{A}}(\q=\K, \omega > 0)$}, centered on an energy 
$\omega_0$ which varies as a function of temperature.
This peak is well described by Voigt profile 
\begin{equation}
	V(\omega, \sigma, \Gamma) =
	\frac{{\sf Re}[w(z)]}{\sigma \sqrt{2 \pi}}	\, ,
\label{eq:voigt}
\end{equation}
where the Faddeeva function
\begin{equation}
	w(z) = e^{-z^2} \text{erfc}(-i z)
\end{equation}
is evaluated for 
\begin{equation}
	z = \frac{(\omega - \omega_0)+i\Gamma}{\sigma \sqrt{2}}	\, .
\end{equation}
The Voigt profile reflects a Lorentzian lineshape 
\begin{equation}
	f(\omega) 
	= \frac{\Gamma}{2\pi} 
		\frac{1}{(\omega - \omega_0)^2 + \Gamma^2}  \; ,
\label{eq:lorentzian}
\end{equation}
appropriate to a single excitation of energy $\omega_0$ and 
inverse lifetime $\Gamma$, convoluted with a Gaussian with 
full--width half--maximum (FWHM) determined by 
$\sigma$ [\Autoref{eq:FWHM}].


Empirical fits of \Autoref{eq:voigt} to simulation data are shown 
with solid lines in \Autoref{fig:T-scaling_Sqw_no_wT}.
The parameters used in MD simulation
completely determine $\sigma$, leaving $\omega_0$, $\Gamma$, 
and the overall normalisation (total spectral weight) as a fit parameters.
The fits found in the way are excellent, confirming that simulations recover 
a single excited mode for $\omega > 0$, with finite, temperature--dependent 
energy and lifetime.


As temperature is reduced, the peak in $S^{\sf MD}_{\rm{A}}(\q=\K, \omega)$ 
migrates to higher energies, and becomes sharper, while retaining its underlying 
Lorentzian structure. 
In both of these respects, for $T \to 0$, simulation results approach 
the \mbox{$T=0$} quantum result, where spectral weight is 
concentrated in a delta function, the $\Gamma \to 0$ limit of \Autoref{eq:lorentzian}.
However at low temperatures, the u3MD results also exhibit a dramatic 
loss of intensity, with integrated spectral weight tending to zero as \mbox{$T \to 0$}. 
And even at $T \sim 0.1$, the difference in intensity is at least a factor 
of $\times\ 100$, reflected in different scales on the axes for 
with $S^{\sf MD}_{\rm{A}}(\q=\K, \omega)$ 
and $S^{\sf QM}_{\rm{A}}(\q=\K, \omega)$.
The reason for this discrepancy, and the way in which it can be 
corrected, will be discussed \Autoref{section:quantum.vs.classical}.


\section{Quantum--Classical correspondence}
\label{section:quantum.vs.classical}

The reason why ``raw'' results of molecular dynamics 
simulations, presented in \Autoref{section:numerics.dynamics}, 
capture the dispersion of quantum excitations, 
while failing to describe their spectral weight, is rooted in the 
classical statistics of the underlying classical Monte Carlo 
simulations.


The equation of motions (EoM) on which the u3MD is based, \Autoref{eq:EoM.u3},
correctly describe the dynamics of a \mbox{spin--1} moment, at a semi--classical level.
And, solved analytically, with appropriate quantization, these EoM yield identical results 
to the linear multiple--Boson expansion developed in 
\Autoref{section:quantum.theory} \cite{Kim-unpublished}.
However, the spectral weight found in u3MD simulation is not subject 
to any quantization condition.  
Instead this is determined by thermal fluctuations, subject to the 
classical statistics of Monte Carlo simulation.
And for this reason, all spectral weight vanishes for $T \to 0$, as thermal 
fluctuations are eliminated, cf. \Autoref{fig:T-scaling_Sqw_no_wT}.


None the less, the fact that spectral weight is concentrated in a single peak,
with Lorentzian lineshape, that becomes arbitrarily sharp 
for $T \to 0$, suggests that low--temperature simulation results 
can be understood within a single--mode approximation.
And this encourages us to believe that it may be possible to ``undo''
the effect of classical statistics, in limit \mbox{$T \to 0$}. 
This line of reasoning, developed below, leads to a simple 
prescription for correcting MD simulation results
\begin{eqnarray}
S^{\sf QM} ({\bf q}, \omega, T=0) = \lim_{T \to 0} \frac{\hbar\omega}{k_B T} S^{\sf MD} ({\bf q}, \omega, T) \; ,
\end{eqnarray}
previously introduced in \Autoref{eq:corrected.Sqw}.
This prescription is shown to restore to perfect agreement with 
zero--temperature quantum results, at a semi--classical level.
%

In what follows, we set out this analysis in more detail.
In \Autoref{section:MD.deconstructed}, we 
``deconstruct'' the dynamical structure factors 
found in u3MD simulations, analysing their intensities in terms of 
excitations with classical statistics, while retaining the quantum 
(more precisely, semi--classical) nature of their dynamics.
Using what we have learned, in \Autoref{section:MD.reconstructed}, 
we show explicitly that u3MD simulation results can be corrected 
to yield dynamical structure factors in agreement with the predictions 
of \Autoref{section:quantum.theory}.
We conclude, in \Autoref{section:cf.QMC}, with a comparison 
of  u3MC and u3MD simulations with published results from 
Quantum Monte Carlo (QMC).

\subsection{Molecular dynamics, deconstructed} 		
\label{section:MD.deconstructed}

\nic{We start by exploring} the relationship between classical 
and quantum theories for fluctuations about FQ order, and their 
implication for the understanding of simulation.
We concentrate on the experimentally--relevant structure factor 
for dipole moments $S_{\sf S}(\q)$, re--deriving the classical 
result quoted in \Autoref{section:classical.theory} 
in a framework which permits direct comparison with the quantum 
result given in \Autoref{section:quantum.theory}.


We take as starting point the quantum theory of excitations about the FQ state,  
\Autoref{eq:H.quantum}, and include a term $\Delta \Ham [{\bf h}]$ 
describing coupling of dipole moments to a transverse field 
\begin{eqnarray}
	\Ham &=& \Ham^\prime_{\sf BBQ} + \Delta \Ham [{\bf h}] \nonumber\\
	&=& E_0 + \Delta E_0 
	+ \sum_{\k} \hbar \epsilon(\k)  \left[ \balphad{\k} \balpha{\k}
	+ \bbetad{\k} \bbeta{\k} \right ] \nonumber\\
	&& - \sum_{\k} \xi_{\sf S} (\k) \left[ i h^x_{\k} ( \bbeta{\k} - \bbetad{-\k} ) 
	+ i h^z_{\k} ( \balphad{-\k} - \balpha{\k} ) \right] \; , \nonumber\\
\label{eq:HamLowTS3}
\end{eqnarray}
where the excitation energy $\epsilon(\k)$ is defined through 
\Autoref{eq:omega.k}, $\xi_{\sf S} (\k)$ is the coherence factor 
defined in \Autoref{eq:coherence.factor}, and all terms at 
cubic and higher order in Bosons have been neglected.
Here and in what follows we restore dimensional constants $\hbar$ and 
$k_B$, which have been set to unity elsewhere.


Recognising \Autoref{eq:HamLowTS3} as the Hamiltonian for a set of independent 
simple harmonic oscillators (SHO), we introduce a new set of coordinates 
\begin{subequations}
\begin{eqnarray}
\balpha{\k} &=& \sqrt{\frac{ m \epsilon(\k) }{2\hbar}} \hat{x}_{1,\k}
		+\frac{i}{\sqrt{2 \hbar m \epsilon(\k) }} \hat{p}_{1,\k}
		\label{eq:alpha_x_p} \; , \\
\bbeta{\k} &=& \sqrt{\frac{ m \epsilon(\k) }{2\hbar}} \hat{x}_{2, \k}
		+\frac{i}{\sqrt{ 2 \hbar m \epsilon(\k) }} \hat{p}_{2, \k}
		\label{eq:beta_x_p} \; ,
\end{eqnarray}
\end{subequations}
satisfying the canonical commutation relation 
\begin{eqnarray}
[ \hat{x}_{\nu,\k}, \hat{p}_{\nu',\k'}] 
	= i \hbar \delta_{\k\k'} \delta_{\nu\nu'}  \; .
\label{eq:commutation.relation}
\end{eqnarray}
with $\nu = 1, 2$. 
Written in terms of these coordinates, the  Hamiltonian
[\Autoref{eq:HamLowTS3}] becomes
\begin{eqnarray}
\Ham &=& E_0 + \Delta E_0 \nonumber\\
	&& + \sum_{\nu, \k} 
	\left[ 
		\frac{m\epsilon(\k)^2}{2} \hat{x}_{\nu,\k}^2 
		+ \frac{1}{2m}  \hat{p}_{\nu,\k}^2 
	\right. \nonumber\\
	&& 	
	\qquad - \left. \sqrt{ \frac{ 2 \xi_{\sf S}^2(\k) }{m \hbar \epsilon(\k)}} 
		\left( h^z_{\k}  \delta_{1,\nu} + h^x_{\k}  \delta_{2,\nu} \right) 
	 \hat{p}_{\nu, \k} \right ]\; , 
\label{eq:HamLowTS4}
\end{eqnarray}


As long as the commutation relation, \Autoref{eq:commutation.relation}, is respected, 
the excitations of \Autoref{eq:HamLowTS4} continue to have well--defined, Bosons statistics.
Meanwhile, the neglect of higher--order terms means that the dynamics of 
these excitations are treated at the level of a semi--classical approximation.
MD simulation, on the other hand, imposes quantum (semi--classical) dynamics on spin 
configurations drawn from classical MC simulation, and so not subject to any 
quantization condition.
And, crucially, the thermal distribution of the states generated by MC simulation
at low temperatures is conditioned by a classical, and not a quantum band 
dispersion [cf. \Autoref{fig:comparison.classical.quantum.dispersion}].


We can model the classical statistics found in MD simulation by ``turning off'' 
the quantization of excitations in \Autoref{eq:HamLowTS4}, 
and treating  $x_{\nu, \k}$ and $p_{\nu, \k}$ as independent, classical, variables.
This will inevitably lead us back to the classical theory developed in \Autoref{section:classical.theory}, 
but expressed in a form that makes it easier to draw conclusions about the relationship 
between classical and quantum results.
Doing so, the partition function 
associated with \Autoref{eq:HamLowTS4} is given by
\begin{align*}
	Z^{\sf CL'} &= \exp{-\beta (E_0 + \Delta E_0) } 
	\prod_{\nu,\k}
	\left[ \left( \int  \dd x_{\nu, \k}\  
	\exp{ -\frac{1}{2} \beta m\epsilon(\k)^2 x_{\nu, \k}^2 }  \right) \right.\\
	& \times \left. \left( \int \dd p_{\nu, \k}\ 
	\exp{ -\frac{\beta}{2m} p_{\nu, \k}^2 } 
	\exp{ \beta \sqrt{ \frac{2 \xi_{\sf S}^2(\k)}{m \hbar \epsilon(\k) } } 
	\left( h^z_{\k}  \delta_{1,\nu} + h^x_{\k}  \delta_{2,\nu} \right) 
	p_{\nu, \k} }  \right)
	\right] \; ,
	\label{eq:Z_ME1} \numberthis
\end{align*}
where $\beta = 1/k_B T$. 
The integrals in \Autoref{eq:Z_ME1} can be evaluated exactly 
[\Autoref{eq:gauss_int_S1}, \Autoref{eq:gauss_int_S2}], to give
\begin{align*}
	Z^{\sf CL'} = \exp{-\beta (E_0 + \Delta E_0) } 
	\prod_{\k}^{N} &\left[\frac{2\pi}{\beta \epsilon(\k)} 
			\exp{\frac{\beta \xi_{\sf S}^2(\k) (h^z_\k)^2 }{\hbar\epsilon(\k) }} \right.\\
			&\times\left.\frac{2\pi}{\beta\epsilon(\k)}
			\exp{\frac{ \beta \xi_{\sf S}^2(\k) (h^x_\k)^2 }{\hbar\epsilon(\k) }} 
	\right] \; . 
\label{eq:Z_ME2} \numberthis
\end{align*}
By construction, this theory now describes excitations subject to the 
classical (i.e. Boltzmann) statistics used in MC simulation.
We are now in a position to calculate equal--time spin correlations 
using the same method as in \Autoref{sec:classical.structure.factors}, i.e. by 
constructing a free energy and differentiating this with respect to $h^\alpha_\k$.
Doing so, we find
\begin{eqnarray}
	S^{\sf CL'}_{\rm{S}}(\q, T) 
		&=& \sum_\alpha\nn{ \hat{S}^\alpha_\q \hat{S}^\alpha_\q }
		= \frac{4 \xi_{\sf S}^2(\q)}{\beta\hbar \epsilon(\q)}  \; ,
\label{eq:S.q.mixed.v1}
\end{eqnarray}
where we have used the fact that $\nn{\hat{S}^\alpha_\q} \equiv 0$ in the FQ state.


\begin{figure*}[t]
	\begin{center}
		\begin{minipage}[b]{\textwidth}
			\subfloat[$\frac{\omega}{T}S^{\sf MD}_{\rm{S}}(\q, \omega), T = 0.05$	  	\label{fig:wTSqw_MC}]{
				\includegraphics[width=0.32\linewidth]{Sim_wTSwq_T005.pdf}
			}
			\subfloat[$\frac{\omega}{T}S^{\sf MD}_{\rm{Q}}(\q, \omega), T = 0.05$		\label{fig:wTQqw_MC}]{
				\includegraphics[width=0.32\linewidth]{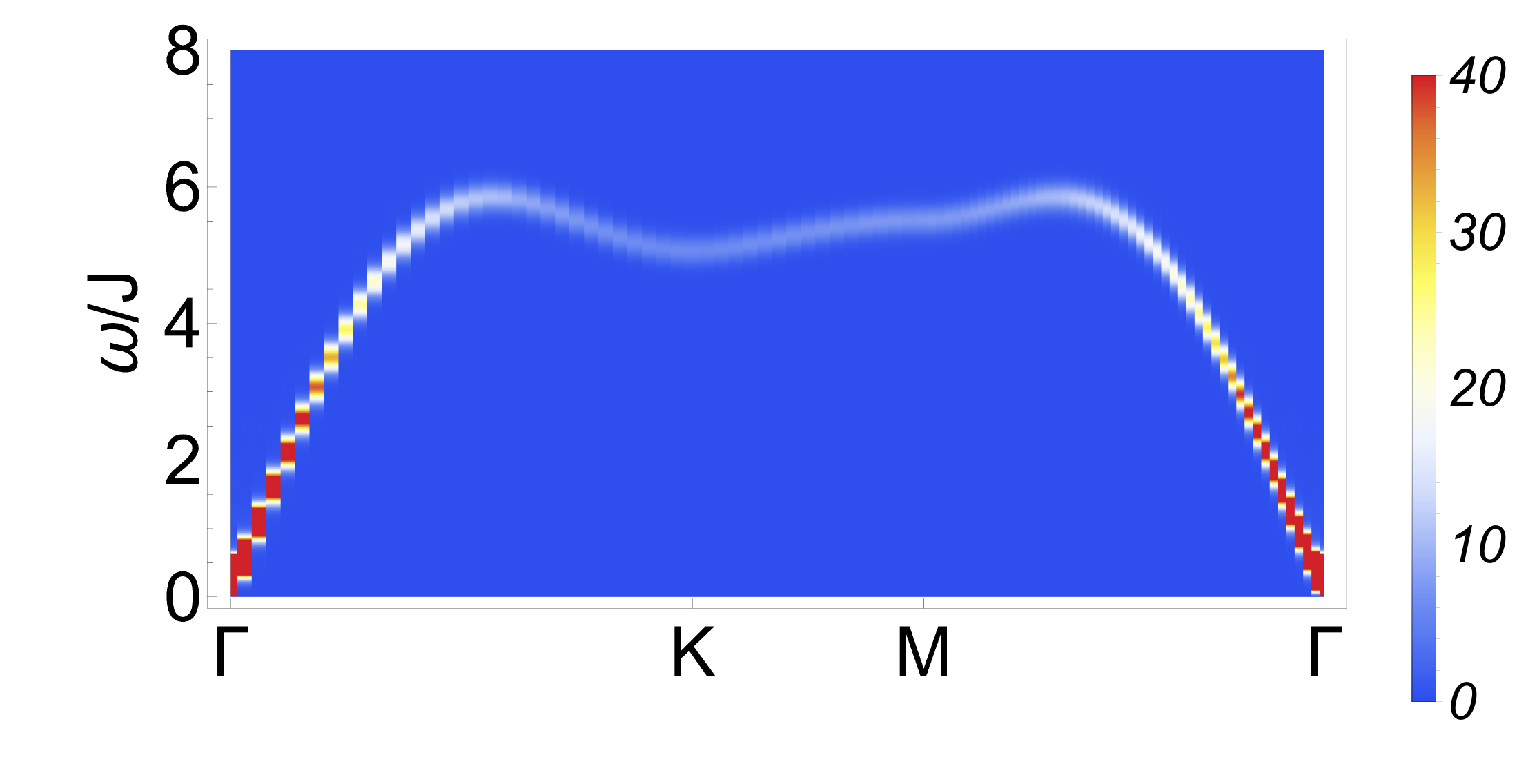}
			}
			\subfloat[$\frac{\omega}{T}S^{\sf MD}_{\rm{A}}(\q, \omega), T  = 0.05$		\label{fig:wTAqw_MC}]{
				\includegraphics[width=0.32\linewidth]{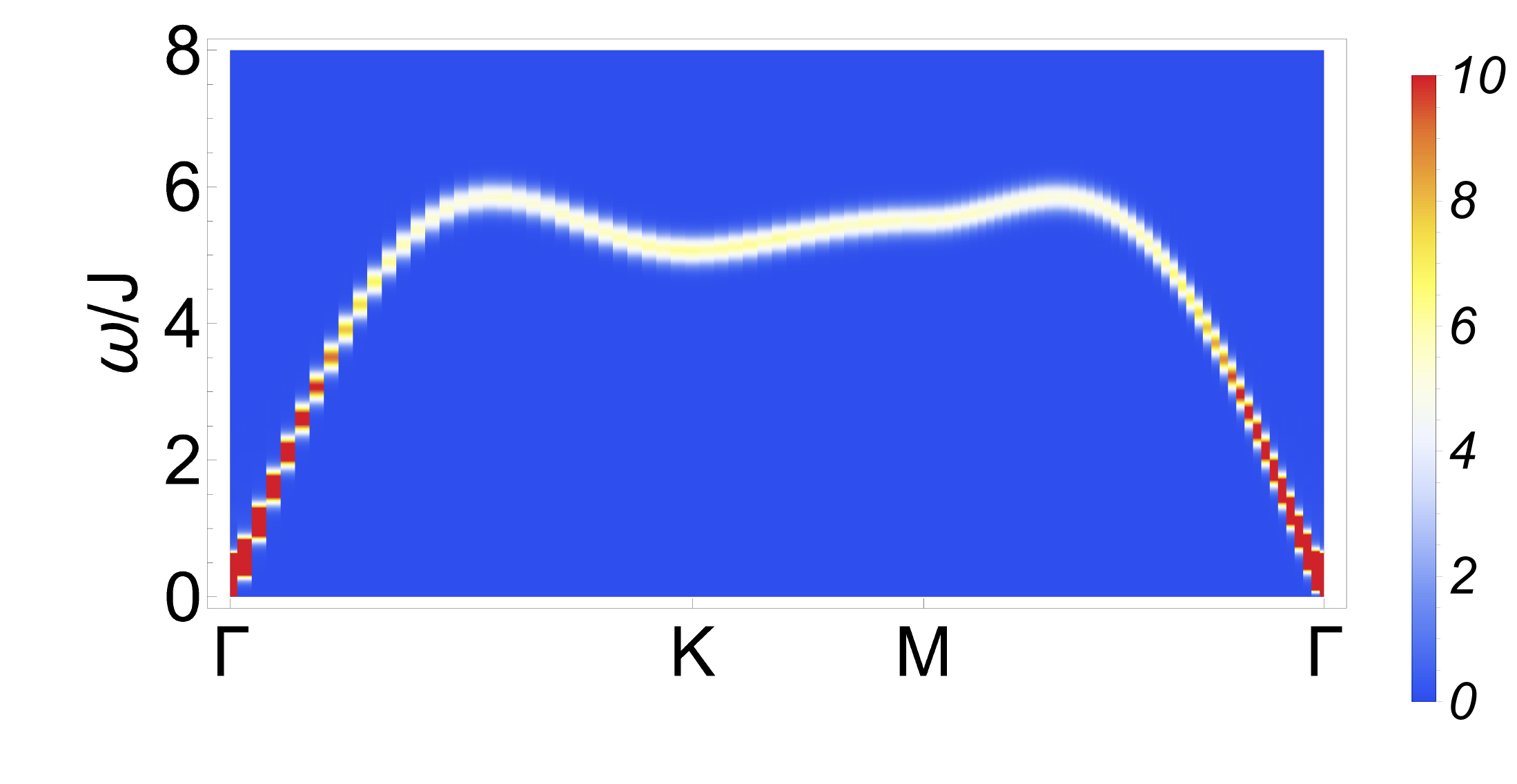}
			}\\		
			\subfloat[$S^{\sf QM}_{\rm{S}}(\q, \omega), T = 0$		\label{fig:Sqw_ana}]{
				\includegraphics[width=0.32\linewidth]{Ana_FW_Sqw.pdf}
			}
			\subfloat[$S^{\sf QM}_{\rm{Q}}(\q, \omega), T = 0 $		\label{fig:Qqw_ana}]{
				\includegraphics[width=0.32\linewidth]{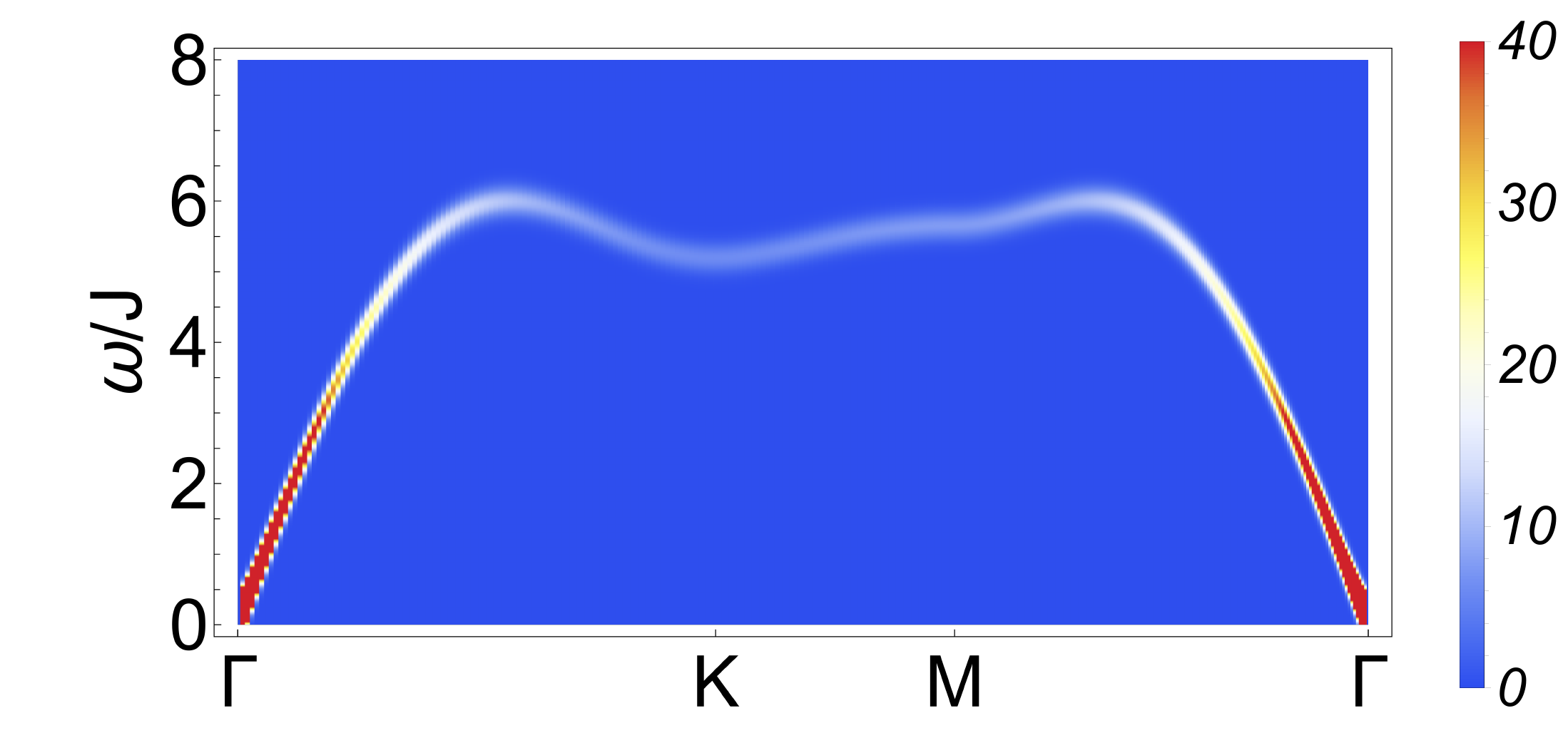}
			}
			\subfloat[$S^{\sf QM}_{\rm{A}}(\q, \omega), T = 0$		\label{fig:Aqw_ana}]{
				\includegraphics[width=0.32\linewidth]{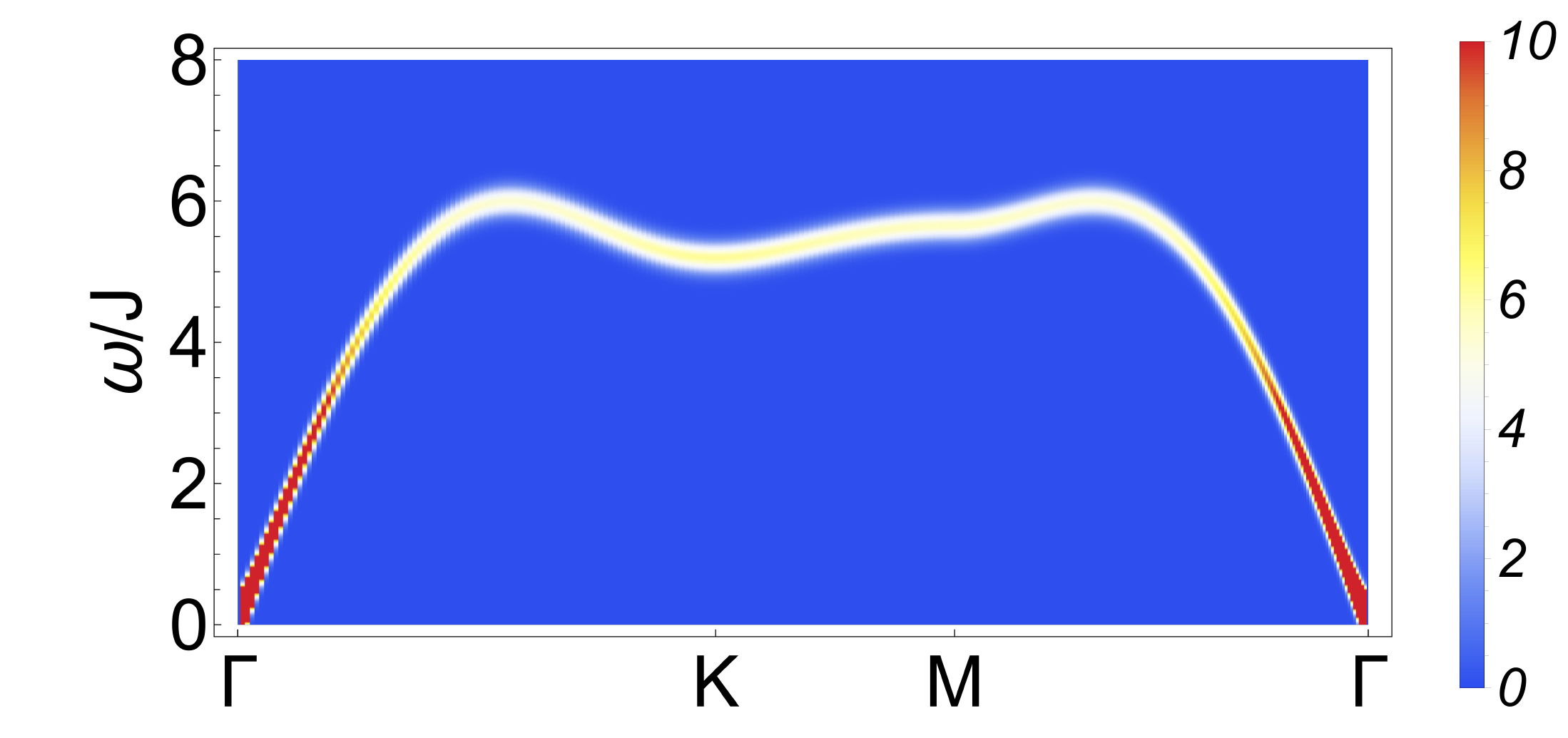}
			} 
		\end{minipage}
	\end{center}
	\caption{
	Comparison between dynamical structure factors found in molecular dynamics 
	(u3MD) simulations of a ferroquadrupolar (FQ) state, and those found in a \mbox{$T=0$} 
	quantum theory.
	\protect\subref{fig:wTSqw_MC}~Simulation results for dynamical structure factor associated with dipole 
	moments, $S^{\sf MD}_{\rm{S}}(\q, \omega)$, corrected for classical statistics, 
	following \protect\Autoref{eq:correcting.for.classical.statistics}.
	\protect\subref{fig:wTQqw_MC}~Equivalent results for quadrupole moments, $S^{\sf MD}_{\rm{Q}}(\q, \omega)$.
	\protect\subref{fig:wTAqw_MC}~Equivalent results for A--matrices, $S^{\sf MD}_{\rm{Q}}(\q, \omega)$.
	\protect\subref{fig:Sqw_ana}~Prediction for $S^{\sf QM}_{\rm{S}}(\q, \omega)$ from \mbox{$T=0$} quantum theory 
	[\protect\Autoref{eq:Sqw_qm}].
	\protect\subref{fig:Qqw_ana}~Equivalent prediction for $S^{\sf QM}_{\rm{Q}}(\q, \omega)$ [\protect\Autoref{eq:Qqw_qm}].
	\protect\subref{fig:Aqw_ana}~Equivalent prediction for $S^{\sf QM}_{\rm{A}}(\q, \omega)$ [\protect\Autoref{eq:Aqw_qm}].
	Simulations were carried out using the u3MD simulation scheme described 
	in \protect\Autoref{sec:u3MD}, for $\Ham_{\sf BBQ}$ [\protect\Autoref{eq:H.BBQ.3}] with 
	parameters \protect\Autoref{eq:simulation.parameters}, at a temperature $T = 0.05\ J$, 
	in a cluster of linear dimension $L = 96$ ($N = 9216$ spins).
	All result have been convoluted with a Gaussian in frequency 
	of \mbox{FWHM = 0.35 J}.
	}
	\label{fig:comparison_dispersion_intensities}
\end{figure*}


The presence of the quantum dispersion $\epsilon(\q)$ and 
coherence factor $\xi_{\sf S} (\q)$ in \Autoref{eq:S.q.mixed.v1}, 
is suggestive of the quantum theory developed in \Autoref{section:quantum.theory}.
And, by direct comparison with \Autoref{eq:S.equal.time.structure.factor}, 
we find
\be
S^{\sf CL'}_{\rm{S}}(\q, T) 
	= 2 \frac{S^{\sf QM}_{\rm{S}}(\q, T=0)}{\beta \hbar \epsilon(\q)} \; ,
\label{eq:intermediate.result}
\ee
a result which holds in the limit of low temperature. 
At the same time, $S^{\sf CL'}_{\rm{S}}(\q)$ must ultimately be equivalent 
to the earlier classical result $S^{\sf CL}_{\rm{S}}(\q)$ [\Autoref{eq:SF_class}].
To this end, we can simplify \Autoref{eq:S.q.mixed.v1} using 
\Autoref{eq:simplification}, to recover 
\begin{eqnarray}
	S^{\sf QM}_{\rm{S}} (\q) 
	= 2 \xi_{\sf S}^2 (\q)
	\; ,
\end{eqnarray}
previously introduced as \Autoref{eq:Sqomega.simplified}. 
It follows that 
\begin{eqnarray}
	S^{\sf CL'}_{\rm{S}}(\q, T) &=& \frac{4}{ \beta(A_\q - B_\q ) } 
	= S^{\sf CL}_{\rm{S}}(\q, T) \; ,
\label{eq:S.q.mixed.v2}
\end{eqnarray}
where this result also holds in the limit of low temperature.  
Combining this with \Autoref{eq:intermediate.result}, we arrive at 
a result which relates classical correlations at finite temperature, 
to those of a quantum system at $T=0$:
\be
S^{\sf QM}_{\rm{S}}(\q, T=0) 
	= \lim_{T \to 0} \frac{ \hbar \epsilon(\q)}{2 k_B T} S^{\sf CL}_{\rm{S}}(\q, T)  \; .
\label{eq:SME_SQM}
\ee


The approach developed above can be generalised from dipole moments 
$\lambda = {\sf S}$, to quadrupole moments, $\lambda = {\sf Q}$ 
and A--matrices, $\lambda = {\sf A}$.
This leads to the general result
\begin{eqnarray}
S^{\sf QM}_\lambda (\q, T=0) 
	= \lim_{T \to 0}  \frac{\hbar \epsilon(\q)}{2 k_B T} S^{\sf CL}_\lambda (\q, T) \; ,
\label{eq:general.result.one}
\end{eqnarray}
where we make explicit the role of temperature,  
and restore dimensional constants $k_B$ and $\hbar$. 
We emphasize that the factor of $\epsilon(\q)$ in \Autoref{eq:general.result.one}
reflects the dispersion for a {\it quantized} excitation [\Autoref{eq:omega.k}], 
and not the eigenvalue of a classical theory.
It is also important to note that quantum mechanics have been treated at 
a semi--classical level, i.e. taking account of quantization, but considering 
only one path in the path integral.
This approximation is, of course, exact for a SHO.


The principle problem encountered in ``raw'' MD results for dynamical structure 
factors, relative to quantum results at low temperatures, was the loss of spectral 
weight at low temperatures [cf. \Autoref{fig:T-scaling_Sqw_no_wT}].
At low temperatures, we can equate $S^{\sf CL}_\lambda (\q, T)$ 
with the structure factor found in MC simulation
\begin{eqnarray}
	\lim_{T \to 0} S^{\sf MC}_\lambda(\q, T) = \lim_{T \to 0} S^{\sf CL}_\lambda (\q, T) \; ,
\end{eqnarray}
permitting us to write 
\begin{eqnarray}
S^{\sf QM}_\lambda (\q, T=0) 
	= \lim_{T \to 0}  \frac{\hbar \epsilon(\q)}{2 k_B T} S^{\sf MC}_\lambda (\q, T) \; .
\label{eq:general.result}
\end{eqnarray}
We can therefore use MC simulation to estimate the total spectral weight 
in a zero--temperature quantum theory, at given $\q$, as long as we had 
prior knowledge of the characteristic energy scale $\epsilon(\q)$.
What remains is to understand the relationship between classical and quantum 
results in the absence of prior knowledge of the dispersion.


\nic{
The effect of MD simulation is to redistribute the spectral weight at 
a given $\q$ over a range of different $\omega$, subject to the sum rule, 
\begin{eqnarray}
	S^{\sf MC}_\lambda(\q, T) = \int^\infty_{-\infty} d\omega\ S^{\sf MD}_\lambda(\q, \omega, T)  \; .
\label{eq:sum.rule.on.S.q.omega}
\end{eqnarray}
To estimate the zero--temperature quantum result 
\mbox{$S^{\sf QM}_\lambda(\q, \omega, T=0)$}, we therefore need to 
construct a model for this redistribution of spectral weight,  
subject to the condition that dynamics are treated at a semi--classical level.
}


\nic{
Here it is instructive to return to the simulation results for fluctuations 
about FQ order, described in \Autoref{section:numerics.dynamics}.
From the ``raw'' results, \Autoref{fig:dispersion_rawMD} 
and \Autoref{fig:T-scaling_Sqw_no_wT}, we learn that 
\begin{enumerate}[label=(\roman*)]
\item for  \mbox{$T \to 0$}, the characteristic energy scale of excitations 
converges on the exact quantum (semi--classical) result 
$\epsilon({\bf q})$ [\Autoref{eq:omega.k}], and 
\item excitations become sharp (resolution limited) for \mbox{$T \to 0$}.
\end{enumerate}
}


\nic{
FQ order, studied here, show a single, two--fold degenerate band of excitations.   
More generally, there may be many different excitations at a given ${\bf q}$.
None the less, at a semi--classical level, (i.e. treated as a set of independent 
oscillators), in a finite--size system, each of these will have a well--defined 
energy.
It is also important to remember that, while only results for positive frequency 
have been plotted in \Autoref{fig:dispersion_rawMD}, MD simulation will return 
solutions at both positive and negative energy, with equal weight 
[\Autoref{eq:symmetry.of.Sqomega}].
}


\begin{figure*}[t]
	\begin{center}
		\begin{minipage}[t]{0.73\textwidth}
			\subfloat[\label{fig:Awq_K_T}]{
				\includegraphics[width=\textwidth]{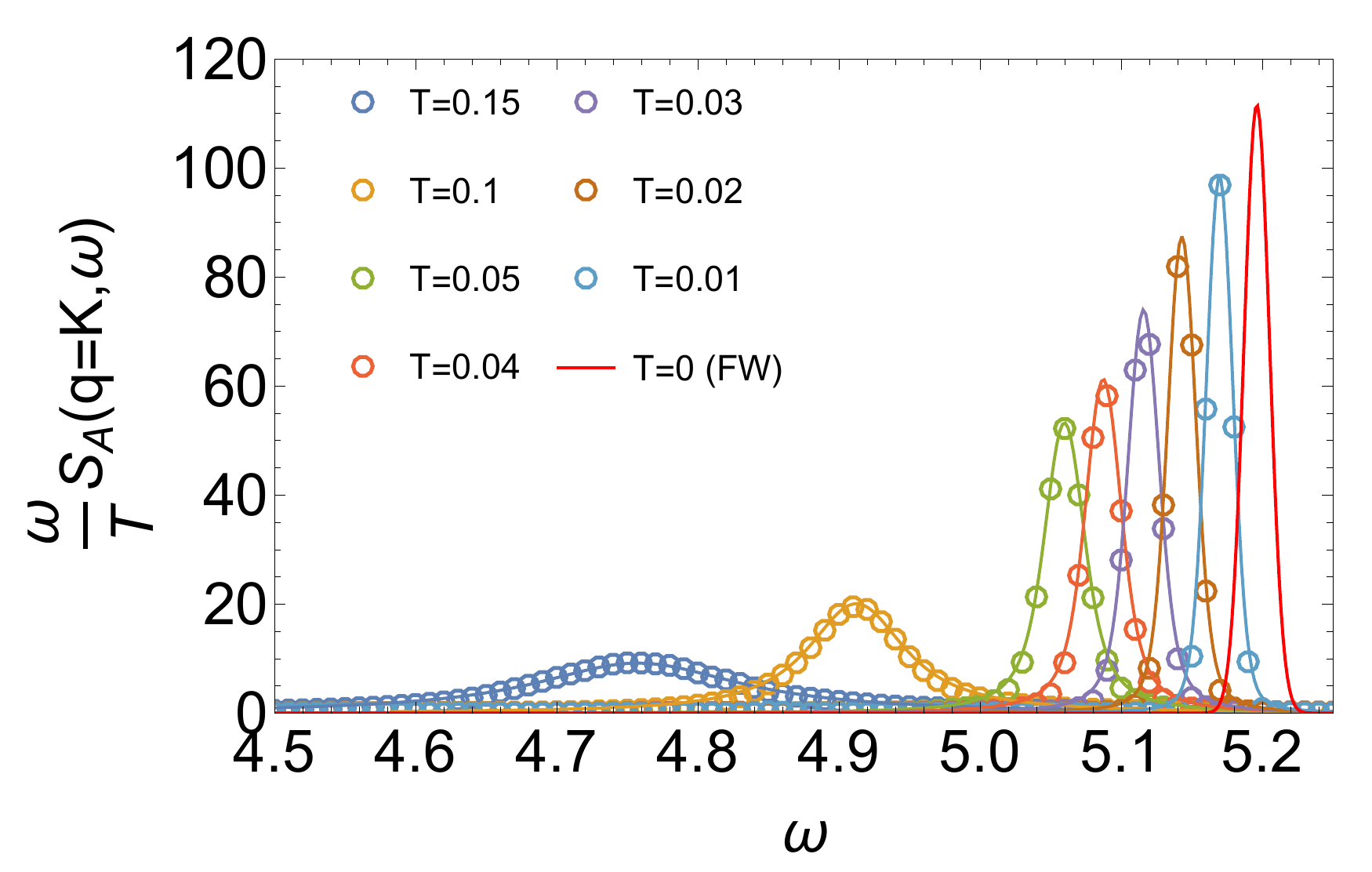}
			}
		\end{minipage}
		\begin{minipage}[t]{0.23\textwidth}
			\subfloat[\label{fig:Awq_K_T_w}]{
				\includegraphics[width=\linewidth]{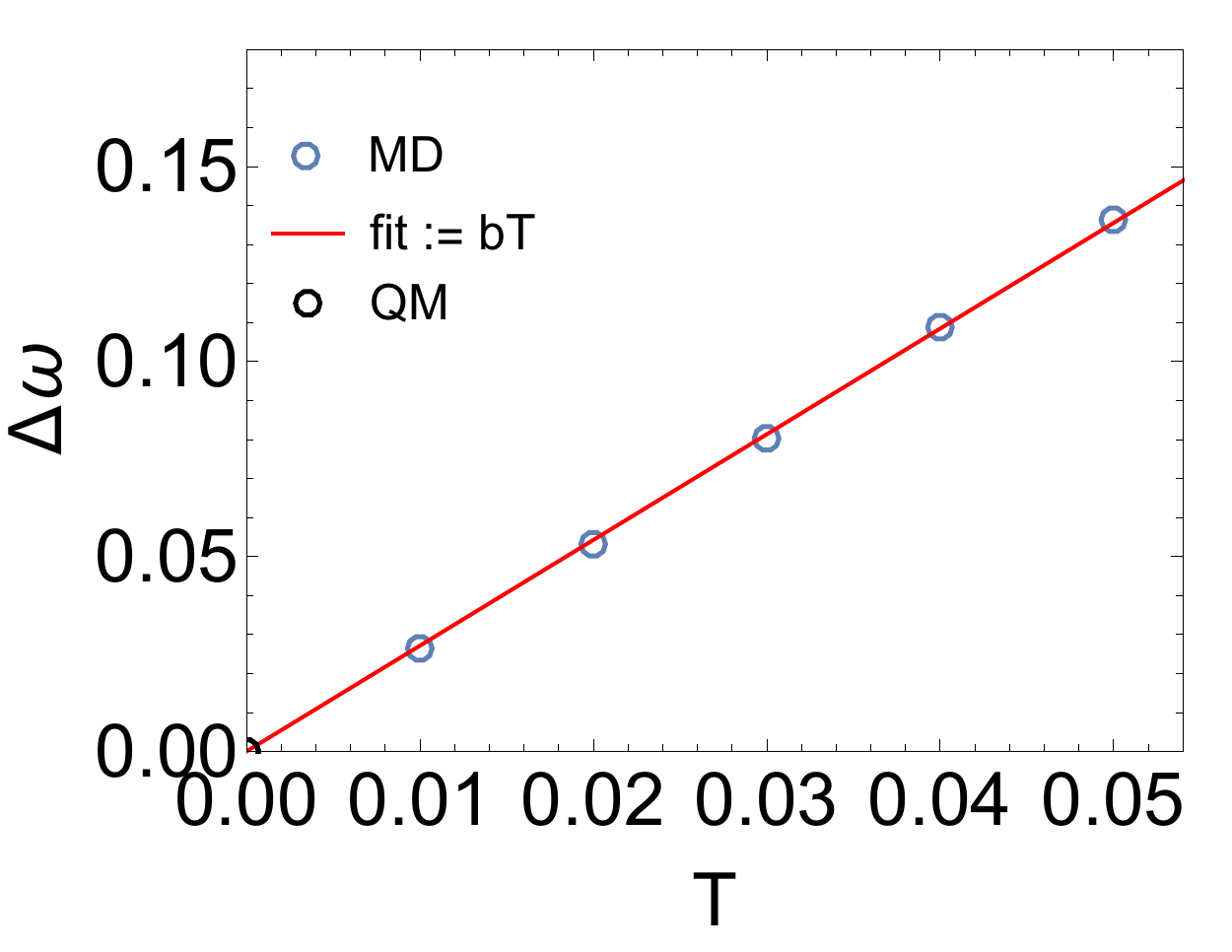}
			}\\
			\subfloat[\label{fig:Awq_K_T_Gamma}]{
				\includegraphics[width=\linewidth]{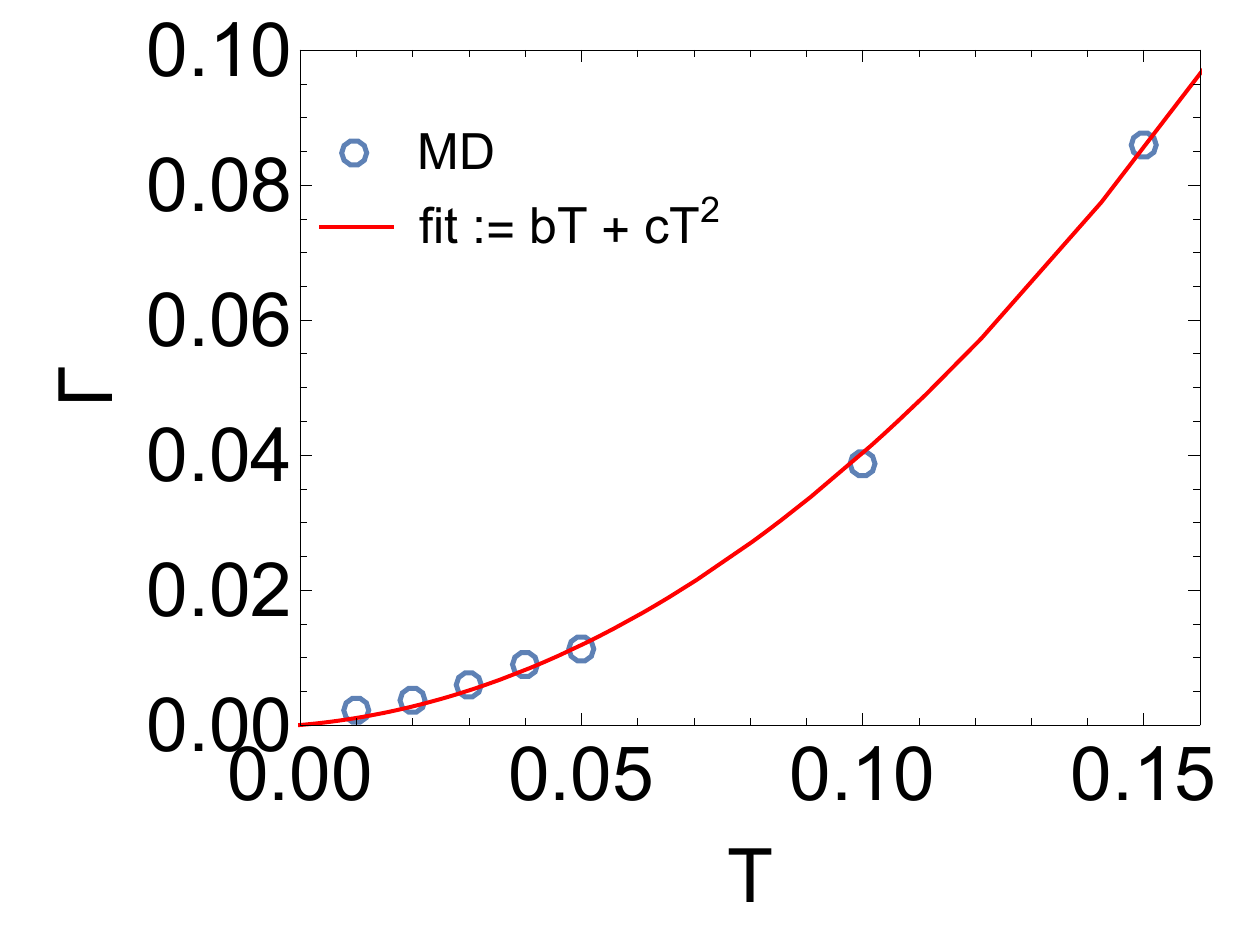}
			}\\
			\subfloat[\label{fig:Awq_K_T_peak}]{
				\includegraphics[width=\linewidth]{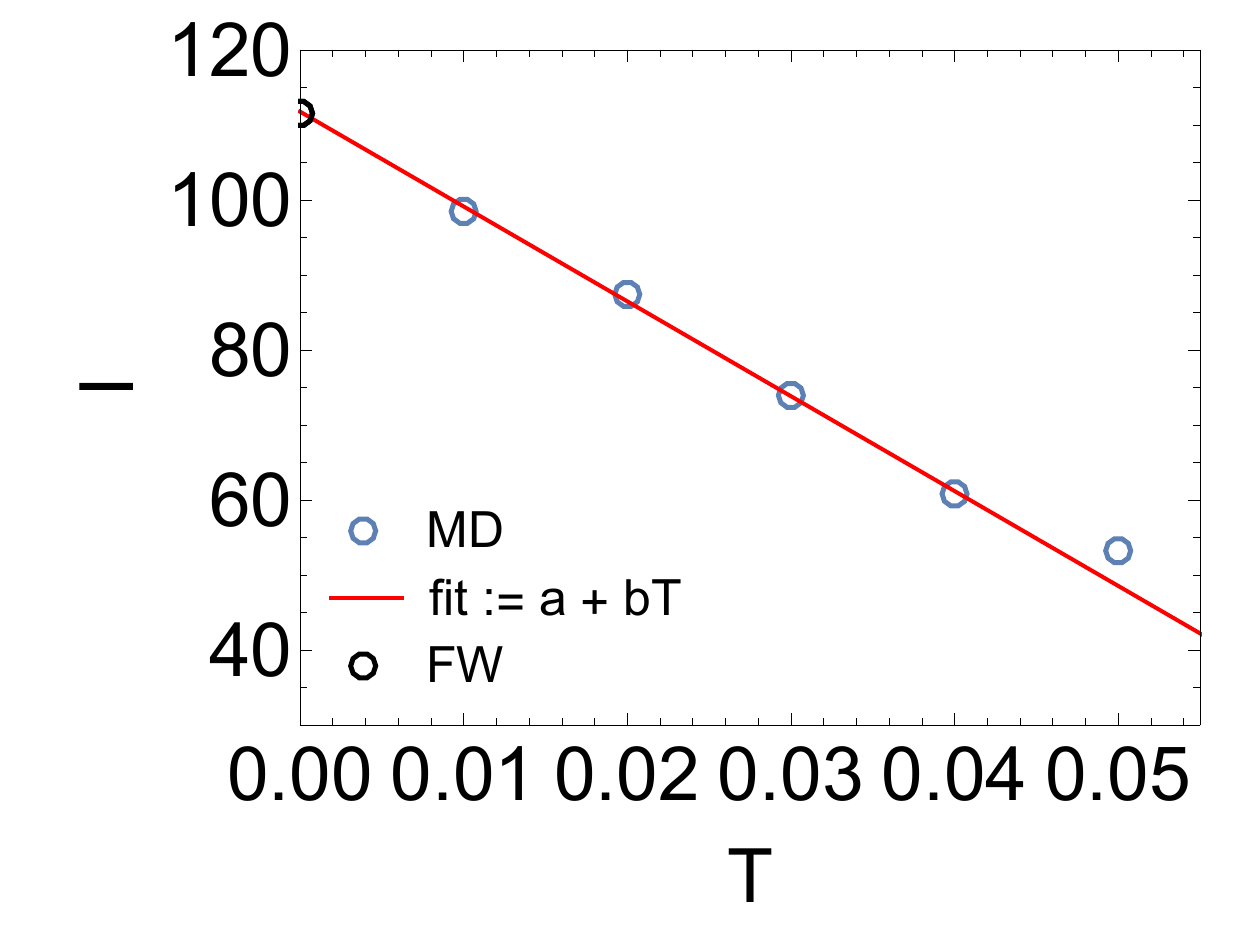}
			}
		\end{minipage}
	\end{center}
	\caption{
	Temperature dependence of results of $U(3)$ Molecular Dynamics (u3MD) 
	simulation after correction for classical statistics, showing convergence 
	on the predictions of a \mbox{$T=0$} quantum theory for \mbox{$T \to 0$}. 
	\protect\subref{fig:Awq_K_T}~u3MD results for dynamical structure 
	factor associated with A--matrices, $S_{\rm{A}}^{\sf MD}(\q, \omega)$, at wavevector 
	$\q=\K$, for temperatures ranging from \mbox{$T = 0.01\ J$} to \mbox{$T = 0.15\ J$}.  
	Simulation results (points), have energy resolution $0.02\ J$, 
	and have been corrected for classical statistics through 
	\protect\Autoref{eq:correcting.for.classical.statistics}.
	Lines show fits to a Voigt profile, \protect\Autoref{eq:voigt}.
	The prediction of the \mbox{$T=0$} quantum theory, \protect\Autoref{eq:Aqw_qm}, 
	convoluted with a Gaussian of \mbox{$\text{FWHM = 0.02\ J}$}, 
	is shown with a solid red line.   
	\protect\subref{fig:Awq_K_T_w}~Shift in peak energy $\Delta\omega(T)$, 
	found from fits to a Voigt profile, showing convergence of peak position 
	on the prediction of \mbox{$T=0$} quantum theory.
	\protect\subref{fig:Awq_K_T_Gamma}~Equivalent results for the inverse lifetime $\Gamma(T)$. 
	%
	\protect\subref{fig:Awq_K_T_peak}~Equivalent results for the peak height, $I(T)$. 
	%
	Parameters for simulations are identical to those used in 
	\protect\Autoref{fig:comparison_dispersion_intensities}.
	}
	\label{fig:T-scaling_Sqw}
\end{figure*}


\nic{
With these assumptions in mind, we model MD simulation results in the limit $T \to 0$
in terms of delta--function peaks at energy $ \omega = \pm \epsilon_\nu (\k) $, 
with spectral weight shared equally between these two peaks, vis
\begin{eqnarray} 
	&& \lim_{T \to 0} S^{\sf MD}_\lambda(\q, \omega, T)  \nonumber \\
	&=& \sum_\nu \left[ A_ {\lambda,\nu} ({\bf q}, T) 
		\delta ( \omega - \epsilon_\nu (\k) )  
		+ A_ {\lambda,\nu} ({\bf q}, T) 
		\delta ( \omega + \epsilon_\nu (\k) )  \right]
		\nonumber \\
	&& \qquad + {\mathcal O}(T^2) \; .
\label{eq:model.of.MD}
\end{eqnarray} 
Here the sum on $\nu$ runs over all eigenmodes of the cluster 
with wavevector ${\bf q}$, and the corresponding spectral weight 
\begin{eqnarray} 
	A_{\lambda,\nu} ({\bf q}, T) 
		= \frac{k_B T}{\hbar \epsilon_\nu ({\bf q})}  
			\xi_{\lambda,\nu}^2 (\q) \; 
\end{eqnarray} 
is defined through a generalized coherence factor  
\begin{eqnarray} 
	\xi_{\lambda,\nu}^2 (\q) \geq 0 \; , 
\end{eqnarray} 
specific to the structure factor in question.
The total spectral weight in these modes is constrained 
through \Autoref{eq:sum.rule.on.S.q.omega}, 
and satisfies 
%
%
\begin{eqnarray}
S^{\sf QM}_\lambda (\q, T=0) 
	&=& \int_{-\infty}^\infty d\omega\ S^{\sf QM}_\lambda (\q, \omega, T=0) 
	\nonumber \\
	&=& \sum_\nu \xi_{\lambda,\nu}^2 (\q) \; ,
\label{eq:general.result.S.q.omega}
\end{eqnarray}
where the sum on $\nu$ runs over the two degenerate branches of FQ excitations.
For the dipolar structure factor, $\lambda = S$, this sum contributes a 
factor $\times2$, and \Autoref{eq:general.result.S.q.omega} can be compared 
directly with \Autoref{eq:Sqomega.simplified}.
}


\nic{
Where the model \Autoref{eq:model.of.MD} holds, no prior knowledge of 
excitation energies $\epsilon_\nu ({\bf q})$ is needed to correct for the 
effect of classical statistics in MD simulation.
And since only positive frequencies, corresponding to transfer 
of energy to the system, are relevant at $T=0$, we can write
\begin{eqnarray}
S^{\sf QM}_\lambda (\q, \omega, T=0) 
	&=& \lim_{T \to 0}  \frac{\hbar \omega}{k_B T} 
		S^{\sf MD}_\lambda (\q, \omega, T) \; \; [\omega > 0] \; .
\nonumber \\
\label{eq:the.big.result}
\end{eqnarray}
Here} we understand that ``{\sf QM}'' should be taken to imply 
``semi--classical'', i.e. pertaining to excitations with quantum 
statistics, treated a Gaussian level of approximation.
Empirical evidence for the validity of \Autoref{eq:the.big.result}
is provided in \Autoref{section:MD.reconstructed}, below.
Equivalent results for a system with many bands can be found in 
in \cite{Pohle2021}.


\nic{
We conclude by noting that the approach of correcting for 
classical statistics by multiplying dynamical structure factors 
by a prefactor \mbox{$\omega/T$} has been anticipated several times 
in the literature of MD simulation, including in studies of the 
spin--1/2 magnet Ca$_{10}$Cr$_{7}$O$_{28}$
\cite{Pohle2017-arXiv,Pohle2021}, 
the spin--1 magnet NaCaNi$_7$O$_7$ \cite{Zhang2019}, 
and dynamical scaling in Yb$_2$Ti$_2$O$_7$ \cite{Scheie2022-arXiv}.
The  factor $\omega/2T$ used in \cite{Pohle2021} reflects 
a different normalisation of MD results.
}

\subsection{Quantum results, reconstructed} 		
\label{section:MD.reconstructed}

Armed with \Autoref{eq:the.big.result}, we are now in a 
position to revisit MD simulation results for excitations about a FQ ground state, 
previously discussed in \Autoref{section:numerics.dynamics}.
In \Autoref{fig:comparison_dispersion_intensities}, we show a comparison 
between MD simulation results, and the predictions of the zero--temperature 
quantum theory developed in \Autoref{section:quantum.theory}.
\nic{
Following \Autoref{eq:the.big.result}, simulation results 
have been corrected by multiplying them by prefactor $\omega/T$, vis
\begin{eqnarray}
	\tilde{S}^{\sf MD}_\lambda (\q, \omega, T) 
		&=& \frac{\omega}{T} S^{\sf MD}_\lambda (\q, \omega, T)  \; ,
\label{eq:correcting.for.classical.statistics}
\end{eqnarray}
}
where the constants $k_B$ and $\hbar$ have again been set to unity.
Results obtained at $T = 0.05\ J$, corrected in this way, are shown in \Autoref{fig:comparison_dispersion_intensities}~\protect\subref{fig:wTSqw_MC}--\protect\subref{fig:wTAqw_MC}.
In this case, u3MD simulations were carried out at a resolution
of $ 0.02\ J$, corrected according to \Autoref{eq:correcting.for.classical.statistics}, 
and then further convoluted with with a Gaussian envelope of 
$\text{FWHM} =  0.33\ J$, so as to achieve a net energy resolution of 
$0.35\ J$, directly comparable to results in \Autoref{section:numerics.dynamics}.

The results 
in \Autoref{fig:comparison_dispersion_intensities}~\protect\subref{fig:wTSqw_MC}--\protect\subref{fig:wTAqw_MC}, 
should be contrasted with the ``raw'' results 
of MD simulation, shown \mbox{\Autoref{fig:dispersion_rawMD}~\protect\subref{fig:Sqw_MC}--\protect\subref{fig:Aqw_MC}}.
Relative to these, corrected results show a far less spectral 
weight at low energies, an effect which is particularly evident for 
$S^{\sf MD}_{\sf} (\q, \omega, T)$.
Meanwhile, for comparison, in \mbox{\Autoref{fig:comparison_dispersion_intensities}
\protect\subref{fig:Aqw_ana}--\protect\subref{fig:Sqw_ana}}, we reproduce equivalent 
results from the \mbox{$T=0$} analytic theory, previously shown in 
\Autoref{fig:comparison.classical.quantum.dispersion}. 
Compared at the level of density plots, the agreement between corrected 
simulation results 
and the zero--temperature quantum prediction 
is essentially perfect, with no visible mismatches in dispersion or intensity.


A more precise comparison between simulation and zero--temperature quantum 
theory can be achieved by plotting \mbox{$\tilde{S}^{\sf MD}_\lambda (\q, \omega, T)$} 
at fixed wavevector $\q$, for a sequence of temperatures converging on \mbox{$T=0$}.
This is accomplished in \Autoref{fig:T-scaling_Sqw}~\subref{fig:Awq_K_T}, where we plot  
results for \mbox{$\tilde{S}^{\sf MD}_{\sf A} (\q=\K, \omega, T)$}, 
for temperatures ranging from \mbox{$T = 0.15\ J$} to \mbox{$T = 0.01\ J$}.
For comparison, we also show the result of the \mbox{$T=0$} analytic theory, 
\Autoref{eq:Aqw_qm}.
Both simulation and analytic prediction have been convoluted
with a Gaussian of FWHM $0.02\ J$.
Plotted in this way, the role of the limit in \Autoref{eq:the.big.result}
becomes clear: MD simulation results corrected using 
\Autoref{eq:correcting.for.classical.statistics} form a sequence 
which converge on the $T=0$ analytic prediction for \mbox{$T \to 0$}. 
%


Having established the validity of \Autoref{eq:the.big.result}, it is  
interesting to examine more precisely the way in which corrected 
simulation results converge on the zero--temperature quantum result.
The dispersing peak in \mbox{$\tilde{S}^{\sf MD}_\lambda (\q, \omega, T)$} 
is \nic{still} well--described by the Voigt lineshape, \Autoref{eq:voigt}, with fits 
shown as solid lines in \Autoref{fig:comparison_dispersion_intensities}~\protect\subref{fig:wTSqw_MC}--\protect\subref{fig:wTAqw_MC}.
Within limits set by the energy resolution of simulations, these fits allow us to extract 
quantitative estimates for the shift in excitation energy $\Delta\omega(T)$ [\Autoref{fig:T-scaling_Sqw}~\subref{fig:Awq_K_T_w}], 
the inverse lifetime of the excitation, $\Gamma(T)$ [\Autoref{fig:T-scaling_Sqw}~\subref{fig:Awq_K_T_Gamma}], 
and the intensity maximum $I(T)$ [\Autoref{fig:T-scaling_Sqw}~\subref{fig:Awq_K_T_peak}], 
as a function of temperature.


We find that the peak position converges linearly on the zero--temperature 
quantum result from below, with 
\begin{eqnarray}
	\Delta\omega(T) &=& b T + \mathcal{O}(T^3) \; , \; 
	\text{[b = 2.71]} \; , 
\end{eqnarray}
Meanwhile, the inverse lifetime of the excitation vanishes (approximately) quadratically 
as $T \to 0$
\begin{eqnarray}
	\Gamma(T) &=& b T + c T^2 + \mathcal{O}(T^3) \; , \; 
		\text{[b = 0.07, c = 3.26]} \; , 
\end{eqnarray}
while the maximum intensity of the peak also converges linearly on
the expected value
\begin{eqnarray}
	I(T) &=& a + b T + \mathcal{O}(T^3) \; , \; 
		\text{[a = 111, b = -1230]} \; ,
\end{eqnarray}
where the coefficient $a$ matches the prediction of the \mbox{$T=0$} quantum theory.


It is possible to construct a diagrammatic expansion for the self energy 
of excitations within the mixed ensemble of MD simulation \cite{owen-unpublished}.
Such calculations lie beyond the scope of this paper but, on general grounds, 
it is possible to offer an interpretation of some of the trends observed in simulation.


At low temperatures, the shift in peak position,  $\Delta \omega(T)$ 
will depend on the density of excitations (one--loop diagram).
Because of the classical statistics of the MC simulation, 
this density is linear in $T$.
Meanwhile, the inverse lifetime, $\Gamma(T)$, will be determined by interactions 
which are present in finite--temperature simulations, 
but absent from the Gaussian--level quantum theory developed 
in \Autoref{section:quantum.theory}.
These processes correspond to Feynman diagrams with a finite 
imaginary part, and will generically have the form of ``bubbles''.
Empirically the dominant low--temperature contribution occurs at 
${\mathcal O}(T^2)$, i.e. at second order in the density of fluctuations.


We leave a more quantitative analysis of these effects for future work.

\subsection{Comparison with the results of QMC simulation}
\label{section:cf.QMC}

In \Autoref{section:MD.reconstructed}, we have explored the correspondence 
between u3MC simulation results at finite temperature, and analytic 
quantum (semi--classical) results at $T=0$.
It is also interesting to consider how they compare with published 
quantum Monte Carlo (QMC) simulation data.

\subsubsection{Ground state}

The spin--1 BBQ model on a triangular lattice [\Autoref{eq:BBQ.model}] 
is accessible to QMC simulation for $J_1 \leq 0$, $J_2 \leq 0$.
(Equivalently, from \Autoref{eq:J.from.theta}, the quadrant $-\pi \leq \theta \leq \pi/2$).
Stochastic series expansion (SSE) methods have been used to obtain results 
for both thermodynamics and dynamics, at finite temperature, across this 
parameter range \cite{Voell2015}.
A more specialized loop--expansion method has also been used to study 
properties at the special point $J_1 = 0$, $J_2 = -1$ 
\mbox{($\theta = -\pi/2$)} \cite{Kaul2012}.


Both QMC results \cite{Voell2015}, and u3MC simulations [\Autoref{sec:phase.diagram}], 
are consistent with a FQ ground state extending from 
the special point \mbox{$\theta = -\pi/2$}, to the $SU(3)$ point, \mbox{$\theta = -3\pi/4$}.
Meanwhile, for $-3\pi/4 < \theta -\pi$, both methods find a FM ground state.
This distribution of FQ and FM ground states
is consistent with mean--field predictions [\Autoref{fig:MF.phase.diagram}], 
results from exact diagonalisation \cite{Lauchli2006}, and more recent calculations 
using tensor--network methods \cite{Niesen2018}.


\begin{figure*}[t]
	\centering
	\subfloat[$c(T)$ as function of $\log T$ \label{fig:heat.capacity1}]{
		\includegraphics[width=0.45\textwidth]{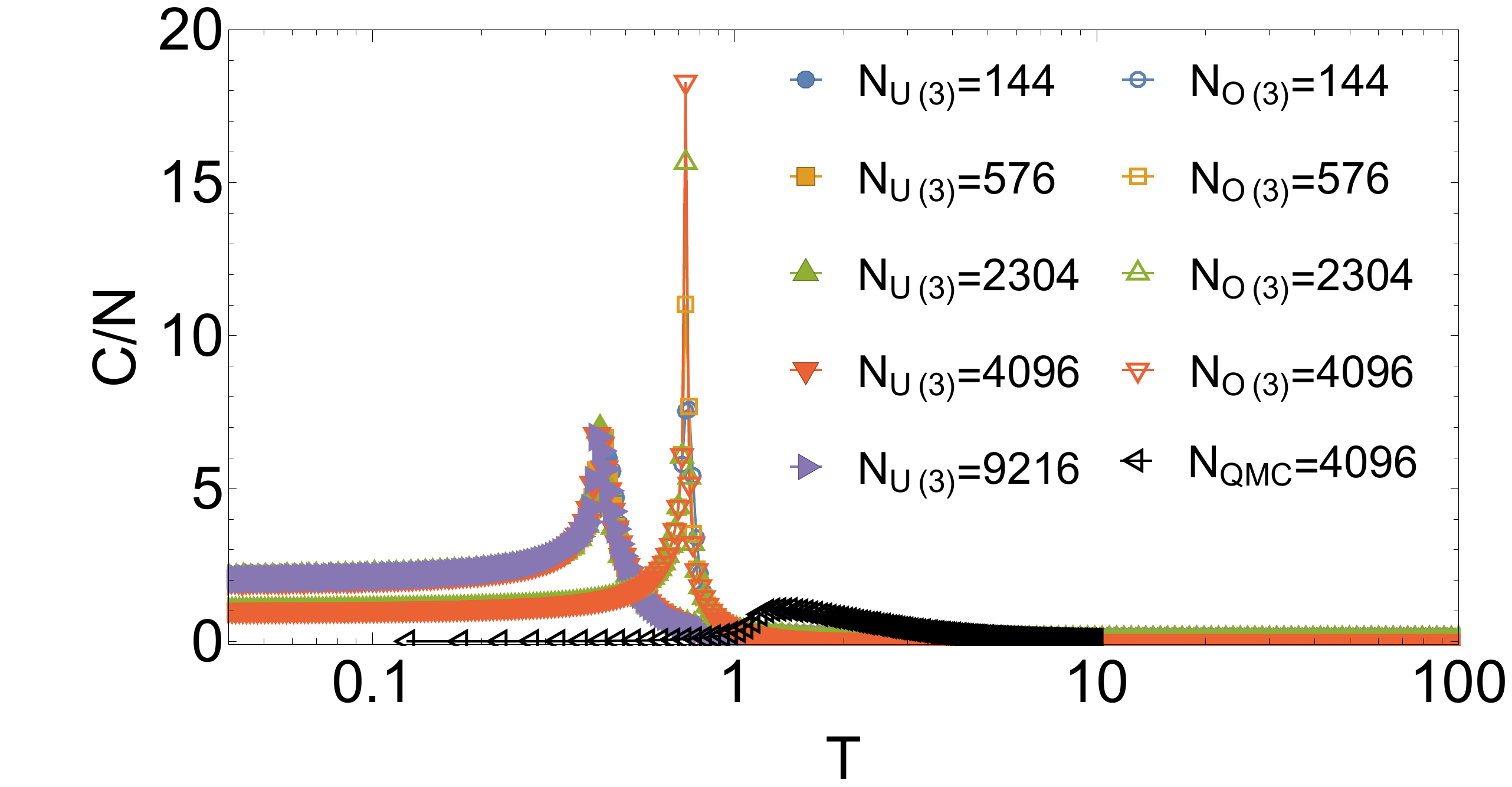}
	}\hspace{1cm}
		\subfloat[$\log c(T)$ as function of $\log T$ \label{fig:heat.capacity2}]{
		\includegraphics[width=0.45\textwidth]{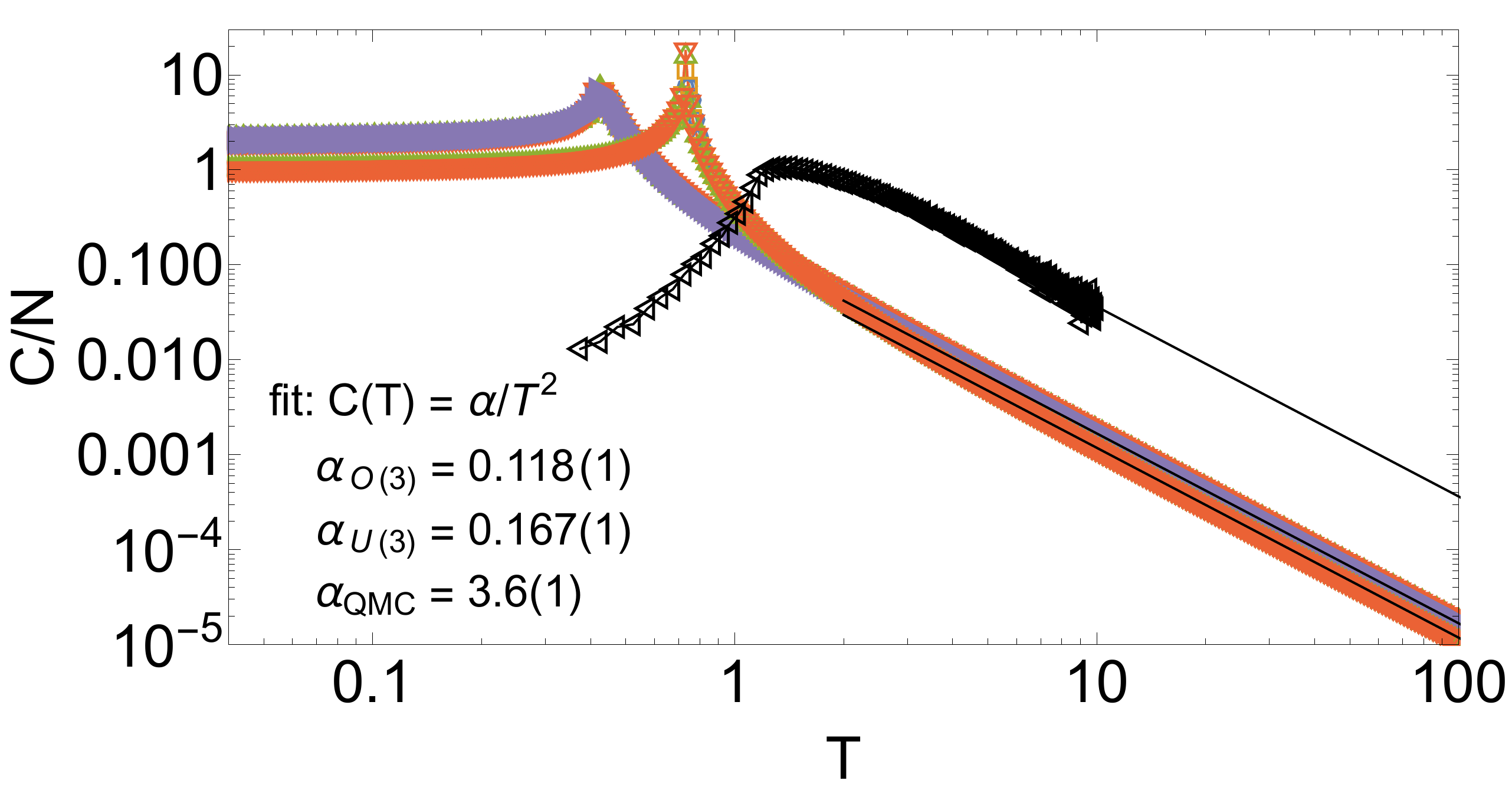}
	}
	\caption{
		Temperature--dependence of heat capacity found in Monte Carlo 
		simulations of the spin--1 bilinear--biquadratic model 
		$\Ham_{\sf BBQ}$ [\protect\Autoref{eq:BBQ.model}].
		\protect\subref{fig:heat.capacity1}~Heat capacity per spin, $c(T)$, plotted as a function of $\log T$.
		Results are shown for simulations using quantum Monte Carlo (QMC), 
		classical Monte Carlo in space of $u(3)$ matrices (u3MC), and 
		classical Monte Carlo in space of $O(3)$ vectors (o3MC).
		The peak in $c(T)$ at intermediate temperatures reflects the onset of 
		fluctuations of ferroquadrupolar (FQ) order.   
		\protect\subref{fig:heat.capacity2}~Equivalent results, plotted on a log--log scale.
		At low temperatures, classical results tend to a constant reflecting 
		the number of generators of excitations about the ground state, while 
		quantum results show a power--law onset $c(T) \propto T^2$.
		At high temperatures, all results scale as $c(T) = \alpha/T^2$.
		Simulations were carried out for parameters $J_1 = 0.0$ and $J_2 = -1.0$, 
		with QMC data taken from \cite{Voell2015}. 
	}\label{fig:heat.capacity3}
\end{figure*}

\subsubsection{Dynamics}

Comparison between semi--classical simulations based on $u(3)$, 
and QMC, is most straightforward for dynamics at low temperatures.
Here, as shown above, the u3MD approach exactly reproduces published results 
from a (Gaussian--level) multiple--boson expansion \cite{Lauchli2006}.
The comparison between QMC and the predictions of this  
multiple--boson expansion is discussed in \cite{Voell2015}.
At a qualitative level, good agreement is found between 
the multiple--Boson expansion 
at $T=0$, and QMC results for $T \ll J$.
It follows that agreement between QMC and u3MD simulations is  
equally good, once the effect of classical statistics have been taken into 
account [\Autoref{section:MD.reconstructed}].


At a quantitative level, QMC results show some differences in values of 
hydrodynamic parameters governing long--wavelength excitations, such as 
the quadrupole--wave velocity \cite{Voell2015}.
At low temperatures, the renormalisation of these parameters is a consequence 
of quantum effects present in QMC, but not accessible within the semi--classical 
description provided by Gaussian--level flavor--wave theory, or u3MD.
None the less, these quantum corrections are small, and become too small to 
measure approaching the $SU(3)$ point \mbox{$\theta = -3\pi/4$}. 


Dynamics at temperatures $T \lesssim J$ have also been explored using QMC 
simulation \cite{Voell2015}.
As temperature is increased, excitations near the top of the band, which have 
predominantly spin--wave character, become heavily damped, and suffer a 
dramatic loss of intensity [cf. results for \mbox{$S_{\sf Q} (\q, \omega)$}   
in \cite{Voell2015}, Fig.~8].
Meanwhile quadrupolar fluctuations near the ordering vector show considerable 
spectral weight at low energy.
The analysis of dynamics across the topological phase transition occurring 
for $T^* \approx 0.4\ J$ lies outside the scope of this paper.
However we note that similar trends in spectral weight are 
observed in u3MD results for \mbox{$S_{\sf Q} (\q, \omega)$}, at temperatures $T \sim J$. 
These will be discussed elsewhere \cite{Pohle-in-preparation}.


For completeness, we note that a phenomenological theory   
of the relaxational dynamics of spin--1 magnets has been introduced in 
\cite{Baryakhtar2013}.
This makes the prediction that long--wavelength quadrupolar waves 
have damping $\propto k^2$.
To the best of our knowledge, this phenomenological approach has yet to be 
used to make quantitative predictions for FQ order in the BBQ model, or compared with 
QMC simulation results.
We have made a preliminary analysis of damping as a function of $\k$, within 
u3MD simulation.
Precise evaluations of the damping of long--lived excitations at low energy 
and temperature is challenging, but initial results are consistent with a damping  
\begin{eqnarray}
	\Gamma(\k) &\propto& k^2 + \mathcal{O}(k^4) \; , \; 
\end{eqnarray}
at fixed temperature $T/J \ll 1$.
We leave the further investigation of this point for future work.

\subsubsection{Thermodynamics}

Probably the most interesting thermodynamic quantity to compare between 
QMC and classical $U(3)$ simulations is the heat capacity.
In \Autoref{fig:heat.capacity3}~\subref{fig:heat.capacity2} we show results of simulations of the 
spin--1 BBQ model [\Autoref{eq:BBQ.model}] carried out 
using QMC, classical MC in the space of A--matrices (u3MC), 
and classical MC carried out in the space of $O(3)$ vectors.
QMC simulation results are taken from \cite{Voell2015}, while 
u3MC results have already been introduced in \Autoref{section:heat.capacity}.
$O(3)$ MC simulations parallel earlier work \cite{Kawamura2007}.
Heat capacity per spin, $c(T)$, is shown plotted on both 
log--linear [\Autoref{fig:heat.capacity3}~\subref{fig:heat.capacity1}] and log--log scales 
[\Autoref{fig:heat.capacity3}~\subref{fig:heat.capacity2}], 
with temperature measured in units of $J$.
We discuss the particulars of different temperature regimes below. 


At low temperatures, \mbox{$T/J \ll 1$}, analytic theory for FQ order predicts 
\begin{eqnarray}
c(T \to 0) = 2 \times \frac{3 \zeta(3)}{\pi v^2}  T^2 + \ldots 
\end{eqnarray}
where  $\zeta(3) \approx 1.2$, $v$ is the velocity of the linearly--dispersing 
Goldstone modes, and the factor $\times 2$ comes from the fact these 
are two--fold degenerate \cite{Lauchli2006}.
This result follows from the Bosonic nature of low--lying excitations.
Fits to QMC simulation, confirm the expected $T^2$ scaling~\footnote{
We note that interacting theory predicts a logarithmic correction to this scaling, 
$c(T) \sim T^2 \ln T$ \cite{Baryakhtar2013}, however published QMC results 
may not extend to sufficiently low temperatures to distinguish this.},
and return a value of $v$ consistent with that found in simulations of dynamics~\cite{Voell2015}.


In contrast, classical MC simulations carried out in the basis of $u(3)$ matrices 
find $c(T \to 0) = 2$ [\Autoref{section:theory.classical.thermodynamics}, \Autoref{section:heat.capacity}].
The profound difference between classical and quantum results for $c(T \to 0)$ 
is a consequence of the fact that, in the absence of quantum statistics, 
entropy is not well posed for $T \to 0$. 
And in this case, the effect of classical statistics cannot be corrected 
as easily as for the semi--classical dynamics discussed above.


At intermediate temperatures, \mbox{$T/J \sim 1$} both classical and quantum 
simulation results for $c(T)$ are dominated by a large peak.
In all three cases, this peak is associated with the onset of fluctuations 
of FQ order.
The peak found in classical simulations, which are carried out for 
much larger systems, is sharp, and can be linked
to the unbinding of $\mathbb{Z}_2$ vortices \cite{Kawamura2007,Pohle-in-preparation}.
Meanwhile, the peak found in QMC is much broader, 
and occurs at a slightly higher temperature.
These differences reflect both different statistics, and the large 
length scales needed to accurately describe a topological 
phase transition.


Finally, we turn to the limit of high temperature, \mbox{$T/J \gg 1$}.
Here results must scale as 
\begin{eqnarray}
c(T \to \infty) 
	= \lim_{T\to \infty} \frac{\langle E^2 \rangle - \langle E \rangle^2}{T^2} 
	=  \frac{\alpha}{T^2} \; ,
\label{eq:c.high.T}
\end{eqnarray}
where $\alpha$ is coefficient depending on model parameters and the ensemble 
of states sampled.
In \Autoref{fig:heat.capacity3}~\subref{fig:heat.capacity2} this behaviour is reflected in parallel lines with gradient 
\begin{eqnarray}
\frac{d \ln c(T)}{d\ln T} = -2 \; ,
\end{eqnarray} 
for \mbox{$T/J \gtrsim 10$}.  
In this high--temperature limit, u3MC results \mbox{($\alpha_{\sf u3MC} = 0.16$)} are  
intermediate between conventional $O(3)$ MC simulations \mbox{($\alpha_{\sf o3MC} = 0.11$)} , 
and QMC \mbox{($\alpha_{\sf QMC} = 3$)} .

It has been argued elsewhere that simulation in the 
space of ${\bf d}$--vectors (vis A--matrices) should yield results equivalent  
to QMC at high temperature \cite{Stoudenmire2009}.
Empirically this is not the case.
And since, at high temperatures, finite--size effects are small, we infer 
that the different values of $\alpha$ found in different simulations  
reflect different asymptotic values of the variance in energy, \Autoref{eq:c.high.T}.
This asymptote, and leading corrections to it, can be calculated within a 
high--temperature series expansion \cite{Oitmaa2006-book}.
We find this expansion takes on a different form for quantum spin--1 
moments and A--matrices, and so will generally lead to different results 
\cite{nic-unpub}.
We leave further analysis of the high--temperature limit for future work.

\section{Generalization to spin--anisotropic interactions}
\label{section:anisotropy}

In the preceding sections of this paper, we have shown it is possible to calculate the 
thermodynamic and dynamical properties of spin--1 magnets through simulations carried out 
in the basis of $u(3)$.
So far, this analysis has been confined to the bilinear--biquadratic (BBQ) model, 
\Autoref{eq:BBQ.model}, which is invariant under $SU(2)$ spin rotations.
Here we show that the same approach can be applied to models with interactions 
anisotropic in spin--space. 


At first sight, this is not a trivial generalization, since the group $U(3)$  
encompasses spins with length $\ne 1$.
We therefore need to show that spin--anisotropic interactions do not 
mix different spin sectors, at the level of individual spin--1 moments.
As we shall see, this condition is satisfied by both u3MC and u3MD simulations,
as long as dynamical simulations are initiated from a valid spin--1 state.


\begin{figure*}[t]
	\begin{center}
		\begin{minipage}[b]{\textwidth}
			\subfloat[$S^{\sf QM}_{\rm{S}}(\q, \omega), T = 0$		\label{fig:Sqw_ana_ani}]{
				\includegraphics[width=0.3\linewidth]{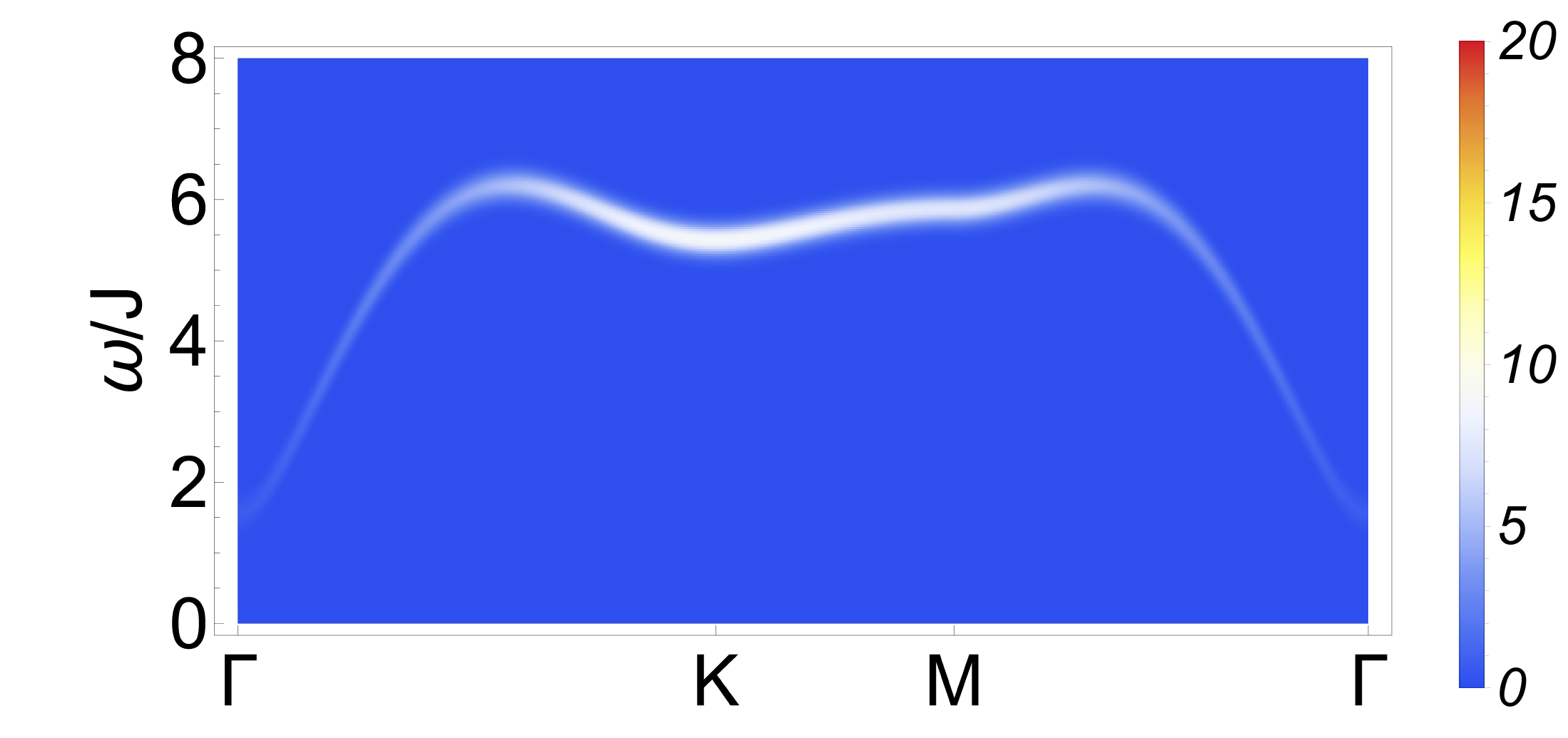}
			}
			\subfloat[$S^{\sf QM}_{\rm{Q}}(\q, \omega), T = 0 $		\label{fig:Qqw_ana_ani}]{
				\includegraphics[width=0.3\linewidth]{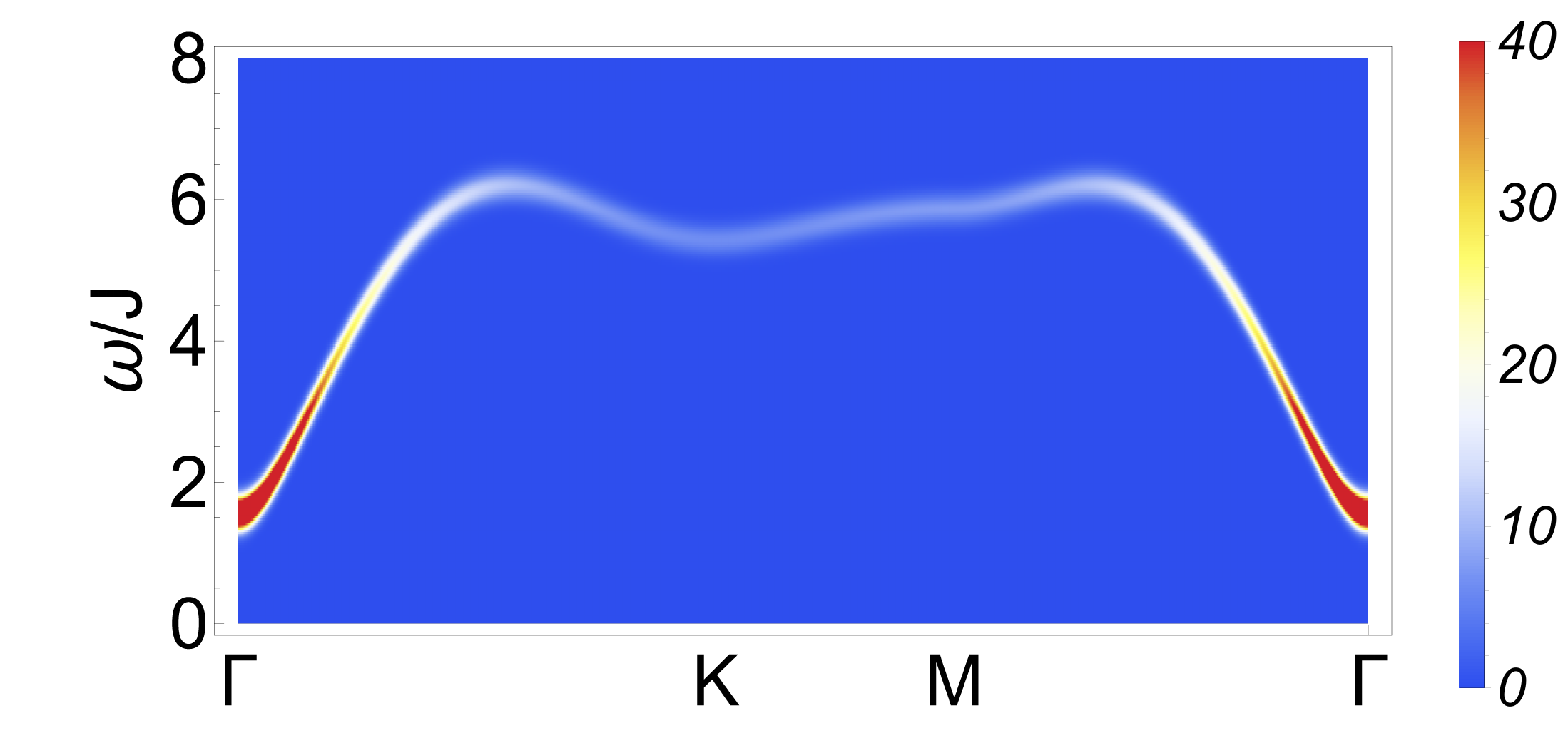}
			}
			\subfloat[$S^{\sf QM}_{\rm{A}}(\q, \omega), T = 0$		\label{fig:Aqw_ana_ani}]{
				\includegraphics[width=0.3\linewidth]{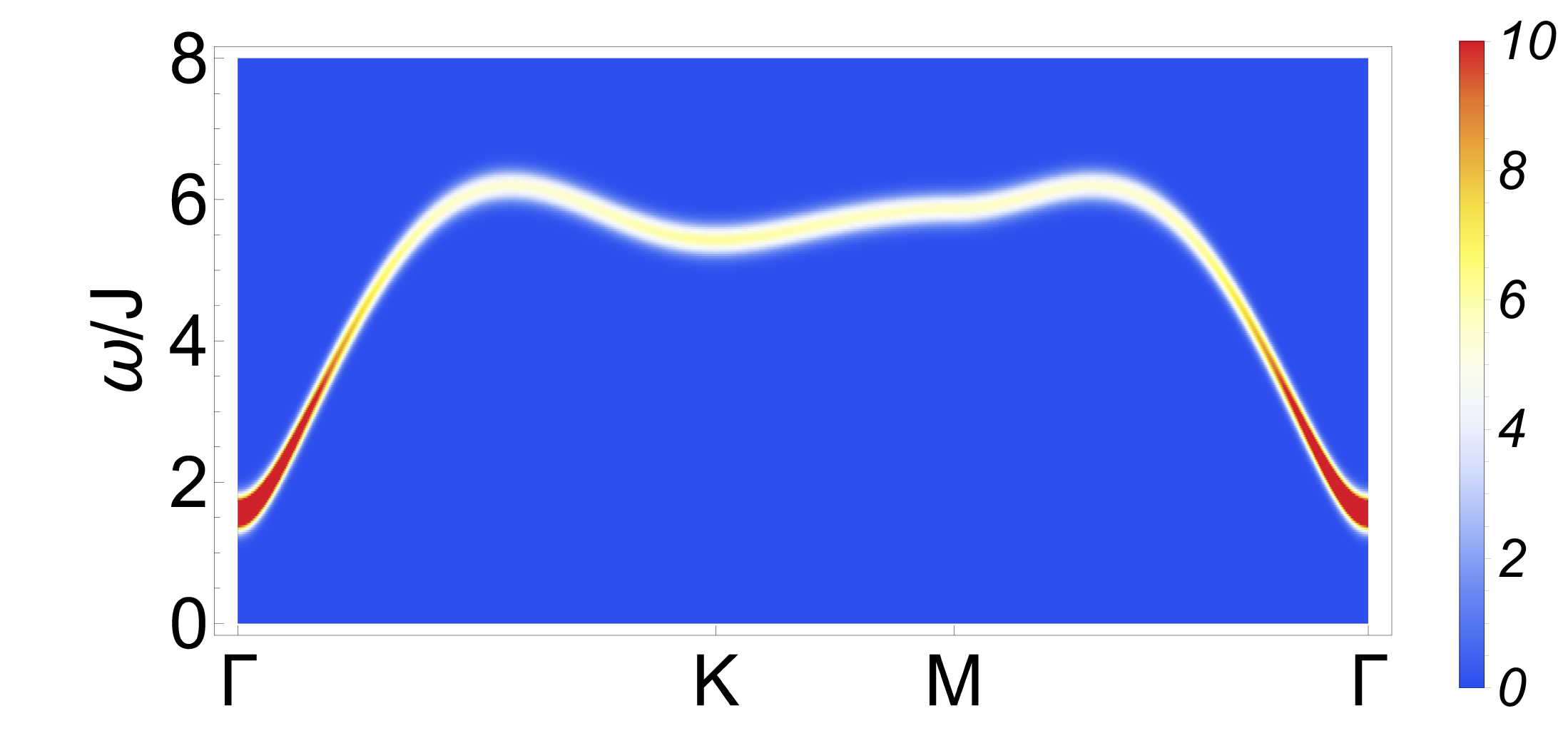}
			}\\
			\subfloat[$\frac{\omega}{T}S^{\sf MD}_{\rm{S}}(\q, \omega), T/J = 0.05$	  	\label{fig:wTSqw_MC_ani}]{
				\includegraphics[width=0.3\linewidth]{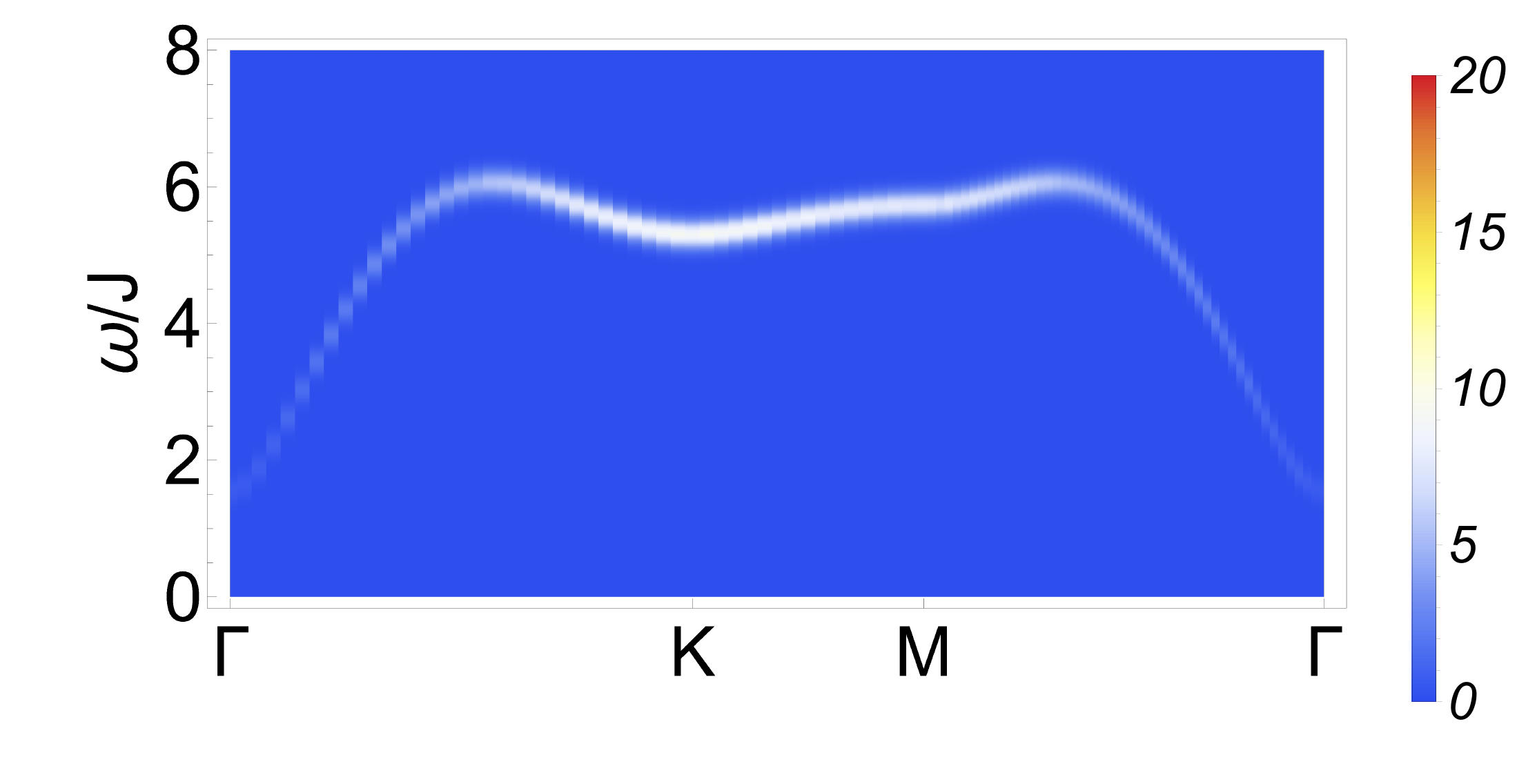}
			}
			\subfloat[$\frac{\omega}{T}S^{\sf MD}_{\rm{Q}}(\q, \omega), T/J = 0.05$		\label{fig:wTQqw_MC_ani}]{
				\includegraphics[width=0.3\linewidth]{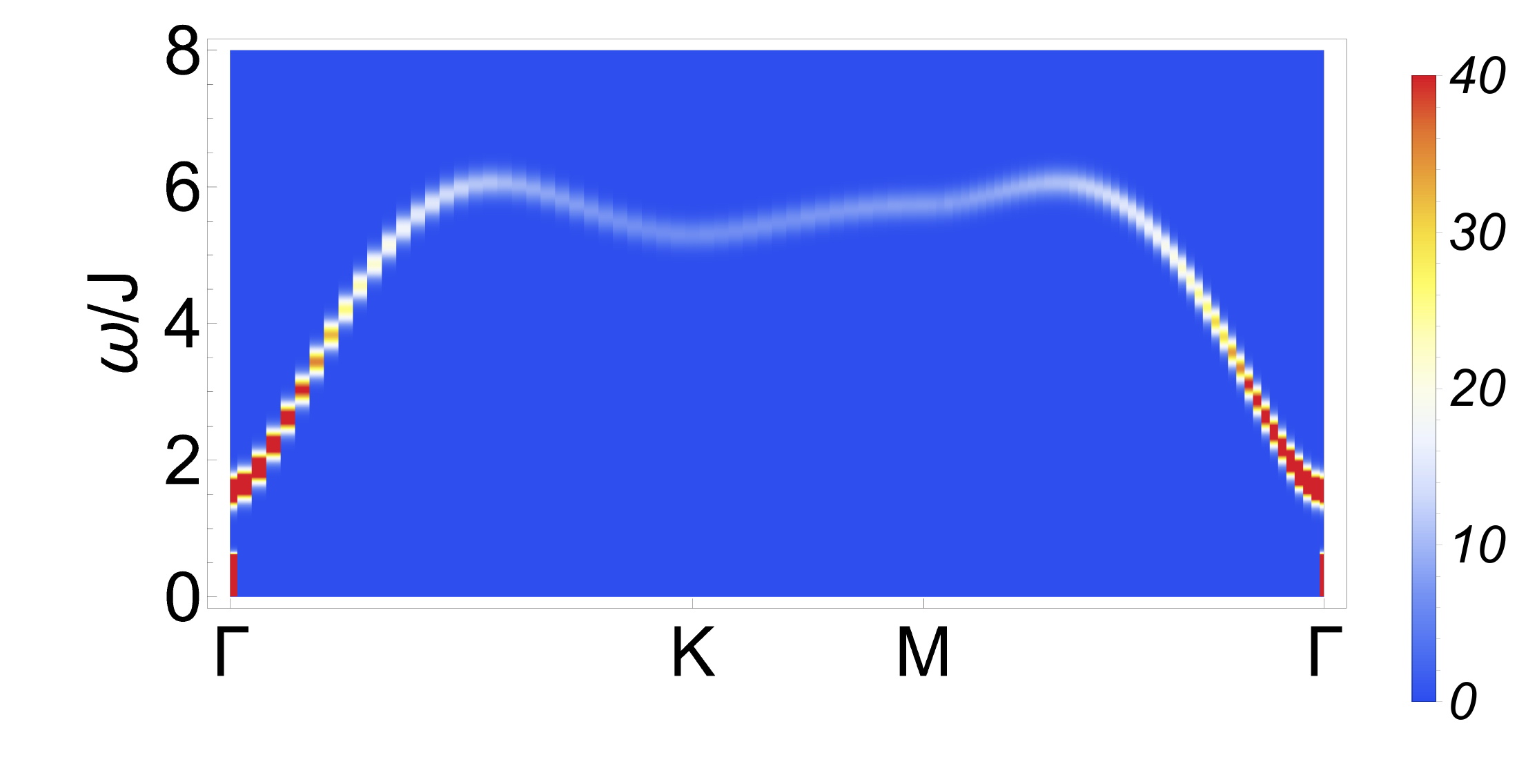}
			}
			\subfloat[$\frac{\omega}{T}S^{\sf MD}_{\rm{A}}(\q, \omega), T/J = 0.05$		\label{fig:wTAqw_MC_ani}]{
				\includegraphics[width=0.3\linewidth]{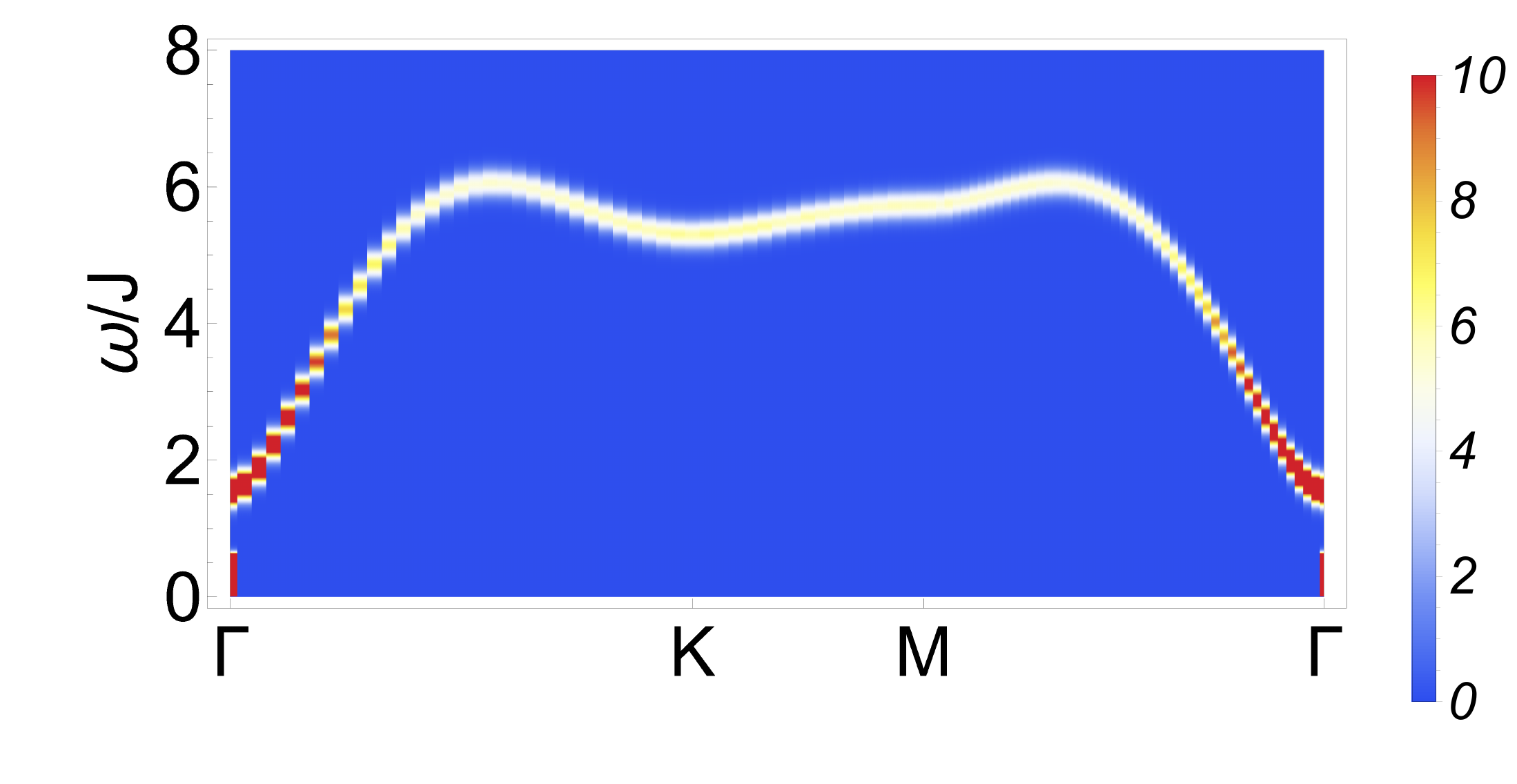}
			}
		\end{minipage}
	\end{center}
	\caption{
		Dynamics of ferroquadrupolar (FQ) state in the spin--1 bilinear--biquadratic (BBQ) 
		model with easy--plane anisotropy.
		\protect\subref{fig:Sqw_ana_ani}~Dynamical structure factor for dipole moments, $S^{\sf QM}_{\rm{S}}(\q, \omega)$, 
		as predicted by zero--temperature quantum theory of 
		\protect\Autoref{sec:SI_anisotropy.analytic}.
		\protect\subref{fig:Qqw_ana_ani}~Equivalent results for quadrupole moments, $S^{\sf QM}_{\rm{Q}}(\q, \omega)$.
		\protect\subref{fig:Aqw_ana_ani}~Equivalent results for A--matrices, $S^{\sf QM}_{\rm{A}}(\q, \omega)$.
		\protect\subref{fig:wTSqw_MC_ani}~Dynamical structure factor $S^{\sf MD}_{\rm{S}}(\q, \omega)$ 
		found in molecular dynamics simulations within $u(3)$ representation (u3MD) .
		\protect\subref{fig:wTQqw_MC_ani}~Equivalent results for $S^{\sf MD}_{\rm{Q}}(\q, \omega)$.
		\protect\subref{fig:wTAqw_MC_ani}~Equivalent results for $S^{\sf MD}_{\rm{A}}(\q, \omega)$.
		Simulations were carried out for $\Ham_{\sf D}$ [\protect\Autoref{eq:Ham_ani}], 
		with parameters \protect\Autoref{eq:anisotropic.parameter.set}, at a 
		temperature \mbox{$T = 0.05\ J$}, for a cluster of linear dimension 
		$L =96$ ($N= 9216$ spins).  
		Numerical results have been multiplied by a prefactor $\omega/T$ to correct 	
		for classical statistics, following \protect\Autoref{eq:correcting.for.classical.statistics}.
		All results have been convoluted with a Gaussian in frequency 
		of \mbox{FWHM = 0.35 J}.		
		}
	\label{fig:comparison_dispersion_intensities_SI_ani_FQ}
\end{figure*}

\subsection{Validity of u3MD approach}
\label{eq:validity.of.approach}

For simulations carried out in a $u(3)$ basis to be valid, it must remain true 
that each site in the lattice is host to a single spin--1 moment.
Once spins are transcribed in terms of generators of $U(3)$, this imposes 
the condition that 
\begin{eqnarray}
	{\rm Tr}\ \hat{\mathcal{A}} \equiv 1 \; ,
\end{eqnarray}
[\Autoref{eq:fix.spin.length}].
This condition is true by construction in u3MC simulation [\Autoref{sec:MC}].
And, in \Autoref{sec:u3.EoM}, we showed that 
\begin{eqnarray}
	\ddp_t \rm{Tr}\ \opa{i}{}{}  \equiv 0\; ,
\label{eq:conservation.of.A}
\end{eqnarray}
for u3MD simulations carried out for the spin--rotationally invariant 
BBQ model [\Autoref{eq:BBQ.model}], implying that spin--length 
is conserved.
We now extend this result to models which break spin--rotation invariance.


We consider most general form of spin--anisotropic Hamiltonian 
allowed for a spin--1 magnet
\begin{eqnarray}
	\Ham_{\Delta} = \sum_{\nn{i,j}} {J^{\alpha\mu}_{\beta\nu}} 
		\opa{i}{\alpha}{\beta}\opa{j}{\mu}{\nu} 
		+ \sum_{i} L^{\alpha}_{\beta} \opa{i}{\alpha}{\beta} \; , 
\label{eq:Ham_A_general}
\end{eqnarray}
where the only restriction placed on the interactions  
$J^{\alpha\mu}_{\beta\nu}$, and single--ion anisotropy $L^{\alpha}_{\beta}$, 
is the requirement that $\Ham_{\Delta}$ be Hermitian.
It follows that 
\begin{align*}
	\ddp_t \opa{i}{\gamma}{\eta}=-i\left[\right.\opa{{i}}{\gamma}{\eta},&\Ham_{\Delta}\left.\right]\\
	=-\frac{i}{2}\sum_{\delta}\left(\right.&J^{\eta\alpha}_{\mu\beta}\opa{{i}}{\gamma}{\mu}+J^{\alpha\eta}_{\beta\mu}\opa{{i}}{\gamma}{\mu}\\
	&-J^{\mu\alpha}_{\gamma\beta}\opa{{i}}{\mu}{\eta}-J^{\alpha\mu}_{\beta\gamma}\opa{{i}}{\mu}{\eta}\left.\right)\opa{{i+\delta}}{\alpha}{\beta}\\
	-\frac{i}{2}\left(\right.L^{\eta}_{\alpha}&\opa{{i}}{\gamma}{\alpha}-L^{\alpha}_{\gamma}\opa{{i}}{\alpha}{\gamma}\left.\right)\; . 
\label{eq:DA_general} \numberthis
\end{align*}
Setting $\eta=\gamma$ and taking the trace, we find
\begin{align*}
	\ddp_t {\rm Tr}\ \mathcal{A}_i = -\frac{i}{2}\sum_{\delta}\left(\right.&J^{\alpha\gamma}_{\beta\mu}\opa{{i}}{\gamma}{\mu}-J^{\alpha\mu}_{\beta\gamma}\opa{{i}}{\mu}{\gamma}\left.\right)\opa{{i+\delta}}{\alpha}{\beta}\\
	-\frac{i}{2}\left(\right.L^{\gamma}_{\alpha}&\opa{{i}}{\gamma}{\alpha}-L^{\alpha}_{\gamma}\opa{{i}}{\alpha}{\gamma}\left.\right)\; ,\label{eq:D_Trace_A}\numberthis
\end{align*}
where we have used the relationship
\be
	J^{\alpha\mu}_{\beta\nu}=J^{\mu\alpha}_{\nu\beta}  \; , 
\label{eq:J_cons}
\ee
which follows from the fact that components of $\mathcal{A}$ 
on different lattice sites commute [\Autoref{eq:ComRelAop}]. 
By rearranging indices on the right hand side of \Autoref{eq:D_Trace_A}, 
we can easily show that 
\begin{eqnarray}
	\ddp_t \rm{Tr}\ \opa{i}{}{}  = 0 \; ,
\end{eqnarray}
as required.


It follows that the trace of $\mathcal{A}$ is conserved within u3MD simulations, 
and therefore that simulations carried for out for arbitrary spin--anisotropic interactions 
respect the constraint on spin--length.
The implication of this result is that solving the $u(3)$ equations of motion, 
\Autoref{eq:EoM.u3}, for a spin--1 state, is exactly equivalent to solving 
the much more complicated equations of motion for spin--1 moments 
found in the algebra $su(3)$ \cite{Balla2014-Thesis,Remund2015-Thesis,Zhang2021}, 
regardless of spin--anisotropy.

\subsection{Application to FQ state with easy--plane anisotropy: analytic theory}
\label{sec:SI_anisotropy.analytic}

For illustration, we now consider the simplest extension of the results developed 
thus far to anisotropic interactions: the spin--1 BBQ model with single--ion, 
easy--plane anisotropy
\begin{eqnarray}
	\Ham_{\sf D} = \Ham_{\sf BBQ} + \Ham_{\sf SI} \; ,
\label{eq:Ham_ani} 
\end{eqnarray}
where $\Ham_{\sf BBQ}$ is defined in \Autoref{eq:BBQ.model}, and 
\begin{eqnarray}
	\Ham_{\sf SI} = \sum_{i} D (\ops{i}{y})^2  \; , \quad [D > 0] \; . 
\label{eq:Ham_SI} 
\end{eqnarray}
This model has previously been studied in \cite{Onufrieva1985}.


Like the BBQ model it descends from, $\Ham_{\sf D}$ supports a FQ ground 
state for a wide range of \mbox{$(J_1, J_2 < 0)$}, and we can easily generalise 
the theory of excitations developed in \Autoref{section:quantum.theory} to 
take account of single--ion anisotropy.
Transcribing $\Ham_{\sf SI}$ in terms of $\mathcal{A}$--matrices, by way of 
\Autoref{eq:dipole.in.terms.of.A}, we find
\begin{align*}
	\Ham_{\sf SI}&=\sum_{i}D \left(-\frac{2}{3}\opa{i}{y}{y}+\frac{1}{3}\opa{i}{x}{x}+\frac{1}{3}\opa{i}{z}{z}+\frac{2}{3}\right) \; . \numberthis 
\label{eq:Ham_SI_A} 
\end{align*}
Written in this form, it is immediately clear that $\Ham_{\sf SI}$ is a special case of 
the single--ion term in \Autoref{eq:Ham_A_general}.


From here, we can use \Autoref{eq:Acond2} to express $\Ham_{\sf SI}$ in terms 
of the Bosonic basis introduced in \Autoref{section:small.fluctuations}.
Its effect is to introduce new diagonal terms in the matrix controlling 
the dispersion of excitations [\Autoref{eq:def_wboson}], vis
\begin{eqnarray}
M_{\k}^{\sf{SI}}&=&\begin{pmatrix}
	A_{\k}+D& -B_{\k} &  0& 0 \\
	-B_{\k} &A_{\k}+D &  0& 0 \\
	0 &  0 & A_{\k}+D &- B_{\k}  \\
	0 & 0 & -B_{\k} & A_{\k}+D\\
\end{pmatrix} \; , \nonumber\\
\label{eq:def_phi_vec_M_k_ani}
\end{eqnarray}
where $A_{\k}$ and $B_{\k}$ are given in \Autoref{eq:DefAB}.
Solving the appropriate eigensystem [\Autoref{eq:quantum.eigensystem}], 
we find two physical branches of excitation, with dispersion 
\be
\epsilon_\k = \sqrt{(A_{\k}+D)^2-B_{\k}^2} \; .
\label{eq:omega_ani_qm}
\ee
It follows that the effect of easy--plane anisotropy is to open a gap 
\be
	\Delta =\sqrt{2 A_{0}D+D^2}\; ,
\label{eq:FQ_SI_gap}
\ee
to the Goldstone modes of FQ order.
This is to be expected since, in the presence of easy--plane anisotropy, 
the FQ ground state does not break spin--rotation symmetry.


It is also straightforward to generalise the calculations of structure
factors described in \Autoref{sec:quantum.structure.factors}.
Results for $S^{\sf QM}_{\rm{A}} (\q, \omega) $ [\Autoref{eq:Aqw_qm}], 
$S^{\sf QM}_{\rm{Q}} (\q, \omega) $ \Autoref{eq:Qqw_qm}] and 
$S^{\sf QM}_{\rm{S}} (\q, \omega)$ [\Autoref{eq:Sqw_qm}]
can be adapted to easy--plane anisotropy through the simple substitution
\begin{eqnarray}
	A_\k \longrightarrow A_\k + D \; .
\label{eq:SI_ani_rule}
\end{eqnarray}
Doing so, and considering parameters 
\begin{eqnarray}
	J_1 =0 \; , \;  J_2=-1.0 \; , \; D=0.2 \; ,  
\label{eq:anisotropic.parameter.set}
\end{eqnarray}
we obtain the predictions shown in 
\Autoref{fig:comparison_dispersion_intensities_SI_ani_FQ}~\subref{fig:Sqw_ana_ani}--\subref{fig:Aqw_ana_ani}.

\subsection{Application to FQ state with easy--plane anisotropy: numerical results}
\label{sec:SI_anisotropy.numerical}

Building on \Autoref{eq:validity.of.approach}, we can also apply the u3MD 
simulation approach to the easy--plane model $\Ham_{\sf D}$ [\Autoref{eq:Ham_ani}].
In \Autoref{fig:comparison_dispersion_intensities_SI_ani_FQ}~\subref{fig:wTSqw_MC_ani}--\subref{fig:wTAqw_MC_ani}, we show 
results obtained for the parameter set \Autoref{eq:anisotropic.parameter.set}.
Once corrected for the effect of classical statistics, through 
\Autoref{eq:correcting.for.classical.statistics}, 
simulations show good agreement with the predictions analytic theory 
developed in \Autoref{sec:SI_anisotropy.analytic}.


These results provide an explicit demonstration of the ability of u3MD simulations to 
describe the excitations of spin--1 models with spin--anisotropic interactions.

\section{Summary, Conclusions and Outlook} 						
\label{section:conclusions}

In this Article, we have introduced a new method for simulating both the thermodynamics 
and dynamics of spin--1 magnets, established the validity of this method 
through detailed comparison with known limits, and used it to obtain a number of new 
results for the spin--1 bilinear--biquadratic (BBQ) model on a triangular lattice.
Several other interesting findings entail.
Foremost among these is an explicit connection between classical simulations at finite 
temperature, and zero--temperature quantum dynamics, treated at a semi--classical level.
Also of interest are a low--temperature expansion for the thermodynamic properties 
of spin--1 magnets, and a novel derivation of a well--established multiple--Boson expansion.


The key to this method, introduced in \Autoref{section:maths.for.spin.1},  
is the representation of spin--1 moments through the algebra $u(3)$. 
Unlike mappings onto an $O(3)$ vector, this approach treats dipole and quadrupole 
moments on an equal footing.
And for this reason, it provides a valid (semi--)classical limit of a spin--1 moment.
From this starting point, we have developed a framework for classical Monte Carlo  
simulation in the space of ``A--matrices'', $\opa{}{\alpha}{\beta}$, which act 
as generators belonging to $u(3)$ (u3MC).
We also derived equations of motion (EoM) for $\opa{}{\alpha}{\beta}$ in a form suitable for 
numerical integration [\Autoref{eq:EoM.u3}].  
These form the basis for a ``molecular dynamics'' scheme for exploring the dynamics of 
spin--1 magnets (u3MD).


The numerical implementation of u3MC and u3MD simulations was described 
in \Autoref{section:numerical.method}.
A Marasaglia--like update in the space of $\opa{}{\alpha}{\beta}$ was introduced, 
and used to develop a MC scheme based on a local Metropolis update.
The resulting u3MC approach was shown to reproduce known results for the 
thermodynamic properties of the spin--1 BBQ model on a triangular lattice, 
and used to derive a finite--temperature phase diagram [\Autoref{fig:MC.phase.diagram}].
Meanwhile, numerical integration of EoM using an RK--4 update was 
shown to conserve the trace $\opa{}{\alpha}{\beta}$, establishing u3MD as valid approach 
for simulating the dynamics of spin--1 magnets.


In order to illustrate the u3MC and u3MD approaches, we then turned to the 
specific example of ferroquadrupolar (FQ) order, as found in the \mbox{spin--1} 
BBQ model on a triangular lattice [\Autoref{fig:FQ.order}].
To better understand simulations, we first developed an analytic theory 
of fluctuations about this state, described in \Autoref{section:classical.theory}.
Treated at a classical level, these fluctuations were shown to form bands with either 
dipolar or quadrupolar character [\Autoref{fig:comparison.classical.quantum.dispersion}~\subref{fig:Ana_Cl_Sqw}--\subref{fig:Ana_Cl_Aqw}], 
which provide the framework for a classical low--temperature (low--T) expansion of the free energy [\Autoref{eq:f0}].
This low--T expansion was used to make predictions for classical thermodynamic properties 
of the BBQ model, for subsequent comparison with u3MC simulation.
 

Next, in \Autoref{section:quantum.theory}, we showed how these fluctuations 
could be quantized, leading to a multiple--Boson expansion of excitations 
about FQ order. 
This theory, which is exactly equivalent to a known ``flavor--wave'' expansion, 
was used to develop zero--temperature quantum predictions for 
dynamical structure factors within a FQ  state, for subsequent comparison 
with u3MD simulation [\Autoref{fig:comparison.classical.quantum.dispersion}~\subref{fig:Ana_FW_Sqw}--\subref{fig:Ana_FW_Aqw}].
 

With this ground work in place, in \Autoref{section:numerics} we explored both the 
predictions of u3MC for the low--temperature thermodynamic properties of the FQ phase, 
and the predictions of u3MD for its dynamics.
u3MC results for heat capacity \mbox{$c(T\to 0)$} [\Autoref{fig:heat.capacity}], 
ordered moment ${\bf Q}$ [\Autoref{fig:ordered.moment}]
and equal time structure factors $S_\lambda (\q)$ [\Autoref{fig:S.of.q}] 
were shown to be in perfect agreement with the predictions of the classical \mbox{low--T} expansion.
Meanwhile ``raw'' u3MD results for dynamics were shown to give a good account of the 
dispersion of excitations, but fail to reproduce their spectral 
weight [\Autoref{fig:dispersion_rawMD}].


The reason for the disagreement between u3MD and the $T=0$ quantum theory 
was identified as coming from classical statistics, inherited from u3MC simulation.
This observation formed the basis for a detailed exploration of quantum--classical 
correspondence within u3MD simulation, building on the analytic theories of classical 
and quantum excitations, and described in \Autoref{section:quantum.vs.classical}.
This analysis leads to a simple, and very general, prescription for correcting MD simulation 
for the effect of classical statistics, in the limit $T\to 0$ [\Autoref{eq:the.big.result}].
Corrected in this way, the predictions of u3MD were shown to perfectly reproduce the 
predictions of zero--temperature quantum theory, considered at a semi--classical level 
[\Autoref{fig:comparison_dispersion_intensities} and \Autoref{fig:T-scaling_Sqw}].
The comparison of u3MC and u3MD results with published QMC simulations of 
the FQ phase of the BBQ model was also discussed.


Up to this point, all results were derived under the assumption of $SU(2)$ symmetry, 
appropriate to the BBQ model.
However many spin--1 magnets display anisotropy in their exchange interactions, 
and at the level of individual ions.
For this reason, in \Autoref{section:anisotropy}, we revisited the derivation 
of the u3MD method, establishing that it remains valid for the most general 
anisotropy permitted for a spin--1 magnet.
To illustrate this result, we demonstrated that u3MD simulations correctly 
describe the dynamics of a FQ state in the presence of single--ion anisotropy, 
perfectly reproducing the predictions of a \mbox{$T=0$} quantum theory 
[\Autoref{fig:comparison_dispersion_intensities_SI_ani_FQ}].


We conclude that the u3MC and u3MD methods introduced in this Article provide 
a reliable guide to the classical thermodynamics, and semi--classical dynamics 
of spin--1 magnets.
This opens many new perspectives for both theory, and the interpretation of experiment.


On the theoretical side, the lack of well--developed methods means that the thermodynamic properties of spin--1 magnets, and in particular their phase transitions, remain relatively unexplored.
This is of particular interest for phases built of on--site quadrupole moments, which 
cannot occur in spin--1/2 magnets, and for orders which support interesting topological 
excitations.
Moreover the possibility of combining u3MC with u3MD means that, where an interesting 
phase transition is identified, the associated dynamics can also be explored.


From this point of view, the phase diagram of the simple 
BBQ model shown in \Autoref{fig:MC.phase.diagram} already poses many interesting questions.
The ordered ground states of this model are already known to support a wide array of 
topological excitations \cite{Ivanov2003,Ivanov2007,Grover2011,Xu2012,Galkina2015}.
These take on particularly interesting form where the model has an 
enlarged, $SU(3)$ symmetry \cite{Ivanov2008,Ueda2016-PRA93}, 
and the range of possibilities becomes still wider in the presence of  
spin--anisotropy \cite{Akagi2021a,Akagi2021b,Zhang2022-arXiv,Amari2022-arXiv}.   
In the context of a two--dimensional model, this presents the opportunity 
to study both the thermodynamics, and the dynamics, of a wide array 
of different topological phase transitions.
We will return to this question elsewhere, in the context of FQ phase of 
the BBQ model \cite{Pohle-in-preparation}.


It would also be interesting to use u3MD to look more deeply into the  
dynamics of spin--1 magnets at finite temperature.
While MD simulation does not respect quantum quantum 
statistics, it does allow for interactions between quasiparticles.
Preliminary analysis of the damping of excitations, described in 
\Autoref{section:quantum.vs.classical} of this Article, suggest that u3MD results 
are consistent with the predictions of hydrodynamic theories, at least at a qualitative level.  
More work would be needed to put these results on a quantitative footing, but this 
remains a promising avenue for future exploration.
To this end, it is worth noting that simulations based on $U(3)$ are still in their infancy, 
and there is considerable room for technical improvement, e.g., in Monte Carlo updates.   
\nic{
And very recently, there have been encouraging developments in the application of $SU(3)$ 
approaches to spin--1 magnets, complimentary to the results of this Article \cite{Dahlbom2022-arXiv,Do2022-arXiv}.
}


\nic{
It is also interesting to speculate about the possible extension of a $U(N)$ approach 
to higher--spin moments.
Paradoxically, while increasing the size of the moment, $S$, 
will suppress quantum fluctuations, it also increases the complexity to the 
problem, through the number of parameters needed to describe a single site, 
and the number of bands of excitations found on a lattice.
For example, the passage from spin--1 to spin--3/2 increases the Hilbert space from 
to $\mathbb{CP}^2$ to $\mathbb{CP}^3$, and the number of parameters needed 
from $4$ to $6$.
It also brings a new piece of physics, octupole moments at the level of a single site, 
and a new algebra, $su(4)$.
This trend continues for larger $S$, with each moment possessing its own, unique, 
semi--classical limit.
Taken appropriately, this limit should become an increasingly good approximation 
as $S \to \infty$.
}


\nic{
For this reason, further development of semi--classical methods for high--spin moments 
makes very good sense.
To this end, we note that coherent--state representation has already been used to derive general 
equations of motion for spin--$S$ moments within the algrebra $su(N)$, 
where \mbox{$N = 2S + 1$} \cite{Zhang2021}.
Quite generally, it is possible to embed such an $su(N)$ algebra within $u(N)$, 
and seek simplification of the algebra representing the original spin--S moment, 
of the type found for spin--1 in this paper.
We leave this as a topic for future study.   
}


On the experimental side, many interesting spin--1 magnets have come to light.
Celebrated examples include the triangular--lattice spin--nematic 
candidate NiGa$_2$S$_4$ \cite{Nakatsuji2005,Nambu2006, Bhattacharjee2006,Tsunetsugu2006,Lauchli2006,Valentine2020}, 
and the pyrochlore spin--liquid candidate 
NaCaNi$_2$F$_7$ \cite{Plumb2019,Zhang2019}.
%
However there are also steady stream of new arrivals, 
and interesting new results for older materials \cite{Bao2021}.
Spin--1 models also arise in the context of cold atoms 
\cite{Demler2002,Imambekov2003,Kurn2013,deForgesdeParny2014,Zibold2016}, 
and as a proxy for describing various forms of quantum liquid crystal, including 
the nematic phases of Fe--based 
superconductors \cite{Fernandes2014,Luo2016,Wang2016,Gong2017,Lai2017}. 
Many aspects of the physics of these systems remain ambiguous, and the ability 
to simulate the dynamics of realistic, microscopic, spin--1 models could prove decisive.


The second major conclusion of this work, is that it is possible 
to correct for the effect of classical statistics in finite--temperature 
``molecular dynamics'' (MD) simulations of magnets, and thereby use them to 
study zero--temperature, quantum (semi--classical) dynamics.
This is a result of broad relevance, applying equally to conventional MD simulations 
in the space of $O(3)$ vectors.
And it goes some way to explaining why, despite its humble classical origins, 
MD simulation has been so successful in describing the dynamics of exotic quantum 
magnets \cite{Taillefumier2017,Samarakoon2017,Pohle2021}.
The deeper exploration of this form of quantum--classical correspondence, 
for example by pairing u3MD with QMC simulations, is another promising avenue 
for future research.
And the $u(3)$ basis which underpins this work could also be used as starting point 
for explicitly quantum calculations, e.g. through the use of variational wave functions
based on tensor--networks.


Consider together, this is an absorbing set of problems, and it will be interesting 
to see how much more can be learned through the numerical simulation 
of spin--1 magnets.

\begin{acknowledgments}

The authors are pleased to acknowledge helpful conversations with 
Yuki Amari, 
Owen Benton, 
Hosho Katsura, 
Yukitoshi Motome, 
Karlo Penc, 
and Mathieu Taillefumier, 
and are grateful to the authors of \cite{Voell2015} for sharing 
numerical values of QMC data for heat capacity.


This work was supported by the Theory of Quantum Matter Unit, OIST, 
JSPS KAKENHI Grants No.~JP17K14352 
and No.~JP20K14411, 
and JSPS Grants-in-Aid for Scientific Research on Innovative Areas 
Topological Materials Science (KAKENHI Grant No.~JP18H04220),  
and Quantum Liquid Crystals (KAKENHI Grant No.~JP20H05154 and JP22H04469).


Numerical calculations were carried out using HPC facilities provided by OIST, 
and the Supercomputer Center of the Institute of Solid State Physics, University of Tokyo.

\end{acknowledgments}

\appendix

\begin{appendices}
	\section{Spin Fluctuation Probability}
	\label{app:Spin_Fluct_Prob}
	In this Appendix, we detail how the spin fluctuation probabilities drawn namely in \Autoref{fig:magnetic.basis},  \Autoref{fig:TR.basis}, \Autoref{fig:FQ.order}, \Autoref{fig:gen_fluct}, \Autoref{fig:basis_x}, and  \Autoref{fig:fluct} are calculated.
	
	Fluctuations around a given state $\ket{\alpha}$ can be calculated by computing its spin fluctuation probability, defined as the spatial probability distribution of the overlapping between the state $\ket{\alpha}$ and the spin coherent state $\ket{\Omega}$. The spin coherent state $\ket{\Omega}$ is obtained by applying a rotation operator in 3 dimensions defined by the angles $\theta$ and $\phi$ on the m=1 state $\ket{1}$:
	\be
	\ket{\Omega}=\mathcal{R}(\theta,\phi)\ket{1}\; .
	\ee
	The spin coherent state represents then a spin pointing in the direction defined by the angles $\theta$ and $\phi$ .
	In the case of a spin 1,  the spin coherent state is expressed as:
	\be
	\ket{\Omega}=\frac{1+\cos{\theta}}{2}\exp{-i\phi}\ket{1}+\frac{\sin{\theta}}{\sqrt{2}}\ket{0}+\frac{1-\cos{\theta}}{2}\exp{-i\phi}\ket{\overline{1}}\; .
	\ee
	The spin fluctuation probability of the state $\ket{\alpha}$ is defined as the norm of the scalar product with the spin coherent state:
	\be
	P(\theta,\phi)_{\ket{\alpha}}=\left| \SP{\alpha}{\Omega}\right|^2\label{eq:FluctProb}\; .
	\ee
	
	\section{Properties of A matrices}
	\label{sec:appA}
	
	Here, we present the fundamental properties of the "A-matrix". From its definition in \Autoref{eq:def_A}, we note that the A-object  is mathematically a (1,1)-tensor, but for simplicity, we might usually refer to it as a  matrix. In this Appendix, we also give the detailed explanations accompanying the symmetry analysis of the BBQ Model [\Autoref{eq:HamA1}]  that we discuss at the end of \Autoref{section:maths.for.spin.1}.
	
	\subsection{Properties of a single A matrix}
	\label{sec:prop_A}
	
	First, we present how \Autoref{eq:LTA} is obtained. \Autoref{eq:LTA} tells us how an object like $\opa{}{\alpha}{\beta}$ would transform under a general linear transformation $\Lambda$.
	To this end, as explained in \Autoref{section:maths.for.spin.1}, we consider a general linear transformation $\Lambda:V\rightarrow V$, such that $\textrm{det}\Lambda\neq 0$, so that $\Lambda$ is invertible, and we define 
	\be
	\tilde{\Lambda}={\Lambda^{-1}}^{T}\; .
	\ee
	Under such a transformation, the basis vector $\V{e}_i$ of the vector space	$V$ will transform according to 
	\be
	\overline{\V{e}}_i={\tilde{\Lambda}_i}^{~j}\V{e}_j\; .
	\ee
	Since the vector $\V{v}=v^i\V{e}_i$ is a mathematical object which existence does not depend on the basis, the components $v^i$ should transform according to 
	\be
	\overline{v}^i={\Lambda}^i_{~j}v^j\; ,
	\ee
	such that the vector $\V{v}=\overline{v}^i\overline{\V{e}}_i=v^i\V{e}_i$ stays invariant.
	It is then also possible to introduce the dual basis  $\{ \V{e}^{\star i}\}$ of the dual vector space $V^{\star}$. The basis vectors can be defined by the relations
	\be 
	{e}^{\star i}(e_j)=\delta^i_j\label{eq:DefDual}\; .
	\ee
	Any element $\V{v}^{\star}$ of $V^{\star}$ can be decomposed as 
	\be
	\V{v}^{\star}=v^{\star}_i  \V{e}^{\star i}\; ,
	\ee
	where the components $v^{\star}_i$ are simply given by the value of the function $\V{v}^{\star}$ on the basis vector $\V{e}_i$ of $V$
	\be
	v^{\star}_i=\V{v}^{\star}(\V{e}_i)	\; .
	\ee
	Under a general transformation $\Lambda$ on the basis vectors  $\V{e}_i$, the dual basis vectors $\V{e}^{\star i}$ will transform according to 
	\be
	\overline{\V{e}}^{\star i}=\Lambda^{i}_{~j}\V{e}^{\star j}\; ,
	\ee
	in order to preserve \Autoref{eq:DefDual}.
	And the component $v^{\star}_i$ will transform as
	\be
	\overline{v}^{\star}_i={\tilde{\Lambda}_i}^{~j}v^{\star}_j \; .
	\ee
	Finally, under such a general transformation $\Lambda$, the component of an object like $\opa{}{\alpha}{\beta}$, which is actually a (1,1)-tensor,  will transform as stated in \Autoref{eq:LTA}.
	
	\subsection{Properties of quadratric terms of A matrices}\label{sec:prop_AA}
	We here show how the products of two objects like $\opa{}{\alpha}{\beta}$ would transform under a linear transformation, in order to analyze the symmetry properties of the BBQ Hamiltonian rewritten in terms of  $\opa{}{}{}$-``matrices'' [\Autoref{eq:HamA1}].
	Again, we emphasis that an object like $\opa{}{\alpha}{\beta}$ is mathematically a tensor, but for simplicity, we might sometimes refer to them as matrices.
	
	Going back to \Autoref{eq:HamA1}, it can easily be seen that the first term  $\opa{i}{\alpha}{\beta}\opa{j}{\beta}{\alpha}$ is $U(3)$ symmetric because both indexes $\alpha$ and $\beta$ are contravariant on one site and covariant on the other. The $\opa{i}{\alpha}{\beta}\opa{j}{\beta}{\alpha}$ will therefore stay invariant under a transformation $U\in U(3)$, for which we have
	\be
	\begin{matrix}
		U\in U(3)\; : & UU^{\dagger}=U^{\dagger}U=\one & ~\Rightarrow~& U^{\dagger}=U^{-1}\; .
	\end{matrix}
	\ee 
	Indeed, under a $U(3)$ symmetry, the first term will transform as 
	\begin{align*}
		(\opa{i}{\alpha}{\beta})^{\mu}_{~\nu}(\opa{j}{\beta}{\alpha})^{\nu}_{~\mu}\rightarrow & U^{\mu}_{~\gamma}{U^{\dagger}}_{~\nu}^{\kappa}(\opa{i}{\alpha}{\beta})^{\gamma}_{~\kappa} U^{\nu}_{~\eta}{U^{\dagger}}_{~\mu}^{\lambda}(\opa{j}{\beta}{\alpha})^{\eta}_{~\lambda}\\
		&=U^{\mu}_{~\gamma}{U^{\dagger}}_{~\mu}^{\lambda}U^{\nu}_{~\eta}{U^{\dagger}}_{~\nu}^{\kappa}(\opa{i}{\alpha}{\beta})^{\gamma}_{~\kappa} (\opa{j}{\beta}{\alpha})^{\eta}_{~\lambda}\\
		&=\delta_{\gamma}^{~\lambda}\delta_{\eta}^{~\kappa}(\opa{i}{\alpha}{\beta})^{\gamma}_{~\kappa} (\opa{j}{\beta}{\alpha})^{\eta}_{~\lambda}\\
		&=(\opa{i}{\alpha}{\beta})^{\gamma}_{~\kappa} (\opa{j}{\beta}{\alpha})^{\kappa}_{~\gamma}\; .\numberthis \label{eq:Hpart1Trans}
	\end{align*}

	The second term in \Autoref{eq:HamA1}, on the other hand,  is not $U(3)$ symmetric,  but it is $O(3)$ symmetric. We can see that under a $U(3)$ transformation, it transforms as:
	\begin{align*}
		(\opa{i}{\alpha}{\beta})^{\mu}_{~\nu}(\opa{j}{\alpha}{\beta})^{\mu}_{~\nu}\rightarrow & U^{\mu}_{~\gamma}{U^{\dagger}}_{~\nu}^{\kappa}(\opa{i}{\alpha}{\beta})^{\gamma}_{~\kappa} U^{\mu}_{~\eta}{U^{\dagger}}_{~\nu}^{\lambda}(\opa{j}{\alpha}{\beta})^{\eta}_{~\lambda}\\
		&=U^{\mu}_{~\gamma}U^{\mu}_{~\eta}{U^{\dagger}}_{~\nu}^{\lambda}{U^{\dagger}}_{~\nu}^{\kappa}(\opa{i}{\alpha}{\beta})^{\gamma}_{~\kappa} (\opa{j}{\alpha}{\beta})^{\eta}_{~\lambda}\; .\numberthis\label{eq:Hpart2Trans1}\\
	\end{align*}
	Clearly, this is not invariant under a $U(3)$ transformation, but it is under a $O(3)$ transformation. If $U=O \in O(3)$, we have
	\be
	\begin{matrix}
		O\in O(3)\; : & OO^{T}=O^{T}O=\one &~\Rightarrow~& O^{T}=O^{-1}\; ,
	\end{matrix}
	\ee
	and under a $O(3)$ transformation, it transforms as:
	\begin{align*}
		(\opa{i}{\alpha}{\beta})^{\mu}_{~\nu}(\opa{j}{\alpha}{\beta})^{\mu}_{~\nu}\rightarrow &O^{\mu}_{~\gamma}O^{\mu}_{~\eta}{O^{T}}_{~\nu}^{\lambda}{O^{T}}_{~\nu}^{\kappa}(\opa{i}{\alpha}{\beta})^{\gamma}_{~\kappa} (\opa{j}{\alpha}{\beta})^{\eta}_{~\lambda}\\
		&={O^T}^{~\mu}_{\gamma}O^{\mu}_{~\eta}{O^{T}}_{~\nu}^{\lambda}{O}_{\nu}^{~\kappa}(\opa{i}{\alpha}{\beta})^{\gamma}_{~\kappa} (\opa{j}{\alpha}{\beta})^{\eta}_{~\lambda}\\
		&=\delta_{\gamma\eta}\delta^{\kappa\lambda}(\opa{i}{\alpha}{\beta})^{\gamma}_{~\kappa} (\opa{j}{\alpha}{\beta})^{\eta}_{~\lambda}\\
		&=(\opa{i}{\alpha}{\beta})^{\gamma}_{~\kappa} (\opa{j}{\alpha}{\beta})^{\gamma}_{~\kappa}\; .\numberthis\label{eq:Hpart2Trans2}
	\end{align*}
	
	The Hamiltonian is therefore overall $O(3)$ symmetric, indeed both terms are invariant under an $O(3)$ symmetry. And in the case of $J_1=J_2$, the second term in \Autoref{eq:HamA1} vanishes, and the Hamiltonian is $U(3)$ symmetric. Therefore, working in $U(3)$ does not change the global symmetry of the Hamiltonian, since $o(3)\simeq su(2)$, and there is an homomorphism from $SU(2)$ into $O(3)$. However, the locally augmented $SU(3)$ symmetry of the Hamiltonian when $J_1=J_2$ is enlarged from $SU(3)$ to $U(3)$.
	
	The Hamiltonian can be rewritten in a more general form as
	\be
	\Ham_{\sf BBQ}=\sum_{\nn{i,j}}{J^{\alpha\mu}_{\beta\nu}}\opa{i}{\alpha}{\beta}\opa{j}{\mu}{\nu}\; ,
	\ee
	with
	\be
	J={\tiny
		\begin{pmatrix}
			\begin{pmatrix}
				J_2 & 0 & 0 \\
				0 & 0 & 0 \\
				0 & 0 & 0 \\
			\end{pmatrix}
			& 
			\begin{pmatrix}
				0 & J_2-J_1 & 0 \\
				J_1 & 0 & 0 \\
				0 & 0 & 0 \\
			\end{pmatrix}
			& 
			\begin{pmatrix}
				0 & 0 & J_2-J_1 \\
				0 & 0 & 0 \\
				J_1 & 0 & 0 \\
			\end{pmatrix}
			\\
			\begin{pmatrix}
				0 & J_1 & 0 \\
				J_2-J_1 & 0 & 0 \\
				0 & 0 & 0 \\
			\end{pmatrix}
			& 
			\begin{pmatrix}
				0 & 0 & 0 \\
				0 & J_2  & 0 \\
				0 & 0 & 0 \\
			\end{pmatrix}
			& 
			\begin{pmatrix}
				0 & 0 & 0 \\
				0 & 0 & J_2-J_1 \\
				0 & J_1 & 0 \\
			\end{pmatrix}
			\\
			\begin{pmatrix}
				0 & 0 & J_1 \\
				0 & 0 & 0 \\
				J_2-J_1 & 0 & 0 \\
			\end{pmatrix}
			& 
			\begin{pmatrix}
				0 & 0 & 0 \\
				0 & 0 & J_1 \\
				0 & J_2-J_1 & 0 \\
			\end{pmatrix}
			& 
			\begin{pmatrix}
				0 & 0 & 0 \\
				0 & 0 & 0 \\
				0 & 0 & J_2  \\
			\end{pmatrix}
			\\
	\end{pmatrix}}\; .\label{eq:J_general_BBQ}
	\ee
	The indexes $\alpha$ and $\beta$ correspond respectively to the line and the row of the table that assigns the designated matrix, whose components are then given by $\mu$ and $\nu$. For example,
	\be
	\begin{matrix}
		J^{1\mu}_{2\nu}=
		\begin{pmatrix}
			0 & J_2-J_1 & 0 \\
			J_1 & 0 & 0 \\
			0 & 0 & 0 \\
		\end{pmatrix}&,~ & 
		{J^{12}_{21}}=J_1 &,~ &
		{J^{11}_{22}}=J_2-J_1\; .
	\end{matrix}
	\ee
	
	The symmetries of the Hamiltonian are now hidden in the symmetries of the tensor $J^{\alpha\mu}_{\beta\nu}$.
	Firstly, we see that the Hamiltonian is $O(3)$ symmetric, because the repeated summed indexes are always either covariant or contravariant.
	The tensor is also symmetric under the exchange ${\alpha\beta}\leftrightarrow{\mu\nu}$
	\be
	J^{\alpha\mu}_{\beta\nu}=J^{\mu\alpha}_{\nu\beta}\; .\label{eq:Jsym1}
	\ee
	\Autoref{eq:Jsym1} expresses the fact that there is actually a tensor product between the two operators $\opa{i}{\alpha}{\beta}$ and $\opa{j}{\mu}{\nu}$ acting on different sites. 
	We also have  ${\alpha}\leftrightarrow{\nu}~\textrm{with}~ {\beta}\leftrightarrow{\mu}$ together, which is just relabeling the indexes.
	
	In the case of $J_1=J_2$, the tensor is also symmetric under the exchanges ${\alpha}\leftrightarrow{\mu}$ and ${\beta}\leftrightarrow{\nu}$ or both
	\be
	J^{\alpha\mu}_{\beta\nu}=J^{\mu\alpha}_{\beta\nu}=J^{\alpha\mu}_{\nu\beta}\; .\label{eq:Jsym2}
	\ee
	It is  also symmetric under the exchanges
	${\beta}\leftrightarrow{\mu}$ and ${\alpha}\leftrightarrow{\nu}$ 
	\be
	J^{\alpha\mu}_{\beta\nu}=J^{\alpha\beta}_{\mu\nu}=J^{\nu\mu}_{\beta\alpha}=J^{\nu\beta}_{\mu\alpha}\; ,\label{eq:Jsym3}
	\ee
	in which case, it can easily be seen that the Hamiltonian is $U(3)$ invariant, since every index is now summed covariantly.
	
	\section{Conventions for the triangular lattice}
	\label{sec:conv_trig_lat}
	In this Appendix, we present the convention that we used to describe the triangular lattice and its reciprocal space.
	We choose the real space lattice vectors, linking a single site unit cell to another, to be
	\be
	\begin{matrix}
		\V{a}=
		\begin{pmatrix}
			1\\
			0
		\end{pmatrix} \; ;&	\V{b}=\frac{1}{2}
		\begin{pmatrix}
			1\\
			\sqrt{3}
		\end{pmatrix} 
	\end{matrix}\; . \label{eq:def_a_b}
	\ee
	The associated vectors in reciprocal space are given by
	\be
	\begin{matrix}
		\V{k}_a=\frac{2\pi}{\sqrt{3}}
		\begin{pmatrix}
			\sqrt{3}\\
			-1
		\end{pmatrix}\; ; &	\V{k}_b=\frac{2\pi}{\sqrt{3}}
		\begin{pmatrix}
			0\\
			2
		\end{pmatrix} 
	\end{matrix}\; .\label{eq:def_ka_kb}
	\ee
	We define the points along the irreducible wedge in reciprocal space to be
	\be
	\begin{matrix}
		\bf{\Gamma}=
		\begin{pmatrix}
			0\\
			0
		\end{pmatrix}\; ; &	\V{K}=\frac{4\pi}{3}
		\begin{pmatrix}
			1\\
			0
		\end{pmatrix} \; ;&	\V{M}=\frac{\pi}{\sqrt{3}}
		\begin{pmatrix}
			\sqrt{3}\\
			1
		\end{pmatrix} \; .
	\end{matrix}\label{eq:K_space_points}
	\ee
	The vectors $\bf{\delta}$ linking the 6 neighboring sites are given by
	\be
	\begin{matrix}
		\bf{\delta}=
		\begin{pmatrix}
			1\\
			0
		\end{pmatrix} \; ;&	\frac{1}{2}
		\begin{pmatrix}
			1\\
			\sqrt{3}
		\end{pmatrix} \; ;&	\frac{1}{2}
		\begin{pmatrix}
			-1\\
			\sqrt{3}
		\end{pmatrix} \; ;\\
		\begin{pmatrix}
			-1\\
			0
		\end{pmatrix} \; ;&	\frac{1}{2}
		\begin{pmatrix}
			-1\\
			-\sqrt{3}
		\end{pmatrix} \; ;&	\frac{1}{2}
		\begin{pmatrix}
			1\\
			-\sqrt{3}
		\end{pmatrix} \; .
	\end{matrix}
	\ee
	For the triangular lattice, the coordination number and the geometrical factor given in \Autoref{eq:def_gamma} yield
	\be
	\begin{matrix}
		z=6&\; ;&	\gamma_{\triangleleft}(\k)=\frac{1}{3}(\cos(k_x)+2\cos(\frac{k_x}{2})\cos(\frac{\sqrt{3}k_y}{2}))
	\end{matrix}\; .\label{eq:def_gamma_tri}
	\ee
	
	The numerical simulations that we present in this Article are all performed on a cluster of sites defined by the  real space basis vectors given by \Autoref{eq:def_a_b} and scaled by $L$, such that $N=L^2$ is the number of lattice sites, with periodic boundary conditions.
	\section{Structure factors classically}
	\label{sec:structure_factors_classical}
	
	We present here in more detail the results obtained in \Autoref{section:theory.classical.thermodynamics} where we introduced a fictive field $\V{h}$ that couples to the moments (dipoles, quadrupoles or A matrices) that we are considering [\Autoref{eq:DeltaH}].
	This allows us to then take the appropriate derivatives of the free energy with respect to the fictive field components evaluated at zero-field, and calculate the desired thermodynamic quantities, such as the structures factors.
	
	The calculation for the structure factors is divided into 2 parts. The first part is valid  for $\q\neq0$. It consists in taking into account up to linear order in the expansion of fluctuations and is presented below.
	We provide details of the calculation for  $\q\neq0$ for dipole moments  in \hyperref[sec:structure_factors_classical_S]{Appendix~}\ref{sec:structure_factors_classical_S}, quadrupole moments in \hyperref[sec:structure_factors_classical_Q]{Appendix~}\ref{sec:structure_factors_classical_Q}, and A--matrices in \hyperref[sec:structure_factors_classical_A]{Appendix~}\ref{sec:structure_factors_classical_A}.
	The second part captures the ground state contribution at $\q=0$ and consists in taking into account up to quadratic order in the expansion of fluctuations. The general steps of the calculation at $\q=0$ are given in  \hyperref[sec:structure_factors_classicalq0]{Appendix~}\ref{sec:structure_factors_classicalq0}. The details at $\q=0$  are provided in \hyperref[sec:structure_factors_classical_Sq0]{Appendix~}\ref{sec:structure_factors_classical_Sq0} for the dipole moments, in \hyperref[sec:structure_factors_classical_Qq0]{Appendix~}\ref{sec:structure_factors_classical_Qq0} for the quadrupole moments,  in \hyperref[sec:structure_factors_classical_Aq0]{Appendix~}\ref{sec:structure_factors_classical_Aq0} for the A--matrices.
	
	We assume that the field dependent part of the Hamiltonian is given by \Autoref{eq:DeltaH},
	and that the moment $ \hat{O}^{\alpha}_{i,\beta} $ can be written down in terms of the fluctuations $\phi_i$. Considering up to second order in fluctuations, the moments $  \hat{O}^{\alpha}_{i,\beta} $ becomes
	\be
	\hat{O}^{\alpha}_{i,\beta}  = {q^{\alpha}_{~\beta}}_{\mu\nu} \phi_{i}^{\mu}\phi_i^{\nu}+{l^{\alpha}_{~\beta}}_{\mu} \phi_{i}^{\mu}+c^{\alpha}_{~\beta} +\O(\phi^3)\; ,
	\ee
	where we implicitly sum over  $\mu$ and  $\nu$, and
	where $ {q^{\alpha}_{~\beta}}_{\mu\nu}$, ${l^{\alpha}_{~\beta}}_{\mu}$ and $c^{\alpha}_{~\beta}$, are respectively the quadratic, linear, and constant coefficients from the expansion of $ \hat{O}^{\alpha}_{i,\beta} $ in terms ot the fluctuations  $\phi_i$. The field depend part of the Hamiltonian then becomes
	\be
	\Delta {\mathcal H} [{\bf{h}}_i] = -\sum_{i} {h}^\alpha_{i,\beta} {q^{\alpha}_{~\beta}}_{\mu\nu}\phi_{i}^{\mu}\phi_i^{\nu}+ {h}^\alpha_{i,\beta}{l^{\alpha}_{~\beta}}_{\mu}\phi_{i}^{\mu}+ {h}^\alpha_{i,\beta}c^{\alpha}_{~\beta}\; ,
	\ee
	where we also implicitly sum over $\alpha,\beta$, and where we neglect terms in ${\mathcal O}(\phi^3)$, which will from now on be disregarded.
	We now perform a Fourier transform according to \Autoref{eq:abFT}, and obtain
	\begin{align*}
		\Delta {\mathcal H} [{\bf{h}}_{\q}] =& -\sum_{\q}\left[{l^{\alpha}_{~\beta}}_{\mu}h^{\alpha}_{\q,\beta}\phi_{-\q}^{\mu}+ \sqrt{N}c^{\alpha}_{~\beta}h^{\alpha}_{\q,\beta}\delta_{\q,0}\right]\\
		&-\sum_{\q}\sum_{\k} \frac{1}{\sqrt{N}} {q^{\alpha}_{~\beta}}_{\mu\nu}h^{\alpha}_{\q,\beta}\phi_{\k}^{\mu}\phi_{-\q-\k}^{\nu}\; ,\numberthis \label{eq:deltaH_2ndO_k}
	\end{align*}
	where $N$ is the number of lattice sites.
	We notice that if we were to include this in the total Hamiltonian \Autoref{eq:Hfield} and write it down in the same form as \Autoref{eq:H.prime},  the interaction matrix $M_{\k}$ would take the same dimension as the number of lattice site, because of the form of quadratic term in \Autoref{eq:deltaH_2ndO_k}. We should then calculate if for a fixed $\q$. Namely for $\q=0$, we get
	\begin{align*}
		\Delta {\mathcal H} [{\bf{h}}_{\q=0}] =& -{l^{\alpha}_{~\beta}}_{\mu}h^{\alpha}_{\q=0,\beta}\phi_{-\q=0}^{\mu}- \sqrt{N}c^{\alpha}_{~\beta}h^{\alpha}_{\q=0,\beta}\\
		&-\sum_{\k}\frac{1}{\sqrt{N}} {q^{\alpha}_{~\beta}}_{\mu\nu}h^{\alpha}_{\q=0,\beta} \phi_{\k}^{\mu}\phi_{-\k}^{\nu}\; .\numberthis \label{eq:deltaH_2ndO_k_q0}
	\end{align*}
	We note that the form of \Autoref{eq:deltaH_2ndO_k_q0} is compatible with the form of \Autoref{eq:H.prime}. Indeed, for $\q=0$, the contribution of the second order in fluctuations will enter the interaction matrix $M_{\k}$, modifying its eigenvalues, i.e., its relation dispersions, which will also depend on the field $\V{h}$, and the interaction matrix $M_{\k}$ can be easily diagonalized.
	Therefore, we decide to only take into account up to second order in fluctuations for $\q=0$, since it is exactly solvable and since we will need it when comparing the ordered moments at $\q=0$, and to neglect them for $\q \neq 0$. To make the fact that we are taking the second order in fluctuations into account only at $\q=0$ more obvious, we write
	\begin{align*}
		\Delta {\mathcal H} &[{\bf{h}}_{\q}] = -\sum_{\q} {l^{\alpha}_{~\beta}}_{\mu}h^{\alpha}_{\q,\beta}\phi_{-\q}^{\mu}\\
		&-\sum_{\q}\left(\sqrt{N}c^{\alpha}_{~\beta}h^{\alpha}_{\q,\beta}+\sum_{\k} \frac{1}{\sqrt{N}} {q^{\alpha}_{~\beta}}_{\mu\nu}h^{\alpha}_{\q,\beta} \phi_{\k}^{\mu}\phi_{-\k}^{\nu}\right)\delta_{\q,0}\; .\numberthis \label{eq:deltaH_2ndO_k_q02}
	\end{align*}
	
	We then rewrite the field depend part of the Hamiltonian such that the Hamiltonian is symmetric in $\vec{\phi_{\q}}$ and $\vec{\phi_{\q}^{T}}$ [\Autoref{eq:phiT}], which will be necessary when calculating the structure factors at $\q=0$. We have
	\be
	\Delta {\mathcal H} [{\bf{h}}_{\q}]=-\sum_{\q}\V{N}_1[{\bf{h}}_{\q}]^{T}\vec{\phi}_{-\q} +\vec{\phi}_{\q}^{T}\V{N}_2[{\bf{h}}_{-\q}]+\tilde{C}[{\bf{h}}_{\q}]\delta_{\q,0}\; , \label{eq:DeltaHexpan}
	\ee
	where we define
	\be
	\begin{matrix}
		\rm{N}_1[{\bf{h}}_{\q}]^{\mu}=\frac{1}{2}{l^{\alpha}_{~\beta}}_{\mu}h^{\alpha}_{\q,\beta}\; , & & \rm{N}_2[{\bf{h}}_{-\q}]^{\mu}=\frac{1}{2} {l^{\alpha}_{~\beta}}_{\mu}h^{\alpha}_{-\q,\beta} \; ,
	\end{matrix}
	\ee
	and
	\begin{align*}
		\tilde{C}[{\bf{h}}_{\q}]&=\sqrt{N}c^{\alpha}_{~\beta}h^{\alpha}_{\q,\beta}+\sum_{\k} \frac{1}{\sqrt{N}}q_{\mu\nu}^{\alpha}h^{\alpha}_{\q=0,\beta} \phi_{\k}^{\mu}\phi_{-\k}^{\nu}\; .\numberthis \label{eq:def_N_C}
	\end{align*}
	$\V{N}_{1,2}[{\bf{h}}_{\q}]$ are n--dimensional vectors whose components depend linearly on the fields $h^{\alpha}_{\q,\beta}$ and represent the linear terms in $\phi_{\q}$ of the moments $\hat{O}_{\q}$. $\tilde{C}[{\bf{h}}_{\q}]$ represents the 0--order term  and the  2\nd--order contribution in $\phi_{\q}$ of the moments $\hat{O}_{\q}$  at $\q=0$.  $\tilde{C}[{\bf{h}}_{\q}]$ is also linear in the fields $h^{\alpha}_{\q,\beta}$. 
	Plugging \Autoref{eq:DeltaHexpan} in \Autoref{eq:Hfield}, using the definition of the partition function in \Autoref{eq:defZ}, and using \Autoref{eq:gauss_int2} to perform the integral, we get
	\begin{align*}
		Z [{\bf{h}}_{\q}]&=\exp{-\beta E_0}\prod_{\q}^{N}\left[\sqrt{\frac{(2\pi)^n}{\beta^n\det (M_{\q})}}\exp{{2\beta}\V{N}_1[{\bf{h}}_{\q}]^{T} M_{\q}^{-1}\V{N}_2[{\bf{h}}_{\q}]}\right.\\
		&\times \left.\exp{\beta(\tilde{C}[{\bf{h}}_{\q}]\delta_{\q,0})}\right]\numberthis\label{eq:Z}
		\; ,
	\end{align*}
	where $E_0$ is given in \Autoref{eq:E0},  and the $n \times n$ square matrix $M_{\q}$ is given by \Autoref{eq:M}. $n$ is the dimension of  $M_{\q}$, i.e., the number of independent classical fluctuations. In our case, we have $n=4$. $N$ is the number of lattice sites.
	
	The free energy then becomes
	\begin{align*}
		F [{\bf{h}}_{\q}]=&-\frac{\log(Z [{\bf{h}}_{\q}])}{\beta }\\
		=&E_0-\sum_{\q}^{N}\tilde{C}[{\bf{h}}_{\q}]\delta_{\q,0}-2\sum_{\q}^{N}\V{N}_1[{\bf{h}}_{\q}]^{T} M_{\q}^{-1}\V{N}_2[{\bf{h}}_{-\q}]\\
		&+\frac{n}{2\beta}\sum_{\q}\log(\frac{\beta}{2\pi})+\frac{1}{2\beta }\sum_{\q}\log({\det}({M}_{\q})) \\
		&+{\mathcal O}(T^2)\; .\numberthis\label{eq:def_f1}
	\end{align*}
	The first derivatives of the free energy with respect to field components ${\bf{h}}_{\q}$ give
	\be
	\nn{\hat{O}^{\mu}_{\q,\nu}}=-\left.\frac{\ddp F}{\ddp h^{\mu}_{\q,\nu}}\right|_{\V{h}=0}=\left.\frac{\ddp \tilde{C}[{\bf{h}}_{\q}]}{\ddp h^{\mu}_{\q,\nu}}\right|_{\V{h}=0}\delta_{\q,0}\; .\label{eq:Ocor1}
	\ee
	The second derivatives of the free energy with respect to field components ${\bf{h}}_{\q}$ correspond to
	\begin{subequations}
		\begin{align}
			&\nn{\hat{O}^{\alpha}_{\q,\beta}\hat{O}^{\mu}_{-\q,\nu}}-\nn{\hat{O}^{\alpha}_{\q,\beta}}\nn{\hat{O}^{\mu}_{-\q,\nu}}=-\left. \frac{1}{\beta}\frac{\ddp^2 F}{\ddp h^{\alpha}_{\q, \beta}\ddp h^{\mu}_{-\q,\nu}}\right|_{\V{h}=0}\\
			&=\left. \frac{2}{\beta }\frac{\ddp^2 }{\ddp h^{\alpha}_{\q,\beta}\ddp h^{\mu}_{-\q,\nu}}\sum_{\q}^{N}(\V{N}_1[{\bf{h}}_{\q}]^{T} M_{\q}^{-1}\V{N}_2[{\bf{h}}_{-\q}])\right|_{\V{h}=0}\; ,\label{eq:Ocor2}
		\end{align}
	\end{subequations}
	where we used the fact that $\tilde{C}[{\bf{h}}_{\q}]$ is linear in the field components $h^{\alpha}_{\q,\beta}$.
	%
	For $\q \neq 0$, it turns out to be more convenient to work with $\tilde{M_{\q}}$ [\Autoref{eq:orthogonal.transformation2_M}] which is diagonal and which inverse then simply holds 
	\be
	(\tilde{M_{\q}}^{-1})^{\lambda\lambda}=\frac{1}{\tilde{M_{\q}}^{\lambda\lambda}}=\frac{1}{\omega_{\q,\lambda}}\; .
	\ee
	We are allowed to do this because for $\q \neq 0$, the interaction matrix stays unchanged. However, we need to be more careful for $\q =0$ as explained in \hyperref[sec:structure_factors_classicalq0]{Appendix~}\ref{sec:structure_factors_classicalq0}.
	Then, $\V{N}_{1,2}[{\bf{h}}_{\q}]$ become $\tilde{\V{N}}_{1,2}[{\bf{h}}_{\q}]$
	\begin{subequations}
		\begin{align}
			\tilde{\V{N}}_1[{\bf{h}}_{\q}]^T&=\V{N}_1[{\bf{h}}_{\q}]^T O \; ,\\
			\tilde{\V{N}}_{2}[{\bf{h}}_{-\q}]&=O^T\V{N}_2[{\bf{h}}_{-\q}]\; ,
		\end{align}\label{eq:Ntilde}
	\end{subequations}
	such that $\tilde{\V{N}}_{1,2}[{\bf{h}}_{\q}]$ corresponds to the linear term when expressing the operators $\hat{O}^{\alpha}_i$ in terms of the fluctuations $\vec{v}_{\q}$ that diagonalize the BBQ Hamiltonian as shown in \Autoref{eq:diagonal.basis}.
	Indeed, we then obtain
	\be
	\Delta {\mathcal H} [{\bf{h}}_{\q}]=-\sum_{\q}\tilde{\V{N}}_1[{\bf{h}}_{\q}]^T\vec{v}_{-\q}+\vec{v}_{\q}^T\tilde{\V{N}}_2[{\bf{h}}_{-\q}] +\tilde{C}[{\bf{h}}_{\q}]\delta_{\q,0}\; ,\label{eq:DeltaH2}
	\ee 
	Therefore, we can simply write 
	\begin{align*}
		F[{\bf{h}}_{\q}]=&E_0+\sum_{\q}^{N}\tilde{C}[{\bf{h}}_{\q}]\delta_{\q,0}-2\sum_{\q}^{N}\sum_{\lambda=1}^{N_{\lambda}}\frac{(\tilde{\V{N}}_1[{\bf{h}}_{\q}]^T)^{\lambda}\tilde{\V{N}}_2[{\bf{h}}_{-\q}]^{\lambda}}{\omega_{\q,\lambda}}\\
		&+\frac{N_{\lambda}}{2\beta }\sum_{\q}\log(\frac{\beta}{2\pi})+\frac{1}{2\beta }\sum_{\q}\sum_{\lambda=1}^{N_{\lambda}}\log(\omega_{\q,\lambda}) +{\mathcal O}(T^2)\; ,\numberthis\label{eq:def_fbis}
	\end{align*}
	where we have used \Autoref{eq:log.det.M} and where $N_{\lambda}=4$ is the number of modes.
	\Autoref{eq:Ocor1} stays unchanged, but \Autoref{eq:Ocor2} takes the simple form given by
	\begin{align}
		&\nn{\hat{O}^{\alpha}_{\q,\beta}\hat{O}^{\mu}_{-\q,\nu}}-\nn{\hat{O}^{\alpha}_{\q,\beta}}\nn{\hat{O}^{\mu}_{-\q,\nu}}=\\
		&\left. \frac{2}{\beta}\frac{\ddp }{\ddp h^{\alpha}_{\q, \beta}\ddp h^{\mu}_{-\q,\nu}}\left(\sum_{\lambda=1}^{N_{\lambda}}\frac{(\tilde{\V{N}}_1[{\bf{h}}_{\q}]^T)^{\lambda}\tilde{\V{N}}_2[{\bf{h}}_{-\q}]^{\lambda}}{\omega_{\q,\lambda}}\right)\right|_{\V{h}=0} \; .\label{eq:Ocor3}
	\end{align}
	%
	The dynamical factor associated with the operator $\hat{O}$ is defined by
	\begin{align*}
		S^{\sf CL}_{\rm{O}}(\q)&=\sum_{\alpha, \beta}\nn{\hat{O}^{\alpha}_{\q,\beta}\hat{O}^{\beta}_{-\q,\alpha}} \numberthis\label{eq:SOF_class1}\\
	\end{align*}
	
	We can generalize a spectral decomposition of the structure factors as
	\begin{eqnarray}
		S^{\sf CL}_{\rm{O}}(\q, \omega) 
		&=& \sum_{\alpha,\beta,\lambda} \nn{\hat{O}^{\alpha}_{\q,\beta}\hat{O}^{\beta}_{-\q,\alpha}}_{\lambda} 
		\delta(\omega - \omega_{\q,\lambda}) \; ,\label{eq:SOF_class_q_w3}
	\end{eqnarray}
	and calculate the following quantity
	\begin{eqnarray}
		S^{\sf CL}_{\rm{O}}(\q, \omega) 
		&=& \sum_{\alpha,\beta,\lambda} \left[ \nn{\hat{O}^{\alpha}_{\q,\beta}}_{\lambda}  \nn{\hat{O}^{\beta}_{-\q,\alpha}}_{\lambda} 
		+ \chi^{\alpha \beta \beta \alpha}_\lambda(\q) 
		\right] \nonumber\\
		&& \qquad \times \delta(\omega - \omega_{\q,\lambda})\; ,\label{eq:SOF_class_q_w4}
	\end{eqnarray}
	where  generalized susceptibility
	\begin{align*}
		&\chi^{\alpha\beta\mu\nu}_{\lambda}(\q)
		=\nn{\hat{O}^{\alpha}_{\q,\beta}\hat{O}^{\mu}_{-\q,\nu}}_{\lambda} - \nn{\hat{O}^{\alpha}_{\q,\beta}}_{\lambda}  \nn{\hat{O}^{\mu}_{-\q,\nu}}_{\lambda} \\
		&=\left. \frac{2}{\beta}\frac{\ddp }{\ddp h^{\alpha}_{\q, \beta}\ddp h^{\mu}_{-\q,\nu}}\left(\frac{(\tilde{\V{N}}_1[{\bf{h}}_{\q}]^T)^{\lambda}\tilde{\V{N}}_2[{\bf{h}}_{-\q}]^{\lambda}}{\omega_{\q,\lambda}}\right)\right|_{\V{h}=0} \; . \label{eq:chi_gen}
		\numberthis
	\end{align*}
	is diagonal in $\lambda$.
	From \Autoref{eq:Ocor1}, we note that the first moments $\nn{\hat{O}^{\alpha}_{\q, \beta}}$ will only contribute at $\q=0$. Therefore, for $\q\neq 0$, we can neglect the $\nn{\hat{O}^{\alpha}_{\q,\lambda}}\nn{\hat{O}^{\beta}_{-\q,\lambda}}$ term and we obtain
	\begin{eqnarray}
		S^{\sf CL}_{\rm{O}} (\q \ne 0) 
		&=&	\sum_{\alpha\beta\lambda}  \chi^{\alpha\beta\beta\alpha}_\lambda(\q) 
		+ \mathcal{O}(T^2) \; ,
		\label{eq:SOF_class_q}
	\end{eqnarray}
	
	\subsection{Structure factors classically at $\q=0$}\label{sec:structure_factors_classicalq0}
	We here show how the calculation for the structure at $\q=0$ is obtained. 
	At $\q=0$, the structure factor associated with the operator $\hat{O}$ is defined by
	\begin{align*}
		S^{\sf CL}_{\rm{O}}(\q=0)&=\sum_{\alpha, \beta}\nn{\hat{O}^{\alpha}_{\q=0,\beta}\hat{O}^{\beta}_{\q=0,\alpha}}\; . \numberthis\label{eq:SOF_class_q0}
	\end{align*}
	The relevant source term is given by \Autoref{eq:DeltaH}.
	By expanding \Autoref{eq:DeltaH} in terms of the fluctuations, we obtained \Autoref{eq:deltaH_2ndO_k_q02}. We see that the contribution of the second order in fluctuations has the same form as the Hamiltonian expressed as \Autoref{eq:H.prime} and  will enter the interaction matrix $M_{\k}$, modifying its eigenvalues, i.e., relation dispersion relations, which will all also depend on the field $\V{h}$.
	We can therefore assume that,  at $\q=0$, the total Hamiltonian [\Autoref{eq:Hfield}] has the following form 
	\begin{align*}
		\Ham &= E_0 + \frac{1}{2} \sum_{\k} 
		\left[ \vec{\phi}_{\k}^T M_{\k}[{\bf{h}}_{\q=0}] \vec{\phi}_{-\k} \right]\\
		&+\sum_{\k} \left[ N^{T}_1[{\bf{h}}_{\k}] \vec{\phi}_{-\k} +\vec{\phi}_{\k}^T N_2[{\bf{h}}_{-\k}]  \right]\delta_{\k,0} \\
		&+\sum_{\k} C[{\bf{h}}_{\q=0}] \delta_{\k,0} 
		+	\O(\phi^3) \; , \numberthis 
		\label{eq:H.q0}
	\end{align*}
	where $C[{\bf{h}}_{\q=0}]$ represents the 0--order term in $\phi_{\q}$ of the moments $\hat{O}_{\q}$. The  2\nd--order contribution at $\q=0$ is now included in $ M_{\k}[{\bf{h}}_{\q=0}] $. As before, $\V{N}_{1,2}[{\bf{h}}_{\k}]$ are n--dimensional vectors whose components depend linearly on the fields $h^{\alpha}_{\k,\beta}$ and represent the linear terms in $\phi_{\k}$ of the moments $\hat{O}_{\k}$.
	Neglecting terms in ${\mathcal O}(\phi^3)$, and using \Autoref{eq:gauss_int2} to perform the integral, we find
	\begin{subequations}
		\begin{align*}
			Z =&  \exp{-\beta E_0} \prod_{\k}^{N}\left[\int \exp{-\beta\frac{1}{2}
				\vec{\phi}_{\k}^T M_{\k}[{\bf{h}}_{\q=0}] \vec{\phi}_{\k}}\right.\\
			& \times \exp{-\beta\left[N^{T}_1[{\bf{h}}_{\k}] \vec{\phi}_{-\k} +\vec{\phi}_{\k}^T N_2[{\bf{h}}_{-\k}]  \right]\delta_{\k,0} }\\
			&\left. \times 	\exp{-\beta C[{\bf{h}}_{\q=0}] \delta_{\k,0} }\right] \dd \vec{\phi}_{\k} \numberthis \\
			=&  \exp{-\beta E_0} \prod_{\k}^{N}\left[\sqrt{\frac{(2\pi)^n}{\beta^n \det M_{\k}[{\bf{h}}_{\q=0}]}}\right.\\
			& \times \left.\exp{2\beta N^{T}_1[{\bf{h}}_{\k}]M_{\k}^{-1}[{\bf{h}}_{\q=0}]N_2[{\bf{h}}_{-\k}] \delta_{\k,0} }\exp{-\beta C[{\bf{h}}_{\q=0}] \delta_{\k,0} } \right]  \numberthis\; , 
			\label{eq:Zq0}
		\end{align*}
	\end{subequations}
	where $E_0$ is defined through \Autoref{eq:E0},  
	and the $n \times n$ matrix $M_{\k}[{h}^\alpha_{\q=0\beta}]$ through \Autoref{eq:H.q0} that includes up to second order in fluctuations. $n$ is the dimension of  $M_{\k}[{h}^\alpha_{\q=0\beta}]$, i.e., the number of independent classical fluctuations. In our case, we have $n=4$. $N$ is the number of lattice sites. 
	It follows that the free energy is 
	\begin{align*}
		F =&-\frac{\log(Z)}{\beta}\\
		=&E_0+\sum_{\k}C[{\bf{h}}_{\q=0}] \delta_{\k,0}\\
		&-\frac{n}{2 \beta}\sum_{\k}\log(\frac{(2\pi)}{\beta})+\frac{1}{2 \beta}\sum_{\k}\sum_{\lambda=1}^{N_{\lambda}}	
		\log(\omega_{\k,\lambda}[{\bf{h}}_{\q=0}])\\
		& - 2\sum_{\k} N^{T}_1[{\bf{h}}_{\k}]M_{\k}^{-1}[{\bf{h}}_{\q=0}]N_2[{\bf{h}}_{-\k}] \delta_{\k,0}  +{\mathcal O}(T^2) \; , \numberthis
		\label{eq:Fq0}
	\end{align*}
	where $\omega_{\k,\lambda}[{\bf{h}}_{\q=0}]$ are the eigenvalues of $M_{\k}[{\bf{h}}_{\q=0}]$, 
	and we have used \Autoref{eq:log.det.M}.
	The moments are given by 
	\begin{eqnarray}
		\nn{\hat{O}^{\alpha}_{\q=0, \beta}} 
		&=&	-\left. \frac{\ddp F}{\ddp h^{\alpha}_{\q=0,\beta}}\right|_{\V{h}=0} \; ,
		\label{eq:Ocorq0}
	\end{eqnarray}
	and 
	\begin{align*}
		\nn{\hat{O}^{\alpha}_{\q=0, \beta}\hat{O}^{\mu}_{\q=0, \nu}} =&	\nn{\hat{O}^{\alpha}_{\q=0, \beta}} 	\nn{\hat{O}^{\mu}_{\q=0, \nu}} \\
		&	-\left. \frac{1}{\beta}\frac{\ddp^2 F}{\ddp h^{\alpha}_{\q=0,\beta} \ddp h^{\mu}_{\q=0,\nu}} \right|_{\V{h}=0} \; .	\label{eq:OOcorq0} \numberthis
	\end{align*}
	Using \Autoref{eq:Fq0}, \Autoref{eq:Ocorq0} yields
	\begin{align*}
		\nn{\hat{O}^{\alpha}_{\q=0, \beta}} 
		&=	-\left. \frac{\ddp C[{\bf{h}}_{\q=0}] }{\ddp h^{\alpha}_{\q=0,\beta}}\right|_{\V{h}=0}\\
		& -\left. \frac{1}{2 \beta}\sum_{\k}\sum_{\lambda=1}^{N_{\lambda}}\frac{1}{\omega_{\k,\lambda}[{\bf{h}}_{\q=0}]} \frac{\ddp \omega_{\k,\lambda}[{\bf{h}}_{\q=0}]}{\ddp h^{\alpha}_{\q=0,\beta}}\right|_{\V{h}=0}\\
		&+\left. 2\frac{\ddp \left[N^{T}_1[{\bf{h}}_{\q=0}]M_{\q=0}^{-1}[{\bf{h}}_{\q=0}]N_2[{\bf{h}}_{\q=0}]\right] }{\ddp h^{\alpha}_{\q=0,\beta}} \right|_{\V{h}=0}\\
		&+{\mathcal O}(T^2)\numberthis \; ,
		\label{eq:Ocorq02}
	\end{align*}
	where the last derivative turns out to be null when evaluated at \mbox{$\V{h}=0$}, for dipoles, quadrupoles and A-matrices. 
	\Autoref{eq:OOcorq0} becomes
	\begin{align*}
		\nn{\hat{O}^{\alpha}_{\q=0, \beta}\hat{O}^{\mu}_{\q=0, \nu}} &
		=	\nn{\hat{O}^{\alpha}_{\q=0, \beta}} 	\nn{\hat{O}^{\mu}_{\q=0, \nu}}\\
		-\frac{1}{\beta}&\left.\frac{\ddp^2 C[{\bf{h}}_{\q=0}]}{\ddp h^{\alpha}_{\q=0,\beta}\ddp h^{\mu}_{\q=0,\nu}}\right|_{\V{h}=0}\\
		+\frac{2}{\beta}&\left. \frac{\ddp^2\left[ N^{T}_1[{\bf{h}}_{\q=0}]M_{\q=0}^{-1}[{\bf{h}}_{\q=0}]N_2[{\bf{h}}_{\q=0}] \right]}{\ddp h^{\alpha}_{\q=0,\beta}\ddp h^{\mu}_{\q=0,\nu}}\right|_{\V{h}=0}\\
		+{\mathcal O}&(T^2) \; , \numberthis \label{eq:OOcorq02}
	\end{align*}
	where the terms including second derivatives of $C[{\bf{h}}_{\q=0}]$ are zero, since $C[{\bf{h}}_{\q=0}]$  is linear in $ {h}^\alpha_{\q=0,\beta}$ by definition.
	Finally, \Autoref{eq:SOF_class_q0} can be calculated by using \Autoref{eq:Ocorq02} and \Autoref{eq:OOcorq02}.
	For each type of moments, dipole, quadrupole or A-matrix, the interaction matrix $M_{\k}[{\bf{h}}_{\q=0}]$, the source terms $N^{T}_1[{\bf{h}}_{\q}]$ and $N_2[{\bf{h}}_{\q}]$ and the constant term will be different. They are given below.
	
	\subsection{Dipole moments: classical structure factor for $\q\neq0$}\label{sec:structure_factors_classical_S}
	First, we consider the structure factor for dipole moments of spin
	\begin{eqnarray}
		S^{\sf CL}_{\rm{S}}(\q) 
		&=& \sum_{\alpha} \nn{\ops{\q}{\alpha} \ops{-\q}{\alpha}}  \; .\label{eq:SS.class_q}
	\end{eqnarray}
	The relevant source term is 
	\begin{eqnarray}
		\Delta\Ham [ {h}^{\alpha}_{i,\beta} ] 
		&=& - \sum_{i,\lambda} h_{i, \beta}^{\alpha}\delta_{\alpha\beta} \ops{i,\lambda}{\beta} \; .
		\label{eq:HamLowTS}
	\end{eqnarray}
	According to \Autoref{eq:dipole.in.terms.of.A} and using \Autoref{eq:A_phi}, we can express the spin dipole components in function of the fluctuations.
	Considering fluctuation terms up to 1$\st$ order and using  \Autoref{eq:orthogonal.transformation},
	the spin dipole moments in terms of the fluctuations diagonalizing the BBQ Hamiltonian are given by
	\begin{align*}
		\ops{i}{x}&=-\sqrt{2}v_{4,i}\; ,\\
		\ops{i}{y}&\simeq 0\; , \numberthis\\
		\ops{i}{z}&=\sqrt{2}v_{2,i}\; .
	\end{align*}
	After performing a Fourier transform, the change in the Hamiltonian due to  $\Delta\Ham [ {h}^{\alpha}_{i,\beta} ] $ [\Autoref{eq:HamLowTS}] yields
	\begin{align*}
		\Delta\Ham [{\bf{h}}_{\q}] =-\sum_{\q}&\left[\frac{\sqrt{2}}{2}h^{z}_{\q}v_{-\q,2}+\frac{\sqrt{2}}{2}h^{z}_{-\q}v_{\q,2}\right. \\
		&\left.-\frac{\sqrt{2}}{2}h^{x}_{\q}v_{-\q,4}-\frac{\sqrt{2}}{2}h^{x}_{-\q}v_{\q,4}\right]\; ,\label{eq:HamLowTS2FT}
	\end{align*}
	and according to \Autoref{eq:DeltaH2}, we get 
	\be
	\tilde{C}[{\bf{h}}_{\q}]=0\; ,\label{eq:C_S}
	\ee
	where we neglected 2$\nd$ order terms in fluctuations, since they only contribute for $\q=0$, and
	\be
	\begin{matrix}
		\tilde{\V{N}}_1[{\bf{h}}_{\q}]^T=\begin{pmatrix}
			0, &
			\frac{\sqrt{2}}{2}h^{z}_{\q},&
			0, &
			-\frac{\sqrt{2}}{2}h^{x}_{\q}
		\end{pmatrix}\; , \\
		\tilde{\V{N}}_2[{\bf{h}}_{-\q}]=\begin{pmatrix}
			0\\
			\frac{\sqrt{2}}{2}h^{z}_{-\q}\\
			0\\
			-\frac{\sqrt{2}}{2}h^{x}_{-\q}
		\end{pmatrix}\; .\label{eq:N_S}
	\end{matrix}
	\ee
	According to \Autoref{eq:Ocor1}, the first moments are given by the first derivative of $ \tilde{C}[{\bf{h}}_{\q}]$ [\Autoref{eq:C_S}] with respect to the fictive field  $\V{h}$. 
	We get
	\begin{align}
		\nn{S^{x}_{\q}}&=\nn{S^{z}_{\q}}=0\; .\label{eq:SxSz}
	\end{align}

	The total structure factor for the dipole moment is given by \Autoref{eq:SS.class_q},
	and using  \Autoref{eq:chi_gen}, \Autoref{eq:SOF_class_q}, and \Autoref{eq:N_S}, we obtain
	\be
	S^{\sf CL}_{\rm{S}}(\q \neq 0)=\frac{2}{\beta\omega_{\q,2}} +\frac{2}{\beta\omega_{\q,4}}+{\mathcal O}(T^2) =\frac{4}{\beta\omega_{\q}^-}+{\mathcal O}(T^2)\label{eq:SF_class1}\; ,
	\ee
	where we used \Autoref{eq:omega_vec_-}.
	Its spectral decomposition  [\Autoref{eq:SOF_class_q_w4}] becomes
	\be
	S^{\sf CL}_{\rm{S}}(\q \neq 0, \omega)=\frac{4}{\beta\omega_{\q}^-}\delta(\omega - \omega_{\q}^-)+{\mathcal O}(T^2)\label{eq:SF_class2} \; .
	\ee
	\subsection{Dipole moments: classical structure factor at $\q=0$}\label{sec:structure_factors_classical_Sq0}
	We now consider the dipole structure factor at the origin of the reciprocal space called the $\Gamma$--point.
	We consider the structure factor for the spin dipole moments
	\begin{eqnarray}
		S^{\sf CL}_{\rm{S}}(\q=0) 
		&=& \sum_{\alpha} \nn{\ops{\q=0}{\alpha} \ops{\q=0}{\alpha}}  \; . \label{eq:Sq0}
	\end{eqnarray}
	We follow the procedure depicted in \hyperref[sec:structure_factors_classicalq0]{Appendix~}\ref{sec:structure_factors_classicalq0}. The relevant source term for dipole moments is given by \Autoref{eq:HamLowTS} that we need to rewrite it in the same form as \Autoref{eq:H.q0}. We use \Autoref{eq:H.prime}--\Autoref{eq:M} for the BBQ Hamiltonian, as well as, \Autoref{eq:A_phi} and \Autoref{eq:dipole.in.terms.of.A} to express \Autoref{eq:HamLowTS} up to second order in terms of the fluctuations. For the total Hamiltonian given in \Autoref{eq:Hfield}, and written in the form of \Autoref{eq:H.q0}, we obtain
	\begin{align*}
		M_{\k}&[{\bf{h}}_{\q=0}] = \\
		&\begin{pmatrix}
			A_{\k} & -B_{\k} &  0& \frac{i}{\sqrt{N}}h^{y}_{\q=0,y} \\
			-B_{\k} &A_{\k} &  -\frac{i}{\sqrt{N}}h^{y}_{\q=0,y}& 0 \\
			0 &  \frac{i}{\sqrt{N}}h^{y}_{\q=0,y} & A_{\k} &- B_{\k}  \\
			-\frac{i}{\sqrt{N}}h^{y}_{\q=0,y} & 0 & -B_{\k} & A_{\k} \\
		\end{pmatrix} \; , \numberthis
		\label{eq:M_Sq0}
	\end{align*}
	\be
	\begin{matrix}
		\V{N}_1[{\bf{h}}_{\k}]^T=\frac{1}{2}\begin{pmatrix}
			-h^{z}_{\k,z}, &
			-h^{z}_{\k,z},&
			h^{x}_{\k,x}, &
			h^{x}_{\k,x}
		\end{pmatrix}\; , \\
		\V{N}_2[{\bf{h}}_{-\k}]=\frac{1}{2}\begin{pmatrix}
			-h^{z}_{-\k,z}\\
			-h^{z}_{-\k,z}\\
			h^{x}_{-\k,x}\\
			h^{x}_{-\k,x}
		\end{pmatrix}\; ,\label{eq:N_Sq0}
	\end{matrix}
	\ee
	\be
	C[ {\bf{h}}_{\k} ] =0\; .\label{eq:C_Sq0}
	\ee
	We diagonalize \Autoref{eq:M_Sq0} to obtain the eigenmodes. We find
	\begin{subequations}
		\begin{align*}
			\omega_{\k}^{+}[{\bf{h}}_{\q=0}]&=\omega_{\k,1}[{\bf{h}}_{\q=0}]=\omega_{\k,3}[{\bf{h}}_{\q=0}]\\
			&=A_{\k}+\sqrt{\left(\frac{h^{y}_{0,y}}{\sqrt{N}}\right)^2+B_{\k}^2}\; ,\numberthis\\
			\omega_{\k}^{-}[{\bf{h}}_{\q=0}]&=\omega_{\k,2}[{\bf{h}}_{\q=0}]=\omega_{\k,4}[{\bf{h}}_{\q=0}]\\
			&=A_{\k}-\sqrt{\left(\frac{h^{y}_{0,y}}{\sqrt{N}}\right)^2+B_{\k}^2}\; .\numberthis
		\end{align*} \label{eq:omega_Sq0}
	\end{subequations}
	We now can calculate the spin dipole moments through \Autoref{eq:Ocorq02}, where  we use \Autoref{eq:C_Sq0}, and \Autoref{eq:omega_Sq0}, and where for the last term, we simply invert \Autoref{eq:M_Sq0} and multiply by the vectors in \Autoref{eq:N_Sq0}.
	We obtain
	\begin{align}
		\nn{S^{x}_{\q=0}}&=	\nn{S^{y}_{\q=0}}=\nn{S^{z}_{\q=0}}=0\; .\label{eq:SxSySz_q0}
	\end{align}
	For the square dipole moments, we use  \Autoref{eq:OOcorq02}. We find
	\begin{subequations}
		\begin{align}
			\nn{S^{x}_{\q=0}S^{x}_{\q=0}}&=\frac{2}{\beta}\frac{1}{A_{\q=0}-B_{\q=0}}=\frac{2}{\beta}\frac{1}{\omega_{0}^{-}}\; ,\\
			\nn{S^{y}_{\q=0}S^{y}_{\q=0}}&=0\; ,\\
			\nn{S^{z}_{\q=0}S^{z}_{\q=0}}&=\frac{2}{\beta}\frac{1}{A_{\q=0}-B_{\q=0}}=\frac{2}{\beta}\frac{1}{\omega_{0}^{-}}\; ,
		\end{align}
	\end{subequations}
	where we used \Autoref{eq:omega_vec_1}.
	
	Finally,  we calculate the dipole structure factor at the $\Gamma$--point  given by \Autoref{eq:Sq0}. We get 
	\begin{eqnarray}
		S^{\sf CL}_{\rm{S}} (\q=0) 
		&=&  \frac{4}{\beta} \frac{1}{\omega^-_0} +{\mathcal O}(T^2)\; .
		\label{eq:classical.structure.factor.Sq0}
	\end{eqnarray}
	Because the $\q=0$ contributions are coming from the ground state and happen for $\omega=0$, the spectral representation of \Autoref{eq:classical.structure.factor.Sq0} yields 
	\begin{eqnarray}
		S^{\sf CL}_{\rm{S}} (\q=0,\omega) 
		&=&  \frac{4}{\beta} \frac{1}{\omega^-_0} \delta(\omega)+{\mathcal O}(T^2)\; .
		\label{eq:classical.structure.factor.Sq0w}
	\end{eqnarray}
	Combining \Autoref{eq:SF_class1} and \Autoref{eq:classical.structure.factor.Sq0}, we get \Autoref{eq:classical.structure.factor.S}.  And considering their respective spectral representation \Autoref{eq:SF_class2} and \Autoref{eq:classical.structure.factor.Sq0w}, we obtain \Autoref{eq:SF_class}.
	
	\subsection{Quadrupole moments: classical structure factor for $\q\neq 0$}\label{sec:structure_factors_classical_Q}
	Next, we consider the structure factor for quadrupole moments of spin
	\begin{eqnarray}
		S^{\sf CL}_{\rm{Q}}(\q) 
		&=& \sum_{\alpha\beta} 
		\nn{\opq{\q}{\alpha\beta} \opq{-\q}{\beta\alpha}} \; ,	\label{eq:QQ}
	\end{eqnarray}
	where the scalar contraction implied by the sum on $\alpha$, $\beta$ respects 
	$SU(2)$ symmetry.
	In this case the source term is 
	\be
	\Delta\Ham [{\bf{h}}_i] = -\sum_{i} h^{\alpha}_{i,\beta} \opq{i}{\alpha\beta} \; . 
	\label{eq:HamLowTQ}
	\ee
	The quadrupole components $\opq{i}{\alpha\beta}$ in the function of  the classical fluctuations can be found using \Autoref{eq:quadrupole.in.terms.of.A} and \Autoref{eq:A_phi}.
	Using \Autoref{eq:orthogonal.transformation}, we can express $\Delta\Ham [{\bf{h}}_i]$ in terms of the fluctuations that diagonalize the BBQ Hamiltonian. After performing a Fourier transform, and rewriting the Hamiltonian in the form of \Autoref{eq:DeltaH2}, we get 
	\be
	\tilde{C}[{\bf{h}}_{\q}]=\sqrt{N}\left(-\frac{4}{3}h_{\q}^{yy}+\frac{2}{3}(h_{\q}^{xx}+h_{\q}^{zz})\right)\; ,\label{eq:C_Q}
	\ee
	where we neglected 2$\nd$ order terms in fluctuations, since they only contribute for $\q=0$, and
	\be
	\begin{matrix}
		\tilde{\V{N}}_1[{\bf{h}}_{\q}]^T=\begin{pmatrix}
			0, &
			\frac{i\sqrt{2}}{2}\xi^1_{\q},&
			0, &
			-\frac{i\sqrt{2}}{2}\xi^1_{\q}
		\end{pmatrix}\; , \\
		\tilde{\V{N}}_2[{\bf{h}}_{-\q}]=\begin{pmatrix}
			0\\
			\frac{i\sqrt{2}}{2}\xi^1_{-\q}\\
			0\\
			-\frac{i\sqrt{2}}{2}\xi^1_{-\q}
		\end{pmatrix}\; ,\label{eq:N_Q}
	\end{matrix}
	\ee
	where
	\be
	\begin{matrix}
		\xi^1_{\q}=(h^{xy}_{\q}+h^{yx}_{\q}) & \; ,& \xi^2_{\q}=(h^{yz}_{\q}+h^{zy}_{\q})\; .
	\end{matrix}
	\ee
	
	The total quadrupole structure factor is given by \Autoref{eq:QQ}.
	According to  \Autoref{eq:chi_gen} and \Autoref{eq:SOF_class_q}, and using \Autoref{eq:N_Q},  we obtain
	\begin{align*}
		S^{\sf CL}_{\rm{Q}}(\q \neq 0)=&\frac{4}{\beta\omega_{\q,1}}+\frac{4}{\beta\omega_{\q,3}}+{\mathcal O}(T^2)=\frac{8}{\beta\omega_{\q}^+}+{\mathcal O}(T^2)\label{eq:QF_class0}\; ,\numberthis
	\end{align*}
	where we used \Autoref{eq:omega_vec_+}.
	and its spectral decomposition  [\Autoref{eq:SOF_class_q_w4}] becomes
	\begin{align*}
		S^{\sf CL}_{\rm{Q}}(\q\neq 0,\omega)=\frac{8}{\beta\omega_{\q}^+}\delta(\omega - \omega_{\q}^+)+{\mathcal O}(T^2)\; . \label{eq:QF_class1}\numberthis
	\end{align*}
	\subsection{Quadrupole moments: classical structure factor at $\q=0$}\label{sec:structure_factors_classical_Qq0}
	We now consider the quadrupole structure factor at the $\Gamma$--point, which is defined as
	\begin{eqnarray}
		S^{\sf CL}_{\rm{Q}}(\q=0) 
		&=& \sum_{\alpha\beta} 
		\nn{\opq{\q=0}{\alpha\beta} \opq{\q=0}{\beta\alpha}} \; . \label{eq:Qq0}
	\end{eqnarray}
	We follow the same procedure as depicted in \hyperref[sec:structure_factors_classicalq0]{Appendix~}\ref{sec:structure_factors_classicalq0}. The relevant source term for quadrupole moments is given by \Autoref{eq:HamLowTQ}. We use \Autoref{eq:M} for the BBQ Hamiltonian as well as \Autoref{eq:A_phi} and \Autoref{eq:quadrupole.in.terms.of.A} to express \Autoref{eq:HamLowTQ} up to second order in terms of the fluctuations. For the total Hamiltonian given by \Autoref{eq:Hfield}, and written in the form of \Autoref{eq:H.q0}, we obtain
	\be
	M_{\k}[{\bf{h}}_{\q=0}] = 
	\begin{pmatrix}
		A_{\k}+\alpha_1 & -B_{\k} &  0& \beta_1\\
		-B_{\k} &A_{\k}+\alpha_1  &   \beta_1& 0 \\
		0 &   \beta_1& A_{\k}+\alpha_2 &- B_{\k}  \\
		\beta_1& 0 & -B_{\k} & A_{\k}+\alpha_2  \\
	\end{pmatrix} \; , 
	\label{eq:M_Qq0}
	\ee
	\be
	\begin{matrix}
		\V{N}_1[{\bf{h}}_{\k}]^T=\frac{-i}{2}\begin{pmatrix}
			-\xi^1_{\k}, &
			\xi^1_{\k},&
			\xi^2_{\k}, &
			-\xi^2_{\k}
		\end{pmatrix}\; , \\
		\V{N}_2[{\bf{h}}_{-\k}]=\frac{i}{2}\begin{pmatrix}
			-\xi^1_{-\k}\\
			\xi^1_{-\k}\\
			\xi^2_{-\k}\\
			-\xi^2_{-\k}
		\end{pmatrix}\; ,\label{eq:N_Qq0}
	\end{matrix}
	\ee
	\be
	C[{\bf{h}}_{\q=0}]=\sqrt{N}\left(\frac{4}{3}h_{\q=0,y}^{y}-\frac{2}{3}(h_{\q=0,x}^{x}+h_{\q=0,z}^{z})\right)\; ,\label{eq:C_Qq0}
	\ee
	where we define
	\be
	\begin{matrix}
		\begin{matrix}
			\alpha_1=\frac{2}{\sqrt{N}}(h_{\q=0,x}^{x}-h_{\q=0,y}^{y})\; ,&\alpha_2=\frac{2}{\sqrt{N}}(h_{\q=0,z}^{z}-h_{\q=0,y}^{y})\; ,
		\end{matrix}\\
		&\\
		\begin{matrix}
			\beta_1=\frac{1}{\sqrt{N}}(h_{\q=0,z}^{x}+h_{\q=0,x}^{z})\;, 
		\end{matrix}\\
		&\\
		\begin{matrix}
			\xi^1_{\k}=(h^{x}_{\k,y}+h^{y}_{\k,x}) & \; ,& \xi^2_{\k}=(h^{y}_{\k,z}+h^{z}_{\k,y})
		\end{matrix}\; .
	\end{matrix}
	\ee
	We diagonalize \Autoref{eq:M_Qq0} to obtain the eigenmodes. We find
	\begin{subequations}
		\begin{align}
			\omega_{\k,1}[{\bf{h}}_{\q=0}]&=A_{\k}+B_{\k}^2+\frac{1}{2}(\alpha_{+}+\Delta)\; ,\\
			\omega_{\k,2}[{\bf{h}}_{\q=0}]&=A_{\k}-B_{\k}^2+\frac{1}{2}(\alpha_{+}+\Delta)\; ,\\
			\omega_{\k,3}[{\bf{h}}_{\q=0}]&=A_{\k}+B_{\k}^2+\frac{1}{2}(\alpha_{+}-\Delta)\; ,\\
			\omega_{\k,4}[{\bf{h}}_{\q=0}]&=A_{\k}-B_{\k}^2+\frac{1}{2}(\alpha_{+}-\Delta)\; ,
		\end{align} \label{eq:omega_Qq0}
	\end{subequations}
	where
	\be
	\begin{matrix}
		\alpha_{+}=\alpha_1+\alpha_2& \; ,& \Delta=\sqrt{(\alpha_1-\alpha_2)^2 +4\beta_1^2}\; .\label{eq:a+_Delta}
	\end{matrix}
	\ee
	
	Finally, we use  \Autoref{eq:Ocorq02} and \Autoref{eq:OOcorq02} to compute the quadrupole structure factor at the $\Gamma$--point given by \Autoref{eq:Qq0}.
	When calculating  \Autoref{eq:Ocorq02} and \Autoref{eq:OOcorq02}, we use \Autoref{eq:C_Qq0} and \Autoref{eq:omega_Qq0}, and for the last term, we simply invert \Autoref{eq:M_Qq0} and multiply by the vectors expressed in \Autoref{eq:N_Qq0}.
	We obtain
	\begin{eqnarray}
		S^{\sf CL}_{\rm{Q}}(\q=0) 
		&=& \frac{8}{\beta} \frac{1}{\omega^+_0} 
		+ \frac{8}{3} N - \frac{8}{\beta}\sum_{\k} \left[ \frac{1}{\omega^+_\k} + \frac{1}{\omega^-_\k} \right] +{\mathcal O}(T^2)\; .
		\label{eq:classical.structure.factor.Qq0}
		\nonumber\\
	\end{eqnarray}
	However, we note that at the $\Gamma$--point, $\omega^+_0=0$. Therefore, in order to get rid of confounding divergent terms, we rewrite the quadrupole structure factor as
	\begin{eqnarray}
		S^{\sf CL}_{\rm{Q}}(\q=0) 
		&=& -\frac{8}{\beta} \frac{1}{\omega^-_0} 
		+ \frac{8}{3} N - \frac{8}{\beta}\sum_{\k\neq0} \left[ \frac{1}{\omega^+_\k} + \frac{1}{\omega^-_\k} \right] +{\mathcal O}(T^2)\; .
		\label{eq:classical.structure.factor.Qq02}
		\nonumber\\
	\end{eqnarray}
	Because the $\q=0$ contributions are coming from the ground state and happen for $\omega=0$, the spectral representation of \Autoref{eq:classical.structure.factor.Qq02} yields 
	\begin{eqnarray}
		S^{\sf CL}_{\rm{Q}}(\q=0,\omega) \nonumber
		&=& -\frac{8}{\beta} \frac{1}{\omega^-_0}\delta(\omega) 
		+ \frac{8}{3} N \delta(\omega) \\
		& & - \frac{8}{\beta}\sum_{\k\neq0} \left[ \frac{1}{\omega^+_\k} + \frac{1}{\omega^-_\k} \right]\delta(\omega)  +{\mathcal O}(T^2)\; .
		\label{eq:classical.structure.factor.Qq0w}
		\nonumber\\
	\end{eqnarray}
	Combining \Autoref{eq:QF_class0} and \Autoref{eq:classical.structure.factor.Qq02}, we obtain \Autoref{eq:classical.structure.factor.Q}. Considering their respective spectral representation given by \Autoref{eq:QF_class1} and \Autoref{eq:classical.structure.factor.Qq0w}, we obtain \Autoref{eq:QF_class}.
	
	\subsection{A-matrices: classical structure factor $\q\neq0$ }\label{sec:structure_factors_classical_A}
	The matrix $\opa{}{\alpha}{\beta}$ is the most fundamental object describing 
	the spins, and its structure factor   
	is defined by 
	\begin{eqnarray}
		S^{\sf CL}_{\rm{A}}(\q) 
		&=& \sum_{\alpha\beta} \nn{\opa{\q}{\alpha}{\beta} \opa{-\q}{\beta}{\alpha}}\; . \label{eq:AA}
	\end{eqnarray}
	We note that the sum on the contracted indices $\alpha, \beta$ preserves 
	the full $U(3)$ symmetry of the representation. 
	The corresponding source term is
	\be
	\Delta\Ham [{\bf{h}}_{i}] 
	= -\sum_{i} h_{i,\beta}^{\alpha} \opa{i}{\alpha}{\beta} \; .
	\label{eq:HamLowTA}
	\ee
	The  components of the  A matrix $\opa{i}{\alpha}{\beta}$ in the function of  the classical fluctuations are given in \Autoref{eq:A_phi}. After expressing them in the function of the fluctuations that diagonalize the BBQ Hamiltonian [\Autoref{eq:orthogonal.transformation}], performing a Fourier transform, and rewriting the total Hamiltonian [\Autoref{eq:Hfield}] according to \Autoref{eq:DeltaH2}, we get 
	\be
	\tilde{C}[{\bf{h}}_{\q}]=\sqrt{N}h^{yy}_{\q}\; ,\label{eq:C_A}
	\ee
	where we neglected 2$\nd$ order terms in fluctuations, since they only contribute for $\q=0$, 
	\be
	\begin{matrix}
		\tilde{\V{N}}_1[{\bf{h}}_{\q}]^T=\begin{pmatrix}
			\frac{i\sqrt{2}}{2}\xi^1_{\q},&
			\frac{i\sqrt{2}}{2}\xi^1_{\q}, &
			-\frac{i\sqrt{2}}{2}\xi^2_{\q}&
			-\frac{i\sqrt{2}}{2}\xi^2_{\q}
		\end{pmatrix}\; , \\
		\tilde{\V{N}}_2[{\bf{h}}_{-\q}]=\begin{pmatrix}
			\frac{i\sqrt{2}}{2}\xi^1_{-\q}\\
			\frac{i\sqrt{2}}{2}\xi^1_{-\q}\\
			-\frac{i\sqrt{2}}{2}\xi^2_{-\q}\\
			-\frac{i\sqrt{2}}{2}\xi^2_{-\q}
		\end{pmatrix}\; ,\label{eq:N_A}
	\end{matrix}
	\ee
	where
	\be
	\begin{matrix}
		\xi^1_{\q}=(h^{xy}_{\q}+h^{yx}_{\q}) & \; ,& \xi^2_{\q}=(h^{yz}_{\q}+h^{zy}_{\q})\; .
	\end{matrix}
	\ee
	
	The total structure factor for A matrices is obtained by computing \Autoref{eq:AA}.
	According to  \Autoref{eq:SOF_class_q}, and \Autoref{eq:chi_gen}, and using \Autoref{eq:N_A},  we obtain
	\begin{align*}
		S^{\sf CL}_{\rm{A}}(\q \neq 0)=&\frac{1}{\beta\omega_{\q,1}}+\frac{1}{\beta\omega_{\q,2}}+\frac{1}{\beta\omega_{\q,3}}+\frac{1}{\beta\omega_{\q,4}}+{\mathcal O}(T^2)\\
		&=\frac{2}{\beta\omega_{\q}^+}+\frac{2}{\beta\omega_{\q}^-}+{\mathcal O}(T^2)\label{eq:AF_class0}\; ,\numberthis
	\end{align*}
	where we used \Autoref{eq:omega_vec_1}.
	It's spectral decomposition is given by
	\begin{align*}
		S^{\sf CL}_{\rm{A}}(\q\neq 0,\omega)=
		&\frac{2}{\beta\omega_{\q}^+}\delta(\omega - \omega_{\q}^+)+\frac{2}{\beta\omega_{\q}^-}\delta(\omega - \omega_{\q}^-)\\
		&+{\mathcal O}(T^2)\label{eq:AF_class1}\; .\numberthis
	\end{align*}
	Again replacing the eigenvalues by their expressions given in \Autoref{eq:omega_vec_1}, we have
	\begin{align*}
		S^{\sf CL}_{\rm{A}}(\q \neq 0,\omega)=&\frac{2}{\beta}\frac{1}{A_{\q}+B_{\q}}\delta(\omega - \omega_{\q}^+)\\
		&+\frac{2}{\beta}\frac{1}{A_{\q}-B_{\q}}\delta(\omega - \omega_{\q}^-)\\
		&+{\mathcal O}(T^2)\; .
		\numberthis\label{eq:AF_class2}
	\end{align*}
	\subsection{A-Matrices: classical structure factor at $\q=0$}\label{sec:structure_factors_classical_Aq0}
	We now consider the structure factor for the A--matrix at the $\Gamma$--point, which is defined as
	\begin{eqnarray}
		S^{\sf CL}_{\rm{A}}(\q=0) 
		&=&  \sum_{\alpha\beta} \nn{\opa{\q=0}{\alpha}{\beta} \opa{\q=0}{\beta}{\alpha}} \; . \label{eq:Aq0}
	\end{eqnarray}
	Again, we follow the procedure depicted in \hyperref[sec:structure_factors_classicalq0]{Appendix~} \ref{sec:structure_factors_classicalq0}. The relevant source term for dipole moments is given by \Autoref{eq:HamLowTA}. We use \Autoref{eq:M} for the BBQ Hamiltonian as well as \Autoref{eq:A_phi} to express \Autoref{eq:HamLowTA} up to second order in terms of the fluctuations. For the total Hamiltonian given in  \Autoref{eq:Hfield}, and written in the form of \Autoref{eq:H.q0}, we obtain
	\be
	M_{\k}[{\bf{h}}_{\q=0}] = 
	\begin{pmatrix}
		A_{\k}-\alpha_1 & -B_{\k} &  0& -\beta_1\\
		-B_{\k} &A_{\k}-\alpha_1  &   -\beta_2& 0 \\
		0 &  - \beta_1& A_{\k}-\alpha_2 &- B_{\k}  \\
		-\beta_2& 0 & -B_{\k} & A_{\k}-\alpha_2  \\
	\end{pmatrix} \; , 
	\label{eq:M_Aq0}
	\ee
	\be
	\begin{matrix}
		\V{N}_1[{\bf{h}}_{\k}]^T=\frac{i}{2}\begin{pmatrix}
			-h^{x}_{\k,y}, &
			h^{y}_{\k,x},&
			h^{y}_{\k,z}, &
			-h^{z}_{\k,y}
		\end{pmatrix}\; , \\
		\V{N}_2[{\bf{h}}_{-\k}]=\frac{-i}{2}\begin{pmatrix}
			-h^{y}_{\k,x}\\
			h^{x}_{\k,y}\\
			h^{z}_{\k,y}\\
			-h^{y}_{\k,z}
		\end{pmatrix}\; ,\label{eq:N_Aq0}
	\end{matrix}
	\ee
	\be
	C[{\bf{h}}_{\q=0}]=-\sqrt{N}h_{\q=0,y}^{y}\; ,\label{eq:C_Aq0}
	\ee
	where we defined
	\be
	\begin{matrix}
		\begin{matrix}
			\alpha_1=\frac{1}{\sqrt{N}}(h_{\q=0,x}^{x}-h_{\q=0,y}^{y})\;, & \alpha_2=\frac{1}{\sqrt{N}}(h_{\q=0,z}^{z}-h_{\q=0,y}^{y})\; ,
		\end{matrix}\\
		&\\
		\begin{matrix}
			\beta_1=\frac{1}{\sqrt{N}}h_{\q=0,x}^{z}\; , &
			\beta_2=\frac{1}{\sqrt{N}}h_{\q=0,z}^{x} \;. 
		\end{matrix}
	\end{matrix}
	\ee
	We diagonalize \Autoref{eq:M_Qq0} to obtain the eigenmodes. We find
	\begin{subequations}
		\begin{align}
			\omega_{\k,1}[{\bf{h}}_{\q=0}]&=A_{\k}-\frac{1}{2}(\alpha_{+}-\Delta^{-})\; , \\
			\omega_{\k,2}[{\bf{h}}_{\q=0}]&=A_{\k}-\frac{1}{2}(\alpha_{+}+\Delta^{-})\; , \\
			\omega_{\k,3}[{\bf{h}}_{\q=0}]&=A_{\k}-\frac{1}{2}(\alpha_{+}-\Delta^{+})\; , \\
			\omega_{\k,4}[{\bf{h}}_{\q=0}]&=A_{\k}-\frac{1}{2}(\alpha_{+}+\Delta^{+})\; , 
		\end{align} \label{eq:omega_Aq0}
	\end{subequations}
	where
	\begin{align*}
		\alpha_{+}&=\alpha_1+\alpha_2\; , \\
		\Delta^{-}&=\sqrt{\alpha_{-}^2+4(B_{\k}^2+\beta_1\beta_2-\sqrt{B_{\k}^2(\alpha_{-}^2+\beta_{-}^2)})}\; , \numberthis\\
		\Delta^{+}&=\sqrt{\alpha_{-}^2 +4(B_{\k}^2+\beta_1\beta_2+\sqrt{B_{\k}^2(\alpha_{-}^2+\beta_{-}^2)})}\; , 
	\end{align*}
	with
	\begin{align*}
		\alpha_{-}&=\alpha_1-\alpha_2\; , \\
		\beta_{-}&=\beta_1-\beta_2\; .\numberthis
	\end{align*}
	
	Finally, we use  \Autoref{eq:OOcorq02} and \Autoref{eq:Ocorq02} to compute the structure factor  for the A--matrix at the $\Gamma$--point given by \Autoref{eq:Aq0}. When calculating  \Autoref{eq:OOcorq02} and \Autoref{eq:Ocorq02}, we use \Autoref{eq:C_Aq0} and \Autoref{eq:omega_Aq0},  and for the last term, we simply invert \Autoref{eq:M_Aq0} and multiply by the vectors in \Autoref{eq:N_Aq0}.
	\begin{align*}
		S^{\sf CL}_{\rm{A}}(\q=0) 
		&=  \frac{2}{\beta} \left[ \frac{1}{\omega^+_0} + \frac{1}{\omega^-_0} \right]\\
		&+ N - \frac{2}{\beta}\sum_{\k} \left[ \frac{1}{\omega^+_\k} + \frac{1}{\omega^-_\k} \right]+{\O}(T^2) \; .\numberthis
		\label{eq:classical.structure.factor.Aq0}
	\end{align*}
	Again, just as for the quadrupole structure factor, we note that at the $\Gamma$--point, $\omega^+_0=0$. Therefore, in order to get rid of confounding divergent terms, we rewrite the structure factor as
	\begin{align*}
		S^{\sf CL}_{\rm{A}}(\q=0) 
		&= N - \frac{2}{\beta}\sum_{\k\neq 0} \left[ \frac{1}{\omega^+_\k} + \frac{1}{\omega^-_\k} \right]+{\mathcal O}(T^2) \; .\numberthis
		\label{eq:classical.structure.factor.Aq02}
	\end{align*}
	Because the $\q=0$ contributions are coming from the ground state and happen for $\omega=0$, the spectral representation of \Autoref{eq:classical.structure.factor.Aq02} yields 
	\begin{eqnarray}
		S^{\sf CL}_{\rm{A}}(\q=0,\omega) \nonumber
		&=& N \delta(\omega) \\
		& & -\frac{2}{\beta}\sum_{\k\neq 0} \left[ \frac{1}{\omega^+_\k} + \frac{1}{\omega^-_\k} \right]\delta(\omega)  +{\mathcal O}(T^2)\; .
		\label{eq:classical.structure.factor.Aq0w}
		\nonumber\\
	\end{eqnarray}
	Combining \Autoref{eq:AF_class0} and \Autoref{eq:classical.structure.factor.Aq02}, we obtain \Autoref{eq:classical.structure.factor.A}. Considering their respective spectral representation given by \Autoref{eq:AF_class1} and \Autoref{eq:classical.structure.factor.Aq0w}, we obtain \Autoref{eq:AF_class}.
	
	\section{Bogolioubov \kim{transformation}}
	\label{sec:bogolioubov_transfomation}
	We here show how the Bogoliubov transformation that we present in \Autoref{section:quantum.theory} is performed.
	
	A Bogolioubov transformation consists in finding new bosons $\opdw{v}{\k}{\alpha}$ and  $\opw{v}{\k}{\alpha}$ expressed in terms of the bosons  $\opdw{w}{\k}{\alpha}$ and  $\opw{w}{\k}{\alpha}$ [\Autoref{eq:def_wboson}] , such that they diagonalize the Hamiltonian
	\be
	\Ham_{\sf BBQ} \sim \sum_{\k}\epsilon_{\k}\opdx{v}{\k}\opx{v}{\k}\; .
	\ee
	Let us assume that the components are given by
	\begin{align*}
		\opw{v}{\k}{\alpha}=&\matU{\k}{\alpha}{\beta}\opw{w}{\k}{\beta}\; ,\\
		\opdw{v}{\k}{\alpha}=&\opdw{w}{\k}{\beta}\matUd{\k}{\alpha}{\beta}\; ,\numberthis\label{eq:U}
	\end{align*}
	where $\matU{\k}{}{}$ is the transformation from basis made out of bosons expressed by time-reversal basis states to the basis in which the Hamiltonian is diagonal.
	Requiring them to have bosonic commutation relations [\Autoref{eq:comrel}], leads to
	\begin{align*}
		\com{\opw{v}{\k}{\alpha},\opdw{v}{\q}{\beta}}&=\com{\matU{\k}{\alpha}{\gamma}\opw{w}{\k}{\gamma},\matUd{\q}{\beta}{\eta}\opdw{w}{\q}{\eta}}\\
		&=\matU{\k}{\alpha}{\gamma}\matUd{\q}{\beta}{\eta}\com{\opw{w}{\k}{\gamma},\opdw{w}{\q}{\eta}}\\
		&=\matU{\k}{\alpha}{\gamma}\matUd{\q}{\beta}{\eta}{\gamma_0}_{\gamma}^{~\eta}\delta_{\k\q}
		\stackrel{!}{=}{\gamma_0}_{\alpha}^{~\beta}\delta_{\k\q}\\
		\Rightarrow ~~~& \matU{\k}{\alpha}{\gamma}{\gamma_0}_{\gamma}^{~\eta}\matUd{\k}{\beta}{\eta}={\gamma_0}_{\alpha}^{~\beta}\\
		\Rightarrow ~~~& {\gamma_0}_{\gamma}^{~\eta}\matUd{\k}{\beta}{\eta}{\gamma_0}_{\beta}^{\alpha}=\matUinv{\k}{\gamma}{\alpha}\; ,\numberthis\label{eq:Uinv}
	\end{align*}
	where ${\gamma_0}$ is defined in \Autoref{eq:gamma_0}, and where we used the fact that 
	\be
	{\gamma_0}={\gamma_0}^{-1}\; .
	\ee
	In the compact form, \Autoref{eq:Uinv} becomes
	\be
	{\gamma_0}\matUd{\k}{}{}{\gamma_0}=\matUinv{\k}{}{}\; .\numberthis\label{eq:Uinv2}
	\ee
	We see that the transformation $\matU{\k}{}{}$ is not unitary, $\matUinv{\k}{}{}\neq\matUd{\k}{}{}$, and that we shall use \Autoref{eq:Uinv2} to find the inverse transformation.
	
	Inverting \Autoref{eq:U} and plugging it into the Hamiltonian leads us to look for a transformation $\matU{\k}{}{}$ such that $\matU{\k}{}{} {\gamma_0} M_{\k} \matUinv{\k}{}{}$ is diagonal. If we define $D_{\k}$ as being a diagonal matrix, we can write
	\begin{align*}
		\matU{\k}{}{}{\gamma_0}_{}^{}{M_{\k}}_{}^{}\matUinv{\k}{}{}&=D_{\k}\\
		\Rightarrow~~~{\gamma_0}_{\nu}^{~\alpha}{M_{\k}}_{\alpha}^{~\beta}\matUinv{\k}{\beta}{i}&=\matUinv{\k}{\nu}{i}{D_{\k}}_{i}^{i} ~~\rm{for}~i=1,2,3,4 \; ,\numberthis\label{eq:Uev}
	\end{align*}
	where we see that $\matUinv{\k}{\nu}{i}$ is an eigenvector of ${\gamma_0}_{}^{}{M_{\k}}_{}^{}$ with eigenvalue ${D_{\k}}_{i}^{i}$. \Autoref{eq:Uev} is rewritten as \Autoref{eq:quantum.eigensystem} in the main text. This means that we need to diagonalize ${\gamma_0} M_{\k}$ and that the corresponding eigenvectors are the column of the matrix $\matUinv{\k}{}{}$.
	
	Finding the Bogolioubov transformation reduces then to find the eigenvalues and eigenvectors of the system in \Autoref{eq:quantum.eigensystem}.
	Since \Autoref{eq:quantum.eigensystem} consists of twice the same system, we only need to solve it once, and we only consider
	\be
	\sigma_z m_{\k}e_i=\epsilon_{\k,i}e_i~~i=1,2\; ,
	\ee
	where
	\be
	\begin{matrix}
		\sigma_z=\begin{pmatrix}
			1 & 0 \\
			0  & -1 \\
		\end{pmatrix} &,~&
		m_{\k}=\begin{pmatrix}
			A_{\k}& -B_{\k} \\
			B_{\k} &-A_{\k}
		\end{pmatrix}
	\end{matrix}\; ,
	\ee
	and where $\sigma_z$ plays the role of $\gamma_0$ but for the two independent subsystems for $\begin{pmatrix}  \ba{-\k}, & \bad{\k}\end{pmatrix}$ and $\begin{pmatrix}\bb{-\k}, & \bbd{\k}\end{pmatrix}$. The eigenvalues $\epsilon_{\k,1/2}$ of $\sigma_z m_{\k}$ are given in \Autoref{eq:omega_qm}.
	The eigenvectors are given by
	\be
	\begin{matrix}
		e_1&=\begin{pmatrix}
			\alpha_1\\
			1
		\end{pmatrix}
		&,~&
		e_2&=\begin{pmatrix}
			\alpha_2\\
			1
		\end{pmatrix}
	\end{matrix}\; ,
	\ee
	in the basis $\{\ba{\k}, \bad{-\k}\}$ and where we define
	\be
	\alpha_i= -\frac{A_{\k}+\epsilon_{\k,i}}{B_{\k}}\; .
	\ee
	The columns of the matrix $\matUinv{\k}{}{}$ are given by the eigenvectors
	\be
	\matUinv{\k}{}{}=\begin{pmatrix}
		\alpha_1 & \alpha_2\\
		1 & 1
	\end{pmatrix}\; .
	\ee
	Using \Autoref{eq:Uinv2}, we can calculate $\matU{\k}{}{}$  as follows:
	\begin{align*}
		\matU{\k}{}{}&=\sigma_z\matUdinv{\k}{}{}\sigma_z=
		\begin{pmatrix}
			\alpha_1 & -1\\
			-\alpha_2 & 1
		\end{pmatrix}\; .\numberthis
	\end{align*}
	Using \Autoref{eq:U}, the new bosons that diagonalize the Hamiltonian are given  by
	\begin{subequations}
		\begin{align}
			\balpha{\k}=\opw{v}{\k}{1}&=\matU{\k}{1}{1}\opw{w}{\k}{1}+\matU{\k}{1}{2}\opw{w}{\k}{2}=\alpha_1\ba{\k} -\bad{-\k} \; ,\\
			\balphad{-\k}=\opw{v}{\k}{2}&=\matU{\k}{2}{1}\opw{w}{\k}{1}+\matU{\k}{2}{2}\opw{w}{\k}{2}=-\alpha_2\ba{\k} +\bad{-\k} \; ,\\
			\balphad{\k}=\opdw{v}{\k}{1}&=\opdw{w}{\k}{1}\matUd{\k}{1}{1}+\opdw{w}{\k}{2}\matUd{\k}{1}{2}=\alpha_1\bad{\k} -\ba{-\k} \; ,\\
			\balpha{-\k}=\opdw{v}{\k}{2}&=\opdw{w}{\k}{1}\matUd{\k}{2}{1}+\opdw{w}{\k}{2}\matUd{\k}{2}{2}=-\alpha_2\bad{\k} +\ba{-\k}\; .
			\numberthis\label{eq:balpha}
		\end{align}
	\end{subequations}
	For instance, we note that we should have $\opw{v}{\k}{1}=\opdw{v}{-\k}{2}$ i.e $\balpha{\k}(\k)=\balpha{-\k}(-\k)$. However, we see that it is not the case
	\be
	\opw{v}{\k}{1}=\alpha_1\ba{\k} -\bad{-\k}\neq -\alpha_2\bad{-\k} +\ba{\k}=\opdw{v}{-\k}{2}\; .
	\ee
	For it to be the case, we see that we need the matrix element of the transformation to be 
	\begin{subequations}
		\begin{align}
			\matU{\k}{1}{1}=\matUd{-\k}{2}{2}\; , \label{eq:Uequality1} \\
			\matU{\k}{1}{2}=\matUd{-\k}{2}{1}\; . \label{eq:Uequality2}
		\end{align}\label{eq:Uequality}
	\end{subequations}
	
	To solve this issue, we can assume that we can multiply the eigenvectors by some parameters, ${a}$ and  ${b}$ for instance, such that \Autoref{eq:Uequality} is satisfied
	\be
	\begin{matrix}
		e_1&= a \begin{pmatrix}
			\alpha_1\\
			1
		\end{pmatrix}
		&,~&
		e_2&= b\begin{pmatrix}
			\alpha_2\\
			1
		\end{pmatrix}
	\end{matrix}\; .
	\ee
	And  $\matUinv{\k}{}{}$ is given by 
	\be
	\matUinv{\k}{}{}=\begin{pmatrix}
		a \alpha_1 & b \alpha_2\\
		a & b
	\end{pmatrix}\; .
	\ee
	Using \Autoref{eq:Uinv2}, we can calculate $\matU{\k}{}{}$ as follows:
	\begin{align}
		\matU{\k}{}{}&=\sigma_z\matUdinv{\k}{}{}\sigma_z=
		\begin{pmatrix}
			a \alpha_1 & -a\\
			-b \alpha_2 & b
		\end{pmatrix}\; .
	\end{align}
	We also have 
	\begin{align}
		\matUd{\k}{}{}&=
		\begin{pmatrix}
			a \alpha_1 & -b \alpha_2 \\
			-a & b
		\end{pmatrix}\; .
	\end{align}
	Note that the coefficients $\alpha_1$ and $\alpha_2$ depend on $\k$ through $\epsilon_{\k}$. However we have $\epsilon_{\k}=\epsilon_{-\k}$ and the dependency in $\k$ for $\alpha_1$ and $\alpha_2$ has been dropped.
	%
	\Autoref{eq:Uequality} implies then 
	\begin{subequations}
		\be
		\begin{matrix}
			\matU{\k}{1}{1}&=&\matUd{-\k}{2}{2}\; ,\\
			a \alpha_1 &=&b \; ,\\
			a\frac{-A_{\k} - \sqrt{A_{\k}^2-B_{\k}^2}}{B_{\k}}&=&b\; ,\\
			\frac{a}{-B_{\k}}&=&\frac{b}{A_{\k} +\sqrt{A_{\k}^2-B_{\k}^2}}\; ,
		\end{matrix}\label{eq:U11U22}
		\ee
		and
		\be
		\begin{matrix}
			\matU{\k}{1}{2}&=&\matUd{-\k}{2}{1}\; ,\\
			-b \alpha_2 &=& -a\; , \\
			-b \frac{-A_{\k} +\sqrt{A_{\k}^2-B_{\k}^2}}{B_{\k}}&=& -a\; ,\\
			\frac{b}{-B_{\k}}&=&\frac{-a}{-A_{\k} +\sqrt{A_{\k}^2-B_{\k}^2}}\; .
		\end{matrix}\label{eq:U12U21}
		\ee
	\end{subequations}
	We see that if we multiply the last line of \Autoref{eq:U12U21}  by  $\frac{-B_{\k}}{A_{\k} +\sqrt{A_{\k}^2-B_{\k}^2}}$ we get
	\begin{subequations}
		\begin{align*}
			\frac{-B_{\k}}{A_{\k} +\sqrt{A_{\k}^2-B_{\k}^2}}&\frac{b}{-B_{\k}}\\
			=&\frac{-B_{\k}}{A_{\k} +\sqrt{A_{\k}^2-B_{\k}^2}}\frac{-a}{-A_{\k} +\sqrt{A_{\k}^2-B_{\k}^2}}\; ,\numberthis\\
			\Rightarrow	&\frac{b}{A_{\k} +\sqrt{A_{\k}^2-B_{\k}^2}}=\frac{a}{-B_{\k}}\numberthis\; .\label{eq:ab_cond}
		\end{align*}
	\end{subequations}
	We note that \Autoref{eq:ab_cond} is exactly the same condition as in the last line of \Autoref{eq:U11U22}. This makes sense, because the 1\st condition, namely $\matU{\k}{1}{1}=\matUd{-\k}{2}{2}$ is correlated the the second one $\matU{\k}{1}{2}=\matUd{-\k}{2}{1}$, as the components $\matU{\k}{1}{1},~\matU{\k}{1}{2}$ are not independent, as they need to be eigenvectors,  and nor are the components $\matUd{-\k}{2}{2},~\matUd{-\k}{2}{1}$. We could also choose the 2$\nd$ solution, as up to scalar multiplication, it gives the same eigenvectors. By normalizing the eigenvectors, we then get rid of this.
	This means, for instance, that we can choose 
	\begin{align}
		a&=-B_{\k}\; ,\\
		b &=A_{\k} +\sqrt{A_{\k}^2-B_{\k}^2} \; .
	\end{align}
	In this case, the eigenvectors become 
	\be
	\begin{matrix}
		e_1&= \begin{pmatrix}
			\Delta_{\k}\\
			-B_{\k}
		\end{pmatrix}
		&,~&
		e_2&=\begin{pmatrix}
			-B_{\k}\\
			\Delta_{\k}
		\end{pmatrix}
	\end{matrix}\; ,
	\ee
	where  $\Delta_{\k}$ is given \Autoref{eq:Delta}.
	And the transformation matrix becomes
	\be
	\matUinv{\k}{}{}=\begin{pmatrix}
		\Delta_{\k} & -B_{\k}\\
		-B_{\k} & \Delta_{\k}
	\end{pmatrix}\label{eq:Uinv3}\; .
	\ee
	Using \Autoref{eq:Uinv2}, we can calculate $\matU{\k}{}{}$ as follows:
	\begin{align}
		\matU{\k}{}{}&=\sigma_z\matUdinv{\k}{}{}\sigma_z=
		\begin{pmatrix}
			\Delta_{\k} & B_{\k}\\
			B_{\k} & \Delta_{\k}
		\end{pmatrix}\; .
	\end{align}
	We also have 
	\begin{align}
		\matUd{\k}{}{}&=
		\begin{pmatrix}
			\Delta_{\k} & B_{\k}\\
			B_{\k} & \Delta_{\k}
		\end{pmatrix}\; .
	\end{align}
	
	Using \Autoref{eq:U}, the new bosons that diagonalize the Hamiltonian are given by
	\begin{subequations}
		\begin{align}
			\balpha{\k}=\opw{v}{\k}{1}=&\matU{\k}{1}{1}\opw{w}{\k}{1}+\matU{\k}{1}{2}\opw{w}{\k}{2}=\Delta_{\k}\ba{\k} + B_{\k}\bad{-\k} \; ,\\
			\balphad{-\k}=\opw{v}{\k}{2}=&\matU{\k}{2}{1}\opw{w}{\k}{1}+\matU{\k}{2}{2}\opw{w}{\k}{2}=B_{\k}\ba{\k} +\Delta_{\k}\bad{-\k} \; ,\\
			\balphad{\k}=\opdw{v}{\k}{1}=&\opdw{w}{\k}{1}\matUd{\k}{1}{1}+\opdw{w}{\k}{2}\matUd{\k}{1}{2}=\Delta_{\k}\bad{\k} +B_{\k} \ba{-\k} \; ,\\
			\balpha{-\k}=\opdw{v}{\k}{2}=&\opdw{w}{\k}{1}\matUd{\k}{2}{1}+\opdw{w}{\k}{2}\matUd{\k}{2}{2}= B_{\k}\bad{\k} +\Delta_{\k}\ba{-\k}\; .
			\numberthis\label{eq:balpha2}
		\end{align}
	\end{subequations}
	We see that now we indeed have $\opw{v}{\k}{1}=\opdw{v}{-\k}{2}$, i.e., $\balpha{\k}(\k)=\balpha{-\k}(-\k)=\balpha{\k}$.
	However, we still need to normalize the new bosons. Indeed they should also satisfy bosonic commutation relations
	\begin{subequations}
		\begin{align*}
			\com{\balpha{\k},\balphad{\k}}&=\com{\frac{1}{\sqrt{N}}(\Delta_{\k}\ba{\k} + B_{\k}\bad{-\k}),\frac{1}{\sqrt{N}}(\Delta_{\k}\bad{\k} +B_{\k} \ba{-\k})}\\
			&=\frac{1}{N}( \Delta_{\k}^2\com{\ba{\k},\bad{\k}} + B_{\k}^2\com{\bad{-\k},\ba{-\k}})\\
			&=\frac{1}{N}( \Delta_{\k}^2- B_{\k}^2)\overset{!}{=}1\numberthis\\
			\Rightarrow ~~~N&= \Delta_{\k}^2- B_{\k}^2\; .\numberthis
		\end{align*}
	\end{subequations}
	Finally, the transformation matrix becomes 
	\be
	\matU{\k}{}{}=\frac{1}{\sqrt{\Delta_{\k}^2- B_{\k}^2}}
	\begin{pmatrix}
		\Delta_{\k} & B_{\k}\\
		B_{\k} & \Delta_{\k}
	\end{pmatrix}\; .
	\ee
	The inverse [\Autoref{eq:Uinv2}] holds
	\be
	\matUinv{\k}{}{}=\sigma_z\matUd{\k}{}{}\sigma_z=\frac{1}{\sqrt{\Delta_{\k}^2- B_{\k}^2}}
	\begin{pmatrix}
		\Delta_{\k} & -B_{\k}\\
		-B_{\k} & \Delta_{\k}
	\end{pmatrix}\; . \label{eq:Uinv4}
	\ee
	By inverting the Bogolibov transformation [\Autoref{eq:U}], we can express the old bosons in terms  of the new Bogolibov bosons using \Autoref{eq:Uinv4},
	\begin{subequations}
		\begin{align}
			\ba{\k}{}=\opw{w}{\k}{1}=&\matUinv{\k}{1}{\beta}\opw{v}{\k}{\beta}=\frac{1}{\sqrt{\Delta_{\k}^2- B_{\k}^2}}(\Delta_{\k}\balpha{\k}-B_{\k}\balphad{-\k})\label{eq:v3_1}\: ,\\
			\bad{-\k}{}=\opw{w}{\k}{2}=&\matUinv{\k}{2}{\beta}\opw{v}{\k}{\beta}=\frac{1}{\sqrt{\Delta_{\k}^2- B_{\k}^2}}(-B_{\k}\balpha{\k}+\Delta_{\k}\balphad{-\k})\; ,\\
			\bad{\k}{}=\opdw{w}{\k}{1}=&\opdw{v}{\k}{\beta}\matUdinv{\k}{\beta}{1}=\frac{1}{\sqrt{\Delta_{\k}^2- B_{\k}^2}}(\Delta_{\k}\balphad{\k}-B_{\k}\balpha{-\k})\; ,\\
			\ba{-\k}{}=\opdw{w}{\k}{2}=&\opdw{v}{\k}{\beta}\matUdinv{\k}{\beta}{1}=\frac{1}{\sqrt{\Delta_{\k}^2- B_{\k}^2}}(-B_{\k}\balphad{\k}+\Delta_{\k}\balpha{-\k})\; . \numberthis\label{eq:v3_4}
		\end{align}\label{eq:v3}
	\end{subequations}
	
	For the other part of the Hamiltonian containing the $\bbd{}$ bosons, the problem is exactly the same, and therefore, we can just use the solutions we found above.
	The eigenvalues $\epsilon_{\k,3}$ and $\epsilon_{\k,4}$ associated to the Bogolibov bosons for the $\begin{pmatrix}\bb{-\k}, & \bbd{\k}\end{pmatrix}$ subsystem are given by \Autoref{eq:omega_qm}.
	To express the old bosons in terms  of the new Bogolibov bosons, we can just use \Autoref{eq:v3}: 
	\begin{subequations}
		\begin{align}
			\bb{\k}{}&=\frac{1}{\sqrt{\Delta_{\k}^2- B_{\k}^2}}(\Delta_{\k}\bbeta{\k}-B_{\k}\bbetad{-\k})\; ,\\
			\bbd{-\k}{}&=\frac{1}{\sqrt{\Delta_{\k}^2- B_{\k}^2}}(-B_{\k}\bbeta{\k}+\Delta_{\k}\bbetad{-\k})\; ,\\
			\bbd{\k}{}&=\frac{1}{\sqrt{\Delta_{\k}^2- B_{\k}^2}}(\Delta_{\k}\bbetad{\k}-B_{\k}\bbeta{-\k})\; ,\\
			\bb{-\k}{}&=\frac{1}{\sqrt{\Delta_{\k}^2- B_{\k}^2}}(-B_{\k}\bbetad{\k}+\Delta_{\k}\bbeta{-\k}) \; .
		\end{align}\label{eq:v3beta}
	\end{subequations}
	
	There is a constant term coming from the Bogoliubov transformation [\Autoref{eq:v3} and \Autoref{eq:v3beta}].
	\begin{subequations}
		For the bosons $\bad{\pm\k}{},\ba{\pm\k}{}$ it holds:
		\be
		\begin{matrix}
			-A_{\k}+\sqrt{A_{\k}^2-B_{\k}^2}=- A_{\k}+\epsilon_{\k,1}\; ,
		\end{matrix}
		\ee
		and for the bosons $\bbd{\pm\k}{},\bb{\pm\k}{}$:
		\be
		\begin{matrix}
			-A_{\k}+\sqrt{A_{\k}^2-B_{\k}^2}=- A_{\k}+\epsilon_{\k,3}\; ,
		\end{matrix}
		\ee
	\end{subequations}
	where we use \Autoref{eq:omega_qm}.
	
	After performing the Bogolibov transformation, the Hamiltonian becomes
	\begin{align*}
		\Ham=&E_0+\frac{1}{2}\left[\sum_{\k}\epsilon_{\k,1}\balphad{\k}\balpha{\k}+\epsilon_{\k,1}\balphad{-\k}\balpha{-\k}- A_{\k}+\epsilon_{\k,1} \right]\\
		&+\left.\epsilon_{\k,3}\bbetad{\k}\bbeta{\k}+\epsilon_{\k,3}\bbetad{-\k}\bbeta{-\k}- A_{\k}+\epsilon_{\k,3}\right]\; ,\numberthis\label{eq:HamBogo}
	\end{align*}
	that we can rewrite as 
	\begin{align*}
		\Ham=&E_0+\left[\sum_{\k}\epsilon_{\k,1}(\balphad{\k}\balpha{\k}+\frac{1}{2})+\epsilon_{\k,3}(\bbetad{\k}\bbeta{\k}+\frac{1}{2})- A_{\k}\right]\; ,\numberthis\label{eq:HamBogo2}
	\end{align*}
	where $E_0$ is given in \Autoref{eq:E0}. \Autoref{eq:HamBogo2} is given in the main text by \Autoref{eq:H.quantum}, where we used the fact that \mbox{$\epsilon_{\k,1}=\epsilon_{\k,3}$} [\Autoref{eq:omega_qm}], since they are the eigenvalues of an identical problem.
	%
	\section{Dynamical structure factors within zero–temperature quantum theory}\label{sec:structure_factors_quantum}
	In this Appendix, we present the outline of the method used to calculate the zero--temperature quantum structure factors in \Autoref{sec:quantum.structure.factors}.
	
	In  \hyperref[sec:structure_factors_quantum_qn0]{Appendix~}\ref{sec:structure_factors_quantum_qn0}, we first present how to calculate dynamical structure factors at finite energy through the explicit calculation of matrix elements within a multiple–Boson expansion, and its application to dipole [\hyperref[sec:structure_factors_quantum_Sq]{Appendix~}\ref{sec:structure_factors_quantum_Sq}], quadrupole [\hyperref[sec:structure_factors_quantum_Qq]{Appendix~}\ref{sec:structure_factors_quantum_Qq}], and A--matrix moments [\hyperref[sec:structure_factors_quantum_Aq]{Appendix~}\ref{sec:structure_factors_quantum_Aq}].
	
	In \hyperref[sec:structure_factors_quantum_q0]{Appendix~}\ref{sec:structure_factors_quantum_q0}, we explain how the calculation for the static structure factors ($\omega = 0$) can also be computed through functional derivatives of the ground–state energy, in order to account for the ground--state and zero--point energy contribution at $\q=0$. We show calculations for the  dipole [\hyperref[sec:structure_factors_quantum_Sq0]{Appendix~}\ref{sec:structure_factors_quantum_Sq0}], quadrupole [\hyperref[sec:structure_factors_quantum_Qq]{Appendix~}\ref{sec:structure_factors_quantum_Qq0}], and A--matrix moments [\hyperref[sec:structure_factors_quantum_Aq0]{Appendix~}\ref{sec:structure_factors_quantum_Aq0}].
	
	\subsection{ Quantum structure factors at general values of $\q$}\label{sec:structure_factors_quantum_qn0}
	The definition of the structure factor is given by \Autoref{eq:SqOF} and its components by
	\be
	S^{\sf QM}_{\rm{O}}(\q, \omega)^{\alpha\mu}_{\beta\nu} 
	= \int_{-\infty}^{\infty} \frac{\dd t}{2\pi} \exp{i\omega t} 
	\Av{ \hat{O}_{\q,\beta}^{\alpha}(t) \hat{O}_{-\q,\nu}^{\mu}(0) }\; .
	\label{eq:SqOF_comp}
	\ee
	where in our case, the averages $\Av{ \hat{O}_{\q,\beta}^{\alpha}(t) \hat{O}_{-\q,\nu}^{\mu}(0) }$ are taken on the ground state.
	We can rewrite the time dependency of $\hat{O}_{\q,\beta}^{\alpha}(t)$ in the Heisenberg picture using the time evolution operator, and we obtain
	\be
	\hat{O}_{\q,\beta}^{\alpha}(t)=\exp{\frac{i\hat{\Ham}t}{\hbar}}\hat{O}_{\q,\beta}^{\alpha}(0)\exp{\frac{-i\hat{\Ham}t}{\hbar}}\; .\label{eq:St}
	\ee
	For a complete basis $\{\ket{\nu}\}$ of Hilbert space,  the closure relation holds
	\be
	\sum_{\nu}\ket{\nu}\bra{\nu}=1\; .\label{eq:close}
	\ee
	
	%
	By using \Autoref{eq:St} and inserting the closure relation, \Autoref{eq:close} twice in \Autoref{eq:SqOF_comp}, we get 
	\begin{align*}
		&S^{\alpha\mu}_{\rm{O},\beta\mu}(\q, \omega)\\
		&=\int_{-\infty}^{\infty}\frac{\dd t}{2\pi}\exp{i\omega t}\langle\exp{\frac{i\hat{\Ham}t}{\hbar}}\sum_{\nu}\ket{\nu}\bra{\nu}\hat{O}_{\q,\beta}^{\alpha}(0)\sum_{\mu}\ket{\mu}\\
		&\times\bra{\mu}\exp{\frac{-i\hat{\Ham}t}{\hbar}}\hat{O}_{-\q,\nu}^{\mu}(0)\rangle +S^{\sf GS}_{\rm{O}}(\q=0, \omega)\\
		&=\sum_{\mu}\bra{0}\hat{O}_{\q,\beta}^{\alpha}(0)\ket{\mu}\bra{\mu}\hat{O}_{-\q,\nu}^{\mu}(0)\ket{0}\delta(\omega-\epsilon_{\mu}) \\
		& +S^{\sf GS}_{\rm{O}}(\q=0, \omega)\; ,\label{eq:SO_qn0} \numberthis 
	\end{align*}
	where we assumed that $\ket{\nu}$ is an eigenstate of the Hamiltonian of energy $E_{\nu}=\hbar\epsilon_{\nu}$ and  used 
	\begin{align*}
		\exp{\frac{i\hat{\Ham}t}{\hbar}}\ket{\nu}&=\sum_{n=0}^{\infty}\frac{(i\hat{\Ham}t)^n}{n\!}\ket{\nu}=
		\sum_{n=0}^{\infty}\frac{(iE_{\nu}t)^n}{n\!}\ket{\nu}\\
		&=\exp{\frac{iE_{\nu}t}{\hbar}}\ket{\nu}=\exp{i\epsilon_{\nu}t}\ket{\nu}\; ,
		\numberthis
	\end{align*}
	and where  $S^{\sf GS}_{\rm{O}}(\q=0, \omega)$ represents the ground state and zero-point energy contribution to the structure factor, as explained below.
	
	In order to compute \Autoref{eq:SO_qn0}, we first note that, in our case, the excited states, for all values of $\k$ 
	\begin{align}
		\ket{\alpha_{\k}}=\balphad{\k}\ket{0_{\alpha}}\; ,\\
		\ket{\beta_{\k}}=\bbetad{\k}\ket{0_{\beta}}\; ,
	\end{align}  
	form a complete basis, where $\ket{0_{\alpha}}$ is the Bogoliubov ground state for the $\balpha{}$ bosons, i.e., $\balpha{\k}\ket{0_{\alpha}}=0$, and similarly for the $\bbeta{}$ bosons. Since the Hilbert space consists of the direct product $\ket{{\alpha}}\oplus\ket{{\beta}}$,we can replace
	\be
	\sum_{\mu}\ket{\mu}\bra{\mu}\rightarrow\sum_{\k}\balphad{\k}\ket{0_{\alpha}}\bra{0_{\alpha}}\balpha{\k} +\sum_{\k}\bbetad{\k}\ket{0_{\beta}}\bra{0_{\beta}}\bbeta{\k}=1\; .\label{eq:ab.close}
	\ee
	in  \Autoref{eq:SO_qn0}.
	However, by replacing \Autoref{eq:close} by \Autoref{eq:ab.close} in \Autoref{eq:SO_qn0}, we account for the 1$\st$ excited states and we therefore disregard the ground state and zero-point energy contribution to the structure factor, which is expressed by the term $S^{\sf GS}_{\rm{O}}(\q=0, \omega)$ in \Autoref{eq:SO_qn0}. The ground state and zero-point energy only contribute at $\q=0$ and $\omega=0$. We present how to calculate it below in \hyperref[sec:structure_factors_quantum_q0]{Appendix~}\ref{sec:structure_factors_quantum_q0}.
	
	\subsection{Dipole moments: quantum structure factor at general values of $\q$}\label{sec:structure_factors_quantum_Sq}
	
	We consider first the dynamical spin structure factor 
	\be
	S^{\sf QM}_{\rm{S}} (\q, \omega) 
	= \int_{-\infty}^{\infty} \frac{\dd t}{2\pi} \exp{i\omega t} 
	\sum_\mu \Av{ \ops{\q}{\mu}(t) \ops{-\q}{\mu}(0) } \; .
	\label{eq:SqSF_app}
	\ee
	Substituting \Autoref{eq:Acond2} in the expression for spin 
	operators, \Autoref{eq:dipole.in.terms.of.A}, and keeping terms to linear order, we find 
	\begin{subequations}
		\begin{align}
			\ops{i}{x}&\simeq i(\bbd{i}-\bb{i})\; ,\\
			\ops{i}{y}&\simeq 0\; ,\\
			\ops{i}{z}&\simeq-i(\bad{i}-\ba{i})\; .
		\end{align}
	\end{subequations}
	Performing a Fourier transform and using the Bogoliubov transformation \Autoref{eq:v_e}, 
	we can express these as 
	\begin{subequations}
		\begin{align}
			\ops{\q}{x}& 
			\simeq i \xi_{\sf{S}}(\q) (\bbetad{-\q}-\bbeta{\q}) \label{eq:Sqm_alpha_x_app} \; , \\
			\ops{\q}{y}&\simeq 0\; ,\\
			\ops{\q}{z}& 
			\simeq- i \xi_{\sf{S}}(\q) (\balphad{-\q}-\balpha{\q})\label{eq:Sqm_alpha_z_app} \; , 
		\end{align}
	\end{subequations}
	where $\xi(\q)$ is the coherence factor
	\be
	\xi_{\sf{S}}(\q) = \frac{\Delta_{\q}+B_{\q}}{\sqrt{\Delta_{\q}^2- B_{\q}^2}}\; .
	\label{eq:coherence.factor_app}
	\ee
	Using \Autoref{eq:SO_qn0}, we can then calculate the structure factor for dipole moments as 
	\begin{align}
		S^{\sf QM}_{\rm{S}} (\q, \omega) 
		&= \sum_{\mu,{\k}}  \norm{ \bra{n_{\k}} \ops{\q}{\mu} \ket{0} }^2
		\delta(\omega-\omega_{n_{\k}})+S^{\sf GS}_{\rm{S}}(\q=0, \omega)\; ,\label{eq:SqSFtot}
	\end{align}
	where $\ket{0}$ is the FQ ground state [\Autoref{eq:FQyGS}], 
	and 
	\be
	\ket{n_{\k}} = \balphad{\k}\ket{0} \otimes \bbetad{\k}\ket{0} \; .\label{eq:nk}
	\ee
	represents the first excited states where $ \otimes$ implies a direct product, as the bosons $\balphad{\q}$ and $\bbetad{\q}$ are independent. By using \Autoref{eq:nk}, we account for the 1$\st$ excited states and we therefore disregard the ground state and zero-point energy contribution to the structure factor which is expressed by the term $S^{\sf GS}_{\rm{O}}(\q=0, \omega)$ in \Autoref{eq:SqSFtot}.

	Finally, we find  
	\be
	S^{\sf QM}_{\rm{S}} (\q, \omega) 
	= 2\frac{\sqrt{A_{\q}+B_{\q}}}{\sqrt{A_{\q}-B_{\q}}} \delta(\omega-\omega_{\q}) +S^{\sf GS}_{\rm{S}}(\q=0, \omega) \; ,
	\label{eq:Sqw_qm_qn0_app}
	\ee
	where we used \Autoref{eq:Delta}.
	Detailed calculations for $\q=0$ contributions to the dipole moment structure factor can be found in \hyperref[sec:structure_factors_quantum_Sq0]{Appendix~}\ref{sec:structure_factors_quantum_Sq0}.
	More precisely, $S^{\sf GS}_{\rm{S}}(\q=0, \omega)$ is given by  \Autoref{eq:SSq0_QM}, which combined with \Autoref{eq:Sqw_qm_qn0_app}  gives the total quantum structure factor for the dipole moments expressed in \Autoref{eq:Sqw_qm}.
	
	\subsection{Quadrupole moments: quantum structure factor at general values of $\q$}\label{sec:structure_factors_quantum_Qq}
	We now consider the dynamical structure factor associated with 
	quadrupole moments 
	\be
	S^{\sf QM}_{\rm{Q}} (\q, \omega) 
	= \int_{-\infty}^{\infty} \frac{\dd t}{2\pi} \exp{i\omega t} 
	\sum_{\mu\nu}  \Av{ \opq{\q}{\mu\nu}(t) \opq{-\q}{\mu\nu}(0) } \; .
	\label{eq:dynamical.quadrupole.structure.factor_app}
	\ee
	Following the same steps as for the spin--structure factor, 
	we use \Autoref{eq:Acond2} to express the quadupole components up to linear order 
	in \Autoref{eq:quadrupole.in.terms.of.A}. We find
	\begin{align}
		\ophb{Q}_i & \cong
		\begin{pmatrix}
			\frac{2}{3} & -\bad{i}-\ba{i} & 0\\
			-\bad{i}-\ba{i} & -\frac{4}{3} & -\bbd{i}-\bb{i} \\
			0 &  -\bbd{i}-\bb{i}  & \frac{2}{3}
		\end{pmatrix} \; ,
		\label{eq:Q_bosons_app}
	\end{align}
	An equivalent calculation of matrix elements in the Bogoliubov basis [\Autoref{eq:v_e}]
	yields
	\begin{align}
		\ophb{Q}_{\q}&\cong\nonumber\\
		&	\small{
			\begin{pmatrix}
				\frac{2}{3}\sqrt{N}\delta(\q)  &	\xi_{\sf{Q}}(\q)(\balphad{-\q}+\balpha{\q}) & 0\\
				\xi_{\sf{Q}}(\q)(\balphad{-\q}+\balpha{\q})& 	-\frac{4}{3}\sqrt{N}\delta(\q)& 		\xi_{\sf{Q}}(\q)(\bbetad{-\q}+\bbeta{\q})\\
				0 &  	\xi_{\sf{Q}}(\q)(\bbetad{-\q}+\bbeta{\q}) & 	\frac{2}{3}\sqrt{N}\delta(\q)
		\end{pmatrix}}\; ,
	\end{align}
	where $N$ is the number of sites and where $\xi_{\sf{Q}}(\q)$ is the coherence factor for quadrupoles defined as
	\be
	\xi_{\sf{Q}}(\q)= \frac{B_{\q}-\Delta_{\q}}{\sqrt{\Delta_{\q}^2- B_{\q}^2}}\; .\label{eq:coherence.factor.quadrupole_app}
	\ee

	Using \Autoref{eq:SO_qn0}, we can then calculate the structure factor for quadrupole moments as defined in \Autoref{eq:dynamical.quadrupole.structure.factor_app}. We obtain
	\begin{align}
		S^{\sf QM}_{\rm{Q}}(\q, \omega)
		=& 4 \frac{\sqrt{A_{\q}-B_{\q}}}{\sqrt{A_{\q}+B_{\q}}} \delta(\omega-\omega_{\q}) +S^{\sf GS}_{\rm{Q}}(\q=0, \omega) \; ,
		\label{eq:Qqw_qm_qn0}
	\end{align} 
	where we used \Autoref{eq:Delta}.
	Detailed calculations for $\q=0$ contributions to the quadrupole moment structure factor can be found in \hyperref[sec:structure_factors_quantum_Qq0]{Appendix~}\ref{sec:structure_factors_quantum_Qq0}.
	More precisely, $S^{\sf GS}_{\rm{Q}}(\q=0, \omega)$ is given by  \Autoref{eq:SQq0_QM}, which combined with \Autoref{eq:Qqw_qm_qn0}  gives the total quantum structure factor for the quadrupole moments expressed in \Autoref{eq:Qqw_qm}. 
	%
	\subsection{A-matrices : quantum structure factors at general values of $\q$}\label{sec:structure_factors_quantum_Aq}
	
	The most fundamental objects in our theory are not dipoles or quadrupoles, 
	but the A--matrices which describe the quantum state of the \mbox{spin--1} moment.
	It is therefore useful to introduce a dynamical structure factor 
	\be
	S^{\sf QM}_{\rm{A}} (\q, \omega) 
	= \int_{-\infty}^{\infty} \frac{\dd t}{2\pi} \exp{i\omega t} 
	\sum_{\mu\nu} \nn{\opa{}{\mu}{\nu}(t) \opa{}{\nu}{\mu}(0)}  \; .
	\label{eq:dynamical.A.matrix.structure.factor_app}
	\ee
	Neglecting 2\nd order and higher terms, \Autoref{eq:Acond2} becomes
	\be
	\ophb{A}_i\simeq
	\begin{pmatrix}
		0& \bad{i} & 0\\
		\ba{i} & 1 & \bb{i}\\
		0 & \bbd{i} & 0
	\end{pmatrix}\; .
	\ee                 
	Once again we can use the Bogoliubov basis [\Autoref{eq:v_e}] to find
	\be
	\ophb{A}_{\q}\simeq
	\begin{pmatrix}
		\begin{array}{c c c c c c}
			\multicolumn{2}{c}{\multirow{2}{*}[-1.5ex]{0}}& \multirow{2}{*}{{\small$\xi_{\sf{A}}^{-}(\q)\balphad{-\q}$}}& \multicolumn{2}{c}{\multirow{2}{*}[-1.5ex]{0}}\\
			& & \multirow{2}{*}{{\small$-\xi_{\sf{A}}^{+}(\q)\balpha{\q} $}}& & \\
			\multirow{2}{*}{{\small$-\xi_{\sf{A}}^{+}(\q)\balphad{-\q}$ }}& \multicolumn{2}{c}{\multirow{2}{*}[-1.5ex]{$\sqrt{N}\delta_{\q, 0}$ }}&	\multirow{2}{*}{{\small$-\xi_{\sf{A}}^{+}(\q)\bbetad{-\q}$}}\\
			\multirow{2}{*}{{\small$+\xi_{\sf{A}}^{-}(\q)\balpha{\q}$}} & & &\multirow{2}{*}{{\small$+\xi_{\sf{A}}^{-}(\q)\bbeta{\q} $}}\\
			\multicolumn{2}{c}{\multirow{2}{*}[-1.5ex]{0}}& \multirow{2}{*}{{\small$\xi_{\sf{A}}^{-}(\q)\bbetad{-\q}$}}& \multicolumn{2}{c}{\multirow{2}{*}[-1.5ex]{0}}\\
			& & \multirow{2}{*}{{\small$-\xi_{\sf{A}}^{+}(\q)\bbeta{\q}$}}& & 
		\end{array}
	\end{pmatrix}
	\ee
	where $N$ is the number of sites and $\xi_{\sf{A}}^{+}(\q)$ and $\xi_{\sf{A}}^{-}(\q)$ are the coherence factors for A--matrices defined as
	\begin{subequations}
		\begin{align}
			\xi_{\sf{A}}^{+}(\q)= \frac{\xi_{\sf{S}}(\q)+\xi_{\sf{Q}}(\q)}{2}\; ,\\
			\xi_{\sf{A}}^{-}(\q)=\frac{\xi_{\sf{S}}(\q)-\xi_{\sf{Q}}(\q)}{2}\; ,
		\end{align}\label{eq:coherence.factor.amatrix_app}
	\end{subequations}
	where $\xi_{\sf{S}}(\q)$ and $\xi_{\sf{Q}}(\q)$ are defined in \Autoref{eq:coherence.factor_app} and \Autoref{eq:coherence.factor.quadrupole_app} respectively.

	Using \Autoref{eq:SO_qn0}, we can then calculate the structure factor for quadrupole moments as defined in \Autoref{eq:dynamical.A.matrix.structure.factor_app}. We obtain
	\begin{align}
		S_{\rm{A}}^{\sf QM}(\q, \omega) 
		=& 2\frac{A_{\q}}{\sqrt{A_{\q}^2-B_{\q}^2}} 
		\delta(\omega-\omega_{\q}) +S^{\sf GS}_{\rm{A}}(\q=0, \omega)\; ,
		\label{eq:Aqw_qm_qn0}
	\end{align}
	where we used \Autoref{eq:Delta}.
	Detailed calculations for $\q=0$ contributions to the A--matrix structure factor can be found in \hyperref[sec:structure_factors_quantum_Aq0]{Appendix~}\ref{sec:structure_factors_quantum_Aq0}.
	More precisely, $S^{\sf GS}_{\rm{A}}(\q=0, \omega)$ is given by  \Autoref{eq:SAq0_QM}, which combined with \Autoref{eq:Aqw_qm_qn0}  gives the total quantum structure factor for the A-matrices expressed in \Autoref{eq:Aqw_qm}. 
	%
	\subsection{ Quantum structure factors: contribution of the ground state at $\q=0$}\label{sec:structure_factors_quantum_q0}
	We here present how to describe the zero-temperature correction of the ground state and the zero-point energy fluctuations' contribution to the quantum structure factors, which is expected to happen at $\q=0$ and $\omega=0$.
	
	We therefore consider the zero--temperature quantum structure factor  at $\q=0$ to be given by 
	\be
	S^{\sf GS}_{\rm{O}}(\q=0)=\sum_{\alpha\beta}\nn{\hat{O}^{\alpha}_{\q=0, \beta}\hat{O}^{\beta}_{\q=0, \alpha}}_{T=0}\; .\label{eq:SOq0_QM}
	\ee
	Calculating the contribution of the ground state and the zero-point energy fluctuations to the zero--temperature quantum structure factor can be achieved by adding a source term to the BBQ Hamiltonian that includes a fictive field $\bf{h}$ coupled to the spin moments, similarly to what we did for the classical case [see \Autoref{sec:classical.structure.factors} and \hyperref[sec:structure_factors_classical]{Appendix~}\ref{sec:structure_factors_classical}]. The structure factors can then be calculated by taking the appropriate derivative of the free energy with respect to the fictive field $\bf{h}$.
	
	We consider the total Hamiltonian to be given by \Autoref{eq:Hfield}, and the source term to be of the form given in \Autoref{eq:DeltaH}. 
	We can then rewrite the operators $\hat{O}^{\alpha}_{\q \beta}$ of \Autoref{eq:DeltaH} in function of the fluctuations orthogonal to the FQ ground state [\Autoref{eq:GS_FQ}]. Refer to \Autoref{section:small.fluctuations} for details on the creation of orthogonal fluctuations.
	Expanding the source term Hamiltonian [\Autoref{eq:DeltaH}] up to second order in bosons, Fourier transforming it and considering its contribution for $\q=0$, we can assume that it takes the following form
	\begin{align*}
		\Delta {\mathcal H} [{\bf{h}}_{\q}]=&C[{\bf{h}}_{\q=0} ] +\frac{1}{2}\sum_{\k}\left[\opdx{w}{\k}{m}_{\k}[{\bf{h}}_{\q=0}]\opx{w}{\k}\right.\\
		&+\left.\left[{\V{N}}[{\bf{h}}_{\k}]^T\opx{w}{\k}+\opdx{w}{\k}{\V{N}}[{\bf{h}}_{\k}]\right]\delta_{\k,0}\right]\; ,\numberthis\label{eq:DeltaH_qm_q0}
	\end{align*}
	where $C[{\bf{h}}_{\q=0} ]$ is the coefficient for the 0$\th$ order term of the source term expended in terms of the fluctuations orthogonal to the FQ ground state,  $\opx{w}{\k}$ represents these fluctuations orthogonal to the FQ ground state and is given \Autoref{eq:def_wboson},  ${m}_{\k}[{\bf{h}}_{\q=0}]$ represents the interaction matrix for 2$\nd$ order terms in fluctuations and depends on ${\bf{h}}_{\q=0}$, and where ${\V{N}}[{\bf{h}}_{\k}]^{\dagger}$ and $ {\V{N}}[{\bf{h}}_{\k}]$ are the coefficients for the linear terms in fluctuations. By definition of the source term Hamiltonian [\Autoref{eq:DeltaH}], all the coefficients $C[{\bf{h}}_{\q=0} ]$, ${\V{N}}_1[{\bf{h}}_{\k}]^T$, $ {\V{N}}_2[{\bf{h}}_{\k}]$, and ${m}_{\k}[{\bf{h}}_{\q=0}]$  depend linearly on the fictive field $\bf{h}$ and will be different whether we are considering dipole, quadrupole, or A-matrix moments for the source term [\Autoref{eq:DeltaH}].
	
	Using \Autoref{eq:HBBQ_qm} for the BBQ Hamiltonian, we can assume that the total Hamiltonian [\Autoref{eq:Hfield}]  in terms of the bosons then take the following form
	\begin{align*}
		\Ham_{}=&E_0+C[{\bf{h}}_{\q=0} ] +\frac{1}{2}\sum_{\k}\left[\opdx{w}{\k}{M}_{\k}[{\bf{h}}_{\q=0}]\opx{w}{\k}\right.\\
		&+\left.\left[{\V{N}}[{\bf{h}}_{\k}]^{\dagger}\opx{w}{\k}+\opdx{w}{\k}{\V{N}}[{\bf{h}}_{\k}]\right]\delta_{\k,0}\right]\; ,\numberthis\label{eq:H_qm_q0}
	\end{align*}
	where $E_0$ is the mean-field ground--state given in \Autoref{eq:E0}, and where ${M}_{\k}[{\bf{h}}_{\q=0}]$ is the interaction matrix for the total Hamiltonian. It includes contributions from the BBQ Hamiltonian and the source term ${m}_{\k}[{\bf{h}}_{\q=0}]$, and, therefore, depends on ${\bf{h}}_{\q=0}$.
	
	Following the method described in \hyperref[sec:bogolioubov_transfomation]{Appendix~}\ref{sec:bogolioubov_transfomation} , we perform a Bogoliubov transformation in order to diagonalize the total Hamiltonian. We assume that the new Bogoliubov bosons $\opx{v}{\k}$ 
	\be
	\opx{v}{\k}=\begin{pmatrix}
		\balpha{\k} \\
		\balphad{-\k} \\
		\bbeta{\k} \\
		\bbetad{-\k} \\
	\end{pmatrix} \; ,
	\label{eq:def_vboson}
	\ee
	are given in terms of the bosons orthogonal to the FQ ground state  $\opx{w}{\k}$ by \Autoref{eq:U}.
	We can then assume that the total Hamiltonian in terms of  the Bogoliubov bosons $\opx{v}{\k}$ becomes
	\begin{align*}
		\Ham_{}=&E_0+\Delta E_0[{\bf{h}}_{\q=0} ]+C[{\bf{h}}_{\q=0} ] \\
		&+\frac{1}{2}\sum_{\k}\left[\epsilon_{\k,\alpha}[{\bf{h}}_{\q=0}]\balphad{\k}\balpha{\k}+\epsilon_{\k,\beta}[{\bf{h}}_{\q=0}]\bbetad{\k}\bbeta{\k}\right.\\
		&+\left.\left[\tilde{\V{N}}[{\bf{h}}_{\k}]^{\dagger}\opx{v}{\k}+\opdx{v}{\k}\tilde{\V{N}}[{\bf{h}}_{\k}]\right]\delta_{\k,0}\right]\; ,\numberthis\label{eq:H_qm_bogo_q0}
	\end{align*}
	where $\epsilon_{\k,\alpha}[{\bf{h}}_{\q=0}]$ and $\epsilon_{\k,\beta}[{\bf{h}}_{\q=0}]$ are the two physical eigenvalues obtained by diagonalizing ${M}_{\k}[{\bf{h}}_{\q=0}]$ [see \Autoref{eq:Uev}], where $\Delta E_0[{\bf{h}}_{\q=0} ]$ is the ground state contribution of the Bogoluibov bosons and where 
	\be
	\begin{matrix}
		\tilde{\V{N}}[{\bf{h}}_{\k}]^{\dagger}={\V{N}}[{\bf{h}}_{\k}]^{\dagger}U^{-1}\; , \\
		\tilde{\V{N}}[{\bf{h}}_{\k}]={U^{\dagger}}^{-1}{\V{N}}[{\bf{h}}_{\k}]\; .\label{eq:Ntilde_q0}
	\end{matrix}
	\ee
	Here, $U$ is the Bogoliubov matrix change defined by  \Autoref{eq:U} and remains to be determined.
	We also note that $U^{-1}$ (and ${U^{\dagger}}^{-1}$) should be calculate from \Autoref{eq:Uinv}.
	
	The canonical partition function is defined by
	\be
	Z=Tr(\exp{-\beta \hat{\Ham}})\; ,
	\ee
	where $\beta$ is defined by \Autoref{eq:def_beta}, and $\hat{\Ham}$ is the operator Hamiltonian.
	However in order to compute the partition function and the free energy, we want to get rid of the linear terms $\tilde{\V{N}}[{\bf{h}}_{\k}]^{\dagger}$ and $\tilde{\V{N}}[{\bf{h}}_{\k}]$, which only contribute for $\k=0$. 
	The partition function is then the one of a set of  independent harmonic oscillators for the $\k\neq0$ terms, but still contains linear terms with respect to the bosons for $\k=0$:
	{\small
		\begin{align*}
			Z=&Tr\left(\exp{-\beta E_0-\beta\Delta E_0[{\bf{h}}_{\q=0} ]-\beta C[{\bf{h}}_{\q=0} ] }\right. \\
			&\times\prod_{\k\neq 0}(\exp{-\beta \epsilon_{\k,\alpha}[{\bf{h}}_{\q=0}]\balphad{\k}\balpha{\k}})(\exp{\beta\epsilon_{\k,\beta}[{\bf{h}}_{\q=0}]\bbetad{\k}\bbeta{\k}})\\
			&\times(\exp{-\frac{1}{2}\beta\epsilon_{\k=0,\alpha}[{\bf{h}}_{\q=0}]\balphad{\k=0}\balpha{\k=0}-\frac{1}{2}\beta n_1[{\bf{h}}_{\k=0}]\balphad{\k=0}-\frac{1}{2}\beta n_2[{\bf{h}}_{\k=0}]\balpha{\k=0}})\\
			&\left.\times(\exp{-\frac{1}{2}\beta\epsilon_{\k=0,\beta}[{\bf{h}}_{\q=0}]\bbetad{\k=0}\bbeta{\k=0}-\frac{1}{2}\beta n_3[{\bf{h}}_{\k=0}]\bbetad{\k=0}-\frac{1}{2}\beta n_4[{\bf{h}}_{\k=0}]\bbeta{\k=0}})\right)\; .\numberthis
	\end{align*}}
	where 
	\be
	\begin{matrix}
		n_1[{\bf{h}}_{\k=0}]=\tilde{N}[{\bf{h}}_{-\k=0}]^{\dagger,2}+	\tilde{N}[{\bf{h}}_{\k=0}]^1 \; , \\
		n_2[{\bf{h}}_{\k=0}]= \tilde{N}[{\bf{h}}_{\k=0}]^{\dagger,1}+	\tilde{N}[{\bf{h}}_{-\k=0}]^2\; ,\\
		n_3[{\bf{h}}_{\k=0}]=\tilde{N}[{\bf{h}}_{-\k=0}]^{\dagger,4}+	\tilde{N}[{\bf{h}}_{\k=0}]^3 \; , \\
		n_4[{\bf{h}}_{\k=0}]= \tilde{N}[{\bf{h}}_{\k=0}]^{\dagger,3}+	\tilde{N}[{\bf{h}}_{-\k=0}]^4\; ,
	\end{matrix}
	\ee
	with $\tilde{\V{N}}[{\bf{h}}_{\k}]^{\dagger,1}$ denoting the first component of $\tilde{\V{N}}[{\bf{h}}_{\k}]^{\dagger}$.
	To do this, we note that we can perform a change a variables by completing the square. For the $\k=0$ term, for the the $\balphad{\k}, \balpha{\k}$ bosons for instance, we have 
	\begin{align*}
		-\frac{1}{2}&\beta\epsilon_{0,\alpha}[{\bf{h}}_{\q=0}]\balphad{0}\balpha{0}-\frac{1}{2}\beta n_1[{\bf{h}}_{\k=0}]\balphad{0}-\frac{1}{2}\beta n_2[{\bf{h}}_{\k=0}]\balpha{0}=\\
		&-\frac{1}{2}\beta\epsilon_{0,\alpha}[{\bf{h}}_{\q=0}](\balphad{0}+\frac{n_1[{\bf{h}}_{\k=0}]}{\epsilon_{0,\alpha}[{\bf{h}}_{\q=0}]}) (\balpha{0}+\frac{n_2[{\bf{h}}_{\k=0}]}{\epsilon_{0,\alpha}[{\bf{h}}_{\q=0}]})\\
		& +\beta\frac{n_1[{\bf{h}}_{\k=0}]n_2[{\bf{h}}_{\k=0}]}{\epsilon_{0,\alpha}[{\bf{h}}_{\q=0}]}  \; ,    \numberthis
	\end{align*}
	We note that we have 
	\be
	n_1[{\bf{h}}_{\k=0}]=n_2[{\bf{h}}_{\k=0}]^{\dagger}\; ,
	\ee
	so that we can define the change of variable
	\be
	\begin{matrix}
		\brhod{\k=0}=\balphad{\k=0}+\frac{n_1[{\bf{h}}_{\k=0}]}{\epsilon_{\k=0,\alpha}[{\bf{h}}_{\q=0}]}	 \; , \\
		\brho{\k=0}=\balpha{\k=0}+\frac{n_2[{\bf{h}}_{\k=0}]}{\epsilon_{\k=0,\alpha}[{\bf{h}}_{\q=0}]}\; ,
	\end{matrix}
	\ee
	which ensures that $\brhod{\k=0}$ and $\brho{\k=0}$ have bosonic commutation relations and are associated with the eigenmode $\epsilon_{\k=0,\alpha}[{\bf{h}}_{\q=0}]$.
	We follow the same argument for the  $\bbetad{\k=0}, \bbeta{\k=0}$ bosons, and get new bosons $\bsigmad{\k=0}$ and $\bsigma{\k=0}$:
	\be
	\begin{matrix}
		\bsigmad{\k=0}=\bbetad{\k=0}+\frac{n_3[{\bf{h}}_{\k=0}]}{\epsilon_{\k=0,\beta}[{\bf{h}}_{\q=0}]}	 \; , \\
		\bsigma{\k=0}=\bbeta{\k=0}+\frac{n_4[{\bf{h}}_{\k=0}]}{\epsilon_{\k=0,\beta}[{\bf{h}}_{\q=0}]}\; ,
	\end{matrix}
	\ee
	associated with the eigenmode $\epsilon_{\k=0,\beta}[{\bf{h}}_{\q=0}]$
	The partition function is then the one of a set of  independent harmonic oscillators. We obtain
	\begin{align*}
		Z=&Tr\left(\exp{-\beta E_0-\beta\Delta E_0[{\bf{h}}_{\q=0} ]-\beta C[{\bf{h}}_{\q=0} ] }\right. \\
		&\times\prod_{\k\neq 0}(\exp{-\beta \epsilon_{\k,\alpha}[{\bf{h}}_{\q=0}]\balphad{\k}\balpha{\k}})(\exp{-\beta\epsilon_{\k,\beta}[{\bf{h}}_{\q=0}]\bbetad{\k}\bbeta{\k}})\\
		&\times(\exp{-\beta \epsilon_{\k=0,\alpha}[{\bf{h}}_{\q=0}]\brhod{\k}\brho{\k}})(\exp{-\beta\epsilon_{\k=0,\beta}[{\bf{h}}_{\q=0}]\bsigmad{\k}\bsigma{\k}})\\
		&\left.\times\exp{\beta\frac{n_1[{\bf{h}}_{\k=0}]n_2[{\bf{h}}_{\k=0}]}{\epsilon_{\k=0,\alpha}[{\bf{h}}_{\q=0}]}  } \exp{\beta\frac{n_3[{\bf{h}}_{\k=0}]n_4[{\bf{h}}_{\k=0}]}{\epsilon_{\k=0,\alpha}[{\bf{h}}_{\q=0}]}}\right)\; .\numberthis
	\end{align*}
	We then perform the trace on the Fock space, and use the fact that the trace is independent of the choice of the basis. This means we can compute it separately for the $\balphad{\k}$ bosons on their respective Fock basis $\ket{n^{\alpha}_{\k}}$ and for the $\brhod{\k=0}$ bosons on its respective Fock basis $\ket{n^{\rho}_{\k=0}}$, and similarly for $\bbetad{\k}$ and $\bsigmad{\k=0}$.
	Taking the trace over the Fock space as explained above, we obtain 
	\begin{align*}
		Z=&\exp{-\beta E_0-\beta\Delta E_0[{\bf{h}}_{\q=0} ]-\beta C[{\bf{h}}_{\q=0} ] } \\
		&\times\prod_{\k}\frac{1}{1-\exp{-\beta\epsilon_{\k,\alpha}[{\bf{h}}_{\q=0}]}} \frac{1}{1-\exp{-\beta\epsilon_{\k,\alpha}[{\bf{h}}_{\q=0}]}}\\
		&\times\exp{\beta\frac{n_1[{\bf{h}}_{\k=0}]n_2[{\bf{h}}_{\k=0}]}{\epsilon_{\k,\alpha}[{\bf{h}}_{\q=0}]}  } \exp{\beta\frac{n_3[{\bf{h}}_{\k=0}]n_4[{\bf{h}}_{\k=0}]}{\epsilon_{\k,\alpha}[{\bf{h}}_{\q=0}]}  }\; .\numberthis
	\end{align*}
	We note here that instead of doing the change of variable for $\k=0$, we could also 
	The free energy is given by
	\begin{align*}
		F &=-\frac{\log(Z)}{\beta}\\
		&=E_0+\Delta E_0[{\bf{h}}_{\q=0} ]+C[{\bf{h}}_{\q=0} ]\\
		&+\frac{1}{ \beta}\sum_{\k}\log(1-\exp{-\beta\epsilon_{\k,\alpha}[{\bf{h}}_{\q=0}]})\\
		&+\frac{1}{ \beta}\sum_{\k}\log(1-\exp{-\beta\epsilon_{\k,\beta}[{\bf{h}}_{\q=0}]})\\
		& - 2\frac{n_1[{\bf{h}}_{\k=0}]n_2[{\bf{h}}_{\k=0}]}{\epsilon_{\k,\alpha}[{\bf{h}}_{\q=0}]} -2\frac{n_3[{\bf{h}}_{\k=0}]n_4[{\bf{h}}_{\k=0}]}{\epsilon_{\k,\alpha}[{\bf{h}}_{\q=0}]} +{\mathcal O}(T^2) \; . \numberthis
		\label{eq:Fq0_QM}
	\end{align*}

	The moments are given by taking the appropriate derivative of the free energy. They are given by the same expression that we obtained for the classical case expressed in \Autoref{eq:Ocorq0} and \Autoref{eq:OOcorq0}. 
	We now note that we are interested in the zero temperature $T=0$ structure factor, and we can disregard the terms with $\frac{1}{ \beta}$ in the free energy.
	\Autoref{eq:Ocorq0} then becomes
	\begin{align*}
		\nn{\hat{O}^{\alpha}_{\q=0, \beta}}_{T=0}
		&=	-\left. \frac{\ddp\left[ \Delta E_0[{\bf{h}}_{\q=0} ] +C[{\bf{h}}_{\q=0} ]\right]}{\ddp h^{\alpha}_{\q=0,\beta}} \right|_{\V{h}=0}\\
		&+2\left. \frac{\ddp\left[ \frac{n_1[{\bf{h}}_{\k=0}]n_2[{\bf{h}}_{\k=0}]}{\epsilon_{\k,\alpha}[{\bf{h}}_{\q=0}]} +\frac{n_3[{\bf{h}}_{\k=0}]n_4[{\bf{h}}_{\k=0}]}{\epsilon_{\k,\alpha}[{\bf{h}}_{\q=0}]}\right]}{\ddp h^{\alpha}_{\q=0,\beta}} \right|_{\V{h}=0}\numberthis \; .
		\label{eq:Ocorq01q0}
	\end{align*}
	We also note that the terms with $n_1[{\bf{h}}_{\k=0}]n_2[{\bf{h}}_{\k=0}]$ and $n_3[{\bf{h}}_{\k=0}]n_4[{\bf{h}}_{\k=0}]$ are at least quadratic (if not of higher order, depending on $\epsilon_{\k}[{\bf{h}}_{\q=0}]$)  in the field components ${\bf{h}}_{\q=0}$, as any of the $n_{i=1,2,3,4}$ is  independently  linear in ${\bf{h}}_{-\k=0}$ by definition. Therefore taking the first derivative of the terms with $n_1[{\bf{h}}_{\k=0}]n_2[{\bf{h}}_{\k=0}]$ and $n_3[{\bf{h}}_{\k=0}]n_4[{\bf{h}}_{\k=0}]$  and evaluating them at zero field will inevitably lead to a null contribution. We then are simply left with
	\begin{align*}
		\nn{\hat{O}^{\alpha}_{\q=0, \beta}}_{T=0}
		&=	-\left. \frac{\ddp \Delta E_0[{\bf{h}}_{\q=0} ] +C[{\bf{h}}_{\q=0} ]}{\ddp h^{\alpha}_{\q=0,\beta}} \right|_{\V{h}=0}\numberthis \; .
		\label{eq:Ocorq0_QM}
	\end{align*}
	The second moments are given by \Autoref{eq:Ocorq0} and disregarding again the term with $\frac{1}{ \beta}$, it simply becomes
	\begin{eqnarray}
		\nn{\hat{O}^{\alpha}_{\q=0, \beta}\hat{O}^{\mu}_{\q=0, \nu}}_{T=0}
		&=&	\nn{\hat{O}^{\alpha}_{\q=0, \beta}}_{T=0}	\nn{\hat{O}^{\mu}_{\q=0, \nu}}_{T=0}\; .	\label{eq:OOcorq0_QM}
		\nonumber\\
	\end{eqnarray}
	We can insert \Autoref{eq:OOcorq0_QM} into \Autoref{eq:SOq0_QM} to calculate the ground state contribution to the quantum structure factor.
	Therefore all we need to do is find the 0$\th$ order contribution of the source term, i.e., find $C[{\bf{h}}_{\q=0} ]$, and compute the zero--point energy of the Bogoliubov transformation $\Delta E_0[{\bf{h}}_{\q=0} ] $.
	
	\subsection{Dipole moments: contribution of the ground state to the quantum structure factor at $\q=0$}\label{sec:structure_factors_quantum_Sq0}
	
	First, we consider the structure factor for dipole moments of spin
	\begin{eqnarray}
		S^{\sf GS}_{\rm{S}}(\q=0) 
		&=& \sum_{\alpha} \nn{\ops{\q=0}{\alpha} \ops{-\q=0}{\alpha}}_{T=0}  \; .\label{eq:Sq0_QM}
	\end{eqnarray}
	The relevant source term is given by \Autoref{eq:HamLowTS}
	According to \Autoref{eq:dipole.in.terms.of.A} and using \Autoref{eq:Acond2},  we can express \Autoref{eq:HamLowTS} in function of fluctuations orthogonal to the FQ ground state [\Autoref{eq:GS_FQ}].
	Considering fluctuation terms up to 2$\nd$ order, we have
	\begin{align*}
		\ops{i}{x}&=-i(\bb{i}-\bbd{i})\; ,\\
		\ops{i}{y}&=i(\bad{i}\bb{i}-\ba{i}\bbd{i})\; , \numberthis\\
		\ops{i}{z}&=-i(\bad{i}-\ba{i})\; .
	\end{align*}
	After performing a Fourier transform, and considering the source term Hamiltonian  $\Delta\Ham [ {h}^{\alpha}_{i,\beta} ] $ [\Autoref{eq:HamLowTS}] at $\q=0$, we have
	\begin{align*}
		\Delta\Ham [{\bf{h}}_{\q}] =-\sum_{\q}&\left[ih^{x}_{-\q}(\bbd{\q}-\bb{\q})-ih^{z}_{-\q}(\bad{\q}-\ba{\q})\right]\delta_{\q,0} \\
		-\sum_{\k}&\frac{i}{\sqrt{N}}h^{y}_{\q=0}(\bad{\k}\bb{\k}-\ba{\k}\bbd{\k})\; .\numberthis \label{eq:Ham_S_q0}
	\end{align*}
	And using \Autoref{eq:HBBQ_qm} for the BBQ Hamiltonian, the total Hamiltonian [\Autoref{eq:Hfield}] in terms of the bosons takes the same form as in \Autoref{eq:H_qm_q0},
	where ${M}_{\k}[{\bf{h}}_{\q=0}]$ is given by
	\begin{subequations}
		\begin{align*}
			M_{\k}[{\bf{h}}_{\q=0}]&=\\
			&\begin{pmatrix}
				A_{\k}& -B_{\k} &  -\frac{i}{\sqrt{N}}h^{y}_{\q=0} & 0 \\
				-B_{\k} &A_{\k} &  0& \frac{i}{\sqrt{N}}h^{y}_{\q=0} \\
				\frac{i}{\sqrt{N}}h^{y}_{\q=0} &  0 & A_{\k} &- B_{\k}  \\
				0 & -\frac{i}{\sqrt{N}}h^{y}_{\q=0} & -B_{\k} & A_{\k}\\
			\end{pmatrix}\; ,\label{eq:def_phi_vec_M_k_Sq0}\numberthis
		\end{align*}
		where $A_{\k}$ and $B_{\k}$ are given in \Autoref{eq:DefAB}
		and where ${\V{N}}[{\bf{h}}_{\k}]$ is given by 
		\be
		\begin{matrix}
			{\V{N}}[{\bf{h}}_{\k}]=i\begin{pmatrix}
				h^{z}_{-\k}\\
				-h^{z}_{\k}\\
				-h^{x}_{-\k}	\\
				h^{x}_{\k}
			\end{pmatrix}\; ,\label{eq:N_S_q0}
		\end{matrix}
		\ee
		and where $ C[{\bf{h}}_{\q=0} ] $ holds
		\be
		C[{\bf{h}}_{\q=0} ] =0\; . \label{eq:C_S_q0}
		\ee
	\end{subequations}
	
	Following the procedure depicted in \Autoref{section:quantum.theory} and detailed in \hyperref[sec:bogolioubov_transfomation]{Appendix~}\ref{sec:bogolioubov_transfomation}, we perform a Bogoliubov transformation and
	the eigenvalues $\epsilon_{\k,\lambda}$ are given by
	\begin{subequations}
		\begin{align}
			\epsilon_{\k,1}[{\bf{h}}_{\q=0}]&=-\epsilon_{\k,2}=+\sqrt{A_{\k}^2-B_{\k}^2} +\frac{1}{\sqrt{N}}h^{y}_{\q=0}\; ,\\
			\epsilon_{\k,3}[{\bf{h}}_{\q=0}]&=-\epsilon_{\k,4}=+\sqrt{A_{\k}^2-B_{\k}^2}-\frac{1}{\sqrt{N}}h^{y}_{\q=0}\; .
		\end{align}\label{eq:omega_qm_S_q0}
	\end{subequations}
	After performing the Bogoliubov transformation, the Hamiltonian can be rewritten as follows:
	\begin{align*}
		\Ham=&E_0+\Delta E_0[{\bf{h}}_{\q=0}]\\
		&+\left[\sum_{\k}\epsilon_{\k,1}[{\bf{h}}_{\q=0}]\balphad{\k}\balpha{\k}+\epsilon_{\k,3}[{\bf{h}}_{\q=0}]\bbetad{\k}\bbeta{\k}\right]\; ,\numberthis\label{eq:HamBogo2_q0}
	\end{align*}
	where $C[{\bf{h}}_{\q=0} ]$ is disregarded since it is null [\Autoref{eq:C_S_q0}], and where $\Delta E_0[{\bf{h}}_{\q=0}]$ is the zero--point energy
	\begin{eqnarray}
		\Delta E_0 [{\bf{h}}_{\q=0}]&=& \frac{1}{2}\sum_\k \left[\epsilon_{\k,1}[{\bf{h}}_{\q=0}]+\epsilon_{\k,3}[{\bf{h}}_{\q=0}]\right]\; .
	\end{eqnarray}
	According to \Autoref{eq:Ocorq0_QM}, the ground state contribution to the first moments yield
	\begin{subequations}
		\begin{align}
			\nn{S^{x}_{\q}}_{T=0}&=\nn{S^{z}_{\q}}_{T=0}=0\; ,\\
			\nn{S^{y}_{\q}}_{T=0}&=\frac{1}{2}\sum_\k \left[\frac{1}{\sqrt{N}}-\frac{1}{\sqrt{N}}\right]=0 \; .
		\end{align}\label{eq:SxSzSyq0}
	\end{subequations}
	And according to \Autoref{eq:OOcorq0_QM} and \Autoref{eq:Sq0_QM}, the spin dipole structure factor at $\q=0$ yields
	\begin{eqnarray}
		S^{\sf GS}_{\rm{S}}(\q=0) 
		&=& 0 \; . \label{eq:SSq0_QM}
	\end{eqnarray}
	Indeed, the ground state is quadrupolar and does not break time--reversal symmetry. Therefore, at zero temperature, the contribution of quantum fluctuations from the zero--point energy should average to zero for the spin dipole moments. 
	The spectral representation of \Autoref{eq:SSq0_QM} is then also trivially null.
	Combining \Autoref{eq:Sqw_qm_qn0_app} and \Autoref{eq:SSq0_QM}, we obtain \Autoref{eq:Sqw_qm}.

	\subsection{Quadrupole moments: contribution of the ground state to the quantum structure factor at $\q=0$}\label{sec:structure_factors_quantum_Qq0}
	We now consider the quadrupole structure factor at the $\Gamma$--point, which is defined as
	\begin{eqnarray}
		S^{\sf GS}_{\rm{Q}}(\q=0) 
		&=& \sum_{\alpha\beta} 
		\nn{\opq{\q=0}{\alpha\beta} \opq{\q=0}{\beta\alpha}} _{T=0}\; . \label{eq:Qq0_QM}
	\end{eqnarray}
	We follow the same procedure as depicted in \hyperref[sec:structure_factors_quantum_q0]{Appendix~}\ref{sec:structure_factors_quantum_q0}. The relevant source term for quadrupole moments is given by \Autoref{eq:HamLowTQ}. We can express \Autoref{eq:HamLowTQ} up to second order in terms of the bosons by using \Autoref{eq:Acond2} and \Autoref{eq:quadrupole.in.terms.of.A}. We use \Autoref{eq:HBBQ_qm} for the BBQ Hamiltonian. We then obtain for the total Hamiltonian given by \Autoref{eq:Hfield}, written in the form of \Autoref{eq:H_qm_q0}, where 
	$M_{\k}[{\bf{h}}_{\q=0}]$ is given by
	\begin{subequations}
		\be
		M_{\k}[{\bf{h}}_{\q=0}]=\begin{pmatrix}
			A_{\k}+\alpha_1 & -B_{\k} &   \beta_1&0\\
			-B_{\k} &A_{\k}+\alpha_1  &   0& \beta_1 \\
			\beta_1 &  0& A_{\k}+\alpha_2 &- B_{\k}  \\
			\beta_1& 0 & -B_{\k} & A_{\k}+\alpha_2  \\
		\end{pmatrix} \; ,\label{eq:M_Qq0_QM}
		\ee
		where ${\V{N}}[{\bf{h}}_{\k}]$ is given by 
		\be
		\begin{matrix}
			{\V{N}}[{\bf{h}}_{\k}]=\begin{pmatrix}
				\xi^1_{-\k}\\
				\xi^1_{\k}\\
				\xi^2_{-\k}\\
				\xi^2_{-\k}\label{eq:N_Q_q0}
			\end{pmatrix}
		\end{matrix}\; ,
		\ee
		and where $ C[{\bf{h}}_{\q=0} ] $ holds
		\be
		C[{\bf{h}}_{\q=0} ] =\sqrt{N}\left(\frac{4}{3}h_{\q=0,y}^{y}-\frac{2}{3}(h_{\q=0,x}^{x}+h_{\q=0,z}^{z})\right)\; ,\label{eq:C_Qq0_QM}
		\ee
	\end{subequations}
	with $A_{\k}$ and $B_{\k}$ being given in \Autoref{eq:DefAB} and with the following definitions
	\be
	\begin{matrix}
		\begin{matrix}
			\alpha_1=\frac{2}{\sqrt{N}}(h_{\q=0,x}^{x}-h_{\q=0,y}^{y})\; ,&\alpha_2=\frac{2}{\sqrt{N}}(h_{\q=0,z}^{z}-h_{\q=0,y}^{y})\; ,
		\end{matrix}\\
		&\\
		\begin{matrix}
			\beta_1=\frac{1}{\sqrt{N}}(h_{\q=0,z}^{x}+h_{\q=0,x}^{z})\;, 
		\end{matrix}\\
		&\\
		\begin{matrix}
			\xi^1_{\k}=(h^{x}_{\k,y}+h^{y}_{\k,x}) & \; ,& \xi^2_{\k}=(h^{y}_{\k,z}+h^{z}_{\k,y})
		\end{matrix}\; .
	\end{matrix}
	\ee
	Following the procedure depicted in \Autoref{section:quantum.theory} and detailed in \hyperref[sec:bogolioubov_transfomation]{Appendix~}\ref{sec:bogolioubov_transfomation}, we perform a Bogoliubov transformation and
the eigenvalues $\epsilon_{\k,\lambda}$ are given by
{\small
\begin{subequations}
			\begin{align*}
				&\epsilon_{\k,1}[{\bf{h}}_{\q=0}]= -\epsilon_{\k,2}[{\bf{h}}_{\q=0}]=\\
				&\sqrt{A_{\k}^2-B_{\k}^2+\beta_1^2+\frac{1}{2} \left(\alpha_{1}^2+\alpha_{2}^2\right)+A_{\k} \alpha^{+}-\Delta  \left(A_{\k}+\frac{\alpha^{+}}{2}\right)}\; ,\numberthis\\
				&\epsilon_{\k,3}[{\bf{h}}_{\q=0}]= -\epsilon_{\k,4}[{\bf{h}}_{\q=0}]=\\
				&\sqrt{A_{\k}^2-B_{\k}^2+\beta_1^2+\frac{1}{2} \left(\alpha_{1}^2+\alpha_{2}^2\right)+A_{\k} \alpha^{+}+\Delta  \left(A_{\k}+\frac{\alpha^{+}}{2}\right)}\; ,\numberthis\\
			\end{align*} \label{eq:omega_qm_Q_q0}
\end{subequations}
}
where $\alpha^{+}$ and $\Delta$ are defined in \Autoref{eq:a+_Delta}.
After performing the Bogoliubov transformation, the Hamiltonian can be rewritten as follows:
	\begin{align*}
		\Ham=&E_0+\Delta E_0[{\bf{h}}_{\q=0}]+C[{\bf{h}}_{\q=0} ]\\
		&+\left[\sum_{\k}\epsilon_{\k,1}[{\bf{h}}_{\q=0}]\balphad{\k}\balpha{\k}+\epsilon_{\k,3}[{\bf{h}}_{\q=0}]\bbetad{\k}\bbeta{\k}\right]\; ,\numberthis\label{eq:HamBogo2_Qq0}
	\end{align*}
	where  $C[{\bf{h}}_{\q=0} ]$ is given in \Autoref{eq:C_Qq0_QM}, and where$\Delta E_0[{\bf{h}}_{\q=0}]$ is the zero--point energy and yields
	\begin{eqnarray}
		\Delta E_0[{\bf{h}}_{\q=0}] &=& \frac{1}{2}\sum_\k \left[\epsilon_{\k,1}[{\bf{h}}_{\q=0}]+\epsilon_{\k,3}[{\bf{h}}_{\q=0}]\right]\; .
	\end{eqnarray}
	According to \Autoref{eq:Ocorq0_QM}, the ground state contribution to the first moments yield
	{\small
		\begin{subequations}
			\begin{align}
				\nn{Q^{xx}_{\q=0}}_{T=0}&=-\frac{2}{3}\sqrt{N}+ \frac{1}{\sqrt{N} }\sum_\k \left[\frac{A_{\k}}{\sqrt{A_{\k}^2-B_{\k}^2}} \right]\; ,\\
				\nn{Q^{xy}_{\q=0,y}}_{T=0}&=\nn{Q^{yx}_{\q=0}}_{T=0}=0\; ,\\
				\nn{Q^{xz}_{\q=0,z}}_{T=0}&=\nn{Q^{zx}_{\q=0}}_{T=0}=0\; ,\\
				\nn{Q^{yy}_{\q=0,y}}_{T=0}&=\frac{4}{3}\sqrt{N}- \frac{1}{\sqrt{N} }\sum_\k \left[\frac{2A_{\k}}{\sqrt{A_{\k}^2-B_{\k}^2}} \right]\; ,\\
				\nn{Q^{yz}_{\q=0,z}}_{T=0}&=\nn{Q^{zy}_{\q=0}}_{T=0}=0\; ,\\
				\nn{Q^{zz}_{\q=0}}_{T=0}&=-\frac{2}{3}\sqrt{N}+ \frac{1}{\sqrt{N}} \sum_\k \left[\frac{A_{\k}}{\sqrt{A_{\k}^2-B_{\k}^2}} \right]\; .
			\end{align}\label{eq:nQq0_QM}
		\end{subequations}
	}
	Before calculating the structure factor, we note that, as given in \Autoref{eq:nQq0_QM}, the first quadrupole moments consist of two terms with different scaling behaviour with respect to the parameter we expand fluctuations about, which is the length of the spin $s$ . Indeed, similarly to multi-boson expansion, or its linear spin-wave version with Holstein--Primakoff bosons or Schwinger bosons in the case of a $su(2)$ representation of the spin, we assume the fluctuation to be sufficiently small compared to the spin length $s$. In other words, $C[{\bf{h}}_{\q=0} ]$  from \Autoref{eq:C_Qq0_QM} and the eigenvalues in \Autoref{eq:omega_qm_Q_q0} scale with $s$ as
	\begin{subequations}
	\be
	C[{\bf{h}}_{\q=0} ]\sim s h^{\mu}_{\q=0\mu} \; ,
	\ee
	\be
	\epsilon_{\k,\lambda}[{\bf{h}}_{\q=0}]\sim s \sqrt{ Const. +\frac{h^{\mu}_{\q=0\mu}}{s^2}+\frac{\O({\bf{h}}_{\q=0}^2)}{s^2} }\; .
	\ee
	\end{subequations}
	Their  derivatives with respect to $h^{\mu}_{\q=0,\mu}$ that enters the quadrupole moments [\Autoref{eq:Ocorq0_QM}] yield
	\begin{subequations}
	\be
	\left.\frac{C[{\bf{h}}_{\q=0} ] }{\ddp h^{\mu}_{\q=0,\mu}}\right|_{\bf{h}=0}\sim s \; ,
	\ee
	\begin{align*}
	\left.\frac{\ddp\epsilon_{\k,\lambda}[{\bf{h}}_{\q=0}] }{\ddp h^{\mu}_{\q=0,\mu}}\right|_{\bf{h}=0}\sim  \left.\frac{s}{\sqrt{ Const.}}\frac{\ddp \frac{h^{\mu}_{\q=0\mu}}{s^2}}{\ddp h^{\mu}_{\q=0,\mu}}\right|_{\bf{h}=0} \sim  \frac{1}{s} \; .
	\end{align*}
\end{subequations}
	This implies that the scaling behaviour of the first quadrupole moments goes as
	\be
	\nn{Q^{\mu\mu}_{\q=0}}_{T=0}\sim s+ \frac{1}{s}\; ,
	\ee
	where $s$ is the length of the spin.
	We now argue that because our approximation is valid up to linear order in $\frac{1}{s}$, i.e., second order in fluctuations,  we can disregard  $\frac{1}{s^2}$ terms.  $\frac{1}{s^2}$ terms are physical but should not enter into our level of approximation. Indeed, one would expect additional contribution to the $\frac{1}{s^2}$ term coming from higher orders in perturbation theory. However, we do not take these into account here and simply consider terms up to $\frac{1}{s}$.
	According to \Autoref{eq:OOcorq0_QM} and \Autoref{eq:Qq0_QM}, the spin quadrupole structure factor at $\q=0$ yields
	\begin{eqnarray}
		S^{\sf GS}_{\rm{Q}}(\q=0)&=& \frac{8}{3}N-8 \sum_\k \left[\frac{A_{\k}}{\sqrt{A_{\k}^2-B_{\k}^2}} \right]\nonumber\\
		&&+\O(\frac{1}{s^2})  \label{eq:SQq0_QM}\; .
	\end{eqnarray}
	Its spectral representation is given by
	\begin{eqnarray}
		S^{\sf GS}_{\rm{Q}}(\q=0,\omega)&=& \left(\frac{8}{3}N-8 \sum_\k \left[\frac{A_{\k}}{\sqrt{A_{\k}^2-B_{\k}^2}} \right]\right)\delta(\omega)\nonumber\\
		&&+\O(\frac{1}{s^2}) \label{eq:SQq0_QM_w}\; .
	\end{eqnarray}
	Combining \Autoref{eq:Qqw_qm_qn0} and \Autoref{eq:SQq0_QM_w}, we obtain \Autoref{eq:Qqw_qm}.
	\subsection{A-matrices: contribution of the ground state to the quantum structure factor at $\q=0$}\label{sec:structure_factors_quantum_Aq0}
	For the quantum zero temperature structure factor for the A--matrices at $\q=0$, we make use of the sum rule given in \Autoref{eq:sum.rule.AQS}.
	This leads to 
	\begin{eqnarray}
		S^{\sf GS}_{\rm{A}}(\q=0)&=&  \frac{1}{4} S^{\sf GS}_{\rm{Q}}(\q=0) 
		+  \frac{1}{2} S^{\sf GS}_{\rm{S}}(\q=0) 
		+  \frac{1}{3}N \delta_{\q,0}\nonumber \\
		&=& N-2 \sum_\k \left[\frac{A_{\k}}{\sqrt{A_{\k}^2-B_{\k}^2}} \right] +\O(\frac{1}{s^2})\; ,\nonumber\\
		\label{eq:SAq0_QM}
	\end{eqnarray}
	where we used \Autoref{eq:SSq0_QM} and \Autoref{eq:SQq0_QM}.
	Its spectral representation yields
	\begin{eqnarray}
		S^{\sf GS}_{\rm{A}}(\q=0, \omega)&=& \left(N-2 \sum_\k \left[\frac{A_{\k}}{\sqrt{A_{\k}^2-B_{\k}^2}} \right]\right)\delta(\omega) \nonumber\\
		&&+\O(\frac{1}{s^2}) \; .\nonumber\\
		\label{eq:SAq0_QM_w}
	\end{eqnarray}
 Combining \Autoref{eq:Aqw_qm_qn0} with \Autoref{eq:SAq0_QM_w}  gives the total quantum structure factor for the A-matrices expressed in \Autoref{eq:Aqw_qm}
	\section{System size dependence of the ordered moments }
	\label{sec:Mermin_Wagner_Theorem}
	We present in this Appendix, the details of the manufacturing of \Autoref{section:ordered.moment}. More precisely, we explain how we fitted the numerical data for the ordered moments and explain how we calculated the ordered moments from the analytical results.
	
	In \Autoref{table:Tintervals}, we show the temperature intervals on which the  corresponding ordered parameters values are used for the fits of the slope $\alpha(L)$ of the ordered parameters in  \Autoref{fig:ordered.moment}~\subref{fig:MW_theorem1}, for different system sizes. 
	
	We also present here how the ordered moments as expressed by \Autoref{eq:OM} and presented in \Autoref{fig:ordered.moment}~\subref{fig:MW_theorem2} are calculated . In order to compute \Autoref{eq:OM}, we need to perform a sum in k--space. We here also show that the sum scales logarithmically with the system size $L$ by explicitly calculating the coefficient $\mu$ correspond to the logarithmic behavior [\Autoref{eq:fitL}]. To do this, we calculate the sum numerically for different system sizes $L$ and fit it according to \Autoref{eq:fitL} (as shown by the orange line in \Autoref{fig:ordered.moment}~\subref{fig:MW_theorem2}). Additionally, we also transform the sum into an integral and extract the logarithmic scaling behaviour.

\begin{center}
\begin{table}[t]
	\begin{tabular}{ |c||c|c|  }
				\hline
				System size L & $T_{min}$ & $T_{max}$\\
				\hline
				L=12 & 0.01 & 0.100177 \\
				L=24 & 0.01 & 0.100177 \\
				L=48 & 0.0252403 & 0.100177 \\
				L=96 & 0.0343658 & 0.100177 \\
				\hline
	\end{tabular}
\caption{Temperature intervals used for fitting the parameter $\alpha(L)$ according to \protect\Autoref{eq:lowTfit2}.}\label{table:Tintervals}
\end{table}
\end{center}
	
The Brillouin zone is turned into a parallelogram of area $\frac{8\pi}{\sqrt{3}}$,
as it is spanned by the  reciprocal vectors $\V{K}_a$ and $\V{K}_b$ given in \Autoref{eq:def_ka_kb}. We then discretized it into $N=L^2$ tiles of dimension $\delta A$ given by
	\be
	\begin{matrix}
		\delta \k_a=\frac{1}{L}\V{K}_a&\; , ~&\delta \k_b=\frac{1}{L}\V{K}_b\; ,
	\end{matrix}
	\ee
	such that 
	\be
	\delta A=\frac{8\pi}{\sqrt{3}L^2}\; .
	\ee
	
	In order to compare numerical with analytical results, we consider
	\be
	\frac{1}{N}\sum_{\k} I_{\k}\Rightarrow N \delta A \sum_{k_x,k_x}  I_{\k}=\int  I_{\k} \dd \k \; ,\label{eq:SumNum}
	\ee
	We can now sum over the k--space, numerically, or integrate, analytically.
	
	In \Autoref{eq:SumNum}, we take as integrant the term expressed as a sum in the result obtained in  \Autoref{eq:OM}, as we wish to compute the temperature--dependent part of the ordered moment given in \Autoref{eq:OM}. Using \Autoref{eq:omega_vec_1}, we obtain
	\be
	I_{\k}= \frac{16 A_{\k}}{\epsilon_{\k,1}^2}\; ,
	\ee
	where $\epsilon_{\k,1}$ is given in \Autoref{eq:omega_qm}.
	We then compute the discrete sum numerically according to \Autoref{eq:SumNum} for the different system sizes, including the ones given in \Autoref{table:Tintervals}. When performing the sum,  we also avoid the origin $\k=(0,0)$, where $\epsilon_{\k,1}$ vanishes, (indeed, $\gamma(0)=1$, and according to \Autoref{eq:DefAB},  $A_0=B_0$) and which is not included in the sum of \Autoref{eq:OM}.   For a specific system size, we then get a number as the results of the discrete sum obtained for that specific system size. These numbers are plotted as the red dots in  \Autoref{fig:ordered.moment}~\subref{fig:MW_theorem2}.
	
	
	According to \Autoref{eq:lowTfit2}, we assume that the system size dependency should be of the form 
	\be
	-\left.\frac{\dd S^{CL}_{\rm{Q}}(\q=\Gamma)}{\dd T}\right|_{T=0}=\frac{1}{N}\sum_{\k} I_{\k} \sim ~ C+\mu \log(L)+\frac{\nu}{L}+\frac{\xi}{L^2} \; .\label{eq:fitL}
	\ee
	We use \Autoref{eq:fitL} to fit the results obtained by computing the discrete sum in \Autoref{eq:SumNum}, i.e, the red dots in  \Autoref{fig:ordered.moment}~\subref{fig:MW_theorem2}. The fit is shown in in  \Autoref{fig:ordered.moment}~\subref{fig:MW_theorem2} by the orange line.
	
	Additionally, we want to investigate how accurate the discrete sum is, compared to the integration, and how it depends on system size.
	If we consider the integral version in 2--dimensions for polar coordinates, we can cut off to some small $k_s=\frac{4\pi}{\sqrt{3}L}$ in order to avoid the origin $\k=(0,0)$ as follows:
	\be
	\int I_{\k}\dd \k=2\pi\int_{0}^{\Lambda} I_{\k}k\dd k=2\pi\int_{0}^{k_s} I_{\k}k \dd k+2\pi\int_{k_s}^{\Lambda} I_{\k}k\dd k\; .
	\ee
	For the FQ state, where we chose, $J_1 = 0.0$ and $J_2 = -1.0$, the coefficients $A_{\k}$ and $B_{\k}$ [\Autoref{eq:DefAB}] and the dispersion relation $\epsilon_{\k,1}$ [\Autoref{eq:omega_qm}] become
	\begin{align*}
		A_{\k}&=z\; ,\\
		B_{\k}&=-z\gamma(\k\; ,)\\
		\epsilon_{\k}^2&=z^2(1-\gamma(\k)^2)\numberthis\; .
	\end{align*}
	For the triangular lattice, the geometrical factor is given by \Autoref{eq:def_gamma_tri}, and for sufficiently small values of k, we can use the Taylor expansion on it. We obtain
	\begin{subequations}
		\begin{align}
			\gamma(\k)&\simeq1-\frac{1}{4}(k_x^2+k_y^2)=1-\frac{1}{4}k^2\; ,\\
			\epsilon_{\k}^2&\simeq z^2\frac{1}{2}k^2\; ,\\
			I_{\k}=\frac{16 A_{\k}}{\epsilon_{\k,1}^2}&\simeq\frac{16 z}{z^2\frac{1}{2}k^2}=\frac{3	2}{zk^2}\; ,\\
			2\pi\int_{k_s}^{k_f} I_{\k}k\dd k&\simeq 2\pi\int_{k_s}^{k_f} \frac{32}{zk^2}k\dd k \; .
		\end{align}
	\end{subequations}
	Since $z=6$ for the triangular lattice, we have
	\begin{align*}
		&\frac{2\pi 16}{3}\int_{k_s}^{k_f} \frac{1}{k}\dd k= \frac{2\pi 16}{3}(\log(k_f)-\log(k_s)) \; ,\\
		&= \frac{2\pi 16}{3}\log(L)+\frac{2\pi 16}{3}\log(k_f)-\frac{2\pi 16}{3}\log(\frac{4\pi}{\sqrt{3}})\; ,\numberthis\label{eq:OM_logL}
	\end{align*}
	where in the last line, we used the fact that we chose to cut off according to $k_s=\frac{4\pi}{\sqrt{3}L}$.
	Before we fit the sum with the expression given by \Autoref{eq:fitL}, we need to account for correction coming from the tiling of the k-space as explained in \Autoref{eq:SumNum}. Therefore, we need to divide by 
	\be
	\delta A*N=\frac{8 \pi^2}{\sqrt{3}}\; .
	\ee
	From \Autoref{eq:fitL}, we can obtain the value for the coefficient $\mu$ for \Autoref{eq:OM_logL}, which we can compare with the fit from the values of the sum calculated numerically as shown in \Autoref{fig:ordered.moment}~\subref{fig:MW_theorem2}:
	\be
	\begin{matrix}
		\mu_{ana}=\frac{\frac{2\pi 16}{3}}{\frac{8 \pi^2}{\sqrt{3}}}=\frac{4}{\sqrt{3}\pi}=0.735&\; , ~ & \mu_{num}=0.735\; .
	\end{matrix}
	\ee
	
	\section{Useful Gaussian integrals}
	\label{sec:table.of.integrals}
	
	We present here useful Gaussian integrals that we used to calculate partition functions for the analytic derivations.
	We namely used the following one--dimensional Gaussian integrals
	\begin{subequations}
		\begin{align}
			\int  \exp{-\frac{1}{2}a x^{2}}\dd x&=\sqrt{\frac{2\pi}{a}}\label{eq:gauss_int_S1}\; ,\\
			\int  \exp{-\frac{1}{2}a x^{2}}\exp{\pm bx}\dd x&=\sqrt{\frac{2\pi}{a}}\exp{\frac{b^2}{2 a}}\; ,\label{eq:gauss_int_S2}\\
			\int  \exp{-\frac{1}{2}a x^{2}}\exp{\pm i b x}\dd x&=\sqrt{\frac{2\pi}{a}}\exp{-\frac{b^2}{2 a}}\; ,\label{eq:gauss_int_Q1}
		\end{align}
		which we can also generalize to a multi-dimensions integral with a source term
		\be
		\int \exp{-\frac{1}{2}x_iA_{ij}x_j+B_ix_i}\dd \vec{x}=\sqrt{\frac{(2\pi)^n}{\det A}}\exp{\frac{1}{2}\vec{B}^{T}A^{-1}\vec{B}}\; .\label{eq:gauss_int}
		\ee
		or more generally, 
		\be
		\int \exp{-\frac{1}{2}x_i{A^{i}}_{j}x^j+B_{1,i}^{T}x^i+x_i B_{2}^{i}}\dd \vec{x}=\sqrt{\frac{(2\pi)^n}{\det A}}\exp{2\vec{B}_1^{T}A^{-1}\vec{B}_2}\; .\label{eq:gauss_int2}
		\ee
	\end{subequations}
	where $n$ is the dimension of of the matrix $\bf{A}$.
	Below, we give the proof for \Autoref{eq:gauss_int2}.
	
	\textit{Proof:}
	We assume $\textbf{A}$ to be a real symmetric $n\times n$-matrix. This means that $\textbf{A}$  is orthogonally diagonalizable, i.e., it is similar to a diagonal matrix $\textbf{D}=\diag(d^{1}_{~1}, \dots, d^{n}_{~n})$
	\be
	\V{A}=\V{S}\V{D}\V{S}^{-1}\label{eq:A_diag_D}\; ,
	\ee
	and the $n\times n$ basis change matrix $\textbf{S}$ is orthogonal.
	The  basis change matrix $\textbf{S}$ then satisfies
	\be
	\V{S}^T=\V{S}^{-1}\label{eq:S_ortho}\; ,
	\ee
	and the old coordinates are related to the new ones by
	\begin{subequations}
		\begin{align}
			\vec{x}&=\V{S}\vec{y}\; ,\\
			\vec{x}^{T}&=\vec{y}\V{S}^T\; ,\\
			\dd \vec{x}&=\det\V{S}\dd \vec{y}=\dd \vec{y}\; ,
		\end{align}
	\end{subequations}
	where in the last line we used the fact that the Jacobian matrix of the map $\vec{x}(\vec{y})\rightarrow\V{S}\vec{y}$ is the matrix $\V{S}$ itself, and that its determinant is 1, since it is an orthogonal matrix.
	The term in the exponential in \Autoref{eq:gauss_int2} then becomes
	\begin{subequations}
		\begin{align}
			E:=&-\frac{1}{2}\vec{x}^{T}\V{A}\vec{x}+\vec{B}_1^T\vec{x}+\vec{x}^{T}\vec{B}_{2}\\
			&=-\frac{1}{2}\vec{y}^{T}\V{D}\vec{y}+\vec{B}_1^T\V{S}\vec{y}+\vec{y}^{T}\V{S}^{-1}\vec{B}_{2}\; .
		\end{align}
		If we expand, we obtain:
		\begin{align}
			E&=\sum_{i}\left[-\frac{1}{2}y_{i} d_{~i}^{i}y^{i}+\sum_{\alpha}\left(B_{1,\alpha}^{T}S^{\alpha}_{~i} y^{i}+y_{i} (S^{-1})^{i}_{~\alpha}B_{2}^{\alpha}\right)\right]\; ,
		\end{align}
	\end{subequations}
	where we used the fact that $\V{D}$ is a diagonal matrix.
	For the $i^{th}$ term, we can complete the square as
	\begin{align*}
		&-\frac{1}{2}y_{i} d_{~i}^{i}y^{i}+\sum_{\alpha}\left(B_{1,\alpha}^{T}S^{\alpha}_{~i} y^{i}+y_{i} (S^{-1})^{i}_{~\alpha}B_{2}^{i}\right)\\
		=&-\frac{1}{2} d_{i}^{i}(y_{i}-\frac{2}{d_{~i}^{i}}\sum_{\alpha}B_{1,\alpha}^{T}S^{\alpha}_{~i})(y^{i}-\frac{2}{d_{i}^{i}}\sum_{\beta}(S^{-1})^{i}_{~\beta}B_{2}^{\beta})\\
		&+2\sum_{\alpha}\sum_{\beta}B_{1,\alpha}^{T}S^{\alpha}_{~i}\frac{1}{d_{i}^{i}}(S^{-1})^{i}_{~\beta}B_{2}^{\beta}\; .\numberthis\label{eq:square_completed}
	\end{align*}
	Using again the fact that $d_{~i}^{i}$ is the $i^{th}$ diagonal term of $\V{D}$, and inverting \Autoref{eq:A_diag_D}, we can rewrite  $\frac{1}{d_{i}^{i}}$ as
	\be
	\frac{1}{d_{i}^{i}}=(D^{-1})^{i}_{~i}=\sum_{\mu,\nu}(S^{-1})^{i}_{~\mu}(A^{-1})^{\mu}_{~\nu}(S)^{\nu}_{~i}\; .\label{eq:1_over_d}
	\ee
	Performing the variable change as 
	\begin{subequations}
		\begin{align}
			z_i&=y_{i}-\frac{2}{d_{~i}^{i}}\sum_{\alpha}B_{1,\alpha}^{T}S^{\alpha}_{~i}\; ,\\
			z^i&=y^{i}-\frac{2}{d_{i}^{i}}\sum_{\beta}(S^{-1})^{i}_{~\beta}B_{2}^{\beta}\; ,\\
			\dd \vec{z}&=\dd \vec{y}\; ,
		\end{align}
	\end{subequations}
	and inserting \Autoref{eq:1_over_d} into the last term of \Autoref{eq:square_completed}, and summing over all the components, we obtain
	\begin{subequations}
		\begin{align}
			E=&-\frac{1}{2}\vec{z}^{T}\V{D}\vec{z}+2\vec{B}_1^T(\V{A}^{-1})\vec{B}_{2}\; .
		\end{align}
	\end{subequations}
	We now have the product of n Gaussian integrals of  the form of \Autoref{eq:gauss_int_S1}
	\begin{align}
		\int& \exp{-\frac{1}{2}x_i{A^{i}}_{j}x^j+B_{1,i}^{T}x^i+x_i B_{2}^{i}}\dd \vec{x}=\nonumber\\
		&=\left[\prod_i\int\exp{-\frac{1}{2}z_i{d^{i}}_{i}z^i}\dd z_i \right]\exp{2\vec{B}_1^T(\V{A}^{-1})\vec{B}_{2}}\nonumber\\
		&=\sqrt{\frac{(2\pi)^n}{\prod_i{d^{i}}_{~i}}}\exp{2\vec{B}_1^T(\V{A}^{-1})\vec{B}_{2}}\; .	
	\end{align}
	We then use the fact that
	\be
	\det\V{A}=\det\V{S}\det\V{D}\det\V{S}^{-1}=\det\V{D}=\prod_i{d^{i}}_{~i}\; ,
	\ee
	to obtain \Autoref{eq:gauss_int2}
\section{Application to an easy--plane ferromagnet}
\label{section:anisotropy_FM}

	We here wish to apply the $u(3)$ formalism and its representation in terms of the A-matrices to the Heisenberg ferromagnetic easy-plane anisotropic model.
	The A-matrices are especially useful to work with on the TR-invariant basis, and relatively easy to use when the ground state is quadrupolar. However, some attention is demanded when working with systems where dipoles rather than quadrupoles order. We demonstrate here how one can carefully apply our method for dipolar ordering. Additionally, as explained in \Autoref{section:anisotropy}, we make the interactions anisotropic.
	We show results for the zero--temperature quantum structure factors for dipole, quadrupole and A-matrix moments applied to the ferromagnetic (FM) state on the triangular lattice for the anisotropic Heisenberg Hamiltonian (BBQ Hamiltonian [\Autoref{eq:BBQ.model}] with anisotropic $J_1$ and $J_2=0$), with single--ion anisotropy.
	

	\begin{figure}[t]
		\centering
		\includegraphics[width=0.48\textwidth]{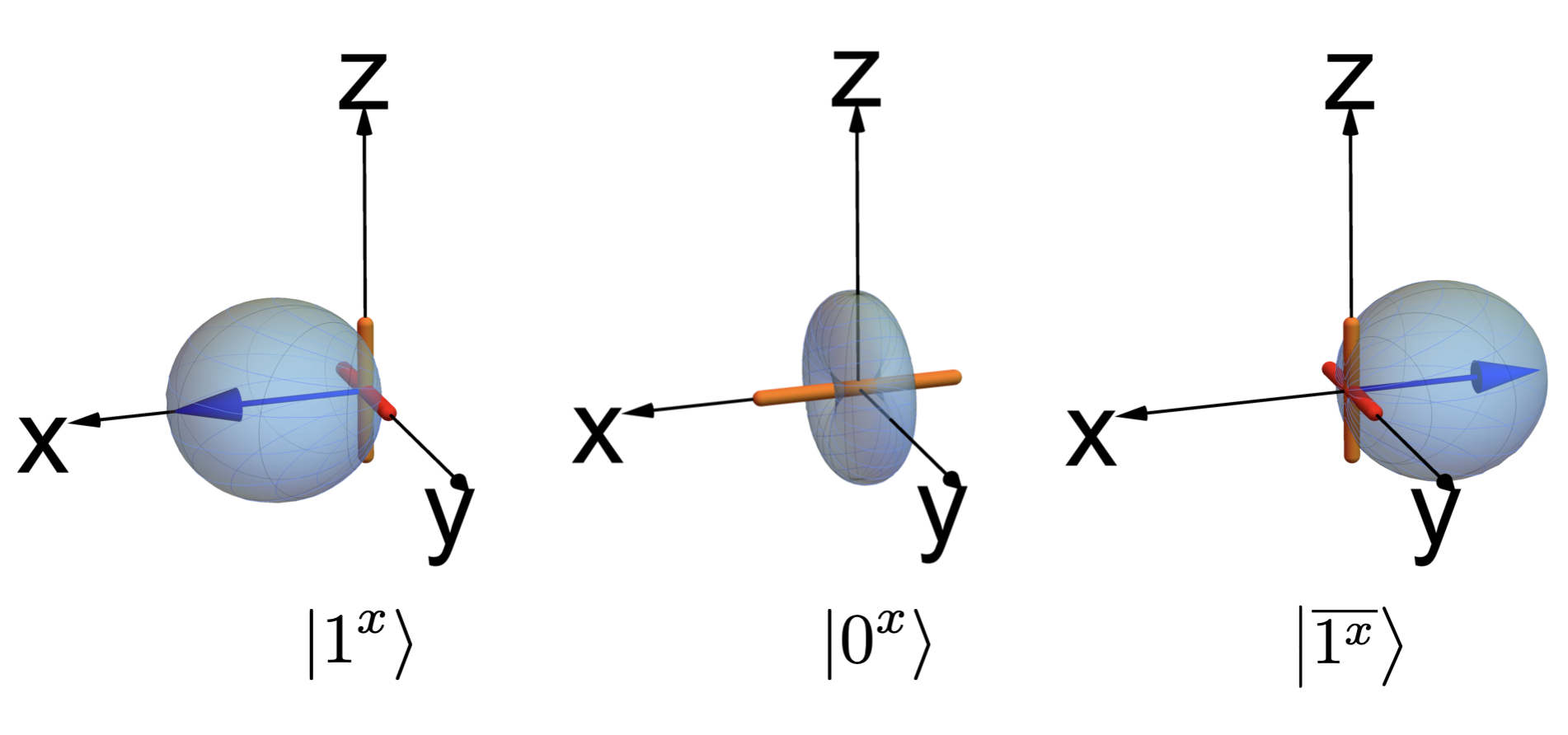}
		\caption{
			Eigenstates of $\ops{i}{x}$ and basis states of $\mathcal{B}^x$
			%
		}
		\label{fig:basis_x}
	\end{figure}
	such that
	\begin{subequations}
		\begin{align}
			\bra{1^x} \ops{i}{x}\ket{1^x}&=1\; ,\\
			\bra{0^x} \ops{i}{x}\ket{0^x}&=0\; ,\\
			\bra{\overline{1^x}} \ops{i}{x}\ket{\overline{1^x}}&=-1\; ,\\
			\bra{\alpha} \ops{i}{\mu}\ket{\alpha}&=0 ~~~~ \textrm{for } \ket{\alpha} \in 	\mathcal{B}^x \textrm{ and }  \mu=y,z\; .
		\end{align}
	\end{subequations}

	
	We consider the following 
	Hamiltonian
	\be
	\Ham=\Ham^{\rm{EP}}+\Ham^{\rm{SI}}\; . \label{eq:Ham_EP_SI_FM}
	\ee
	$\Ham^{\rm{EP}}$ represents the Heisenberg Hamiltonian for spin-1 with easy--plane Heisenberg anisotropic exchange couplings $\textbf{J}$
	\be
	\Ham^{\rm{EP}}=\sum_{\nn{i,j}}\left[\opsb{i}\cdot\textbf{J}\cdot\opsb{j}\right]\; , \label{eq:Ham_EP_FM}
	\ee
	where the spin dipole operator $\opsb{i}$ is defined in \Autoref{eq:Sdipole}, and where $\textbf{J}$ corresponds to the usual nearest neighbor spin-spin coupling tensor.
	$\Ham^{\rm{SI}}$ accounts for single--ion anisotropy and is given by
	\be
	\Ham^{\rm{SI}}=\sum_{i}\opsb{i}\textbf{D}\opsb{i}\label{eq:Ham_SI_FM}\; ,
	\ee
	where $\textbf{D}$ corresponds to the usual single site spin-spin coupling tensor. 
	We assume  the spin-spin coupling tensors $\textbf{J}$ and $\textbf{D}$ to only have diagonal components:
	\be
	\Ham^{\rm{EP}}=\sum_{\nn{i,j}}\left[J^{xx}\ops{i}{x}\ops{j}{x}+J^{yy}\ops{i}{y}\ops{j}{y}+J^{zz}\ops{i}{z}\ops{j}{z}\right]\; , \label{eq:Ham_EP_FM_2}
	\ee
	\be
	\Ham^{\rm{SI}}=\sum_{i}\left[D^{xx}\ops{i}{x}\ops{i}{x}+D^{yy}\ops{i}{y}\ops{i}{y}+D^{zz}\ops{i}{z}\ops{i}{z}\right]\; . \label{eq:Ham_SI_FM_2}
	\ee
	We also assume the coupling constants to be negative and the order to be ferromagnetic:
	\be
	J^{\alpha\alpha}<0\; ,\label{eq:J_FM}
	\ee
	\be
	D^{\alpha\alpha}<0\; .\label{eq:D_FM}
	\ee
	We can assume the ground state to be a state with the spin pointing somewhere in the xy-plane, and we can choose it to be pointing along the x-axis:
	\be
	\ket{GS}=\ket{1^x}\; .\label{eq:gs_x}
	\ee
	As a basis, we choose the eigenstates of $\ops{i}{x}$:
	\begin{align}
		\mathcal{B}^x&=\left\lbrace\ket{1^x},\ket{0^x},\ket{\overline{1^x}}\right\rbrace\; ,\label{eq:Bx}
	\end{align}
	as represented in \Autoref{fig:basis_x}, 
	

	Even though the  A-matrices are deeply linked to the time-reversal (TR) invariant basis, we will here mostly focus on the basis $\mathcal{B}^{x}$ [\Autoref{eq:Bx}] and then transform the required quantities accordingly.
	%
	

\begin{figure}[t]
		\centering
		\includegraphics[width=0.45\textwidth]{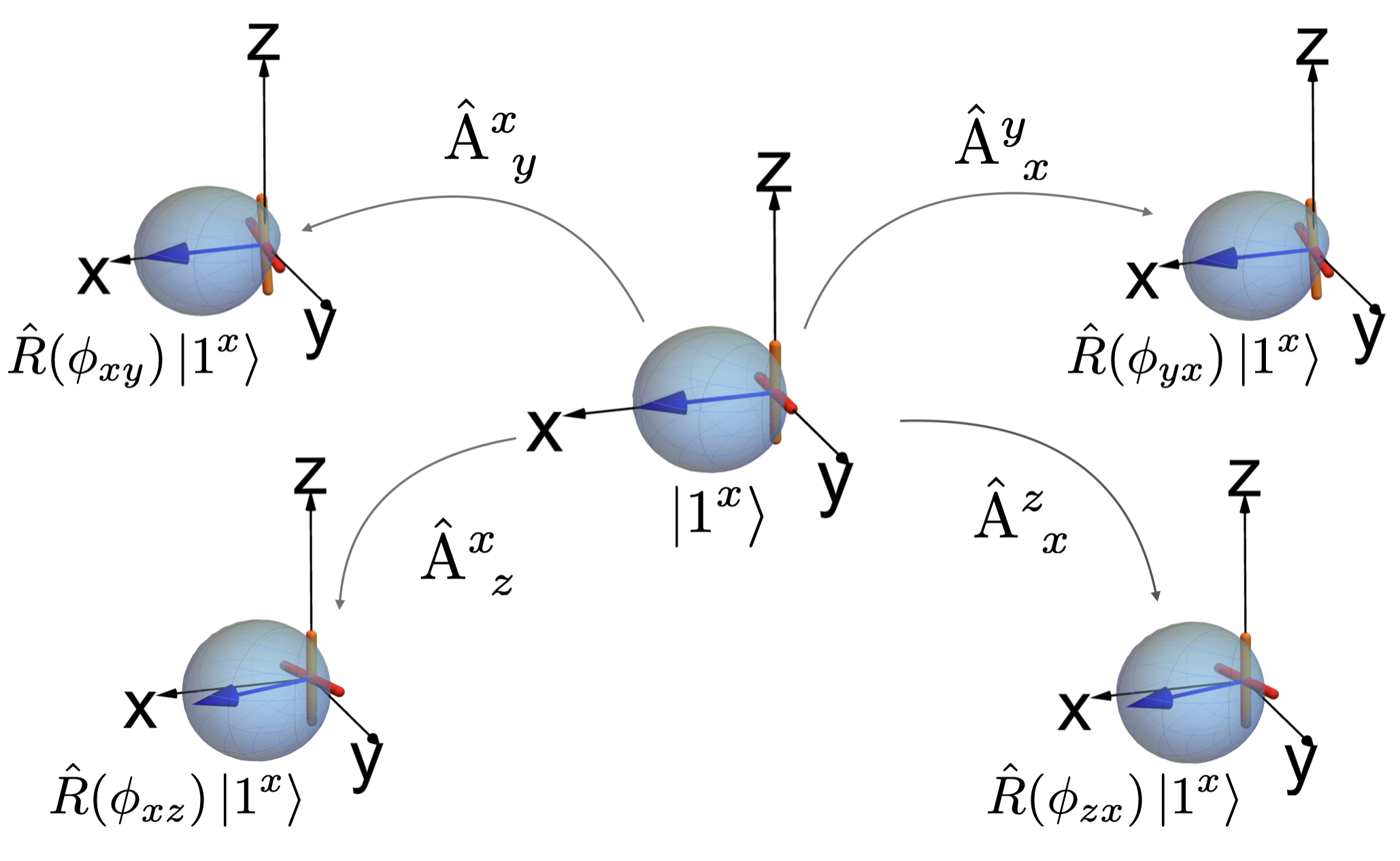}
		\caption{
			Fluctuations created by the generators $\opa{}{x}{y}$, $\opa{}{x}{z}$, $\opa{}{y}{x}$, and $\opa{}{z}{x}$ , according to \protect\Autoref{eq:defR}, for an angle $\phi_{\alpha\beta}=\frac{\pi}{8}$.
		%
		}
		\label{fig:fluct}
\end{figure}


	To do that, we remember that the spin dipole moments can be rewritten in terms of the A matrices [\Autoref{eq:dipole.in.terms.of.A}], expressed in the time-reversal (TR) invariant basis [\Autoref{eq:TRB}] as shown in \Autoref{fig:TR.basis}.
	Using \Autoref{eq:dipole.in.terms.of.A}, the terms of the easy--plane anisotropic Hamiltonian [\Autoref{eq:Ham_EP_FM_2}], in terms of the A-matrices, becomes
	\begin{subequations}
		\begin{align}
			\ops{i}{x}\ops{j}{x}&=-(\opa{i}{y}{z}-\opa{i}{z}{y})(\opa{j}{y}{z}-\opa{j}{z}{y})\; ,\\
			\ops{i}{y}\ops{j}{y}&=-(\opa{i}{z}{x}-\opa{i}{x}{z})(\opa{j}{z}{x}-\opa{j}{x}{z})\; ,\\
			\ops{i}{z}\ops{j}{z}&=(\opa{i}{x}{y}-\opa{i}{y}{x})(\opa{j}{x}{y}-\opa{j}{y}{x})\; .
		\end{align} \label{eq:SS_EP}
	\end{subequations}
	For the single ion terms, we use \Autoref{eq:DefQ} and \Autoref{eq:quadrupole.in.terms.of.A} to rewrite the terms $\ops{i}{\alpha}\ops{i}{\alpha}$ of the single--ion anisotropic Hamiltonian [\Autoref{eq:Ham_SI_FM_2}] in the function of the A-matrices, as
	\begin{subequations}
		\begin{align}
			\ops{i}{x}\ops{i}{x}&=-\frac{2}{3}\opa{i}{x}{x}+\frac{1}{3}\opa{i}{y}{y}+\frac{1}{3}\opa{i}{z}{z}+\frac{2}{3}\; ,\\
			\ops{i}{y}\ops{i}{y}&=-\frac{2}{3}\opa{i}{y}{y}+\frac{1}{3}\opa{i}{x}{x}+\frac{1}{3}\opa{i}{z}{z}+\frac{2}{3}\; ,\\
			\ops{i}{z}\ops{i}{z}&=-\frac{2}{3}\opa{i}{z}{z}+\frac{1}{3}\opa{i}{y}{y}+\frac{1}{3}\opa{i}{x}{x}+\frac{2}{3}\; .
		\end{align} \label{eq:SS_SI}
	\end{subequations}
	Using \Autoref{eq:SS_EP} and \Autoref{eq:SS_SI}, the total Hamiltonian [\Autoref{eq:Ham_EP_SI_FM}] in terms of the A-matrices then becomes
	\begin{align*}
		\Ham=\sum_{\nn{i,j}}&\left[-J^{xx}(\opa{i}{y}{z}-\opa{i}{z}{y})(\opa{j}{y}{z}-\opa{j}{z}{y}))\right.\\
		&\left.-J^{yy}(\opa{i}{z}{x}-\opa{i}{x}{z})(\opa{j}{z}{x}-\opa{j}{x}{z})\right.\\
		&\left.-J^{zz}(\opa{i}{x}{y}-\opa{i}{y}{x})(\opa{j}{x}{y}-\opa{j}{y}{x})\right]\\
		+\sum_{i}&\left[D^{xx}(-\frac{2}{3}\opa{i}{x}{x}+\frac{1}{3}\opa{i}{y}{y}+\frac{1}{3}\opa{i}{z}{z}+\frac{2}{3})\right.\\
		&\left.+D^{yy}(-\frac{2}{3}\opa{i}{y}{y}+\frac{1}{3}\opa{i}{x}{x}+\frac{1}{3}\opa{i}{z}{z}+\frac{2}{3})\right.\\
		&\left.+D^{zz}(-\frac{2}{3}\opa{i}{z}{z}+\frac{1}{3}\opa{i}{y}{y}+\frac{1}{3}\opa{i}{x}{x}+\frac{2}{3})\right]\; . \label{eq:HamU3_EP_SI}\numberthis
	\end{align*}
	
	If we define the basis change $\Lambda_3$ to be the basis change matrix between $\mathcal{B}_{2}$ and $\mathcal{B}^{x}$, such that  if a state $\ket{\phi}_{\mathcal{B}_{2}}$ is given in the TR invariant basis $\mathcal{B}_{2}$, in the basis $\mathcal{B}^{x}$ , its components are given by
	\be
	\ket{\phi}_{\mathcal{B}^x}= \Lambda_3\ket{\phi}_{\mathcal{B}_{2}}\; . \label{eq:Def_lambda3}
	\ee
	We found that the basis change matrix  $\Lambda_3$ yields
	\be
	\Lambda_3=\begin{pmatrix}
		0 &  \frac{1}{\sqrt{2}} &  \frac{i}{\sqrt{2}}\\
		-i & 0 & 0\\
		0 & \frac{1}{\sqrt{2}} &- \frac{i}{\sqrt{2}}
	\end{pmatrix}\; .\label{eq:lambda2}
	\ee
	An operator $\oph{O}_{\mathcal{B}_{2}}$ given in the TR invariant basis $\mathcal{B}_{2}$
	is expressed as
	\be
	\oph{O}_{\mathcal{B}^x}= \Lambda_3\oph{O}_{\mathcal{B}_{2}} \Lambda_3^{\dagger} \; , \label{eq:O_x_TR}
	\ee
	in the basis $\mathcal{B}^{x}$.
	
	We will start working the basis $\mathcal{B}^{x}$, where everything is simple, since the ground state is one of the basis states and the orthogonal fluctuations can be expressed in terms of the other orthogonal basis states.
	Indeed, the ground state matrix takes the simple form
	\be
	{\bf{A}_0}_{\mathcal{B}^x}=
	\begin{pmatrix}
		1& 0 & 0 \\
		0 & 0 & 0 \\
		0& 0 & 0 \\
	\end{pmatrix}\; ,\label{eq:A0}
	\ee
	since the ground state is simply the state  $\ket{1^x}$ [\Autoref{eq:gs_x}]
	or expressed in terms of director components
	\be
	{\textbf{d}^{\dagger}_0}_{\mathcal{B}^x}=\begin{pmatrix}1\\0\\0\end{pmatrix}\; .
	\ee
	We can generate orthogonal fluctuations by application of the exponential map given in \Autoref{eq:defR}.
	The new state describing the fluctuations around the ground state is given by
	\be
	{\textbf{d}^{\dagger}}(\phi)=\hat{R}(\phi){\textbf{d}^{\dagger}_0}\; .
	\ee
	The A matrix  transforms according to \Autoref{eq:NewA}.
	Only the generators $\opa{}{x}{x}$, $\opa{}{x}{y}$, $\opa{}{x}{z}$, $\opa{}{y}{x}$, and $\opa{}{z}{x}$ will have non zero contribution when applied to the ground state matrix [\Autoref{eq:A0}]. \Autoref{fig:fluct} represents the action of the generators on the ground state. We can see, for example, that the generator $\opa{}{x}{y}$, will create a fluctuation along $\ket{0^x}$, i.e. an $\bad{}$ boson, and will induce the new state to exhibit some quadrupolar features.
	

Using the constraint on the trace of A-matrices [\Autoref{eq:traceA.on.A.matrix}], we express the contribution from $\opa{}{x}{x}$ in terms of the others components in order to ensure the length of the spin to be $S=1$ (which is equivalent to constraining the trace of A to be equal to 1), so that we properly restrict to $su(3)$ and make sure that we are correctly representing a spin-1. We obtain
	\begin{align}
		\textbf{A}(\phi)_{\mathcal{B}^x}&=\begin{pmatrix}
			1 -\phi_{xy} \phi_{yx}-\phi_{xz} \phi_{zx}& i \phi_{xy} & i \phi_{xz} \\
			-i \phi_{yx} & \phi_{xy} \phi_{yx} & \phi_{xz} \phi_{yx} \\
			-i \phi_{zx} & \phi_{xy} \phi_{zx} & \phi_{xz} \phi_{zx} \\
		\end{pmatrix}\; .
	\end{align}
	We can then easily introduce bosonic fluctuations  by 
	\begin{subequations}
		\begin{align}
			i \phi_{xy}&=\ba{}\; ,\\
			-i \phi_{yx}&=\bad{}\; ,\\
			i \phi_{xz} &= \bb{}\; ,\\
			-i \phi_{zx} &= \bbd{}\; ,
		\end{align}
	\end{subequations}
	such that we get
	\be
	\ophb{A}_{\mathcal{B}^x}=
	\begin{pmatrix}
		1-\bad{i}\ba{i}- \bbd{i}\bb{i}&\ba{i} & \bb{i}\\
		\bad{i} &\bad{i}\ba{i} &\bad{i} \bb{i}\\
		\bbd{i} & \ba{i}\bbd{i} & \bbd{i}\bb{i}
	\end{pmatrix}\; .\label{eq:A_bos_tilda}
	\ee
	According to \Autoref{eq:O_x_TR}, the A matrices expressed in the TR invariant basis $\mathcal{B}_{2}$ are given by 
	\begin{widetext}
		\begin{align*}
			\ophb{A}_{B_{2}}&=\Lambda_3^{\dagger}\ophb{A}_{\mathcal{B}^x}\Lambda_3\\
			&=\begin{pmatrix}
				\bad{i}\ba{i} &\frac{i}{\sqrt{2}} \bad{i} +\frac{i}{\sqrt{2}}\bad{i}\bb{i} &-\frac{1}{\sqrt{2}} \bad{i} +\frac{1}{\sqrt{2}}\bad{i}\bb{i}\\
				-\frac{i}{\sqrt{2}} \ba{i} -\frac{i}{\sqrt{2}}\ba{i}\bbd{i} &  \frac{1}{2} + \frac{1}{2}\bbd{i}+ \frac{1}{2}\bb{i}- \frac{1}{2}\bad{i}\ba{i}  & \frac{i}{2}+ \frac{i}{2}\bbd{i}- \frac{i}{2}\bb{i}-\frac{i}{2}\bad{i}\ba{i}-i \bbd{i}\bb{i}\\
				-\frac{1}{\sqrt{2}} \ba{i} +\frac{1}{\sqrt{2}}\ba{i}\bbd{i} &-\frac{i}{2}- \frac{i}{2}\bb{i}+ \frac{i}{2}\bbd{i}+ \frac{i}{2}\bad{i}\ba{i}+i \bbd{i}\bb{i} &  \frac{1}{2} - \frac{1}{2}\bbd{i}- \frac{1}{2}\bb{i}- \frac{1}{2}\bad{i}\ba{i} 
			\end{pmatrix}\; .\numberthis\label{eq:A_TR}
		\end{align*}
	\end{widetext}
	Inserting \Autoref{eq:A_TR} into \Autoref{eq:SS_SI}, we get the single--ion terms in the function of the bosons
	\begin{align*}
		\ops{i}{x}\ops{i}{x}&=1-\bad{i}\ba{i}\; ,\\
		\ops{i}{y}\ops{i}{y}&=\frac{1}{2}(1+\bad{i}\ba{i}-\bbd{i}-\bb{i})\; ,\numberthis \label{eq:SS_SI_boson}\\
		\ops{i}{z}\ops{i}{z}&=\frac{1}{2}(1+\bad{i}\ba{i}+\bbd{i}+\bb{i})\; .
	\end{align*}
	We notice that if $D^{yy}$ is not equal to $D^{zz}$, then the Hamiltonian [\Autoref{eq:HamU3_EP_SI}] has single bosons terms, meaning that the state about which we expanded the fluctuations is not the ground state any more.
	Therefore, to be consistent with the easy-plane FM order and the ground state [\Autoref{eq:gs_x}],  we choose 
	\be
	D^{\perp}=D^{yy}=D^{zz}  \textrm{~with~} \norm{D^{\perp}}<\norm{D^{xx}}\; .
	\ee

	After inserting \Autoref{eq:A_TR} into the total Hamiltonian [\Autoref{eq:HamU3_EP_SI}], only keeping fluctuations
	up to 2nd order, and performing a Fourier transform, the Hamiltonian [ \Autoref{eq:HamU3_EP_SI}] becomes
	\begin{align*}
		\Ham=&\frac{1}{2}\sum_{{\k}}\left[\begin{pmatrix}
			\bad{{\k}}, &\ba{-{\k}}
		\end{pmatrix}
		\begin{pmatrix}
			A_{\k} & B_{\k} \\
			B_{\k}  & A_{\k} \\
		\end{pmatrix}
		\begin{pmatrix}
			\ba{{\k}}\\
			\bad{-{\k}}\\
		\end{pmatrix}\right.\\
		&\left.+\begin{pmatrix}
			\bbd{{\k}}, &\bb{-{\k}}
		\end{pmatrix}
		\begin{pmatrix}
			C_{\k} & 0 \\
			0  & C_{\k} \\
		\end{pmatrix}
		\begin{pmatrix}
			\bb{{\k}}\\
			\bbd{-{\k}}\\
		\end{pmatrix}\right]\\
		&+\frac{1}{2}NzJ^{xx}+N(D^{xx}+D^{\perp})\; ,\numberthis
		\label{eq:HamU3FT_SI}
	\end{align*}
	where
	\be\begin{matrix}
		A_{\k}=-J^{xx}z+\frac{1}{2}z(J^{yy}+J^{zz})\gamma(\k)+(D^{\perp}-D^{xx})\; ,\\
		B_{\k}=\frac{1}{2}z(J^{zz}-J^{yy})\gamma(\k)\; ,\\
		C_{\k}=-2zJ^{xx}\; .
	\end{matrix}\label{eq:Def_A_B_C_FM}
	\ee
	Similarly to the FQ case, we need to solve an eigensystem  analogous to \Autoref{eq:quantum.eigensystem}. The dispersion relations for $\bad{\k}$ and $\ba{\k}$ can be found by imposing them to have bosonic commutation relations [\Autoref{eq:Uinv}], and diagonalizing
	\be
	\sigma_z 	\begin{pmatrix}
		A_{\k} & B_{\k} \\
		B_{\k}  & A_{\k} \\
	\end{pmatrix}=
	\begin{pmatrix}
		1 & 0 \\
		0  & -1 \\
	\end{pmatrix}
	\begin{pmatrix}
		A_{\k} & B_{\k} \\
		B_{\k}  & A_{\k} \\
	\end{pmatrix}=
	\begin{pmatrix}
		A_{\k} & B_{\k} \\
		-B_{\k}  & -A_{\k} \\
	\end{pmatrix}\; ,
	\ee
	where the multiplication by $ \sigma_z$ imposes the bosonic commutation relations.
	The eigenvalues $\epsilon_{\k}$ are given by 
	\be
	\begin{matrix}
		\epsilon_{\k,1}=+\sqrt{A_{\k}^2-B_{\k}^2} &,~&\epsilon_{\k,2}=-\sqrt{A_{\k}^2-B_{\k}^2}
	\end{matrix}\label{eq:omega_a_U3_SI}\; .
	\ee
	The dispersion relations for the $\bbd{\k}$ and $\bb{\k}$ are obtained by diagonalizing
	\be
	\sigma_z 		\begin{pmatrix}
		C_{\k} & 0 \\
		0  & C_{\k} \\
	\end{pmatrix}=
	\begin{pmatrix}
		C_{\k} & 0 \\
		0  & -C_{\k} \\
	\end{pmatrix}\; .
	\ee
	The eigenvalues $\epsilon_{\k}$ are given by 
	\be
	\begin{matrix}
		\epsilon_{\k,3}=-C_{\k} &,~&\epsilon_{\k,4}=+C_{\k}
	\end{matrix}\label{eq:omega_b_U3_SI}\; .
	\ee
	Because the coupling constants are negative, the physical results are
	\be
	\begin{matrix}
		\epsilon_{\k,1}=+\sqrt{A_{\k}^2-B_{\k}^2} &,~&\epsilon_{\k,3}=-C_{\k}=2z\norm{J^{xx}}
	\end{matrix}\label{eq:omega_2_U3_SI}\; ,
	\ee
	where $A_{\k}$, $ B_{\k}$, and  $C_{\k}$ are given in \Autoref{eq:Def_A_B_C_FM}.
	
	Following the same procedure as for the FQ state in \Autoref{section:quantum.theory}, we calculate dynamical structure factors for the anisotropic FM case. We start by finding the Bogoliubov transformation that diagonalizes \Autoref{eq:HamU3FT_SI}. Following the steps given in \hyperref[sec:bogolioubov_transfomation]{Appendix~}\ref{sec:bogolioubov_transfomation}, we get
	\begin{subequations}
		\begin{align}
			\ba{\k}{}=&\frac{1}{\sqrt{\Delta_{\k}^2- B_{\k}^2}}(\Delta_{\k}\balpha{\k}-B_{\k}\balphad{-\k})\; ,\label{eq:Bogo_FM_a_1}\\
			\bad{-\k}{}=&\frac{1}{\sqrt{\Delta_{\k}^2- B_{\k}^2}}(-B_{\k}\balpha{\k}+\Delta_{\k}\balphad{-\k})\; ,\\
			\bad{\k}{}=&\frac{1}{\sqrt{\Delta_{\k}^2- B_{\k}^2}}(\Delta_{\k}\balphad{\k}-B_{\k}\balpha{-\k})\; ,\\
			\ba{-\k}{}=&\frac{1}{\sqrt{\Delta_{\k}^2- B_{\k}^2}}(-B_{\k}\balphad{\k}+\Delta_{\k}\balpha{-\k})\; , \label{eq:Bogo_FM_a_4}
		\end{align}\label{eq:Bogo_FM_a}
	\end{subequations}
	and
	\begin{subequations}
		\begin{align}
			\bb{\k}{}&=\bbeta{\k}\; ,\\
			\bbd{-\k}{}&=\bbetad{-\k}\; ,\\
			\bbd{\k}{}&=\bbetad{\k}\; ,\\
			\bb{-\k}{}&=\bbeta{-\k}\; ,
		\end{align}\label{eq:Bogo_FM_b}
	\end{subequations}
	where $\Delta_{\k}$ is given in \Autoref{eq:Delta}, and where $A_{\k}$ and $ B_{\k}$ are given in \Autoref{eq:Def_A_B_C_FM}.
	
	We follow now the calculations outlined in \Autoref{sec:structure_factors_quantum_qn0} in order to calculate the quantum structure factors.
	Since we are working in the Bogoliubov representation, the ground state $\ket{\rm{GS}}$ is the vacuum state $\ket{\rm{vac}}$ for the Bogoliubov bosons. The structure factors are given by \Autoref{eq:SqOF}. We calculate $\bra{\rm{vac}}\hat{O}_{\q}^{\alpha}\ket{\mu}$ with \mbox{$\ket{\mu}= \balphad{\k}\ket{\rm{vac}}\oplus\bbetad{\k}\ket{\rm{vac}}$} and
	$\hat{O}_{\q}^{\alpha}=\ops{\k}{\alpha}$ with $\alpha=x,y,z$, for the dipole structure factor for instance.
	Using \Autoref{eq:dipole.in.terms.of.A}, \Autoref{eq:quadrupole.in.terms.of.A} and \Autoref{eq:A_TR}, we can rewrite the spin dipole, the spin quadropole, and the A-matrix operators in terms of the bosons up to linear order, and after performing a Fourier transform, we can rewrite them in terms of the Bogoliubov bosons using \Autoref{eq:Bogo_FM_a} and \Autoref{eq:Bogo_FM_b}. This allows to easily calculate the structure factors [\Autoref{eq:SqOF}].


\begin{figure*}[h]
\begin{center}
\begin{minipage}[b]{\textwidth}
				\subfloat[$~~~~~~~~~~~S^{\sf QM}_{\rm{A}}(\q, \omega)~~~~~~~~~~~$ $\left(\right. J^{xx}=J^{yy}=J^{zz}\left.\right)$	 	\label{fig:Aqw_FM_iso}]{
					\includegraphics[width=0.3\linewidth]{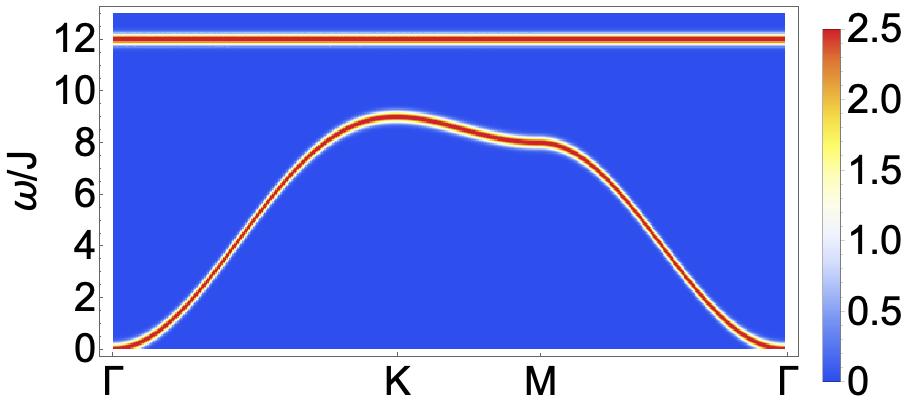}
				}
				\subfloat[$~~~~~~~~~~~S^{\sf QM}_{\rm{Q}}(\q, \omega)~~~~~~~~~~~$ $\left(\right. J^{xx}=J^{yy}=J^{zz} \left.\right)$			\label{fig:Qqw_FM_iso}]{
					\includegraphics[width=0.3\linewidth]{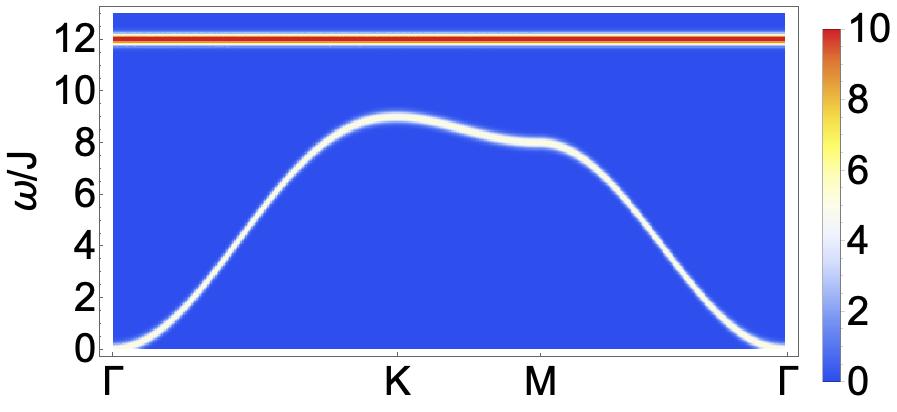}
				}
				\subfloat[$~~~~~~~~~~~S^{\sf QM}_{\rm{S}}(\q, \omega)~~~~~~~~~~~$ $\left(\right. J^{xx}=J^{yy}=J^{zz} \left.\right)$			\label{fig:Sqw_FM_iso}]{
					\includegraphics[width=0.3\linewidth]{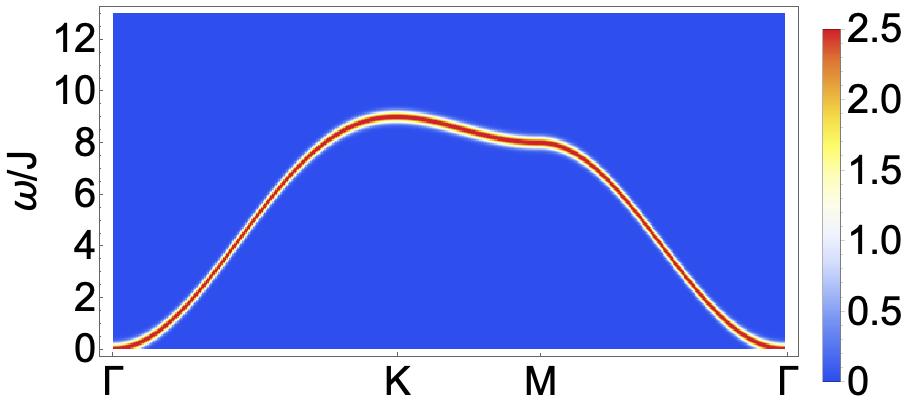}
				}\\
				\subfloat[$~~~~~~~~~~~S^{\sf QM}_{\rm{A}}(\q, \omega)~~~~~~~~~~~$ $\left(\right. J^{yy}=J^{zz}=0.8J^{xx} \left.\right)$			\label{fig:Sqw_FM_A_EP}]{
					\includegraphics[width=0.3\linewidth]{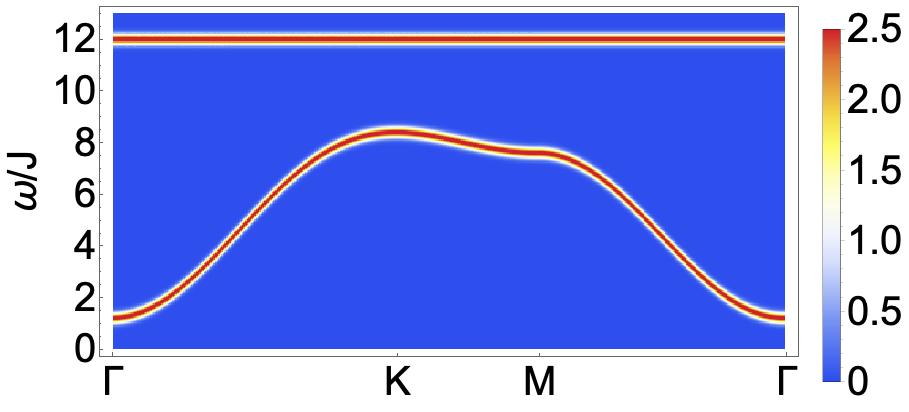}
				}
				\subfloat[$~~~~~~~~~~~S^{\sf QM}_{\rm{Q}}(\q, \omega)~~~~~~~~~~~$ $\left(\right. J^{yy}=J^{zz}=0.8J^{xx} \left.\right)$			\label{fig:Sqw_FM_Q_EP}]{
					\includegraphics[width=0.3\linewidth]{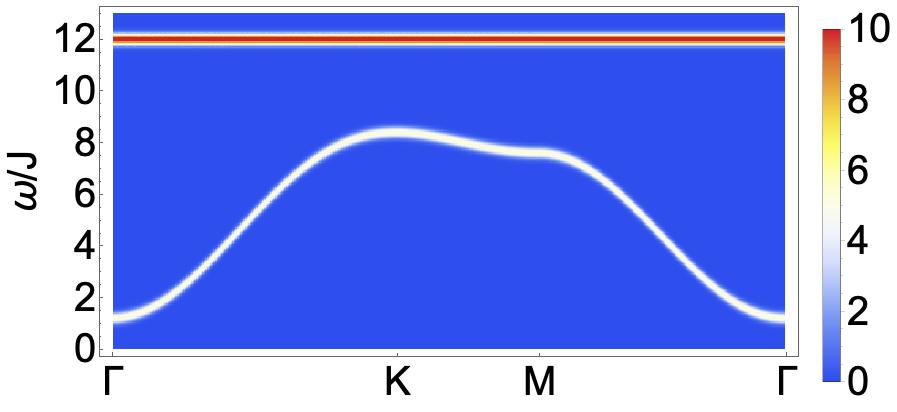}
				}
				\subfloat[$~~~~~~~~~~~~S^{\sf QM}_{\rm{S}}(\q, \omega)~~~~~~~~~~~~$ $ \left(\right.J^{yy}=J^{zz}=0.8J^{xx} \left.\right)$  		\label{fig:Sqw_FM_S_EP}]{
					\includegraphics[width=0.3\linewidth]{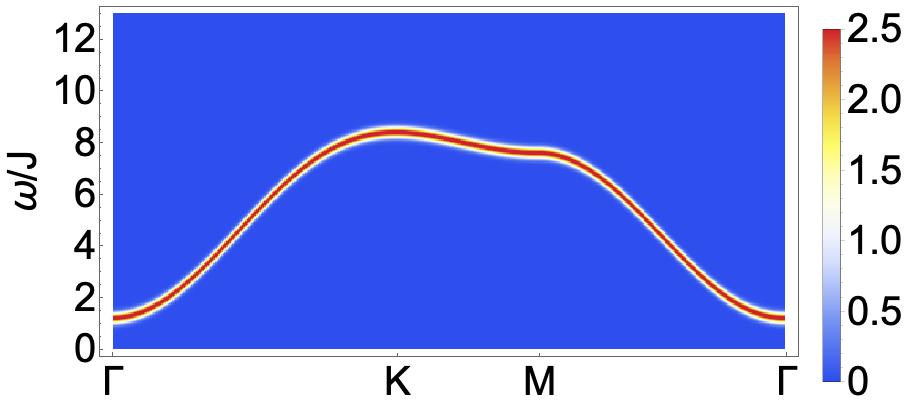}
				}\\
				\subfloat[$~~~~~~~~~~~~~S^{\sf QM}_{\rm{A}}(\q, \omega)~~~~~~~~~~~~~$ $ \left(\right. D^{\perp}=0.5D^{x}\left.\right)$			\label{fig:Sqw_FM_A_SI}]{
					\includegraphics[width=0.3\linewidth]{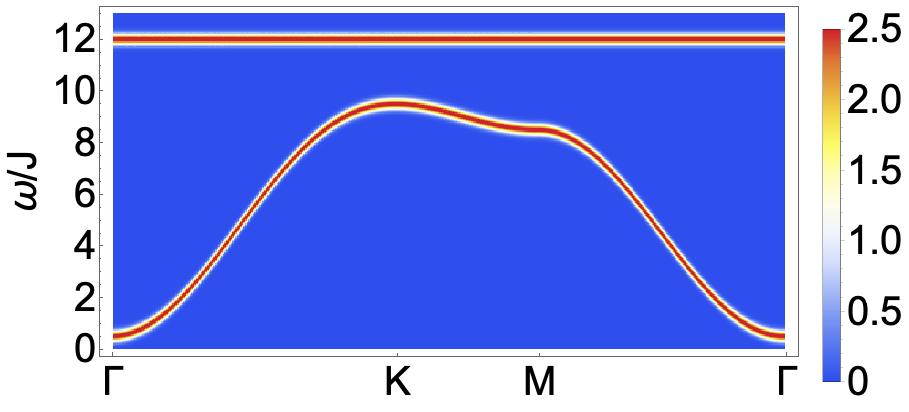}
				}
				\subfloat[$~~~~~~~~~~~~~S^{\sf QM}_{\rm{Q}}(\q, \omega)~~~~~~~~~~~~~$ $ \left(\right. D^{\perp}=0.5D^{x}\left.\right)$			\label{fig:Sqw_FM_Q_SI}]{
					\includegraphics[width=0.3\linewidth]{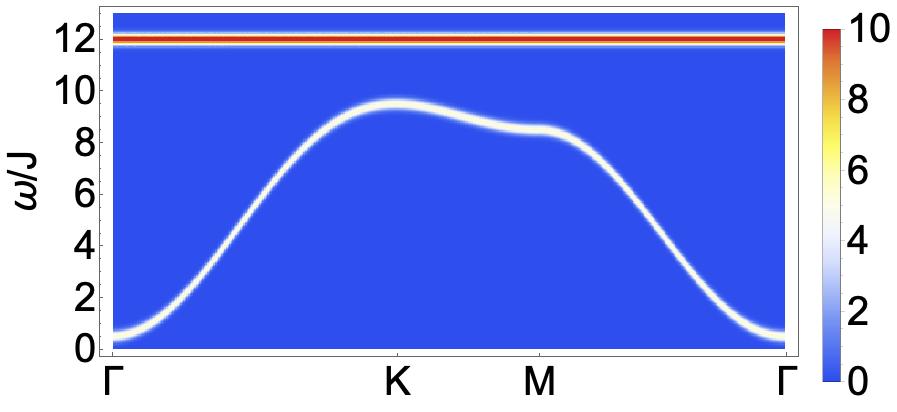}
				}
				\subfloat[$~~~~~~~~~~~~~~S^{\sf QM}_{\rm{S}}(\q, \omega)~~~~~~~~~~~~~~$ $  \left(\right. D^{\perp}=0.5D^{x}\left.\right)$  		\label{fig:Sqw_FM_S_SI}]{
					\includegraphics[width=0.3\linewidth]{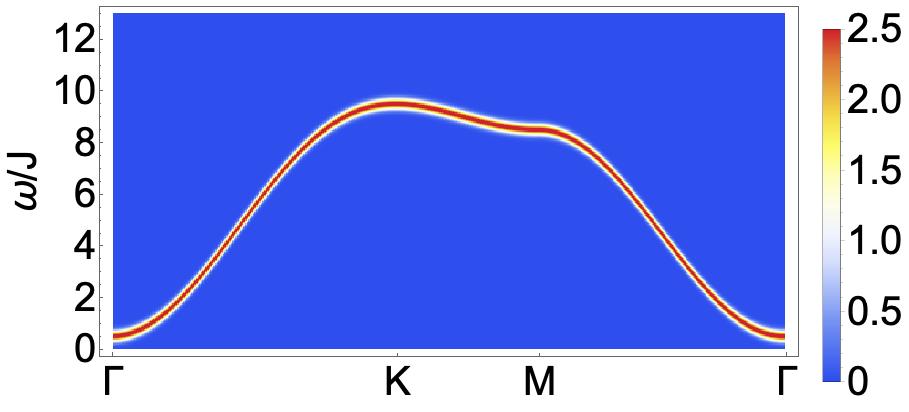}
				}\\
				\subfloat[$~~~~~~~~~~~~~S^{\sf QM}_{\rm{A}}(\q, \omega)~~~~~~~~~~~~~$ $  \left(\right.J^{yy}=D^{x}=J^{xx} \left.\right. ,$ $\left.\right.J^{zz}=D^{\perp}=0.8J^{xx}\left.\right)$
				\label{fig:Sqw_FM_A_EP_SI}]{
					\includegraphics[width=0.3\linewidth]{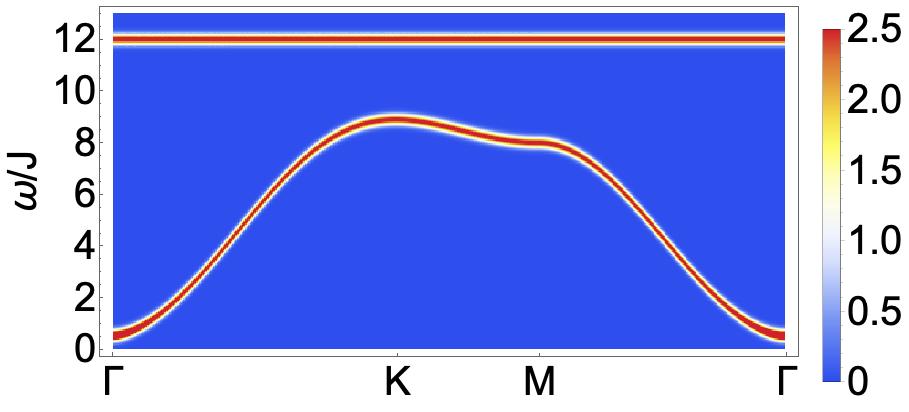}
				}
				\subfloat[$~~~~~~~~~~~~~S^{\sf QM}_{\rm{Q}}(\q, \omega)~~~~~~~~~~~~~$ $  \left(\right.J^{yy}=D^{x}=J^{xx}\left.\right. ,$ $\left.\right.J^{zz}=D^{\perp}=0.8J^{xx}\left.\right)$			\label{fig:Sqw_FM_Q_EP_SI}]{
					\includegraphics[width=0.3\linewidth]{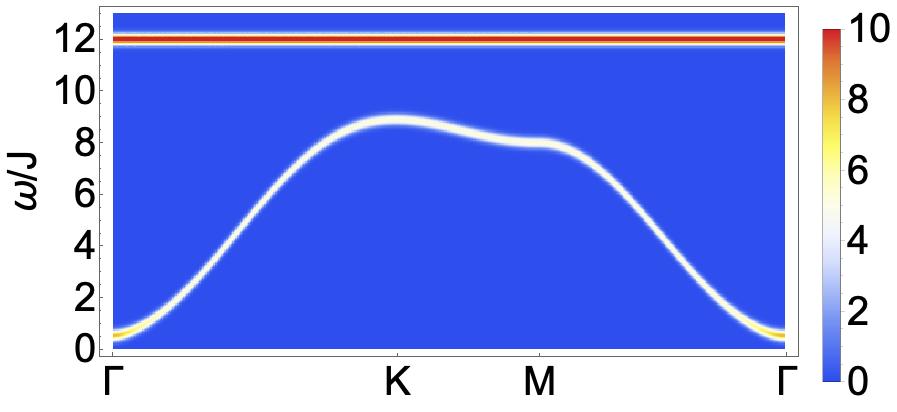}
				}
				\subfloat[$~~~~~~~~~~~~~S^{\sf QM}_{\rm{S}}(\q, \omega)~~~~~~~~~~~~~$ $ \left(\right.J^{yy}=D^{x}=J^{xx}\left.\right. ,$ $\left.\right.J^{zz}=D^{\perp}=0.8J^{xx}\left.\right)$
				\label{fig:Sqw_FM_S_EP_SI}]{
					\includegraphics[width=0.3\linewidth]{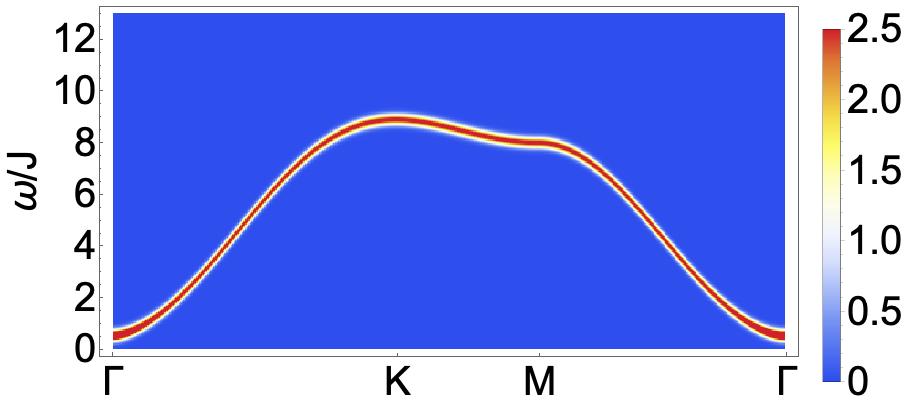}
				}
	\end{minipage}
\end{center}
\caption{
Dynamical structure factors obtained from zero-T quantum calculations 
for $S_{\rm{A}}(\q, \omega)$ (A-matrices consisting of a mixture of dipolar and 
quadrupolar moments), $S_{\rm{Q}}(\q, \omega)$  (quadrupolar moments) and 
$S_{\rm{S}}(\q, \omega)$  (dipolar moments) for the ferromagnetic (FM) phase 
of the BBQ model on the triangular lattice (\protect\Autoref{eq:HamA1}) with 
$J_1$ being considered as Heisenberg anisotropic exchange interactions 
[\protect\Autoref{eq:Ham_EP_FM_2}] and $J_2=0$, and with an additional single 
ion anisotropic exchange Hamiltonian [\protect\Autoref{eq:Ham_SI_FM_2}].
\protect\subref{fig:Aqw_FM_iso}--\protect\subref{fig:Sqw_FM_iso} 
Dynamical structure factors obtained for the isotropic FM state of the Heisenberg 
Hamiltonian [\protect\Autoref{eq:Ham_EP_FM_2}] where $J^{xx}=J^{yy}=J^{zz}=-1$ 
without any single--ion anisotropy [\protect\Autoref{eq:Ham_SI_FM_2}], $D^{\perp}=D^{x}=0$.
\protect\subref{fig:Sqw_FM_A_EP}--\protect\subref{fig:Sqw_FM_S_EP} 
Dynamical structure factors obtained for the easy-plane anisotropic FM state of the 
Heisenberg Hamiltonian [\protect\Autoref{eq:Ham_EP_FM_2}] where 
$J^{yy}=J^{zz}=0.8J^{xx}$ and $J^{xx}=-1$ without any single--ion anisotropy 
[\protect\Autoref{eq:Ham_SI_FM_2}], $D^{\perp}=D^{x}=0$.
\protect\subref{fig:Sqw_FM_A_SI}--\protect\subref{fig:Sqw_FM_S_SI} 
Dynamical structure factors obtained for the isotropic FM state of the 
Heisenberg Hamiltonian [\protect\Autoref{eq:Ham_EP_FM_2}] where 
$J^{yy}=J^{zz}=J^{xx}=-1$ with single--ion anisotropy [\protect\Autoref{eq:Ham_SI_FM_2}], 
$D^{\perp}=0.5D^{x}$ and $D^{x}=J^{xx}$.
\protect\subref{fig:Sqw_FM_A_EP_SI}--\protect\subref{fig:Sqw_FM_S_EP_SI}
Dynamical structure factors obtained for the easy-plane anisotropic FM state 
of the Heisenberg Hamiltonian [\protect\Autoref{eq:Ham_EP_FM_2}] 
where $J^{zz}=0.8J^{xx}$ and $J^{yy}=J^{xx}=-1$  with single--ion anisotropy
[\protect\Autoref{eq:Ham_SI_FM_2}], $D^{\perp}=0.8D^{x}$ and $D^{x}=J^{xx}$.
}
\label{fig:comparison_dispersion_intensities_FM_ani}
\end{figure*}

	
	Using \Autoref{eq:SO_qn0}, the dynamical spin dipole structure factor, defined by \Autoref{eq:SqSF_app}, is given by
	\begin{align}
		S^{\sf FM}_{\rm{S}}(\q, \omega)&=\frac{A_{\q}}{\sqrt{A_{\q}^2-B_{\q}^2}}\delta(\omega-\epsilon_{\q,1})+S^{\sf GS_{\sf{FM}}}_{\rm{S}}(\q=0, \omega)\; .\label{eq:Sqw_qm_FM}
	\end{align}
	The dynamical spin quadrupole structure factor, as given by \Autoref{eq:dynamical.quadrupole.structure.factor_app}, yields
	\begin{align*}
		S^{\sf FM}_{\rm{Q}}(\q, \omega)&=2\frac{A_{\q}}{\sqrt{A_{\q}^2-B_{\q}^2}}\delta(\omega-\epsilon_{\q,1})+4\delta(\omega-\epsilon_{\q,3})\\
		&+S^{\sf GS_{\sf{FM}}}_{\rm{Q}}(\q=0, \omega)\; .\numberthis\label{eq:Qqw_qm_FM}
	\end{align*}
	
	The total dynamical factor for the $\opa{}{}{}$ operators defined in \Autoref{eq:dynamical.A.matrix.structure.factor_app} becomes
	\begin{align*}
		S_{\rm{A}}^{\sf FM}(\q,\omega)&=\frac{A_{\q}}{\sqrt{A_{\q}^2-B_{\q}^2}}\delta(\omega-\epsilon_{\q,1})+\delta(\omega-\epsilon_{\q,3})\\
		&+S^{\sf GS_{\sf{FM}}}_{\rm{A}}(\q=0, \omega)\; ,\numberthis\label{eq:Aqw_qm_FM}
	\end{align*}
	where we explicitly summed over the indexes $\alpha$ and $\beta$ and where the terms of the form $S^{\sf GS_{\sf{FM}}}_{\rm{O}}(\q=0, \omega)$ represent the ground state  and zero--point energy contribution to the structure factors at $\q=0$, but are not calculated here, for simplicity reasons.
	For these 3 results, \Autoref{eq:Sqw_qm_FM}, \Autoref{eq:Qqw_qm_FM}, and \Autoref{eq:Aqw_qm_FM}, we used \Autoref{eq:Delta}, and $\epsilon_{\q,1}$, and $\epsilon_{\q,3}$ are given in \Autoref{eq:omega_2_U3_SI}.

	
We also check that the sum rule \Autoref{eq:sum.rule.AQS} is indeed satisfied after noticing that the constant terms in \Autoref{eq:sum.rule.AQS} would only contribute for $\q=0$ and at equal time, and can therefore be neglected. 
These results are identical to results that one can obtain by performing a conventional multi-bosons expansion.
	
	In \Autoref{fig:comparison_dispersion_intensities_FM_ani}, we show results for the dynamical structure factors [\Autoref{eq:Sqw_qm_FM}, \Autoref{eq:Qqw_qm_FM}, and \Autoref{eq:Aqw_qm_FM}] for the ferromagnetic state for the anisotropic Heisenberg Hamiltonian with single--ion anisotropy [\Autoref{eq:Ham_EP_SI_FM}] on the triangular lattice.
	We first notice that the quadrupolar band $\epsilon_{\q,3}$, which corresponds to the $\Delta S=2$ excitation band associated with the $\bbetad{\k}$ boson,  is gapped and non--dispersive. Because it essentially corresponds to the excitation obtained by applying the lowering operator $S^{+}$ twice, it is quadrupolar in nature and will only contribute to the quadrupolar structure factor channel. Moreover, such a quadrupolar excitation from a FM ground state has a finite energy cost, and it also doesn't have any neighboring quadrupoles to interact with, so it is therefore localized.
	The isotropic FM Heisenberg case without single--ion anisotropy is presented in \Autoref{fig:comparison_dispersion_intensities_FM_ani}~\subref{fig:Aqw_FM_iso}--\subref{fig:Sqw_FM_iso}.
	As shown in \Autoref{fig:comparison_dispersion_intensities_FM_ani}~\subref{fig:Sqw_FM_A_EP}--\subref{fig:Sqw_FM_S_EP}, we note that the introduction of easy--plane anisotropy with $J^{yy}=J^{zz}\neq J^{xx}$ creates a gap and lifts the dispersion relation according to \Autoref{eq:omega_2_U3_SI} and \Autoref{eq:Def_A_B_C_FM}.
	In \Autoref{fig:comparison_dispersion_intensities_FM_ani}~\subref{fig:Sqw_FM_A_SI}--\subref{fig:Sqw_FM_S_SI}, we see that introducing single--ion anisotropy with $D^{\perp}\neq D^{xx}$ also creates a gap and lifts the dispersion relation again according to \Autoref{eq:omega_2_U3_SI} and \Autoref{eq:Def_A_B_C_FM}.
	In \Autoref{fig:comparison_dispersion_intensities_FM_ani}~\subref{fig:Sqw_FM_A_EP_SI}--\subref{fig:Sqw_FM_S_EP_SI}, we display the interplay of easy-plane and single--ion anisotropy.

	
\newpage
	
\end{appendices}
\newpage
\bibliography{paper}

\end{document}